\documentclass[notitlepage,aps,prc,groupedaddress]{revtex4-1}
\usepackage{graphicx}
\usepackage{amssymb,mathtools,float,subfig,hyperref,appendix,braket,enumerate}
\newcommand{\bvec}[1]{\boldsymbol{#1}}
\newcommand{\rb}[1]{\left(#1\right)}
\newcommand{\sqb}[1]{\left[#1\right]}
\newcommand{\abs}[1]{\left|#1\right|}
\newcommand{\floor}[1]{\left\lfloor#1\right\rfloor}
\newcommand{\dd}{\,\mathrm{d}}
\newcommand{\units}[1]{\,\mathrm{#1}}
\newcommand{\pp}{{\prime \prime}}
\newcommand{\up}[1]{{^{#1}}\!}
\newcommand{\tspace}[1]{@{\hskip #1}}

\def\Re{\operatorname{Re}}
\def\Im{\operatorname{Im}}
\begin{document}
\title{Solution of the Bethe-Goldstone Equation without Partial Wave Decomposition}
\author{L. White}
\email[Email:]{whit2657@vandals.uidaho.edu}
\author{F. Sammarruca}
\email[Email:]{fsammarr@uidaho.edu}
\affiliation{Physics Department, University of Idaho, Moscow, ID 83844-0903, U.S.A}
\date{\today}
\begin{abstract}
We present a method for solving the nucleon-nucleon scattering equation without the use of a partial wave expansion 
of the scattering amplitude. After verifying the accuracy of the numerical solutions, we proceed to apply the method 
to the in-medium scattering equation (the Bethe-Goldstone equation) in three dimensions. 
A focal point is a study of Pauli blocking effects calculated in the (angle-dependent) three-dimensional formalism as compared to the           
usual spherical approximation. We discuss our results and their implications.                       
\end{abstract}
\maketitle
\section{Introduction}
Infinite nuclear matter, with equal or unequal concentrations of protons and neutrons, is a 
convenient theoretical laboratory to explore the nucleon-nucleon (NN) interaction in the many-body environment. 
The Bethe-Goldstone equation~\cite{BLM,Bethe56,Goldstone,Bethe71} was developed to describe NN scattering in
a dense hadronic medium through the inclusion of two main effects: 1) corrections of the single-particle energies
to account for the presence of the medium, and 2) the Pauli blocking mechanism, which prevents scattering into occupied 
states. Within the Dirac-Brueckner-Hartree-Fock (DBHF) approach, an additional ``non conventional" medium 
effect comes in through the use of the (density-dependent) nucleon effective mass in the nucleon Dirac spinors. 

The main purpose of this paper is to present a method for the solution of the {\it in-medium} 
scattering equation without the use of partial wave decomposition and to discuss the significance of our results.
Although considerable work can be found in literature on solutions of the NN scattering
equation in three-dimensional {\it free-space} (see for instance, Refs.~\cite{RK,FEG,CPW,Veerasamy:2012sp}), to the best of our knowledge no such calculation has been
reported for in-medium scattering. 

There are advantages to the use of a three-dimensional formalism. First, the computational effort is the same
regardless of the energy, whereas the number of partial waves to be included for satisfactory convergence is well known to grow 
with energy. Second, and most important for our purposes, the Pauli operator can be handled exactly, avoiding
the usual spherical or angle-average approximation which becomes necessary in a partial wave (angle-independent)
framework. 

Studies on the impact of non-spherical components in the Pauli operator have been reported earlier (see for instance Ref.~\cite{Sam2000} and references therein). Remaining within a partial wave formalism, non-spherical components can be included by calculating the matrix elements of the exact Pauli operator. In turn, those allow transitions between states of different total angular momentum $J$, with the resulting partial
wave matrix elements depending on the magnetic quantum number $M$. The presence of states with $J\neq J^{\prime}$ and the dependence on $M$ can be cumbersome and inconvenient, particularly if a large number of partial waves needs to be included. One possibility is to limit the inclusion of non-spherical components to a few partial waves. On the other hand, the issue of the importance of non-spherical components in the Pauli operator can be 
settled in a more definite way by a calculation such as the one reported in this paper. Given that Pauli blocking is perhaps the 
single most important medium effect in nuclear matter, it's a worthwhile effort. 

Our framework is meson theory. 
Although NN potentials based on chiral effective theory have recently become popular, it must be kept in mind 
that a chiral expansion is valid only for low momenta (up to approximately $250 \units{MeV}$ in terms of laboratory kinetic 
energy), whereas relativistic meson theory is a more appropriate framework if one wishes to consider a broad range 
of momenta and densities. 

The organization of this paper is as follows. In Sect.~\ref{Thompson}, we discuss the analytical aspects of our 
approach to the solution of the three-dimensional equation (i.e. choice of basis and partial decoupling of the scattering equations). A detailed description of the NN potential used as our input is found in Appx.~\ref{OBEP}. There, we provide complete expressions for the relativistic one-boson-exchange (OBE) helicity amplitudes in three-dimensional space, with pseudovector coupling for the exchange of pseudoscalar mesons,          
and for the general case of baryons with different masses. This information, which we could not find in the literature,
can be useful to the reader as it can be applied, for instance, to develop a nucleon-hyperon (pseudovector) meson-theoretic potential. 
Furthermore, the presence of two different nucleon masses makes these potentials suitable for 
applications in isospin-asymmetric matter, where neutrons and protons acquire different effective masses. We conclude the section with a brief description of the formalism necessary to connect with the partial wave representation and 
the construction of physical states in the three-dimensional approach. 

In Sect.~\ref{IIB} we incorporate the effects of the Pauli blocking operator, which is                                 
beautifully simple in three-dimensional space. Then we proceed to the solution of the scattering equation.                               

Results are presented in Sect.~\ref{III}, 
where we first verify the accuracy of our numerical solution. We accomplish this by transforming our output into the familiar $LSJ$ formalism and comparing with existing partial wave solutions obtained with the same input. This is done successfully. We then proceed to explore the impact of using the exact or spherical Pauli operator. Possible implications of those effects are discussed. The paper concludes with Sect.~\ref{IV}. 

\section{Formalism}
\subsection{Free-Space Nucleon-Nucleon Scattering in Three-Dimensional Space} \label{Thompson}
Before confronting the in-medium scattering equation, we consider the equation in free-space. Once the necessary tools have been worked out for free-space, natural modifications can be made to account for the presence of the medium.
\subsubsection{The Thomson equation in a helicity basis} \label{thompson_heli}
Two nucleon scattering is described covariantly by the  Bethe-Salpeter equation~\cite{BS}. Being a four-dimensional integral equation, it's difficult to solve~\cite{FT}, so it's customary to resort to relativistic three dimensional-reductions. One such three-dimensional reduction yields the Thompson equation, which is the one we adopt here. In operator form the Thompson equation reads $T=V+VG_oT$, where $T$, $V$, and $G_o$ are the $T$-matrix, the NN potential, and the two-nucleon propagator, respectively. After casting the operator equation into a momentum and total isospin basis we obtain~\cite{Thompson}
\begin{equation}
T^I(\bvec{q}^\prime,\bvec{q}) = V^I(\bvec{q}^\prime,\bvec{q}) + \lim_{\epsilon \to 0} \int_{\mathbb{R}^3}V^I(\bvec{q}^\prime,\bvec{q}^\pp)\frac{m^2}{E_{q^\pp}^2}\frac{1}{2(E_q - E_{q^\pp}+ i \epsilon)} T^I(\bvec{q}^\pp,\bvec{q})\frac{ \dd^3 q^\pp}{(2\pi)^3} \; ,
\label{Eq:Thompson_full}
\end{equation}
with $E_p = \sqrt{p^2 + m^2}$ and $m$ the nucleon mass, which we take to be the average of the proton and neutron mass. The $T$-matrix, $T^I(\bvec{q}^\prime,\bvec{q}) \equiv \braket{\bvec{q}^\prime I|T|\bvec{q} I}$, as well as the NN potential, $V^I(\bvec{q}^\prime,\bvec{q}) \equiv \braket{\bvec{q}^\prime I|V|\bvec{q} I}$, are written in terms of the momentum and the (conserved) total isospin. $\bvec{q}$, $\bvec{q}^\prime$, and $\bvec{q^\pp}$ are the initial, final, and intermediate relative momentum.                                  

Multiplying the equation from the left by $\frac{m}{E_{q^{\prime}}}$ and
from the right by $\frac{m}{E_{q}}$ and defining
\begin{align}
\hat{V} & = \frac{m}{E_{q^{\prime}}} V \frac{m}{E_{q}} \; , \nonumber \\
\hat{T} & = \frac{m}{E_{q^{\prime}}} T \frac{m}{E_q} \; , 
\end{align}
we can write the Thompson equation in a more convenient form
\begin{equation}
\hat{T}^I(\bvec{q}^\prime,\bvec{q}) = \hat{V}^I(\bvec{q}^\prime,\bvec{q}) + \lim_{\epsilon \to 0} \int_{\mathbb{R}^3}\hat{V}^I(\bvec{q}^\prime,\bvec{q}^\pp)\frac{1}{2(E_q - E_{q^\pp}+ i \epsilon)} \hat{T}^I(\bvec{q}^\pp,\bvec{q}) \dd^3 q^\pp \; ,
\label{Eq:Thompson_mod}
\end{equation}
where we have absorbed the $1/(2 \pi)^3$ factor into the NN potential, which       
is described in Appx.~\ref{OBEP}. Next we introduce a helicity basis. A helicity ket is defined as an eigenstate of $(\bvec{\sigma} \cdot \hat{\bvec{p}}) \ket{\lambda} = 2 \lambda \ket{\lambda}$, where $\hat{\bvec{p}}$ is a unit momentum vector and $\bvec{\sigma} = (\sigma_x,\sigma_y,\sigma_z)$ the spin operator. Physically, the helicity is the spin projection along the direction of the momentum. Utilizing a helicity basis along with its completeness relation we obtain
\begin{equation}
\braket{\lambda_1^\prime \lambda_2^\prime|\hat{T}^I(\bvec{q}^\prime,\bvec{q})|\lambda_1 \lambda_2} = \braket{\lambda_1^\prime \lambda_2^\prime|\hat{V}^I(\bvec{q}^\prime,\bvec{q})|\lambda_1 \lambda_2} + \sum_{\lambda_1^\pp, \lambda_2^\pp = \pm } \int_{\mathbb{R}^3} \frac{\braket{\lambda_1^\prime \lambda_2^\prime|\hat{V}^I(\bvec{q}^\prime,\bvec{q}^\pp)|\lambda_1^\pp \lambda_2^\pp} \braket{\lambda_1^\pp \lambda_2^\pp|\hat{T}^I(\bvec{q}^\pp,\bvec{q})|\lambda_1 \lambda_2}}{2( E_q - E_{q^\pp} + i \epsilon)} \dd^3 q^\pp \; ,
\label{Eq:Thompson_heli}
\end{equation}
where for brevity we suppressed $\lim_{\epsilon \to 0}$ and denoted $\pm \frac{1}{2}$ by $\pm$. 

Note that our choice of basis is different from both Ref.~\cite{RK} and Ref.~\cite{FEG}, where states of total helicity are employed. We find that uncoupled-helicity states, $\ket{\lambda_1 \lambda_2}$, are a more convenient and transparent basis because they connect to the NN potential straightforwardly, since the NN potential is constructed in terms of solutions of the single-nucleon Dirac equation [see Eq.~(\ref{Eq:helicity})]. 

As it stands, a three-dimensional integral needs to be performed. Fortunately, the azimuthal degree of freedom can be removed. This is accomplished by applying to both sides of Eq.~(\ref{Eq:Thompson_heli}) the operator $\frac{1}{2 \pi} \int_0^{2 \pi} \dd \phi^\prime$~\cite{RK}
\begin{align}
\frac{1}{2 \pi} \int_0^{2 \pi} & \braket{\lambda_1^\prime \lambda_2^\prime|\hat{T}^I(\bvec{q}^\prime,\bvec{q})|\lambda_1 \lambda_2} \dd \phi^\prime = \frac{1}{2 \pi} \int_0^{2 \pi} \braket{\lambda_1^\prime \lambda_2^\prime|\hat{V}^I(\bvec{q}^\prime,\bvec{q})|\lambda_1 \lambda_2} \dd \phi^\prime \nonumber \\
& + \sum_{\lambda_1^\pp, \lambda_2^\pp = \pm } \int_{\mathbb{R}^3} \rb{\frac{1}{2 \pi} \int_0^{2 \pi} \braket{\lambda_1^\prime \lambda_2^\prime|\hat{V}^I(\bvec{q}^\prime,\bvec{q}^\pp)|\lambda_1^\pp \lambda_2^\pp} \dd \phi^\prime} \frac{\braket{\lambda_1^\pp \lambda_2^\pp|\hat{T}^I(\bvec{q}^\pp,\bvec{q})|\lambda_1 \lambda_2}}{2( E_q - E_{q^\pp} + i \epsilon)} \dd^3 q^\pp \; ,
\end{align}
and observing that the azimuthal dependence of $\hat{V}$ occurs in factors of $\cos(\phi^\prime-\phi)$ and $\sin(\phi^\prime-\phi)$. This symmetry carries over to $\hat{T}$ and is due to rotational invariance. We will revisit this point more rigorously in Sect.~\ref{pwd}. Exploiting this observation,   
we obtain 
\begin{align}
\frac{1}{2 \pi} \int_0^{2 \pi} & \braket{\lambda_1^\prime \lambda_2^\prime|\hat{T}^I(\bvec{q}^\prime,\bvec{q})|\lambda_1 \lambda_2}|_{\phi=0} \dd \phi^\prime = \frac{1}{2 \pi} \int_0^{2 \pi} \braket{\lambda_1^\prime \lambda_2^\prime|\hat{V}^I(\bvec{q}^\prime,\bvec{q})|\lambda_1 \lambda_2}|_{\phi=0} \dd \phi^\prime \nonumber \\
& + \sum_{\lambda_1^\pp, \lambda_2^\pp = \pm } \int_0^\infty \int_0^\pi \rb{\frac{1}{2 \pi} \int_0^{2 \pi} \braket{\lambda_1^\prime \lambda_2^\prime|\hat{V}^I(\bvec{q}^\prime,\bvec{q}^\pp)|\lambda_1^\pp \lambda_2^\pp}|_{\phi^\pp=0} \dd \phi^\prime} \frac{{q^\pp}^2 \sin \theta^\pp}{2( E_q - E_{q^\pp} + i \epsilon)} \nonumber \\
& \times 2 \pi  \rb{ \frac{1}{2 \pi} \int_0^{2 \pi} \braket{\lambda_1^\pp \lambda_2^\pp|\hat{T}^I(\bvec{q}^\pp,\bvec{q})|\lambda_1 \lambda_2}|_{\phi=0} \dd \phi^\pp} \dd \theta^\pp \dd q^\pp \; .
\label{Eq:Thompson_phi0}
\end{align}
To complete the removal of the azimuthal degree of freedom, we introduce the following definitions~\cite{RK}
\begin{subequations}
\begin{align}
\braket{\lambda_1^\prime \lambda_2^\prime|t^I(\tilde{q}^\prime,\tilde{q})|\lambda_1 \lambda_2} & \equiv \frac{1}{2 \pi} \int_0^{2 \pi} \braket{\lambda_1^\prime \lambda_2^\prime|\hat{T}^I(\bvec{q}^\prime,\bvec{q})|\lambda_1 \lambda_2}|_{\phi=0} \dd \phi^\prime, \label{Eq:PIT} \\
\braket{\lambda_1^\prime \lambda_2^\prime|v^I(\tilde{q}^\prime,\tilde{q})|\lambda_1 \lambda_2} & \equiv \frac{1}{2 \pi} \int_0^{2 \pi} \braket{\lambda_1^\prime \lambda_2^\prime|\hat{V}^I(\bvec{q}^\prime,\bvec{q})|\lambda_1 \lambda_2}|_{\phi=0} \dd \phi^\prime,
\label{Eq:PIV}
\end{align}
\label{Eq:PI}
\end{subequations}
with $\tilde{q} \equiv (q,\theta)$ and similarly for primed coordinates. It should be pointed out that even though the three-dimensional potential $\braket{\lambda_1^\prime \lambda_2^\prime|\hat{V}^I(\bvec{q}^\prime,\bvec{q})|\lambda_1 \lambda_2}$, is complex, the $\phi$-integrated NN potential $\braket{\lambda_1^\prime \lambda_2^\prime|v^I(\tilde{q}^\prime,\tilde{q})|\lambda_1 \lambda_2}$, is real, as the $\phi$-integrated imaginary part vanishes due to the $\cos(\phi^\prime-\phi)$ and $\sin(\phi^\prime-\phi)$ factors. Using Eqs.~(\ref{Eq:Thompson_phi0}) and~(\ref{Eq:PI}) we obtain the $\phi$-integrated Thompson equation
\begin{align}
& \braket{\lambda_1^\prime \lambda_2^\prime|t^I(\tilde{q}^\prime,\tilde{q})|\lambda_1 \lambda_2} = \braket{\lambda_1^\prime \lambda_2^\prime|v^I(\tilde{q}^\prime,\tilde{q})|\lambda_1 \lambda_2} \nonumber \\
& \hspace{1cm} + \sum_{\lambda_1^\pp, \lambda_2^\pp = \pm } \pi \int_0^\infty \int_0^\pi \frac{\braket{\lambda_1^\prime \lambda_2^\prime|v^I(\tilde{q}^\prime,\tilde{q}^\pp)|\lambda_1^\pp \lambda_2^\pp} \braket{\lambda_1^\pp \lambda_2^\pp|t^I(\tilde{q}^\pp,\tilde{q})|\lambda_1 \lambda_2}}{E_q - E_{q^\pp} + i \epsilon} {q^\pp}^2 \sin \theta^\pp \dd \theta^\pp \dd q^\pp \; .
\label{Eq:Thompson_final}
\end{align}

Equation~(\ref{Eq:PI}) is consistent with the $\phi$-average procedure in Ref.~\cite{RK}. 
In the past, slightly different definitions have been used to integrate out the azimuthal dependence, see for instance the method of Ref.~\cite{FEG}. There, the initial momentum is taken along the $z$-axis (that is, $\theta=0$). While convenient in some ways, this choice is not compatible with Eq.~(\ref{Eq:PI}), because some of the helicity matrix elements vanish (if $\theta=0$) when integrated over the azimuthal angle. Therefore, in our calculations we don't take $\theta$ equal to a fixed value. Instead, we compute the solution over the $q^\prime \times \theta^\prime \times \theta$ grid.
\subsubsection{Partially decoupling the system of integral equations}
The $\phi$-integrated Thompson equations are a set of sixteen coupled Fredholm integral equations of the second kind for each isospin. Due to parity and isospin conservation, only six amplitudes are independent
\begin{align}    
\braket{++|t^I|++} &= \braket{--|t^I|--} \; , \nonumber \\
\braket{++|t^I|--} &= \braket{--|t^I|++} \; , \nonumber \\
\braket{+-|t^I|+-} &= \braket{-+|t^I|-+} \; , \nonumber \\
\braket{+-|t^I|-+} &= \braket{-+|t^I|+-} \; , \nonumber \\
\braket{++|t^I|+-} &= -\braket{++|t^I|-+} = \braket{--|t^I|+-} = -\braket{--|t^I|-+} \; , \nonumber \\
\braket{+-|t^I|++} &= -\braket{-+|t^I|++} = \braket{+-|t^I|--} = -\braket{-+|t^I|--} \; .    
\label{Eq:symmetries}
\end{align}
For the six independent amplitudes we choose
\begin{alignat}{3}
t^I_1 & \equiv \braket{++|t^I|++} \; , & \quad t^I_2 & \equiv \braket{++|t^I|--} \; , & \quad t^I_3 & \equiv \braket{+-|t^I|+-} \; , \nonumber \\
t^I_4 & \equiv \braket{+-|t^I|-+} \; , & t^I_5 & \equiv \braket{++|t^I|+-} \; , & t^I_6 & \equiv \braket{+-|t^I|++} \; .
\label{Eq:sixamplitudes}
\end{alignat}
Due to the symmetries of the three-dimensional NN potential, we find that the following linear combinations,
\begin{equation}
\up{0}t^I \equiv t^I_1 - t^I_2 \; ,
\quad
\up{1}t^I \equiv t^I_3 + t^I_4 \; ,
\quad
\up{12}t^I \equiv t^I_1 + t^I_2 \; ,
\quad
\up{34}t^I \equiv t^I_3 - t^I_4 \; ,
\quad
\up{55}t^I \equiv 2t^I_5 \; ,
\quad
\up{66}t^I \equiv 2t^I_6 \; ,
\label{Eq:linearcombinations}
\end{equation}
partially decouple the system. As it turns out, the spin triplet amplitudes $^{12}t^I,^{34}t^I,^{55}t^I$, and $^{66}t^I$ remain coupled, whereas the spin singlet amplitude $^0t^I$ and the spin triplet amplitude $^1t^I$ are uncoupled. Note that all the formulas above are applicable to the NN potential.       

\begin{figure}[H]
\begin{center}
\subfloat{
\includegraphics[width=6cm]{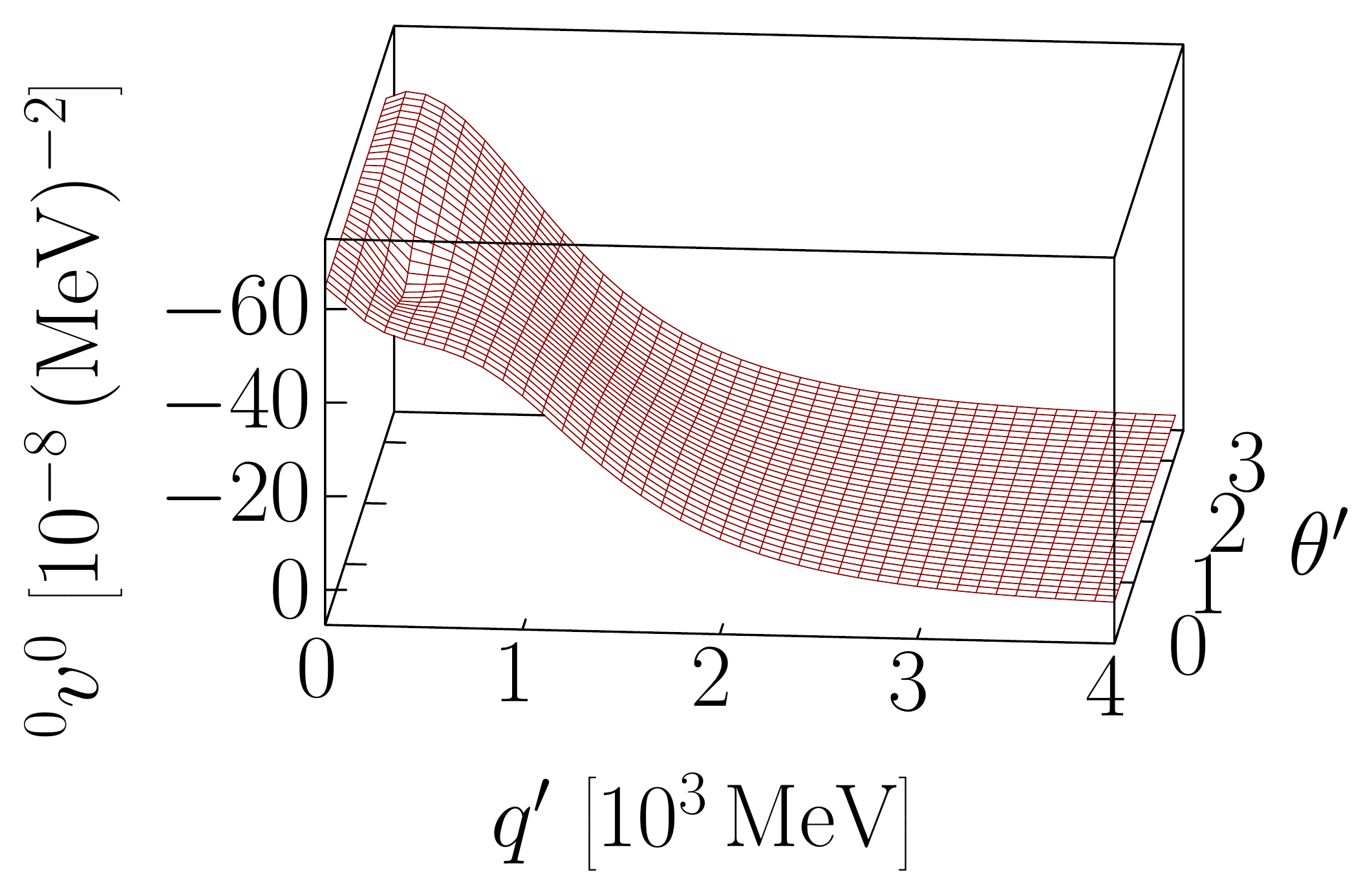}
}
\subfloat{
\includegraphics[width=6cm]{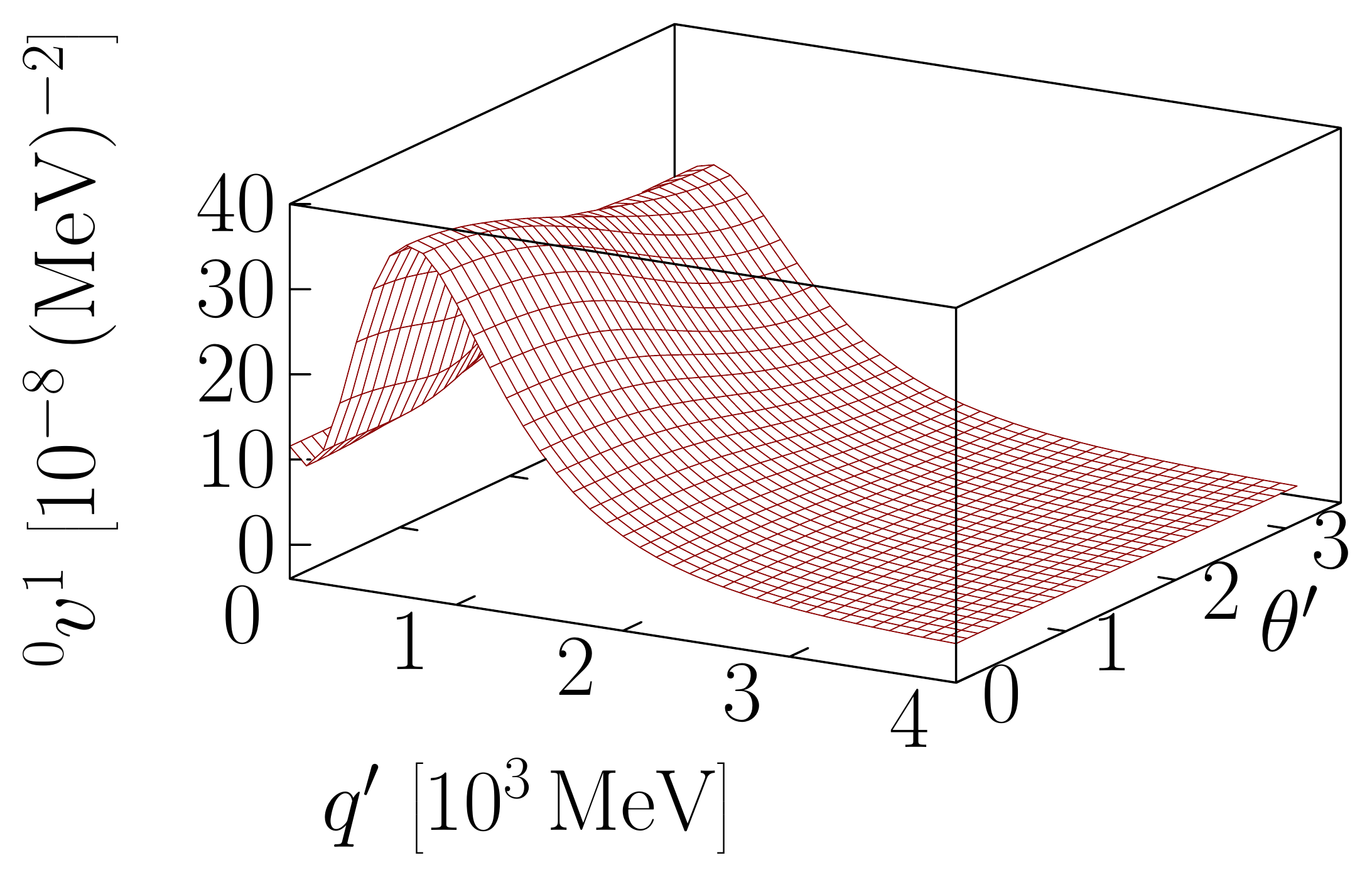}
}
\subfloat{
\includegraphics[width=6cm]{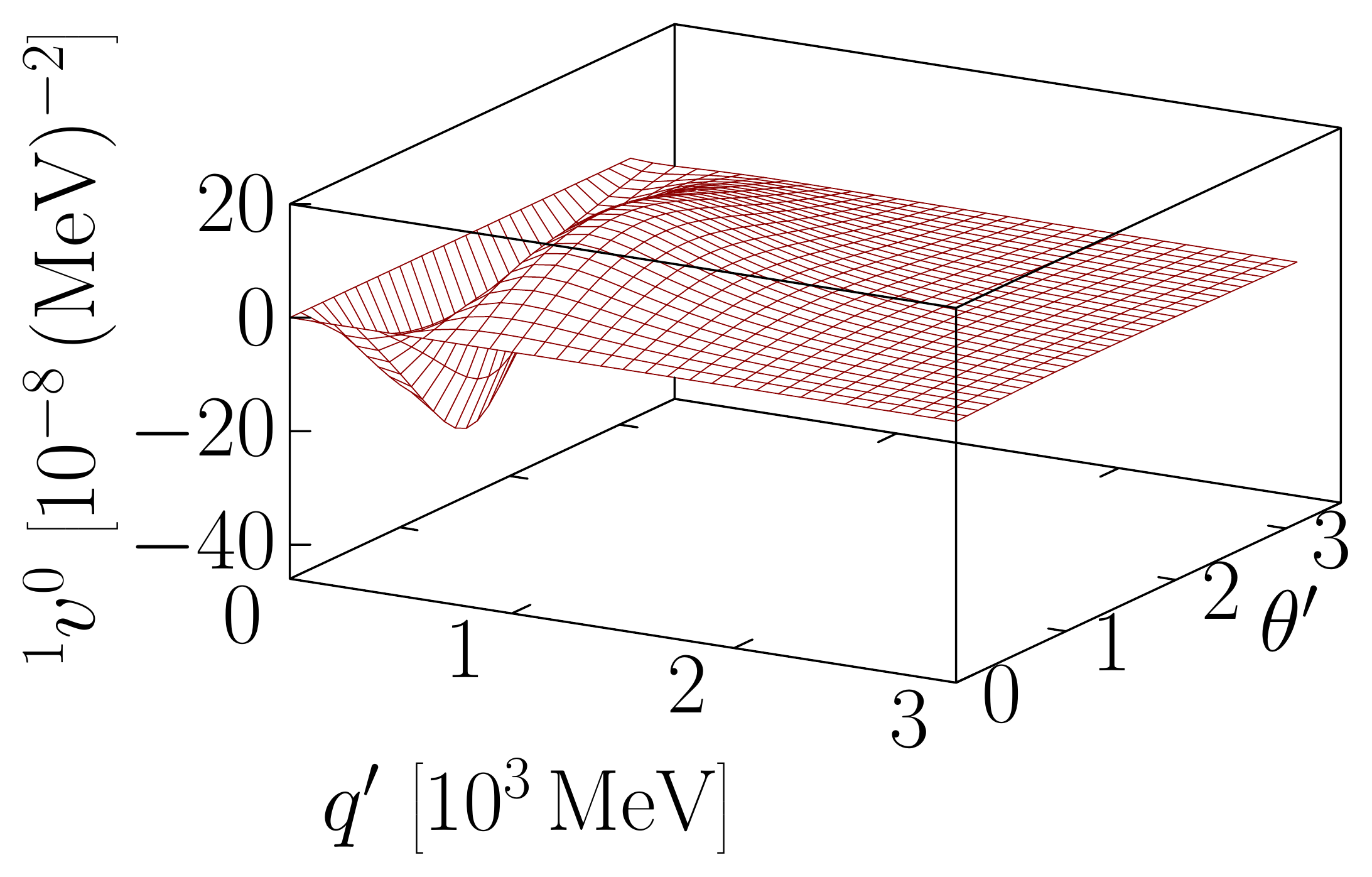}
}
\\
\subfloat{
\includegraphics[width=6cm]{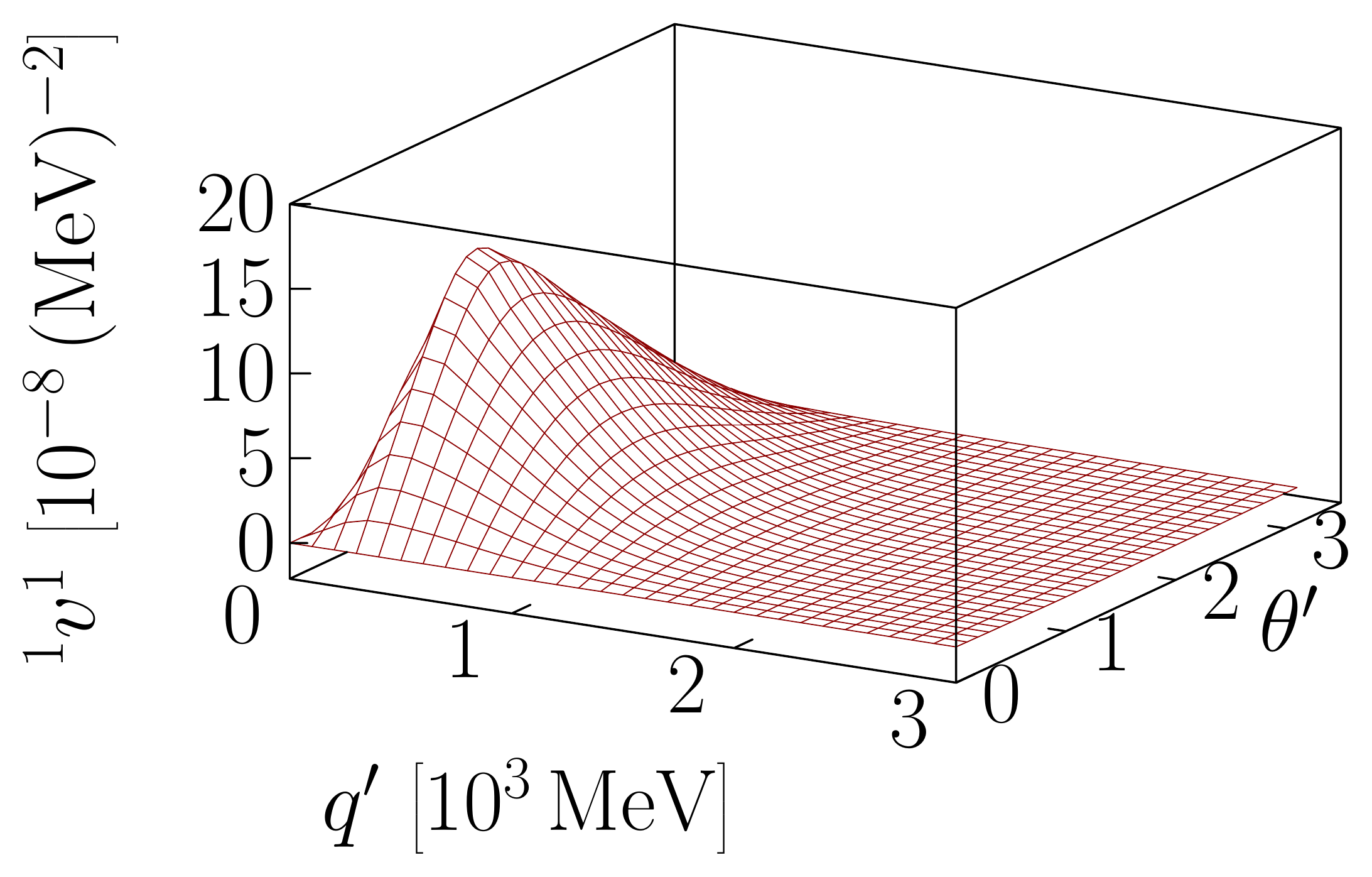}
}
\subfloat{
\includegraphics[width=6cm]{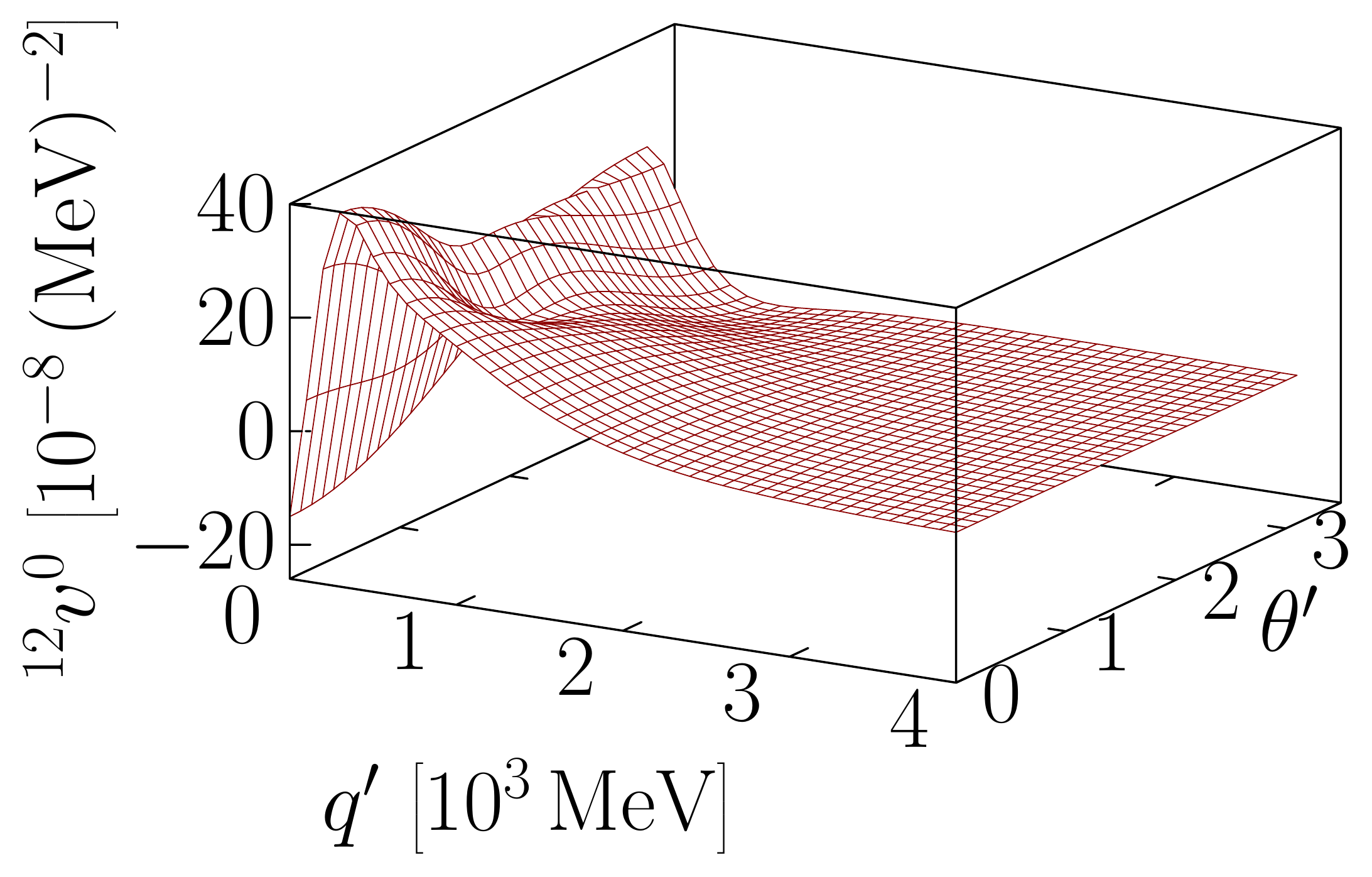}
}
\subfloat{
\includegraphics[width=6cm]{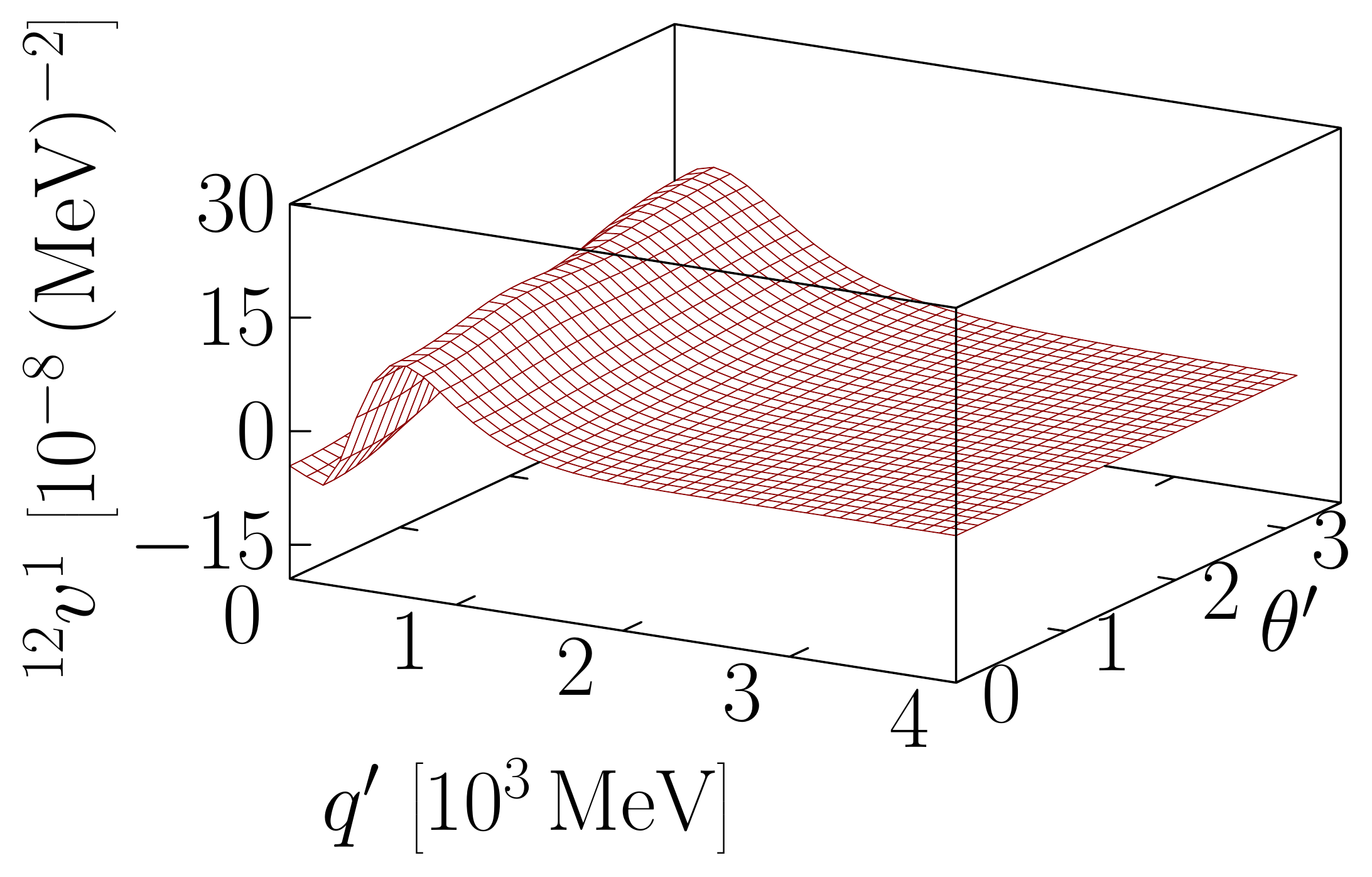}
}
\end{center}
\caption{(Color online) $\phi$-integrated Bonn B potentials as a function of $\tilde{q}^\prime = (q^\prime,\theta^\prime)$. The potentials are evaluated at $\theta = \arccos(0.5)$ and $q=306.42 \units{MeV}$.}
\label{Fig:2d_potential}
\end{figure}

To get a feel for the behavior of the $\phi$-integrated NN potentials which enter the kernel of the equation, we plot in Fig.~\ref{Fig:2d_potential} the $\phi$-integrated Bonn B potentials: $\up{n}v^I$ for $n=0,1,12$ in the notation of Eq.~(\ref{Eq:linearcombinations}). The plots reveal potentials that need a momentum of at least $4000 \units{MeV}$ to approach zero. These observations are insightful with respect to the expected convergence properties of the integral equation. 
The formal numerical solution of the $\phi$-integrated Thompson equation is rather tedious and is developed in Appx.~\ref{systemofeq}. The general idea is to use the linear combinations given in Eq.~(\ref{Eq:linearcombinations}), to obtain six (for each isospin) Fredholm integral equations of the second kind. Then, using Nystrom's method~\cite{nr} or matrix inversion~\cite{Haf} we convert the system of integral equations into a system of matrix equations and invert.
\subsubsection{Connection with partial wave decomposition and construction of physical states} \label{pwd}
A common method for solving Eq.~(\ref{Eq:Thompson_heli}) involves partial wave decomposition~\cite{Mac93}. Although in this paper we avoid that method, we utilize the partial-wave solution for comparison purposes. The expansion of $\hat{T}^I(\bvec{q}^\prime,\bvec{q})$ in a partial wave helicity basis \cite{JW59,BrJack} is given by 
\begin{equation}
\braket{\lambda_1^\prime \lambda_2^\prime|\hat{T}^I(\bvec{q}^\prime,\bvec{q})|\lambda_1 \lambda_2} = \sum_{JM} \frac{2J+1}{4 \pi} D^J_{M \Lambda^\prime}(\phi^\prime,\theta^\prime,-\phi^\prime)^\ast \braket{\lambda_1^\prime \lambda_2^\prime|\hat{T}^{IJ}(q^\prime,q)|\lambda_1 \lambda_2} D^J_{M \Lambda}(\phi,\theta,-\phi) \; ,
\label{Eq:transformation}
\end{equation}
where the Wigner $D$-matrix $D^J_{M \Lambda}(\alpha,\beta,\gamma) = e^{-iM \alpha} d^J_{M \Lambda}(\beta) e^{-i \Lambda \gamma}$ includes the reduced rotation matrix $d^J_{M \Lambda}(\beta)$ with $\Lambda \equiv \lambda_1-\lambda_2$ and an analogous definition for the primed coordinate. The partial wave amplitudes, denoted by $\hat{T}^{IJ}(q^\prime,q)$ (with a similar decomposition done for the NN potential), are the solutions of the partial wave decomposed Eq.~(\ref{Eq:Thompson_heli}).              
 We choose the partial wave helicity amplitudes consistently with those defined in Eq.~(\ref{Eq:sixamplitudes}) and denote them as
\begin{equation}
\hat{T}_n^{IJ}(q^\prime,q) \quad \text{with} \quad n=1,2,3, \dots, 6 \; . 
\label{Eq:sixamplitudesJ}
\end{equation}
It can then be shown that the following linear combinations,
\begin{align}
\intertext{spin singlet $S=0$,} 
\up{0}\mathcal{T}^{IJ} & \equiv \hat{T}^{IJ}_1 - \hat{T}^{IJ}_2 \; , \\
\intertext{spin triplet $S=1$,}  
\up{1}\mathcal{T}^{IJ} & \equiv \hat{T}^{IJ}_3 - \hat{T}^{IJ}_4 \; ,       
\quad
\up{12}\mathcal{T}^{IJ} \equiv \hat{T}^{IJ}_1 + \hat{T}^{IJ}_2 \; ,
\quad
\up{34}\mathcal{T}^{IJ} \equiv \hat{T}^{IJ}_3 + \hat{T}^{IJ}_4 \; ,
\quad
\up{55}\mathcal{T}^{IJ} \equiv 2 \hat{T}^{IJ}_5 \; ,
\quad
\up{66}\mathcal{T}^{IJ} \equiv 2 \hat{T}^{IJ}_6 \; ,
\label{Eq:pwdcoupled}
\end{align}
partially decouple the (partial wave decomposed) scattering equation \cite{Mac87}.
The last four amplitudes in Eq.~\ref{Eq:pwdcoupled} represent coupled triplet states.              
Again, all of the formulas above have similar expressions for the NN potential. We now apply $\frac{1}{2 \pi} \int_0^{2 \pi} \dd \phi^\prime$ on both sides of Eq.~(\ref{Eq:transformation}) to obtain                       
\begin{equation}
\braket{\lambda_1^\prime \lambda_2^\prime|t^I(\tilde{q}^\prime,\tilde{q})|\lambda_1 \lambda_2} = \sum_J \frac{2J+1}{4 \pi} d^J_{0 \Lambda^\prime}(\theta^\prime) \braket{\lambda_1^\prime \lambda_2^\prime|\hat{T}^{IJ}(q^\prime,q)|\lambda_1 \lambda_2} d^J_{0 \Lambda}(\theta) \; ,
\label{Eq:phi_transformation}
\end{equation}
where we made use of Eq.~(\ref{Eq:helicityrot}). Then, recalling the definitions given in Eq.~(\ref{Eq:sixamplitudesJ}), we obtain the transformation from partial waves into the (angle-dependent) $t$-matrix 
\begin{alignat}{3}
\up{0}t^{I} & = \sum_J \frac{2J+1}{4 \pi} d^J_{00}(\theta^\prime) d^J_{00}(\theta) \up{0} \mathcal{T}^{IJ}(q^\prime,q) \; , & \quad \up{1}t^{I} & = \sum_J \frac{2J+1}{4 \pi} d^J_{01}(\theta^\prime) d^J_{01}(\theta) \up{1} \mathcal{T}^{IJ}(q^\prime,q) \; , \nonumber \\
\up{12}t^{I} & = \sum_J \frac{2J+1}{4 \pi} d^J_{00}(\theta^\prime) d^J_{00}(\theta) \up{12} \mathcal{T}^{IJ}(q^\prime,q) \; , & \up{34}t^{I} & = \sum_J \frac{2J+1}{4 \pi} d^J_{01}(\theta^\prime) d^J_{01}(\theta) \up{34} \mathcal{T}^{IJ}(q^\prime,q) \; , \nonumber \\
\up{55}t^{I} & = \sum_J \frac{2J+1}{4 \pi} d^J_{00}(\theta^\prime) d^J_{01}(\theta) \up{55} \mathcal{T}^{IJ}(q^\prime,q) \; , & \up{66}t^{I} & = \sum_J \frac{2J+1}{4 \pi} d^J_{01}(\theta^\prime) d^J_{00}(\theta) \up{66} \mathcal{T}^{IJ}(q^\prime,q) \; ,
\label{Eq:tmatunphy} 
\end{alignat}
where we used the relation $d^J_{0,-1}(\theta)=-d^J_{01}(\theta)$. As it stands, our angle-dependent solutions contain unphysical states. On the other hand, the well-known antisymmetry requirement for the NN system 
imply that only even or odd values of $J$ are allowed in a particular state of definite spin and isospin. 
Thus, starting with Eq.~(\ref{Eq:tmatunphy}) and making use of the identities 
 $(-1)^J d^J_{00}(\theta^\prime) = d^J_{00}(\pi-\theta^\prime)$ and $(-1)^{J+1} d^J_{01}(\theta^\prime) = d^J_{01}(\pi-\theta^\prime)$, we can, in each case, identify the appropriate combination of the direct and the exchange terms which must enter the antisymmetrized amplitudes.                                                   
For those, we obtain:                                   
\begin{align}
\up{0}t_a^I(\tilde{q}^\prime,\tilde{q}) \equiv \up{0}t^{\overset{1}{0}}(\tilde{q}^\prime,\tilde{q}) \pm \up{0}t^{\overset{1}{0}}(-\tilde{q}^\prime,\tilde{q}) & = 2 \sum_{J= \substack{\mathrm{even} \\ \mathrm{odd}}} \frac{2J+1}{4 \pi} d^J_{00}(\theta^\prime) d^J_{00}(\theta) \up{0} \mathcal{T}^{\overset{1}{0}J}(q^\prime,q) \; , \label{Eq:partialwave2d0} \\
\up{1}t_a^I(\tilde{q}^\prime,\tilde{q}) \equiv \up{1}t^{\overset{1}{0}}(\tilde{q}^\prime,\tilde{q}) \pm \up{1}t^{\overset{1}{0}}(-\tilde{q}^\prime,\tilde{q}) & = 2 \sum_{J= \substack{\mathrm{odd} \\ \mathrm{even}}} \frac{2J+1}{4 \pi} d^J_{01}(\theta^\prime) d^J_{01}(\theta) \up{1} \mathcal{T}^{\overset{1}{0}J}(q^\prime,q) \; ,\\
\up{12}t_a^I(\tilde{q}^\prime,\tilde{q}) \equiv \up{12}t^{\overset{1}{0}}(\tilde{q}^\prime,\tilde{q}) \pm \up{12}t^{\overset{1}{0}}(-\tilde{q}^\prime,\tilde{q}) & = 2 \sum_{J= \substack{\mathrm{even} \\ \mathrm{odd}}} \frac{2J+1}{4 \pi} d^J_{00}(\theta^\prime) d^J_{00}(\theta) \up{12} \mathcal{T}^{\overset{1}{0}J}(q^\prime,q) \; , \label{Eq:partialwave2d12}\\
\up{34}t_a^I(\tilde{q}^\prime,\tilde{q}) \equiv \up{34}t^{\overset{1}{0}}(\tilde{q}^\prime,\tilde{q}) \mp \up{34}t^{\overset{1}{0}}(-\tilde{q}^\prime,\tilde{q}) & = 2 \sum_{J= \substack{\mathrm{even} \\ \mathrm{odd}}} \frac{2J+1}{4 \pi} d^J_{01}(\theta^\prime) d^J_{01}(\theta) \up{34} \mathcal{T}^{\overset{1}{0}J}(q^\prime,q) \; ,\\
\up{55}t_a^I(\tilde{q}^\prime,\tilde{q}) \equiv \up{55}t^{\overset{1}{0}}(\tilde{q}^\prime,\tilde{q}) \pm \up{55}t^{\overset{1}{0}}(-\tilde{q}^\prime,\tilde{q}) & = 2 \sum_{J= \substack{\mathrm{even} \\ \mathrm{odd}}} \frac{2J+1}{4 \pi} d^J_{00}(\theta^\prime) d^J_{01}(\theta) \up{55} \mathcal{T}^{\overset{1}{0}J}(q^\prime,q) \; , \\
\up{66}t_a^I(\tilde{q}^\prime,\tilde{q}) \equiv \up{66}t^{\overset{1}{0}}(\tilde{q}^\prime,\tilde{q}) \mp \up{66}t^{\overset{1}{0}}(-\tilde{q}^\prime,\tilde{q}) & = 2 \sum_{J= \substack{\mathrm{even} \\ \mathrm{odd}}} \frac{2J+1}{4 \pi} d^J_{01}(\theta^\prime) d^J_{00}(\theta) \up{66} \mathcal{T}^{\overset{1}{0}J}(q^\prime,q) \; , \label{Eq:partialwave2d66}
\end{align}
where one must read across the top (or bottom) to associate the correct sign with the appropriate $J$ values (even or odd) and isospin ($0$ or $1$). We also used the shorthand notation for the exchange amplitude $\up{n}t^I(-\tilde{q}^\prime,\tilde{q})=\up{n}t^I(q^\prime,\pi-\theta^\prime,q,\theta) \text{ for } n=0,1,12,34,55,66$. 

Even more common to describe the NN system is the $\ket{LSJ}$ basis because these states are traditionally related to phase-shift analyses. In this basis the physical states can simply be selected using the constraint that $L+S+I$ must be odd. To compare with the familiar partial wave states, we first invert Eq.~(\ref{Eq:phi_transformation}) with the help of the orthogonality relation
\begin{equation} 
\int_0^\pi d^{J^\prime}_{0 \Lambda}(\theta) d^J_{0 \Lambda}(\theta) \sin \theta \dd \theta = \frac{2}{2J+1} \delta_{JJ^\prime } \; ,
\end{equation}
to obtain
\begin{equation}
\braket{\lambda_1^\prime \lambda_2^\prime|\hat{T}^{IJ}(q^\prime,q)|\lambda_1 \lambda_2} = \pi(2J+1) \int_0^\pi \int_0^\pi d^J_{0\Lambda^{\prime}}(\theta^\prime) \braket{\lambda_1^\prime \lambda_2^\prime|t^I(\tilde{q}^\prime,\tilde{q})|\lambda_1 \lambda_2} d^J_{0\Lambda}(\theta) \sin \theta^\prime \sin \theta \dd \theta \dd \theta^\prime \; .
\label{Eq:inverse_transformation}
\end{equation}
At this point, an elementary unitary transformation takes us into the $\ket{LSJ}$ partial wave basis. For explicit formulas see Ref.~\cite{Mac87,Mac93}.
\subsection{Solving the Bethe-Goldstone equation in three dimensions}                   
\label{IIB}
\subsubsection{The Bethe-Goldstone equation in a helicity basis}
\label{IIB1} 
In the nuclear matter frame, in analogy with the free-space case and following 
steps similar to Eqs.~(\ref{Eq:Thompson_full}-\ref{Eq:Thompson_mod}), the Bethe-Goldstone equation can be written
as                
\begin{equation}
\hat{G}^I(\bvec{q}^\prime,\bvec{q},\bvec{P},k_F) = \hat{V}^I(\bvec{q}^\prime,\bvec{q}) + \lim_{\epsilon \to 0} \int_{\mathbb{R}^3} \frac{\hat{V}^I(\bvec{q}^\prime,\bvec{q}^\pp) Q(\bvec{q}^\pp,\bvec{P},k_F) \hat{G}^I(\bvec{q}^\pp,\bvec{q},\bvec{P},k_F)}{(e^\ast (\bvec{P},\bvec{q}) - e^\ast (\bvec{P},\bvec{q}^\pp) + i \epsilon)} \dd^3 q^\pp,
\label{Eq:BG_nm}
\end{equation}
where $Q$ stands for the Pauli operator, which suppresses scattering into states below the Fermi momentum, and 
the asterix signify in-medium energies. Depending on the approach one takes, the NN potential may or may not 
be medium-modified through the use of effective masses in the Dirac spinors. 
In Eq.~(\ref{Eq:BG_nm}), we have defined                                                                  
\begin{equation} 
e^\ast (\bvec{P},\bvec{q})=
\epsilon^\ast (\bvec{P}+\bvec{q})  +
\epsilon^\ast (\bvec{P}-\bvec{q}) \; .   
\end{equation} 
The single-particle energy, $\epsilon^\ast$, contains kinetic and potential energy               
\begin{equation}
\epsilon^\ast(\bvec{P} + \bvec{q}) = T(\bvec{P} + \bvec{q}) +                                       
U(\bvec{P} + \bvec{q}) =                                       
E^\ast +U_V \; ,                      
\end{equation}
where $E^\ast = \sqrt{(\bvec{P}+\bvec{q})^2 + (m^\ast)^2}$, and the 
last step is a consequence of the self-consistent determination of the nuclear matter
potential and its parametrization in terms of scalar and vector potentials, 
$U_S=m^\ast - m$ and $U_V$ \cite{Sam10}. 

On the other hand, we are interested in the scattering of two nucleons in the medium  at some positive energy      
and in their center-of-mass system (which makes the comparison with free-space scattering more straightforward). 
For such a case, $\bvec{P} =0$ in the energies, although the Pauli operator 
still depends on the relative velocity between the two frames, as the momenta 
$\bvec{P} \pm \bvec{q}$ are the ones to be compared with the Fermi momentum. 
Thus, in the center-of-mass system, and ignoring medium effects other than Pauli blocking, the equation 
reads 
\begin{equation}
\hat{G}^I(\bvec{q}^\prime,\bvec{q},\bvec{P},k_F) = \hat{V}^I(\bvec{q}^\prime,\bvec{q}) + \lim_{\epsilon \to 0} \int_{\mathbb{R}^3} \frac{\hat{V}^I(\bvec{q}^\prime,\bvec{q}^\pp) Q(\bvec{q}^\pp,\bvec{P},k_F) \hat{G}^I(\bvec{q}^\pp,\bvec{q},\bvec{P},k_F)}{2(E^\ast_q - E^\ast_{q^\pp} + i \epsilon)} \dd^3 q^\pp.
\label{Eq:BG_full}
\end{equation}

The Pauli operator for symmetric nuclear matter is defined as
\begin{equation}
Q(\bvec{q}^\pp,\bvec{P},k_F) \equiv \Theta( \abs{\bvec{P} + \bvec{q}^\pp} - k_F ) \Theta( \abs{\bvec{P} - \bvec{q}^\pp} - k_F ) \; ,
\label{Eq:Q}
\end{equation}
where $\Theta$ is the Heaviside step function, $\bvec{P}$ is one half the center of mass momentum,                    
$\bvec{P} \pm \bvec{q}$ is the momentum of the two particles in the nuclear matter rest frame,
and $k_F$ is the Fermi momentum, related to the nucleon density by $\rho=\frac{2k_F^3}{3 \pi^2}$. Clearly, the free-space equation is recovered by using free-space energies and setting  
$Q$=1. 

From Eq.~(\ref{Eq:BG_full}) we obtain the corresponding $\phi$-integrated Bethe-Goldstone equation
\begin{align}
& \braket{\lambda_1^\prime \lambda_2^\prime|g^I(\tilde{q}^\prime,\tilde{q})|\lambda_1 \lambda_2} = \braket{\lambda_1^\prime \lambda_2^\prime|v^I(\tilde{q}^\prime,\tilde{q})|\lambda_1 \lambda_2} \nonumber \\
& \hspace{1cm} + \sum_{\lambda_1^\pp, \lambda_2^\pp = \pm } \pi \int_0^\infty \int_0^\pi \frac{\braket{\lambda_1^\prime \lambda_2^\prime|v^I(\tilde{q}^\prime,\tilde{q}^\pp)|\lambda_1^\pp \lambda_2^\pp} Q(\tilde{q}^\pp,P,k_F) \braket{\lambda_1^\pp \lambda_2^\pp|g^I(\tilde{q}^\pp,\tilde{q})|\lambda_1 \lambda_2}}{E^\ast_q - E^\ast_{q^\pp} + i \epsilon} {q^\pp}^2 \sin \theta^\pp \dd \theta^\pp \dd q^\pp \; .
\label{Eq:BG_final}
\end{align}
It's important to choose a frame such that $\bvec{P}$ points along the $z$-axis, since this allows $Q$ to become independent of $\phi^\pp$ because we can set $\phi^\pp = 0$ inside $Q$ without loss of generality. In this paper we have chosen $\bvec{P}=q \hat{\bvec{z}}$ and $k_F=1.4 \units{fm^{-1}}$, or 
$\rho$=0.185 fm$^{-3}$, close to normal matter density. Thus, we will suppress the dependence of $g^I$ on those variables.             
\subsubsection{The Pauli operator and the spherical approximation.}
The Pauli operator and its effect on Eq.~(\ref{Eq:BG_full}) is the focal point of this paper. Mathematically, $Q$ restricts the $\theta^\pp$ integration to 
\begin{equation}
\abs{\cos \theta^\pp} < a \; , \quad \text{where} \quad a \equiv \frac{P^2+{q^\pp}^2-k_F^2}{2Pq^\pp} \; ,
\end{equation}
as can be easily shown from Eq.~(\ref{Eq:Q}). As already discussed in the Introduction, non-spherical components can be included by evaluating the matrix elements of $Q$ and including them in a partial wave scattering equation. This method generates couplings between intermediate states with different total angular momentum as well as dependence on the magnetic quantum number. On the other hand, the three-dimensional solution requires some initial effort, but the inclusion of the exact Pauli operator is then absolutely straightforward. Another common method, which avoids the latter approaches, is to use the partial wave scattering equation along with the so called spherical or angle-averaged Pauli operator $\bar{Q}$ (see Ref.~\cite{Haf} and references therein)
\begin{equation}
Q(\bvec{q}^\pp,\bvec{P},k_F) \approx \bar{Q}(q^\pp,P,k_F) = \frac{\int Q(\bvec{q}^\pp,\bvec{P},k_F) \dd \Omega^\pp }{ \int \dd \Omega^\pp } = \frac{1}{2} \int_{-a}^a \dd( \cos \theta^\pp) = \frac{P^2+{q^\pp}^2-k_F^2}{2Pq^\pp} \; , 
\label{Eq:Q_avg}
\end{equation}
unless it's equal to zero or one. 
In the next section we explore the differences (or similarities) resulting from using the exact Pauli operator in a three-dimensional calculation, \textit{or} the spherical Pauli operator in a partial wave calculation.

For the sake of generality, we note that the above Pauli operator can be extended to the case of two different Fermi
momenta, $k_{F1}$ and $k_{F2}$. This makes it suitable for an isospin-asymmetric nuclear matter calculation. All that needs to be done is to modify the angular integration to implement the restrictions
\begin{align}
& \abs{\bvec{P} + \bvec{q}^\pp}  > k_{F1} \quad \text{and} \quad  \abs{\bvec{P} - \bvec{q}^\pp} > k_{F2} \implies \nonumber \\ 
& \quad Q(\bvec{q}^\pp,\bvec{P},k_{F1},k_{F2}) \equiv \Theta( \abs{\bvec{P} + \bvec{q}^\pp} - k_{F1} ) \Theta( \abs{\bvec{P} - \bvec{q}^\pp} - k_{F2} ) \; ,
\label{Eq:Q2}
\end{align}
which again, is easily implemented into our three-dimensional formalism.
\section{Results and Discussion}
\label{III}
As an initial check of our formalism, we calculate the $t$-matrix and transform it into the $\ket{LSJ}$ basis via Eq.~(\ref{Eq:inverse_transformation}). Our comparisons are displayed in Tables~\ref{Table:freespaceLSJ_1} and \ref{Table:freespaceLSJ_2}, where we show $LSJ$ on-shell matrix elements at laboratory energies equal to $50,100,200,$ and $300 \units{MeV}$. (The laboratory energy $E_{Lab}$, is related to the on-shell center-of-mass momentum $q$ by $E_{Lab} = \frac{2q^2}{m}$.) We use the familiar spectroscopic notation for partial waves, e.g., for coupled states $\up{(2S+1)}L_J^\prime$-$\up{(2S+1)}L_J$ refers to $\braket{L^\prime SJ|\hat{T}|LSJ}$.

Looking at the tables in terms of relative error (with the partial wave solution taken to be exact),  
the majority of our results in Tables~\ref{Table:freespaceLSJ_1} and \ref{Table:freespaceLSJ_2} have errors less than $\approx 0.1 \%$. Coupled states have slightly larger errors. For instance, the real part of $\up{3}S_1$ at $300 \units{MeV}$ has an error of $\approx 3 \%$. Only the $\up{1}P_1$ case has consistently larger discrepancies, the largest being $\approx 7 \%$ and occurring in the imaginary part at $300 \, \units{MeV}$. The $\up{1}P_1$ state is a central partial wave with both spin and isospin equal to zero and notoriously problematic, due to large attraction at short range. Thus, some larger discrepancy may be expected. Nevertheless, the worst case we have observed still yields reasonable agreement: $-1.33\times 10^{-7} \units{MeV^{-2}}$ {\it vs.} $-1.43 \times 10^{-7} \units{MeV^{-2}}$.

Before moving on to showing our main results, it's useful to recall that 
Eq.~(\ref{Eq:transformation}) implies rotational invariance (hence conservation of total angular
momentum). While the angle-average calculation clearly maintains rotational invariance,
this symmetry is broken when handling the Pauli operator exactly, due to the directional dependencies introduced. Thus, when entering the medium, we stay with the direct output of our three-dimensional equation, antisymmetrized as displayed in the LHS of Eqs.~(\ref{Eq:partialwave2d0})-(\ref{Eq:partialwave2d66}). The other element of the comparison consists of three-dimensional solutions constructed
from the (Pauli-modified, but rotationally invariant) partial waves as shown in the RHS of 
Eqs.~(\ref{Eq:partialwave2d0})-(\ref{Eq:partialwave2d66}). 

We perform calculations as described in Appx.~\ref{systemofeq} for several initial momenta. Because we wish to
highlight the impact of Pauli blocking in the two different approaches (exact \textit{vs.} angle-averaged),
we apply no other medium effects at this time and thus the matrix elements
can be quite different than those from a realistic Brueckner or Dirac-Brueckner calculation (although we
may refer to our Pauli-modified calculation as a $g$-matrix calculation).
We will show a representative set of amplitudes from Eqs.~(\ref{Eq:partialwave2d0})-(\ref{Eq:partialwave2d66}).                                                                                                   

\begin{figure}[H]
\begin{center}
\subfloat{
\includegraphics[width=6cm]{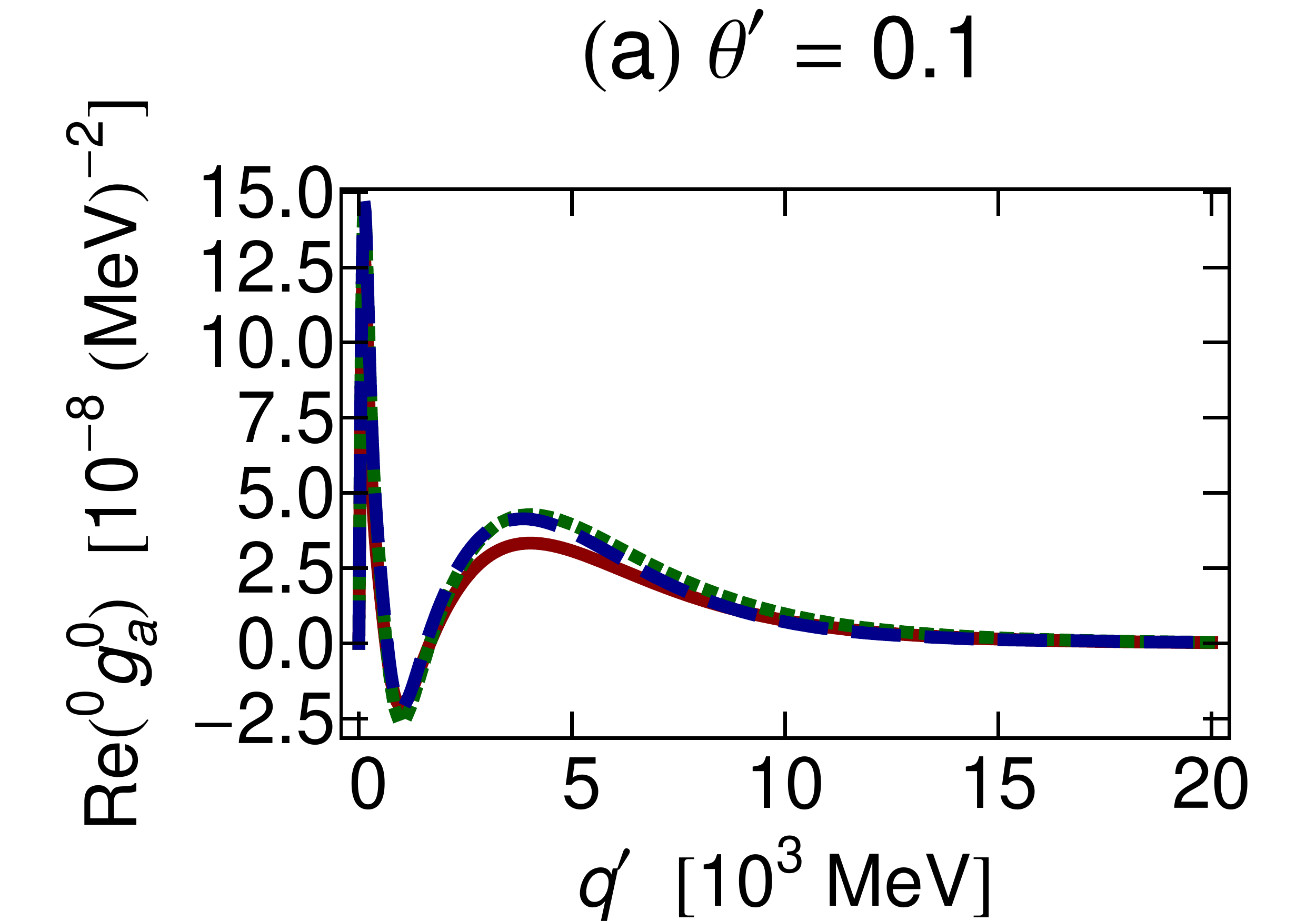}
}
\subfloat{
\includegraphics[width=6cm]{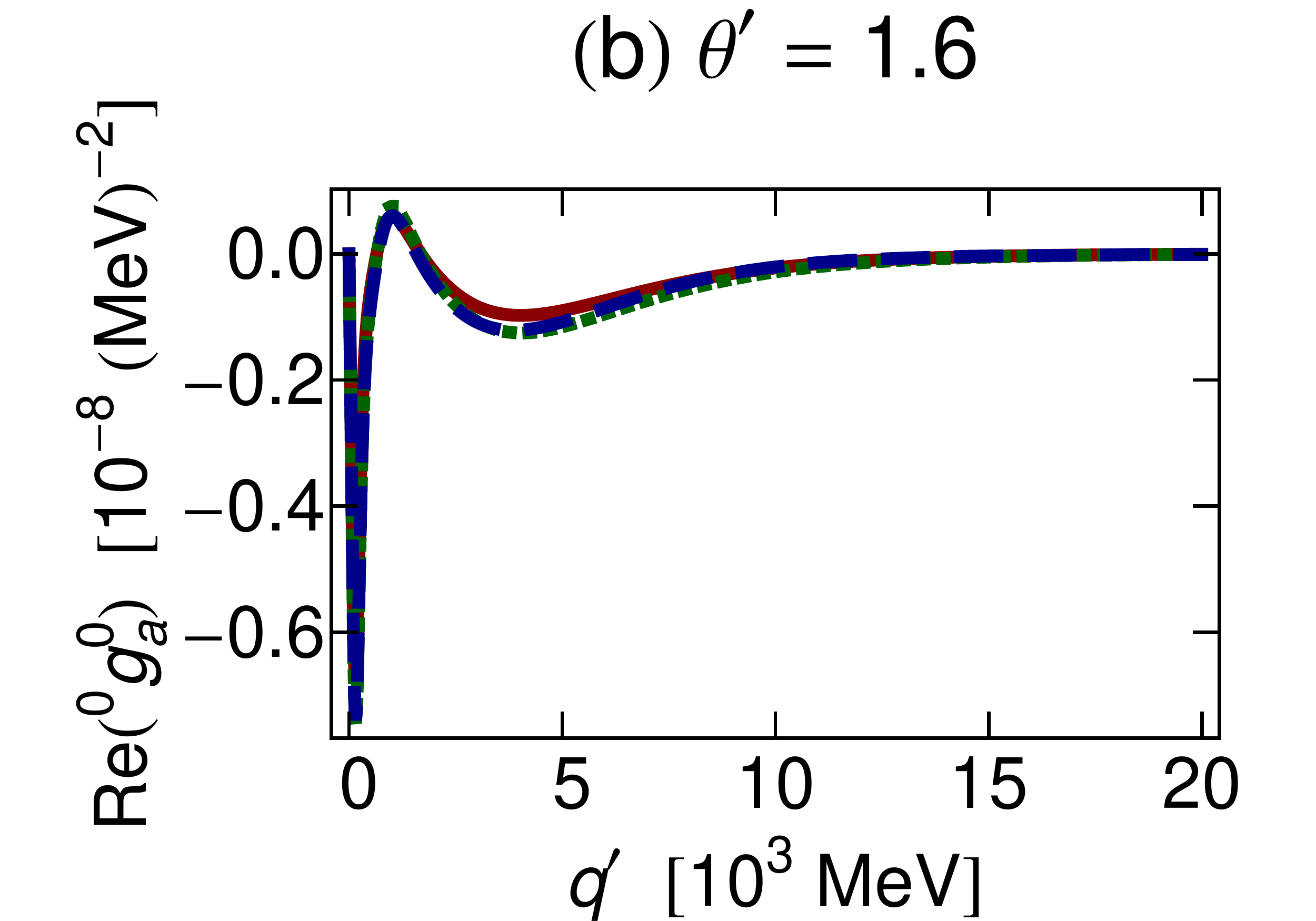}
}
\subfloat{
\includegraphics[width=6cm]{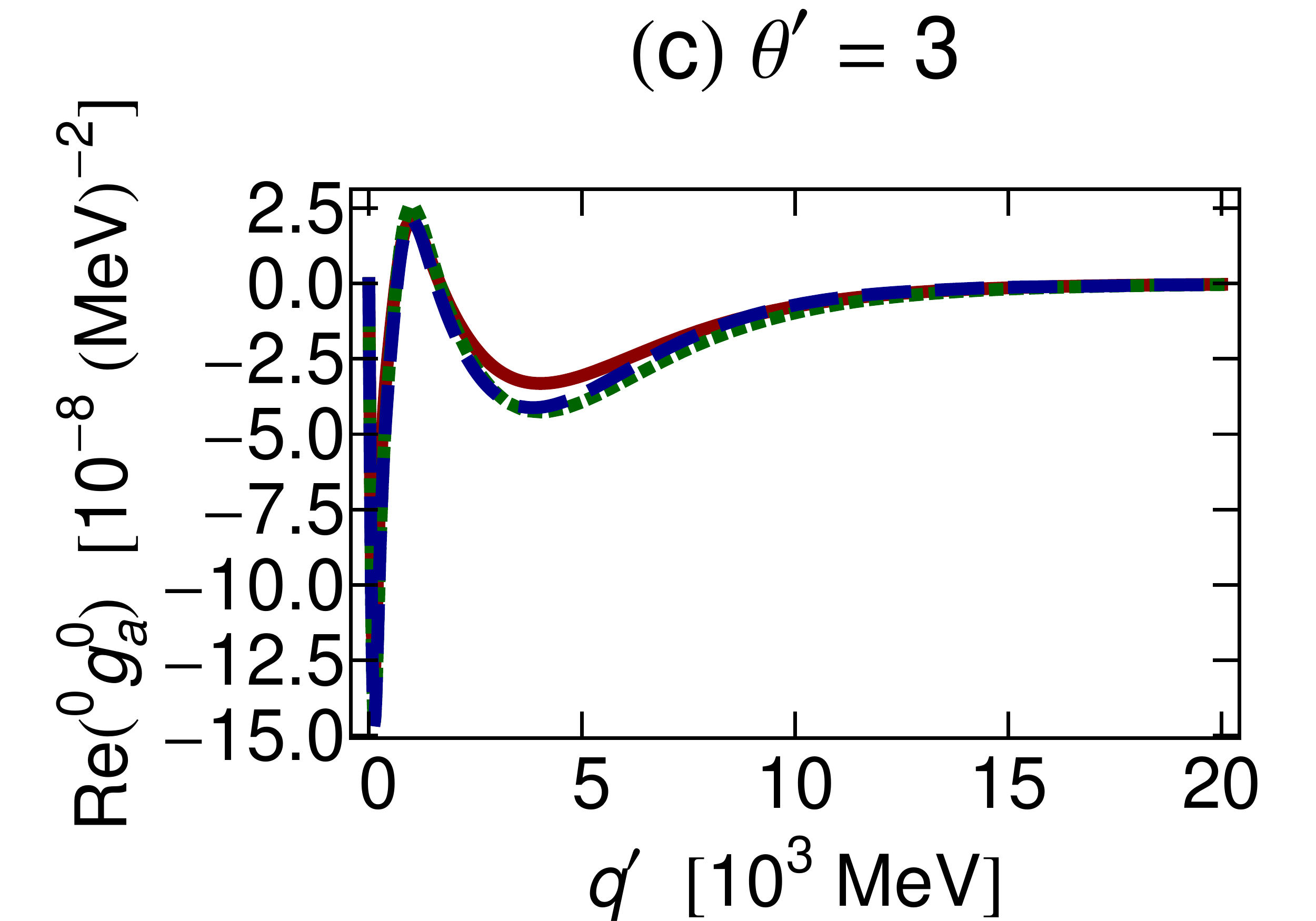}
}
\\
\subfloat{
\includegraphics[width=6cm]{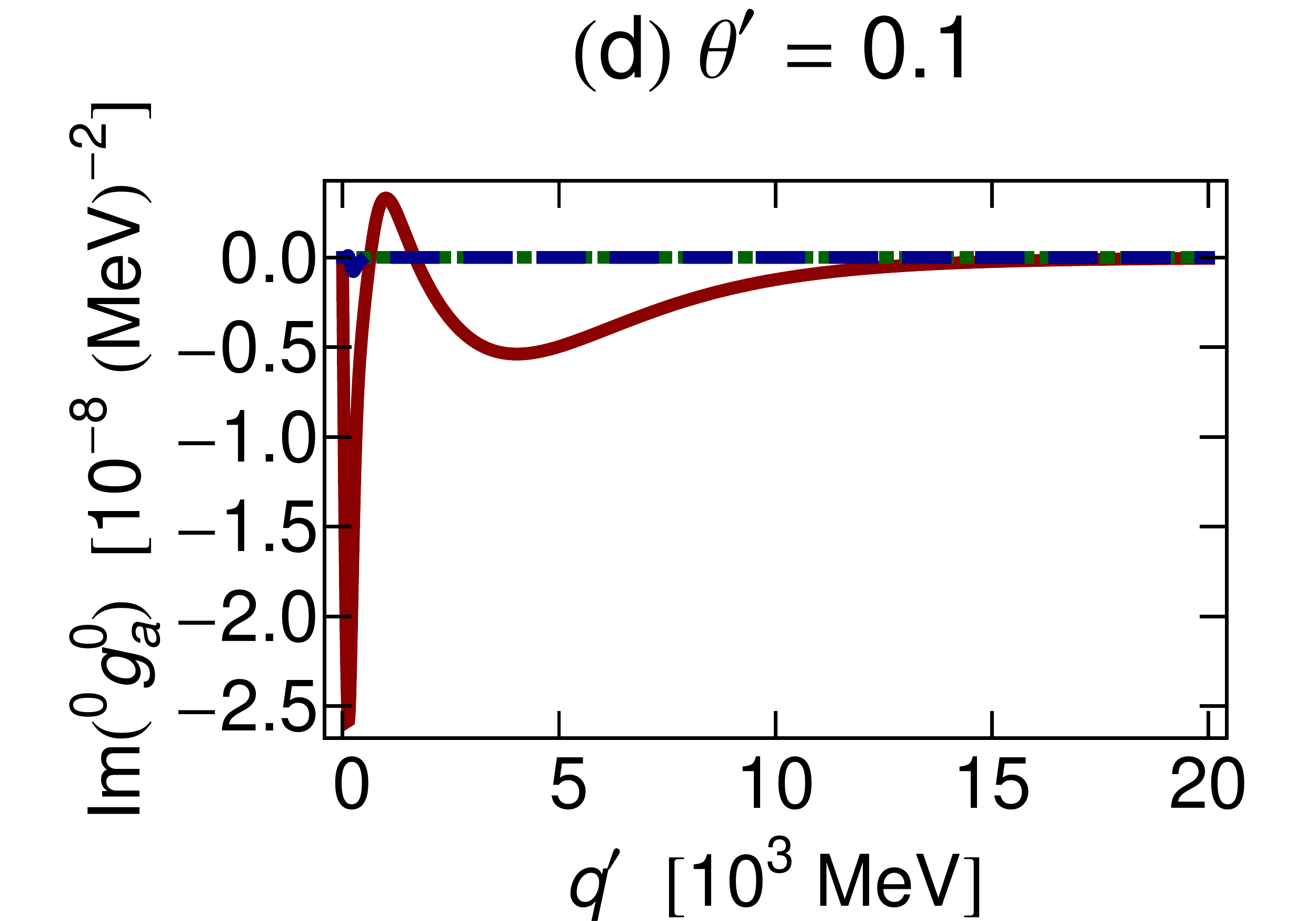}
}
\subfloat{
\includegraphics[width=6cm]{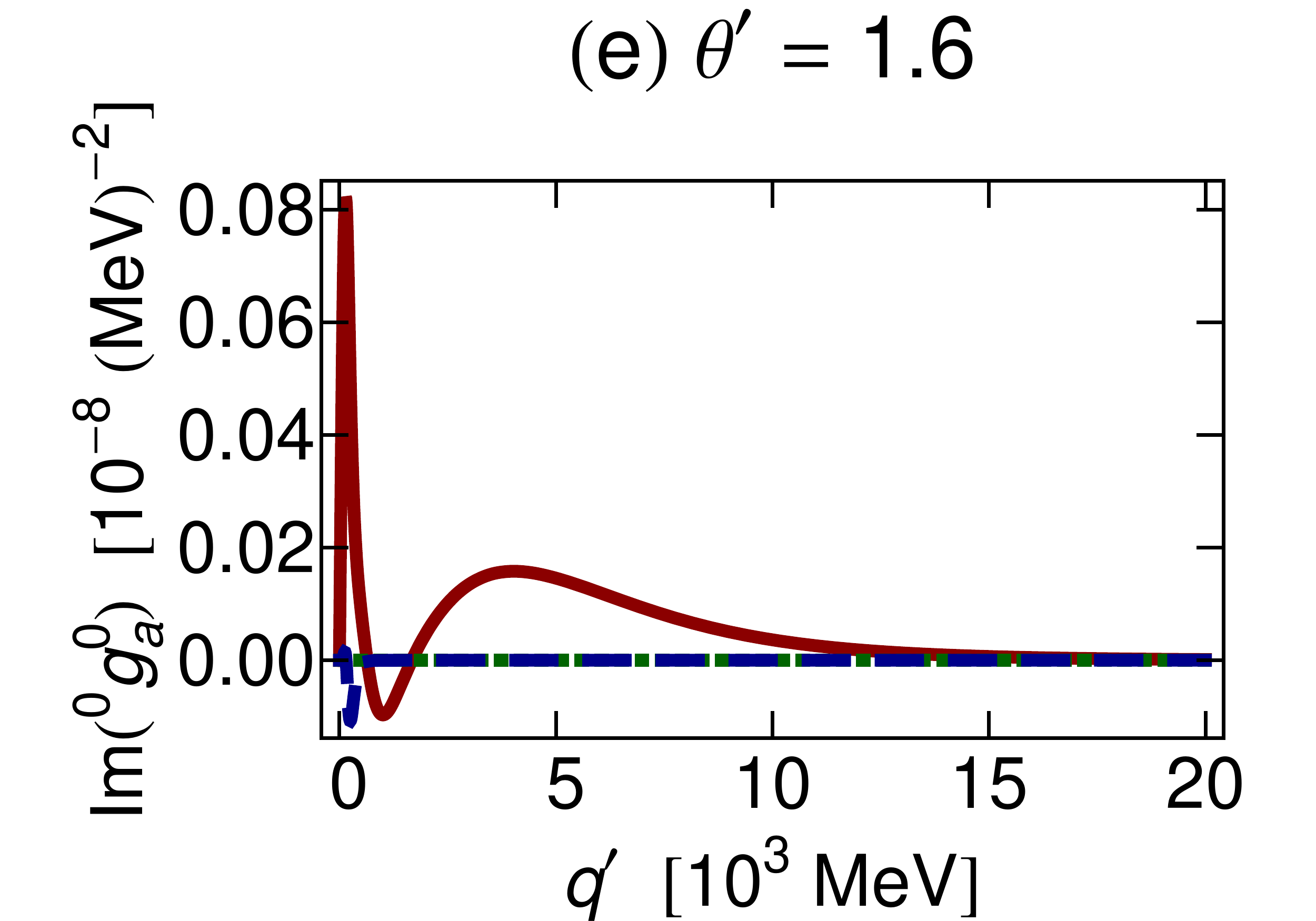}
}
\subfloat{
\includegraphics[width=6cm]{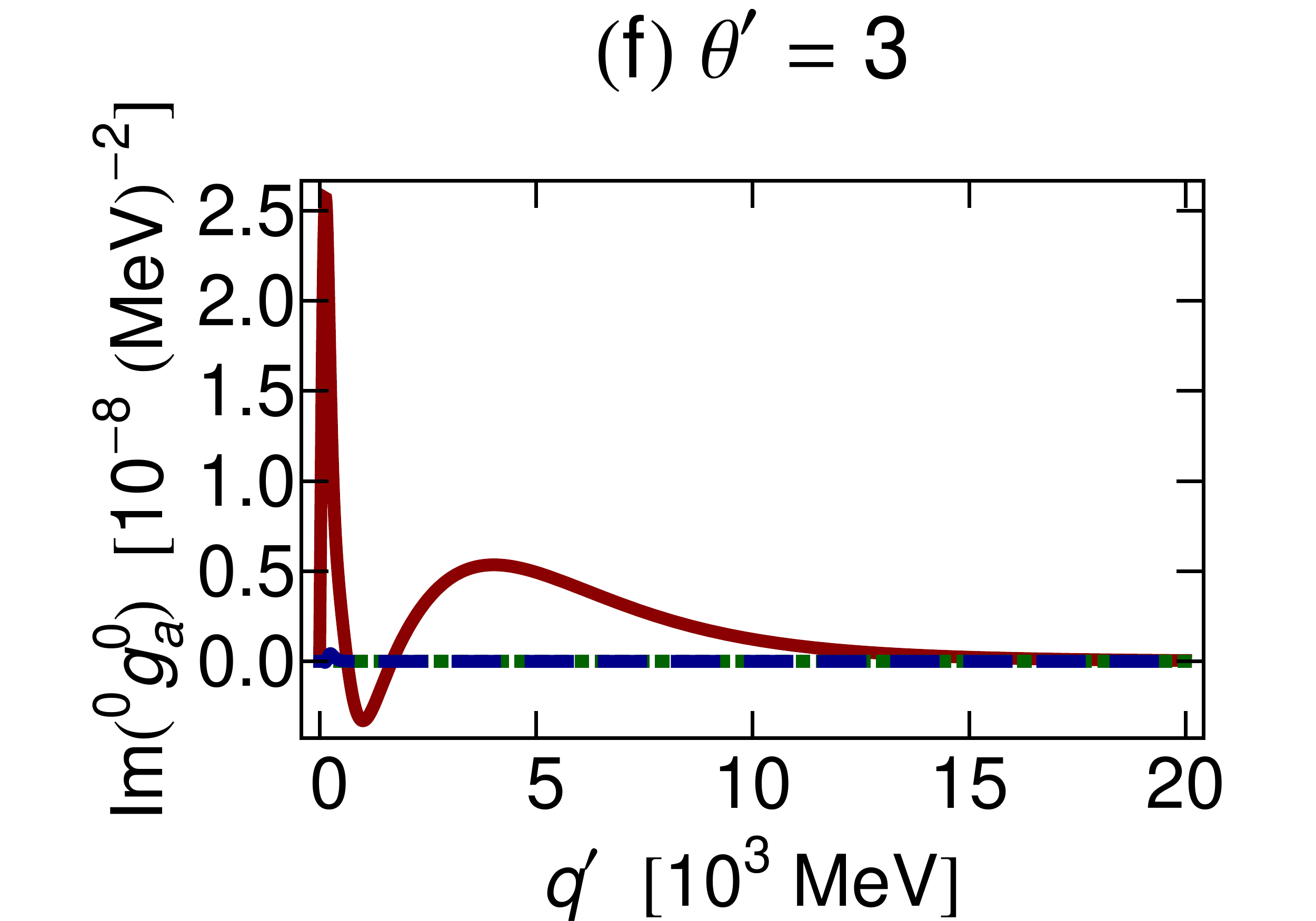}
}
\\
\subfloat{
\includegraphics[width=6cm]{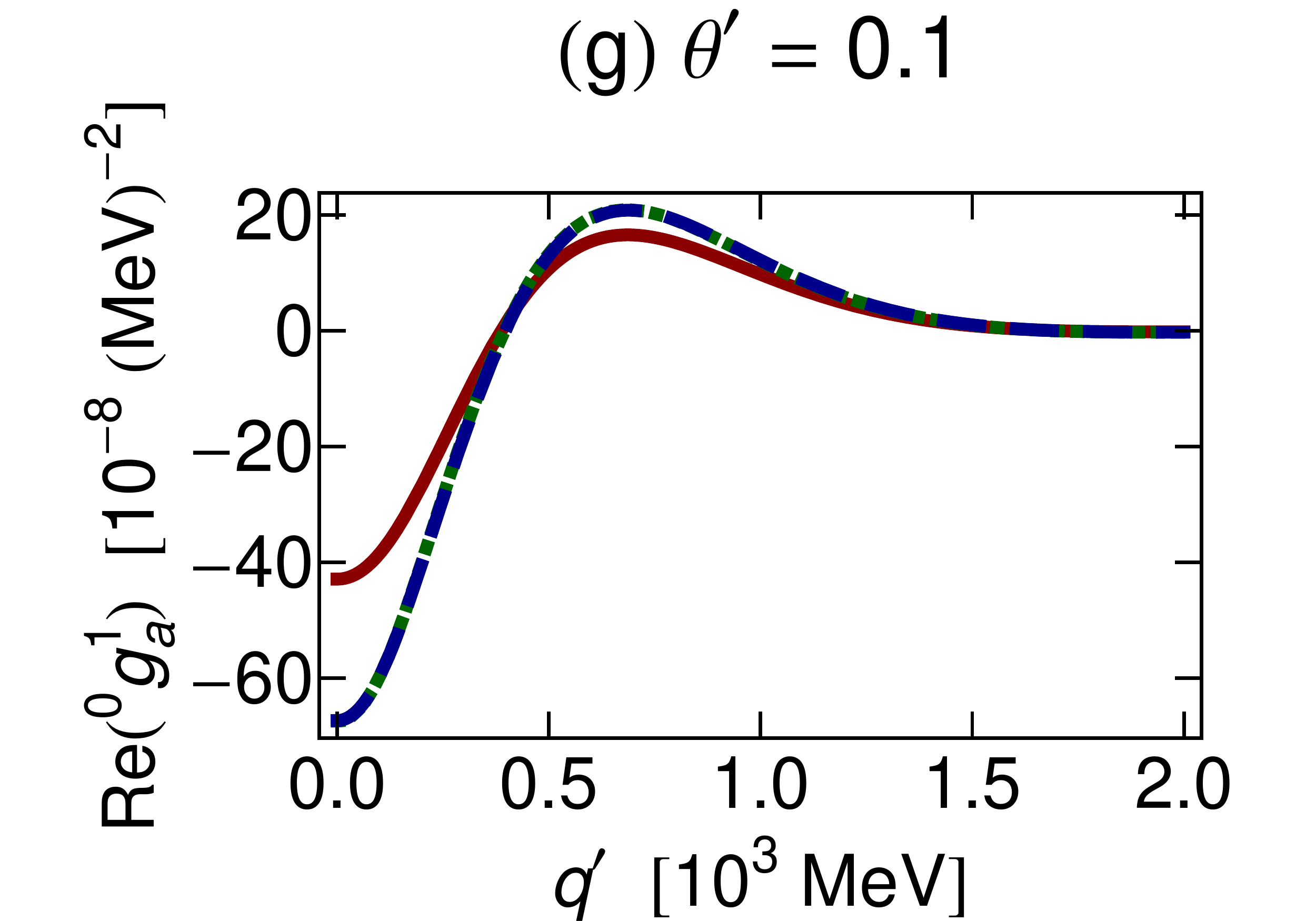}
}
\subfloat{
\includegraphics[width=6cm]{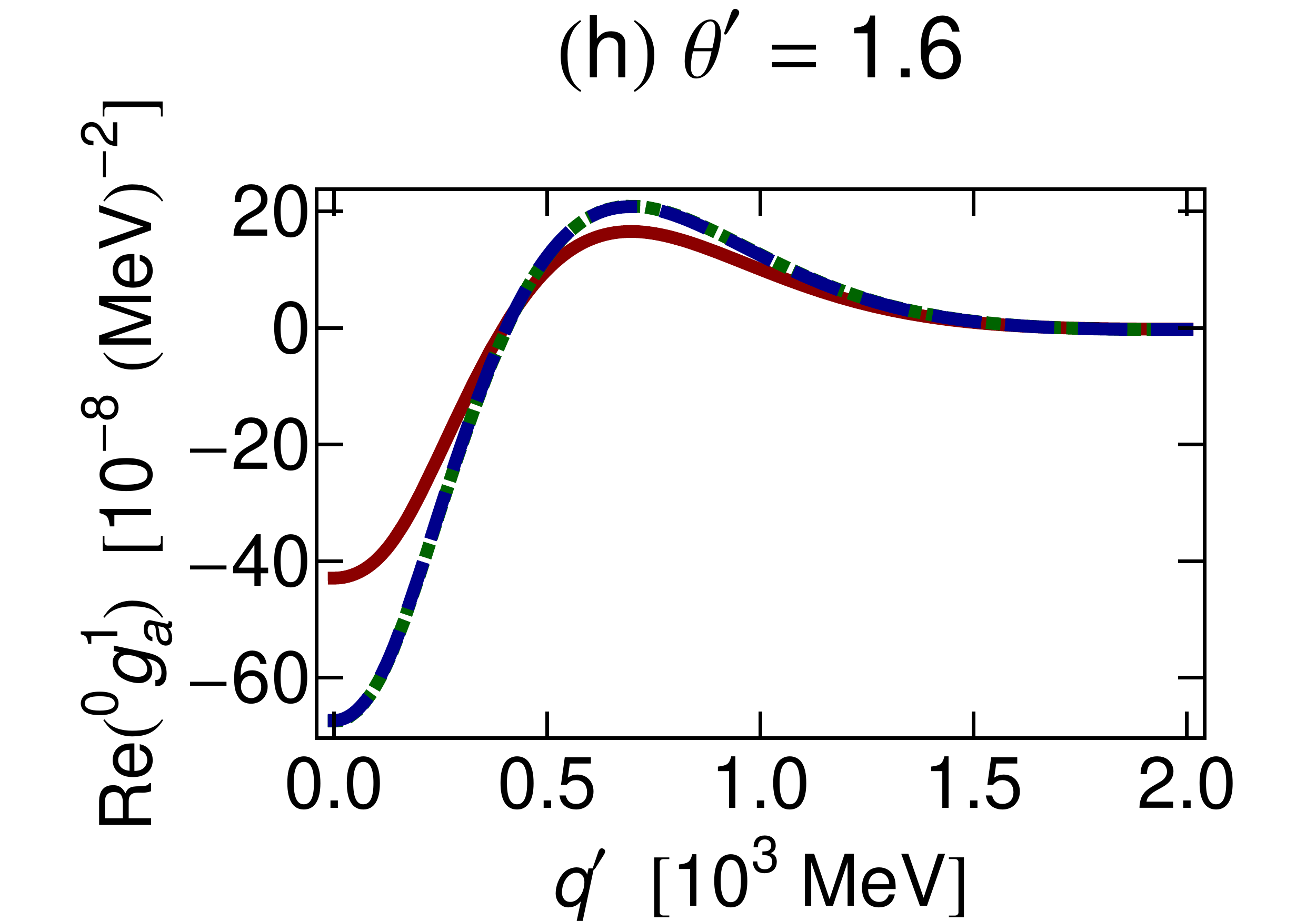}
}
\subfloat{
\includegraphics[width=6cm]{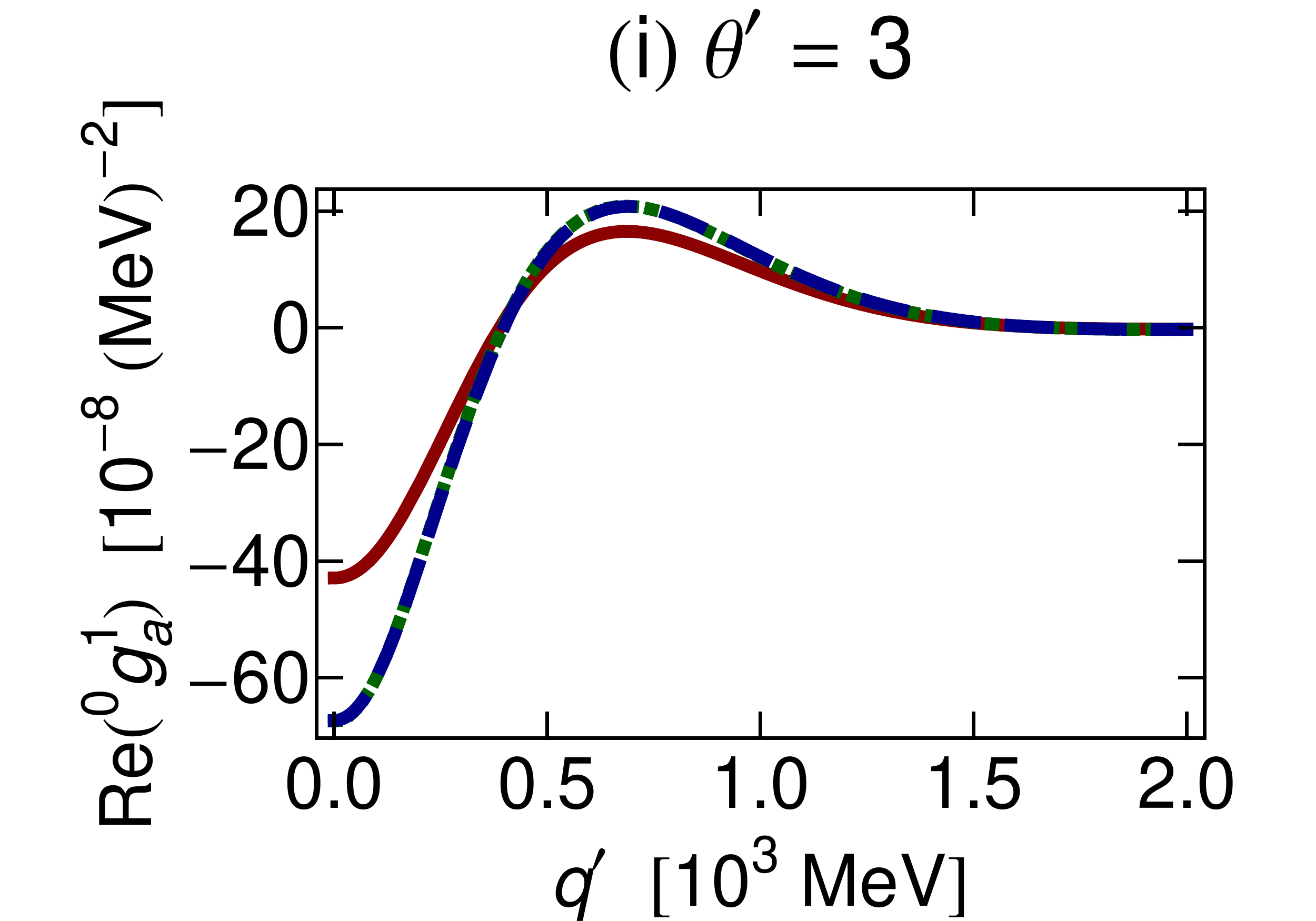}
}
\\
\subfloat{
\includegraphics[width=6cm]{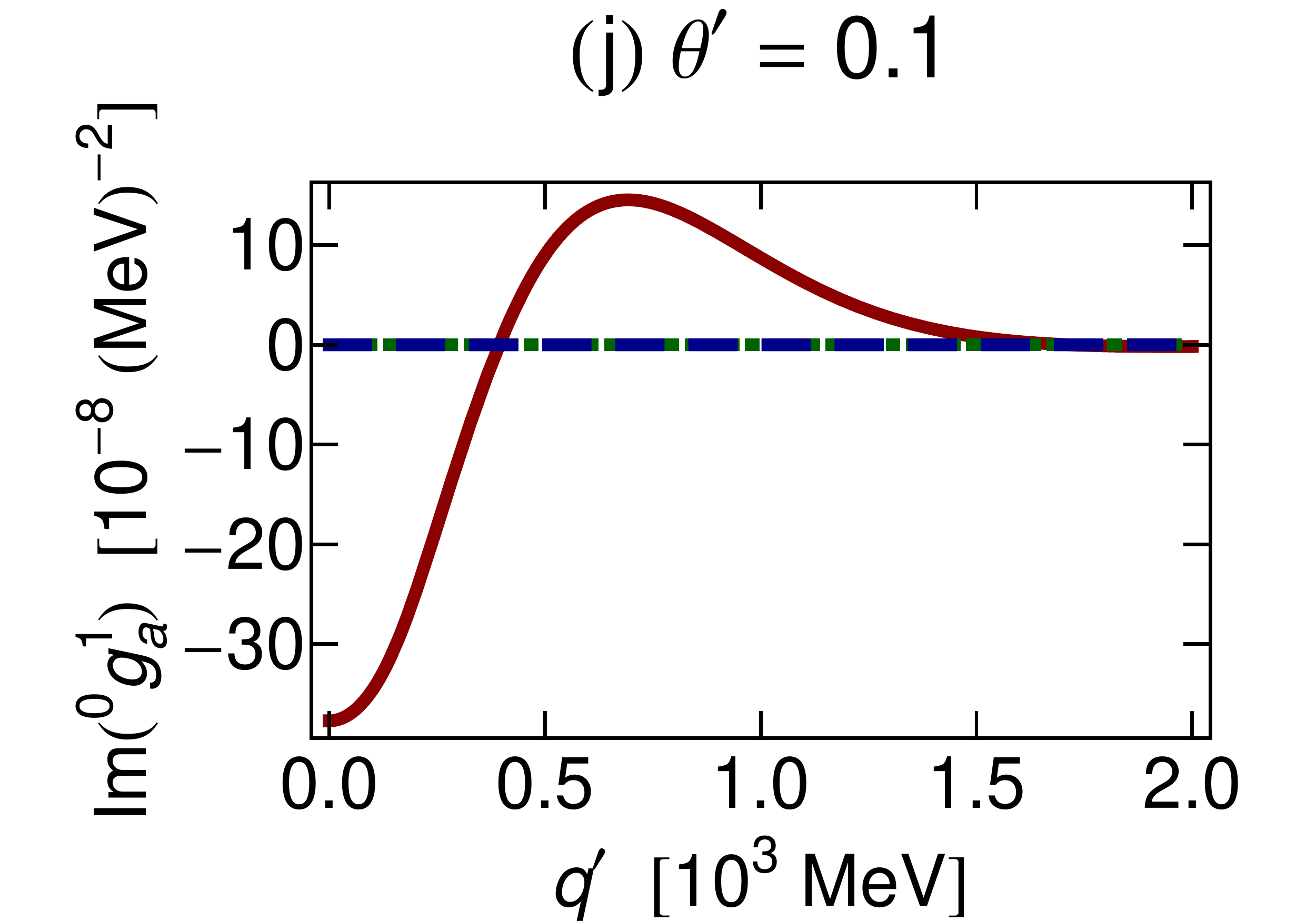}
}
\subfloat{
\includegraphics[width=6cm]{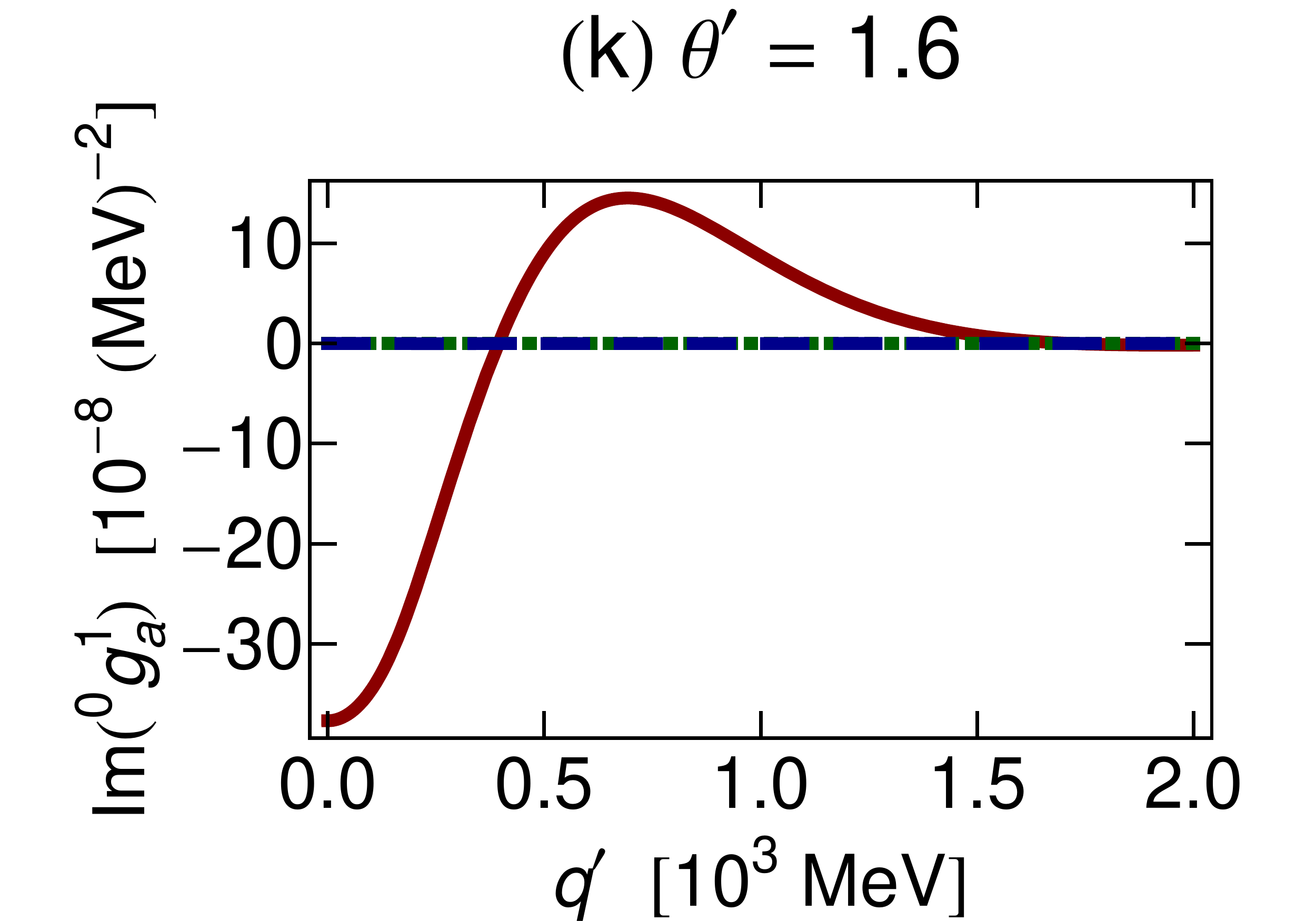}
}
\subfloat{
\includegraphics[width=6cm]{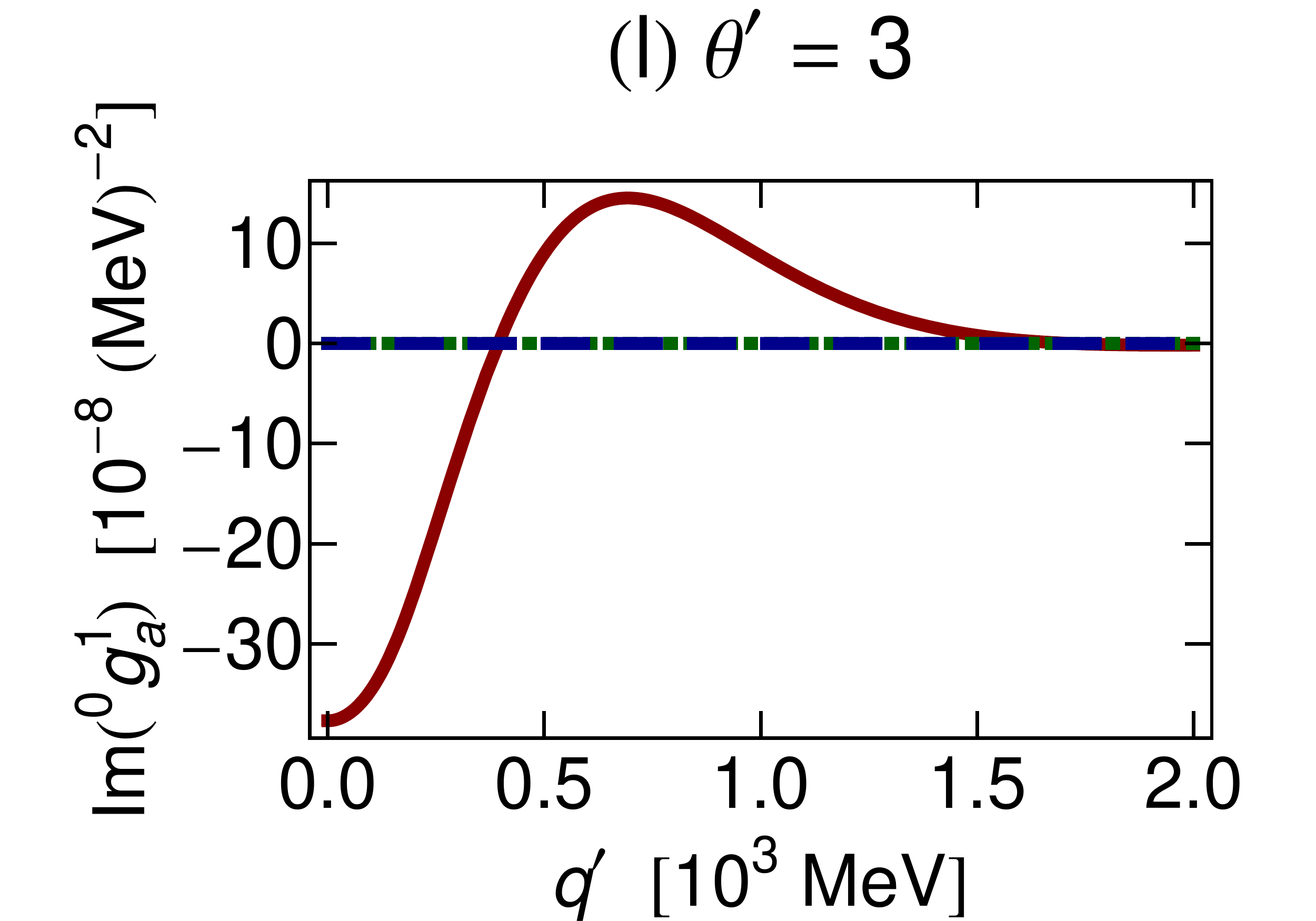}
}
\end{center}
\caption{(Color online) Real and imaginary parts of $\up{0}g^I_a$ and $\up{0}t^I_a$ (both isospins) as a function of $q^\prime$. We set $\theta=\arccos(0.5)$, $q=153.21 \units{MeV}$, and $\theta^\prime = 0.1, 1.6, 3$. The solid (red) curve is the free-space calculation while the dotted (green) and dashed (blue) curves are the exact and spherical Pauli operator calculations respectively.}
\label{Fig:g_plots50_0}
\end{figure}

\begin{figure}[H]
\begin{center}
\subfloat{
\includegraphics[width=6cm]{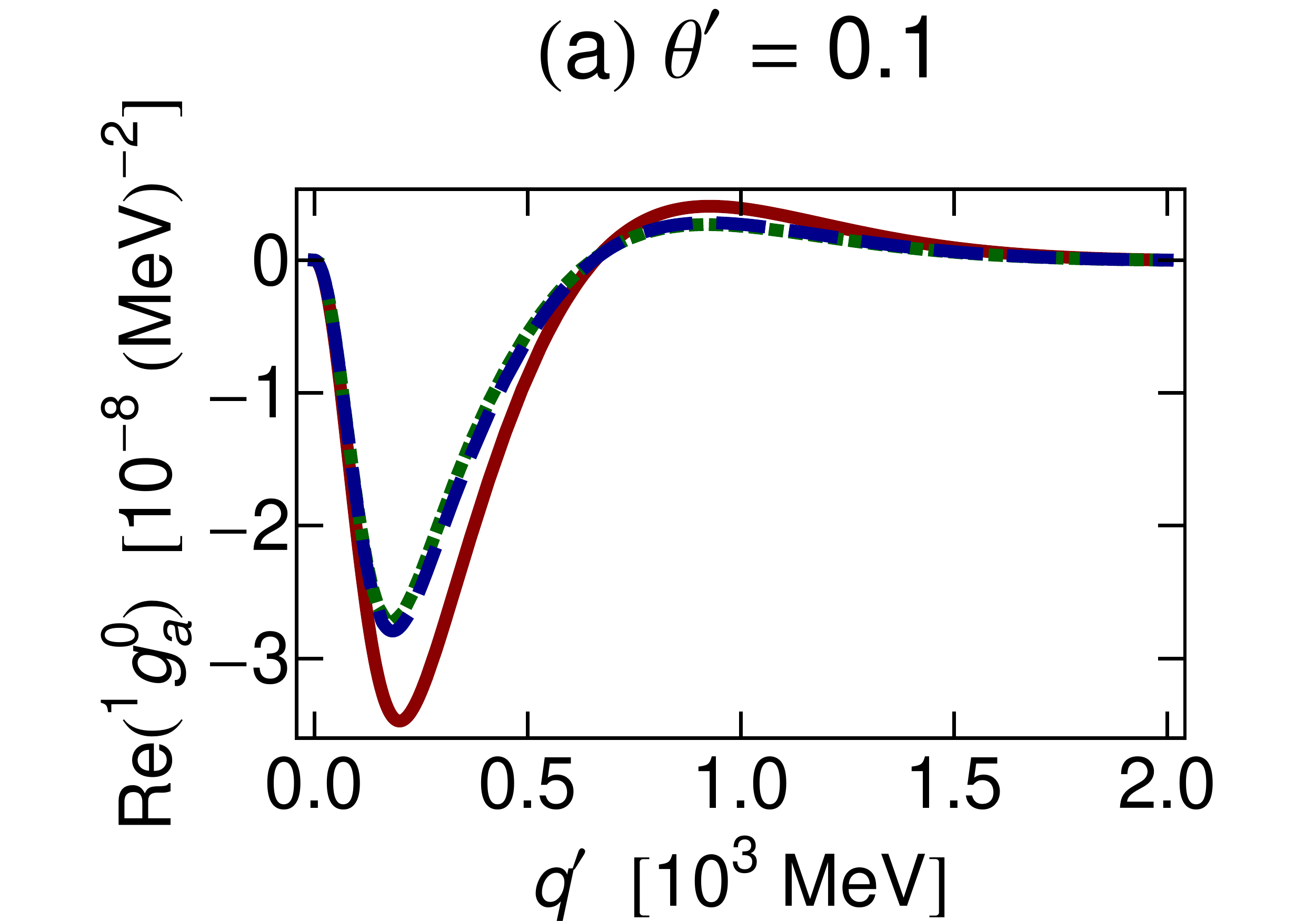}
}
\subfloat{
\includegraphics[width=6cm]{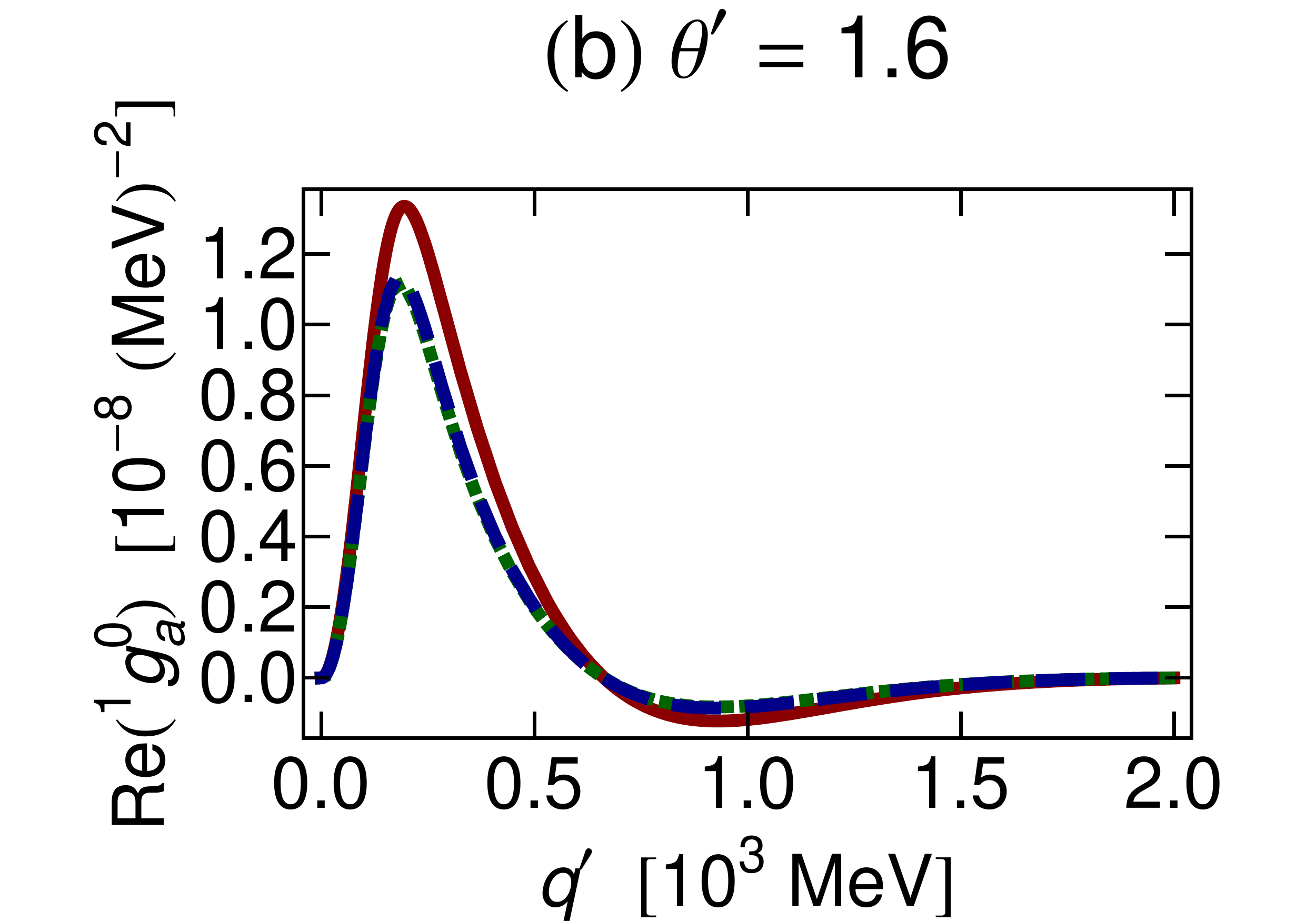}
}
\subfloat{
\includegraphics[width=6cm]{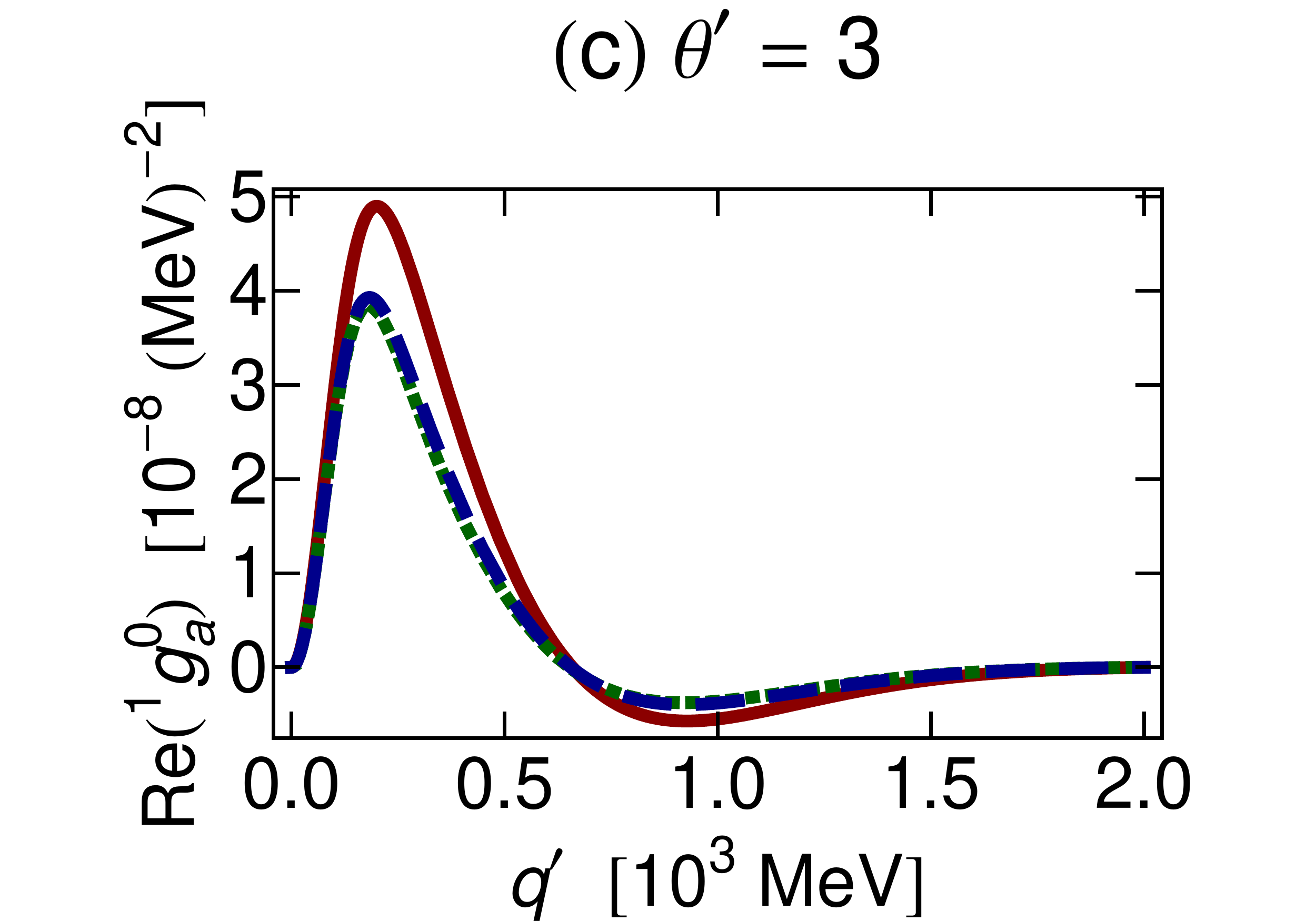}
}
\\
\subfloat{
\includegraphics[width=6cm]{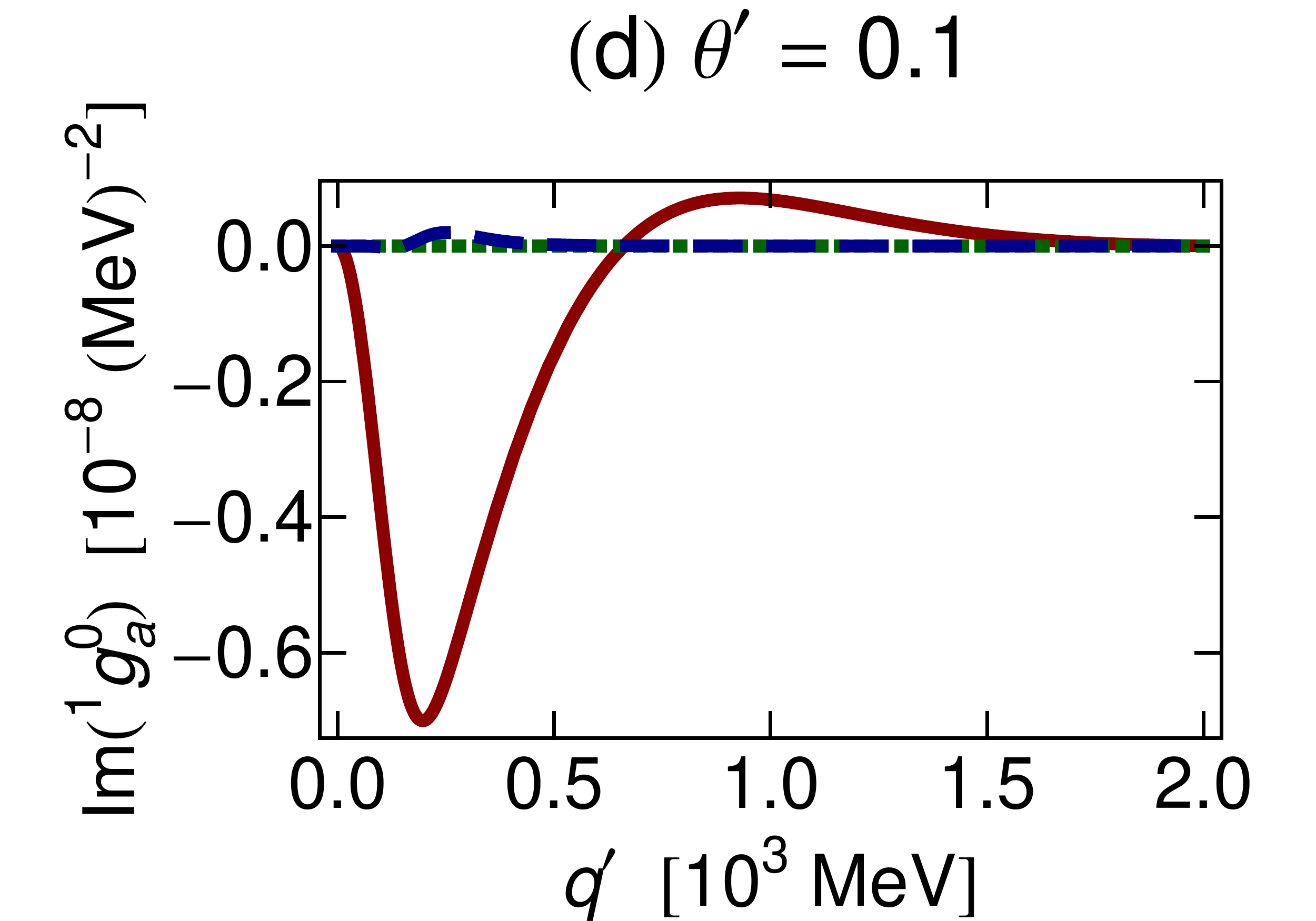}
}
\subfloat{
\includegraphics[width=6cm]{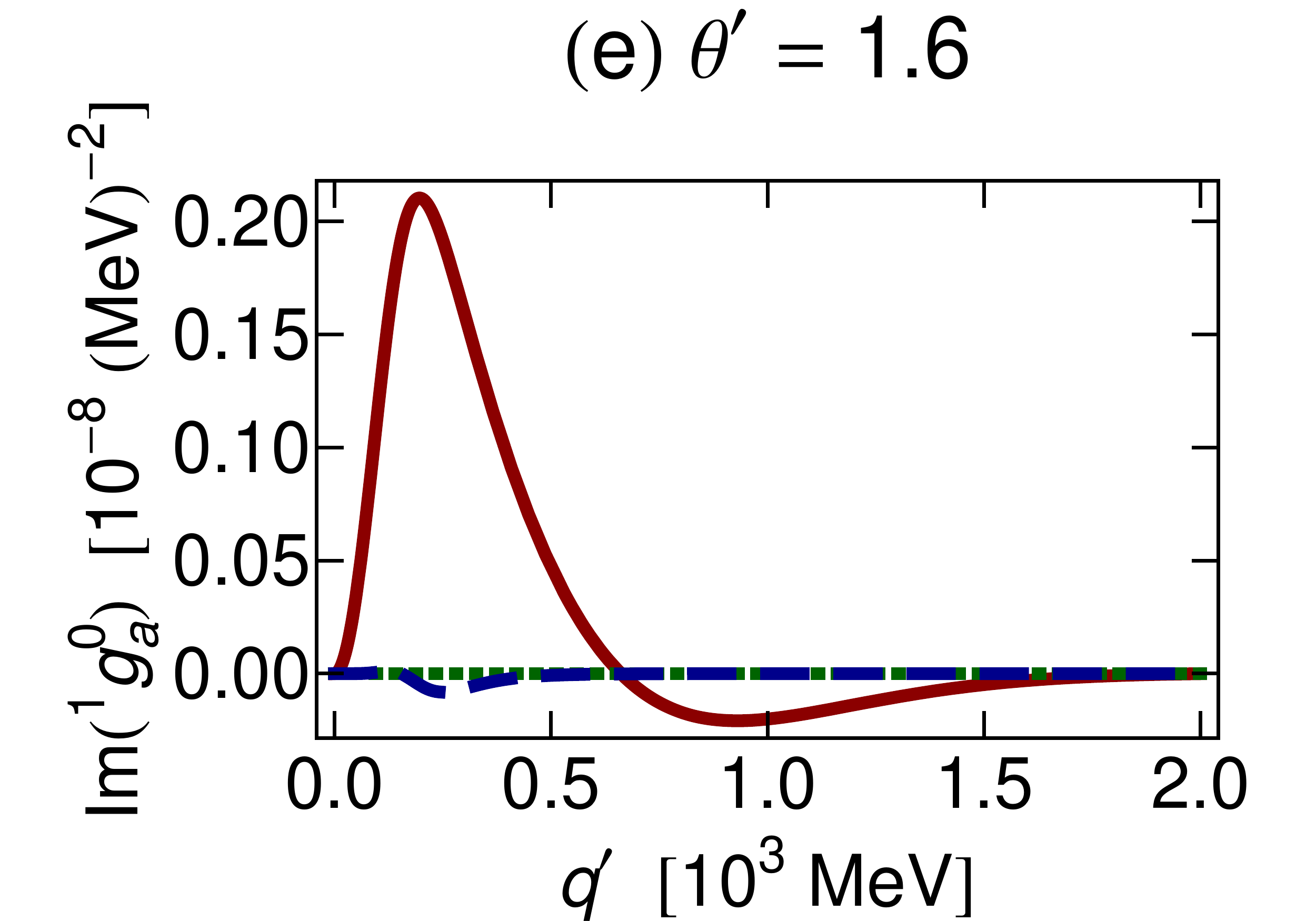}
}
\subfloat{
\includegraphics[width=6cm]{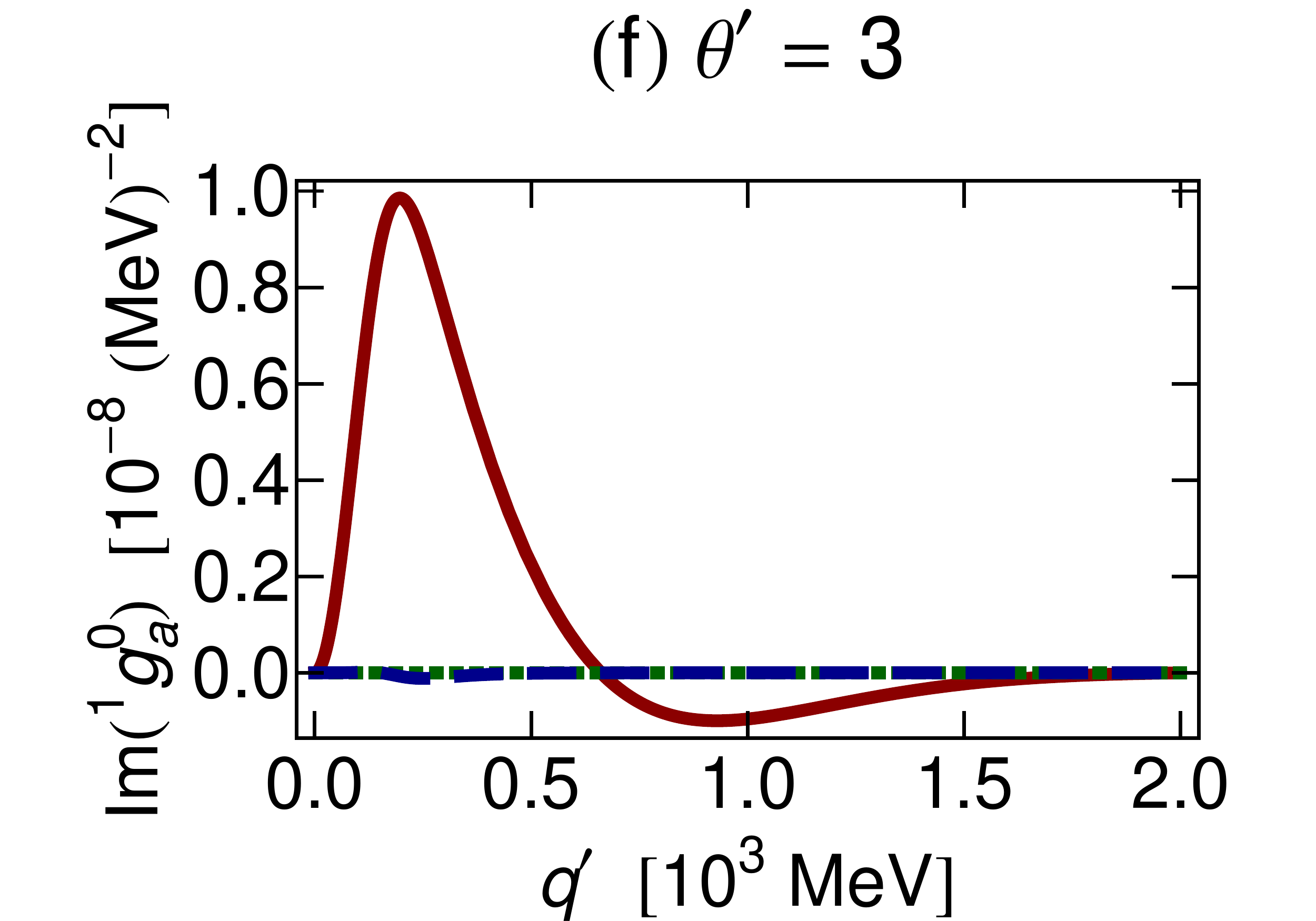}
}
\\
\subfloat{
\includegraphics[width=6cm]{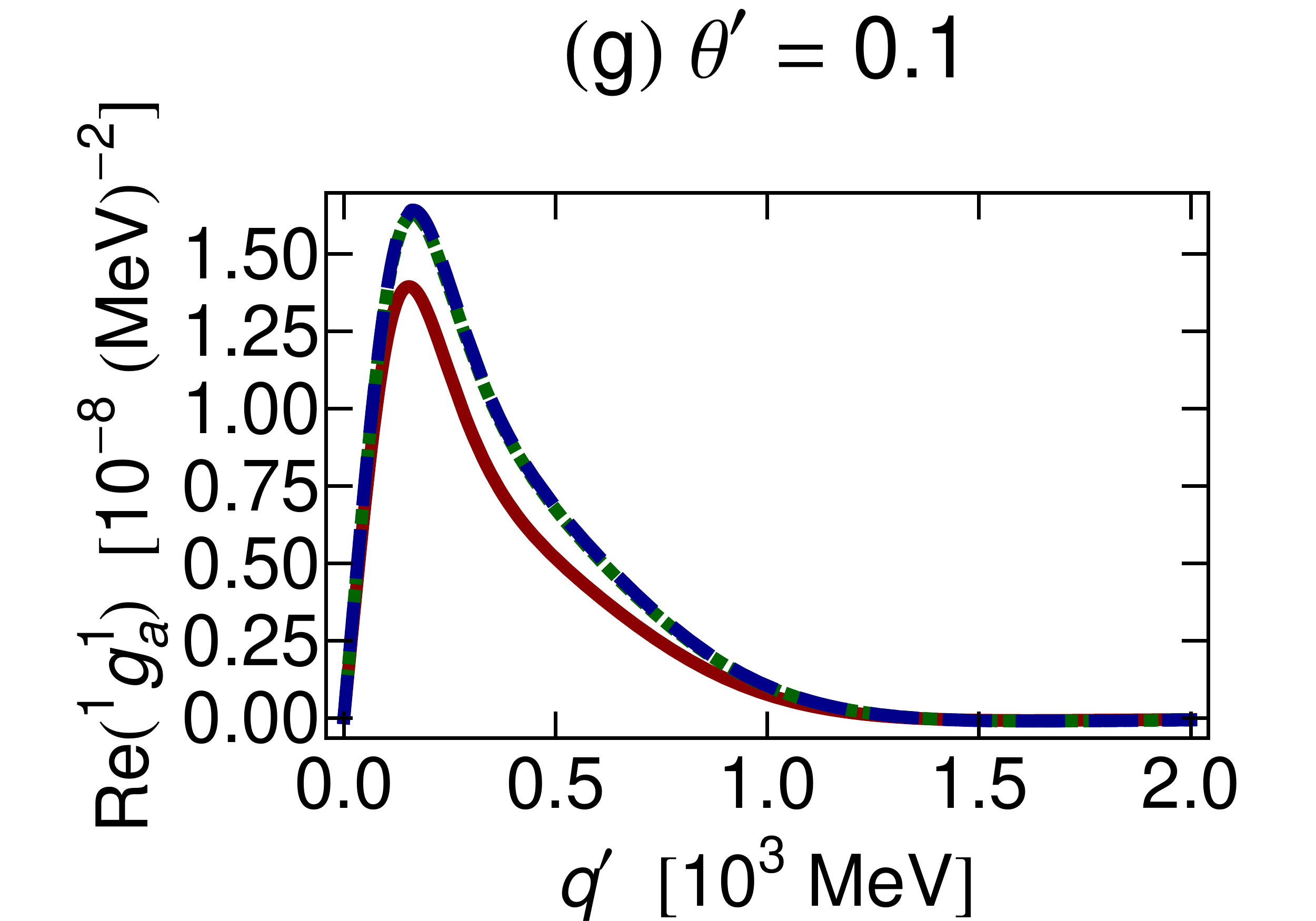}
}
\subfloat{
\includegraphics[width=6cm]{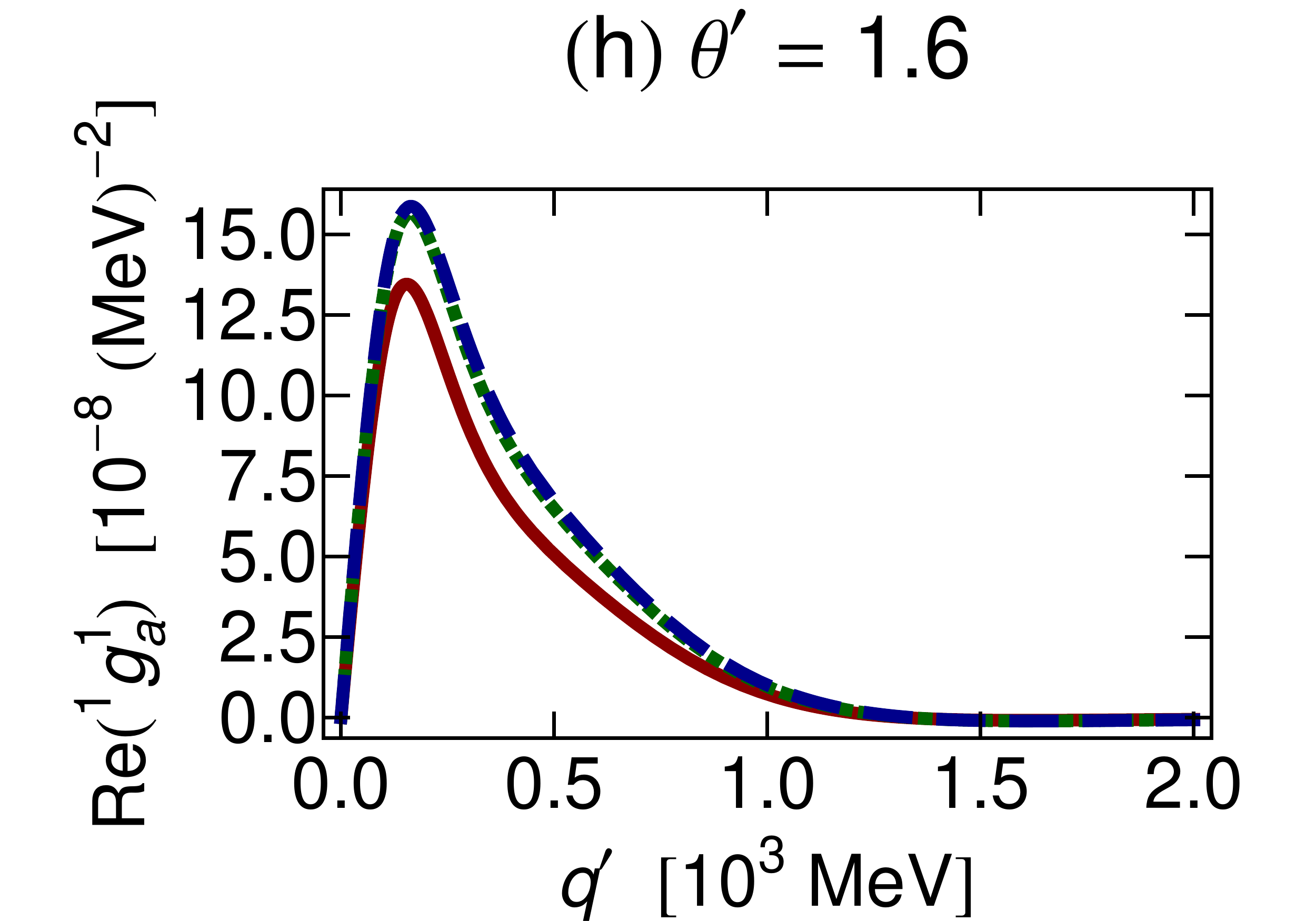}
}
\subfloat{
\includegraphics[width=6cm]{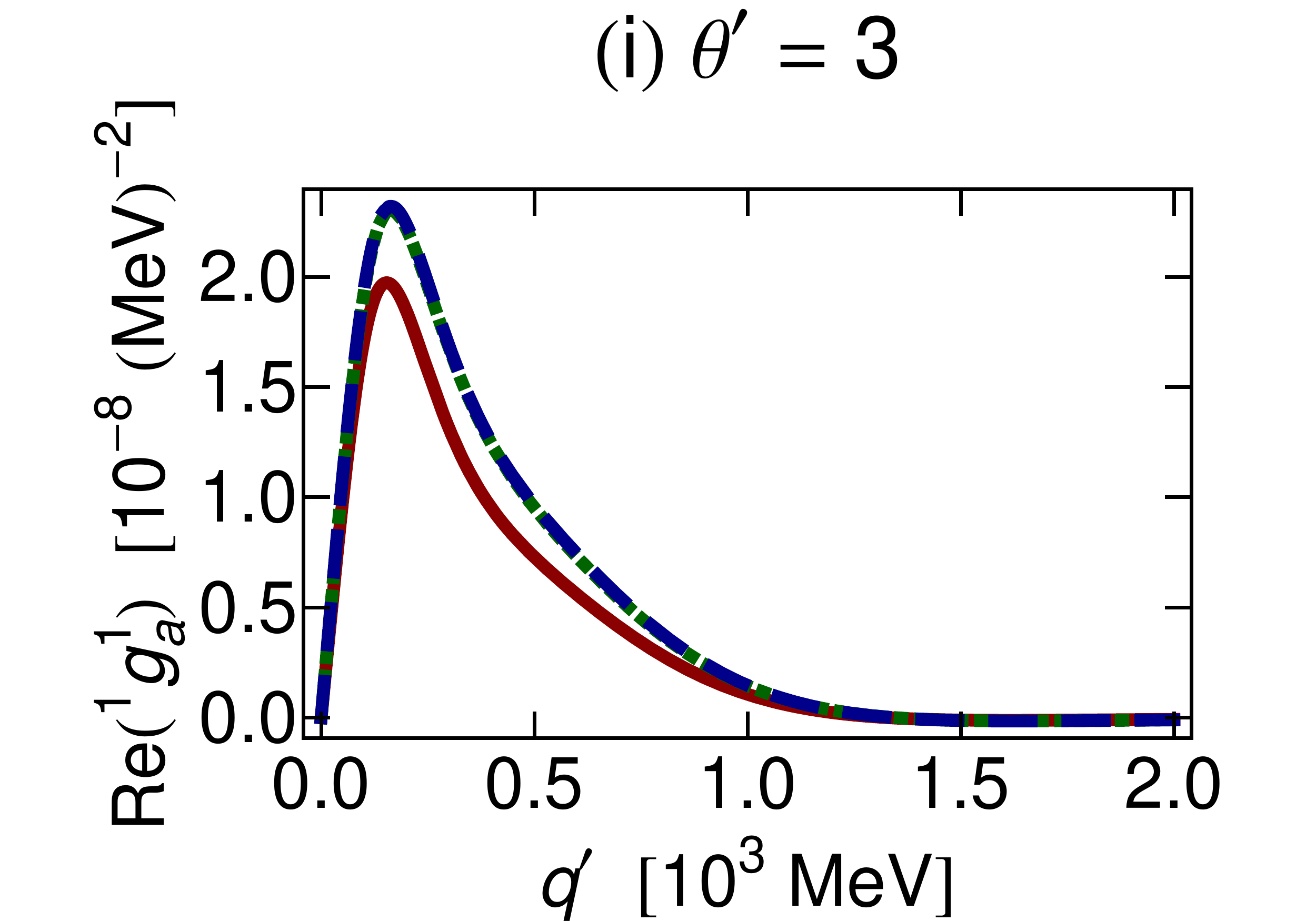}
}
\\
\subfloat{
\includegraphics[width=6cm]{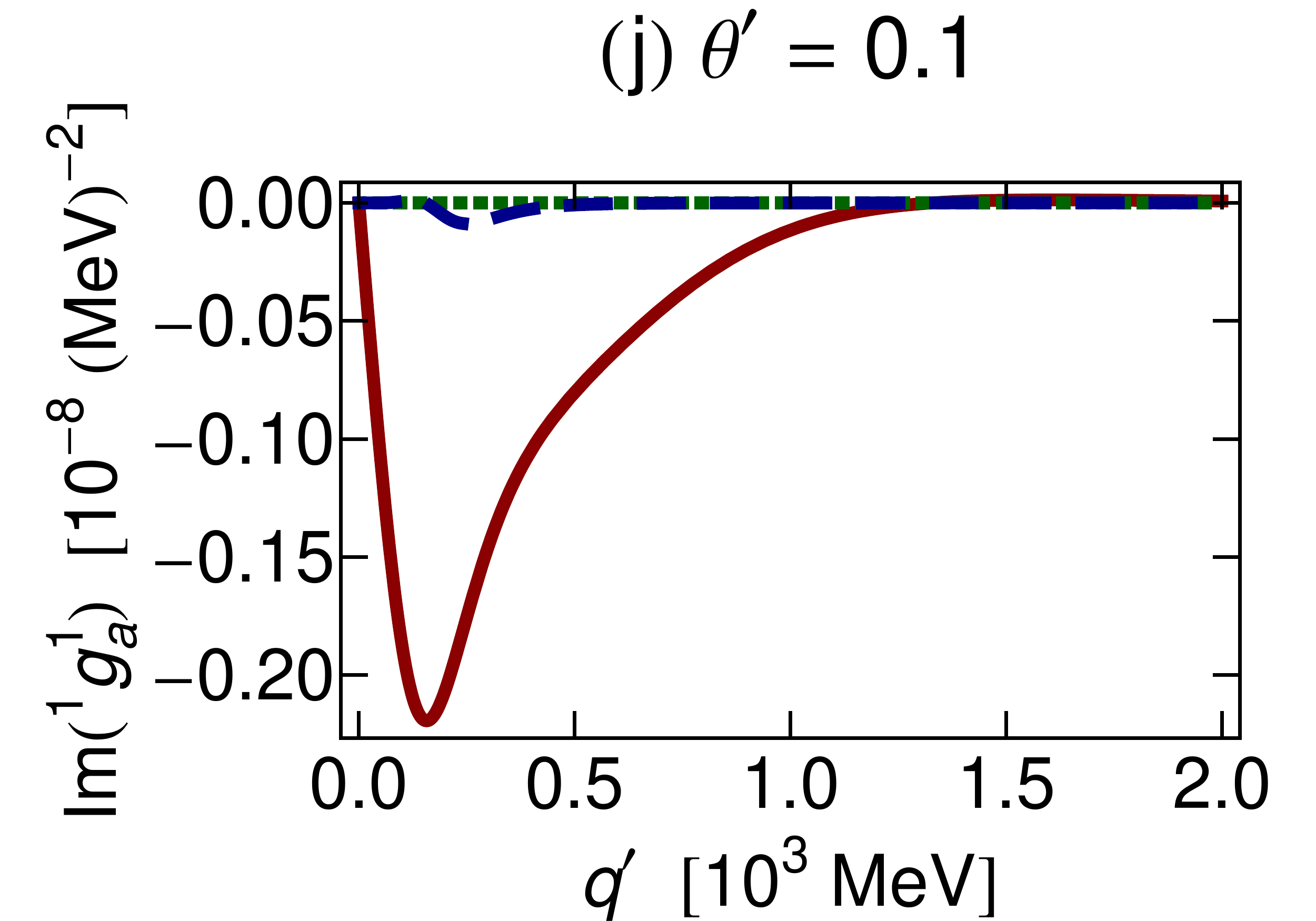}
}
\subfloat{
\includegraphics[width=6cm]{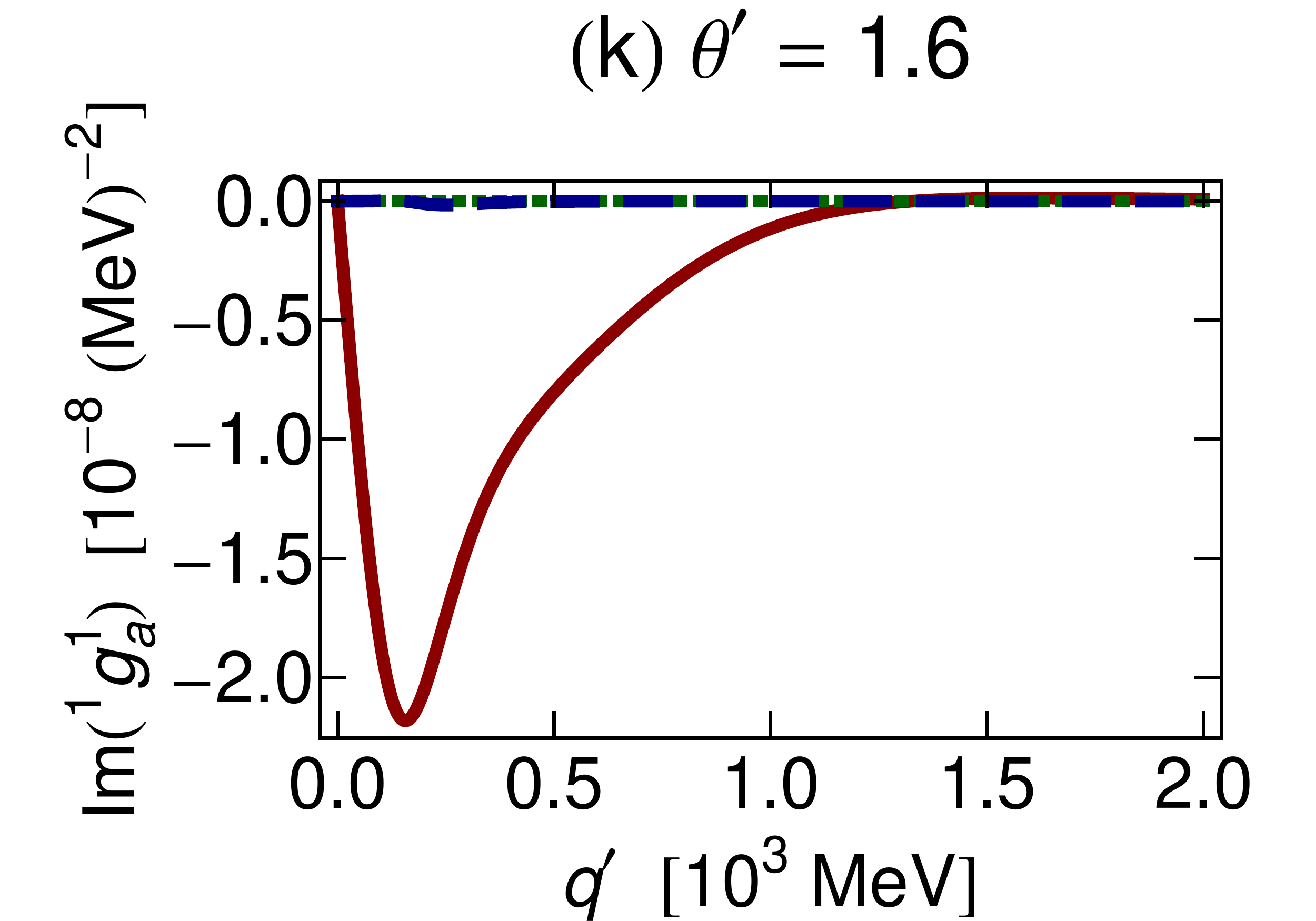}
}
\subfloat{
\includegraphics[width=6cm]{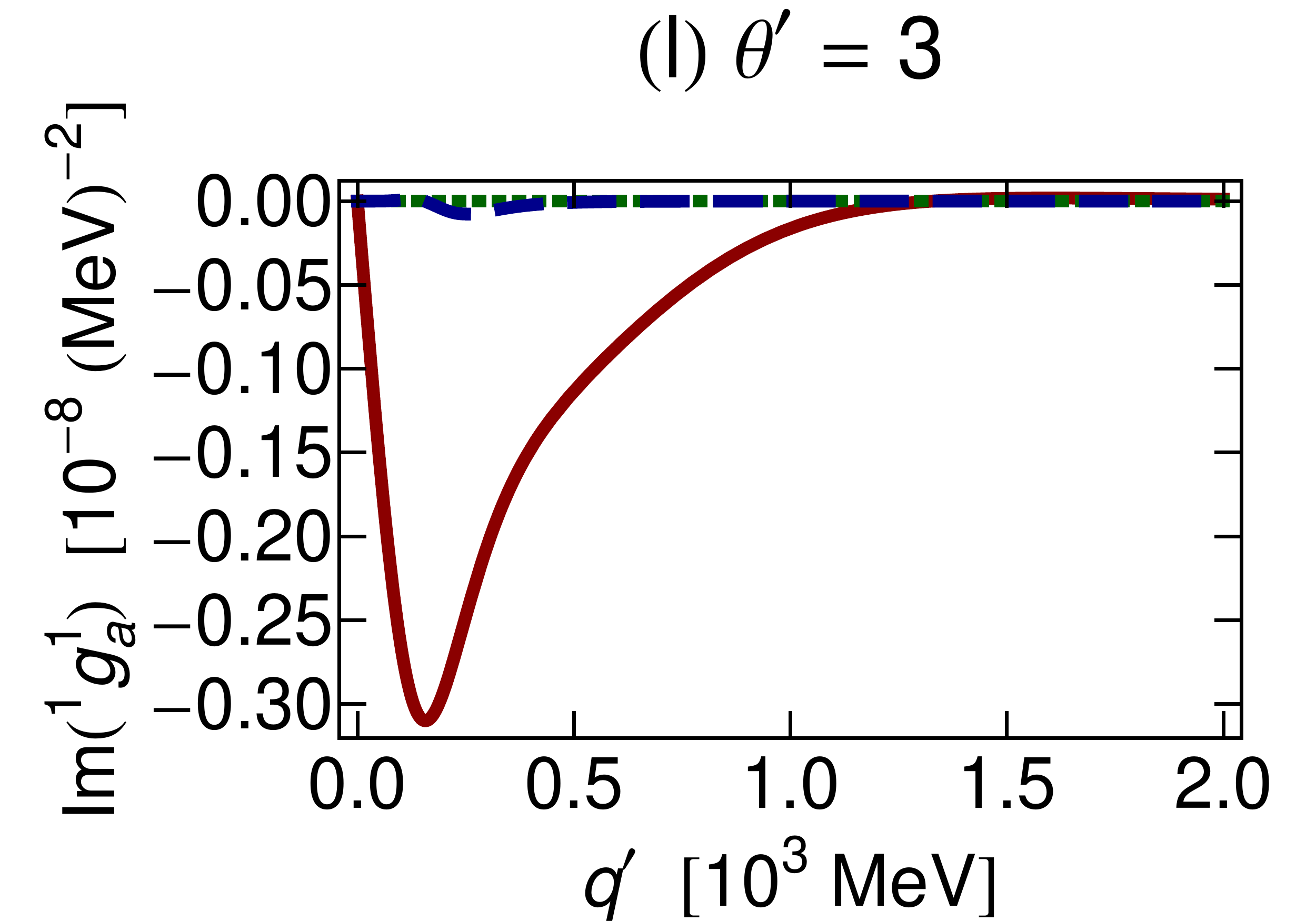}
}
\end{center}
\caption{(Color online) Same as Fig.~\ref{Fig:g_plots50_0} but for $\up{1}g^I_a$ and $\up{1}t^I_a$.}
\label{Fig:g_plots50_1}
\end{figure}

\begin{figure}[H]
\begin{center}
\subfloat{
\includegraphics[width=6cm]{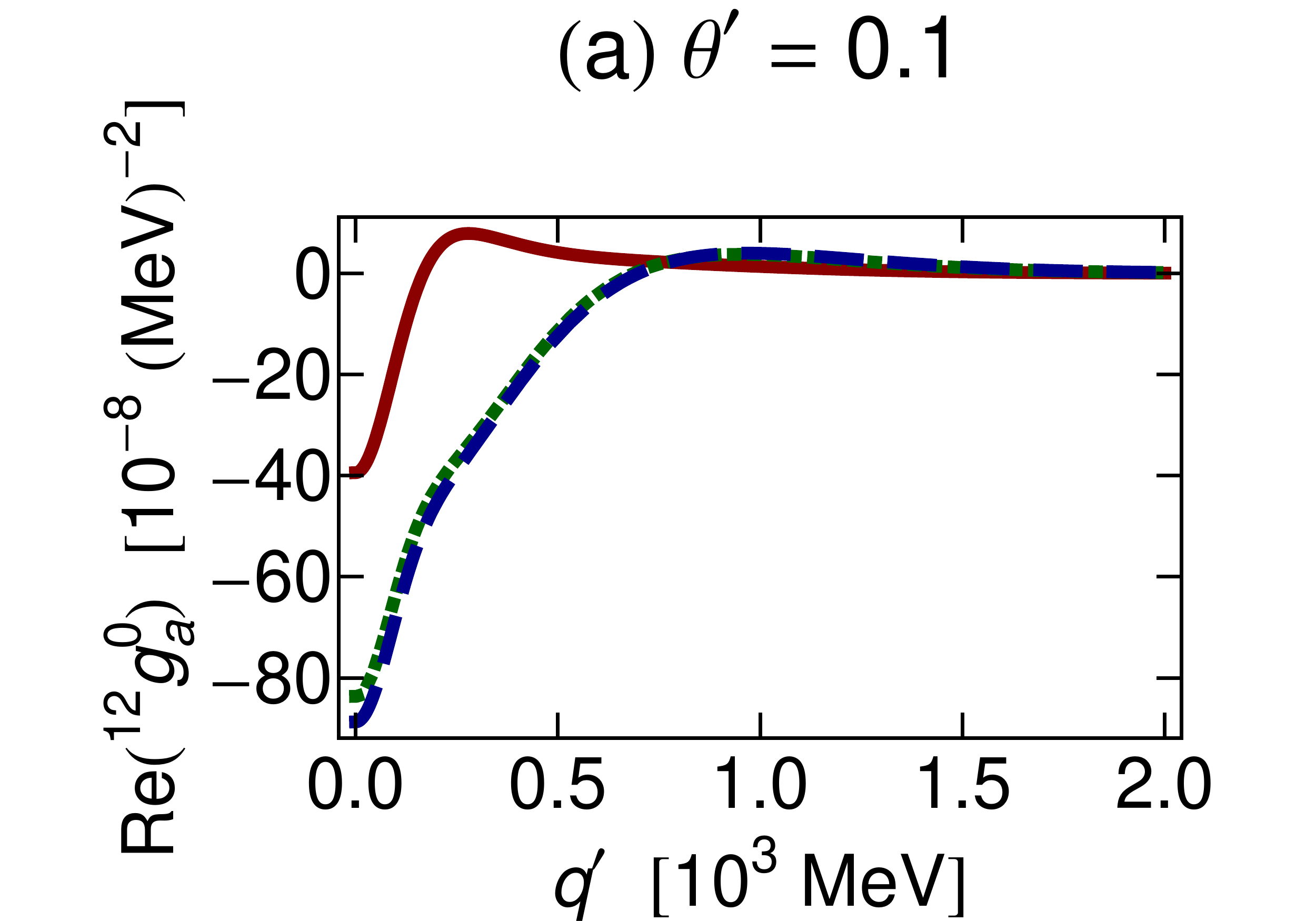}
}
\subfloat{
\includegraphics[width=6cm]{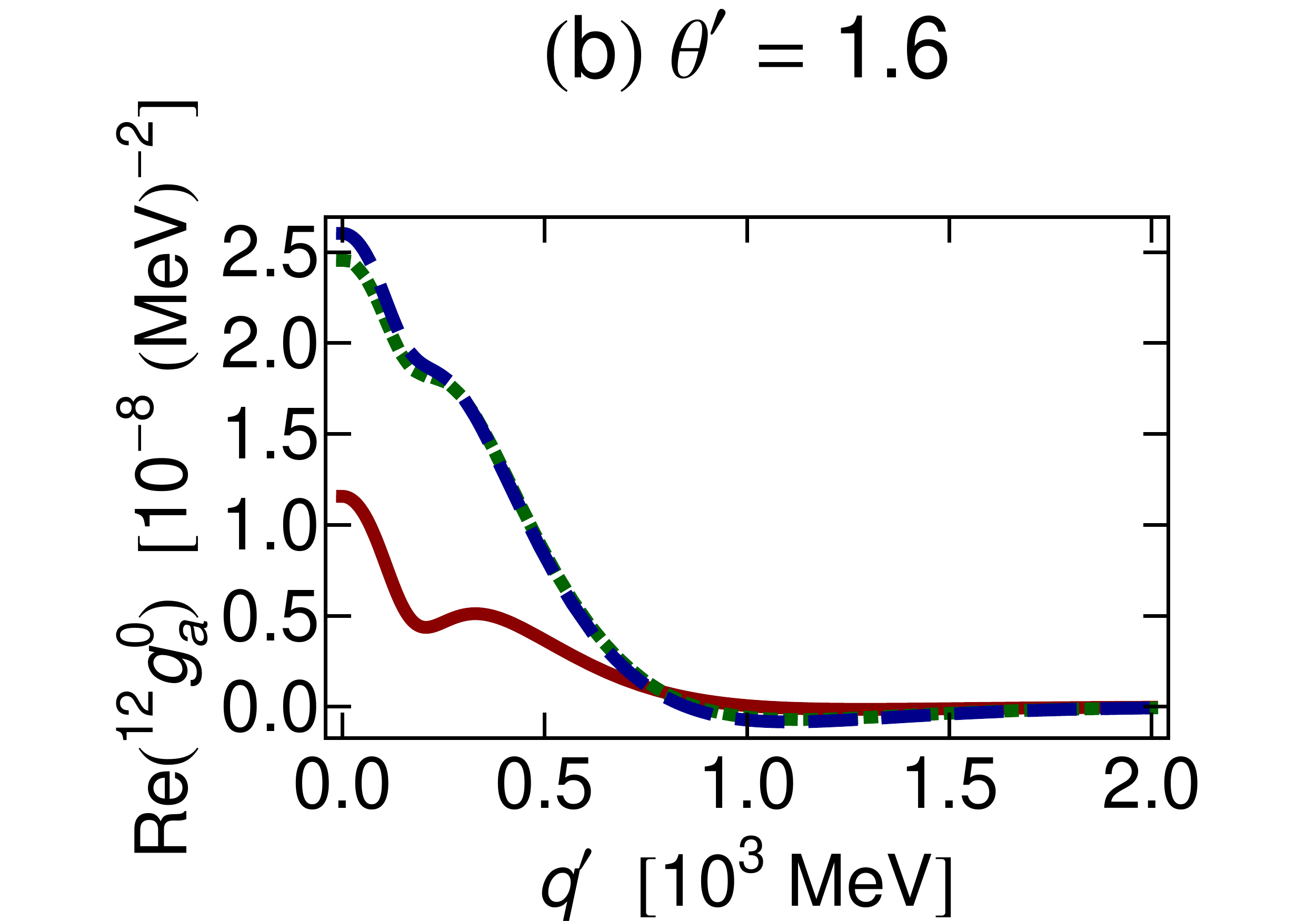}
}
\subfloat{
\includegraphics[width=6cm]{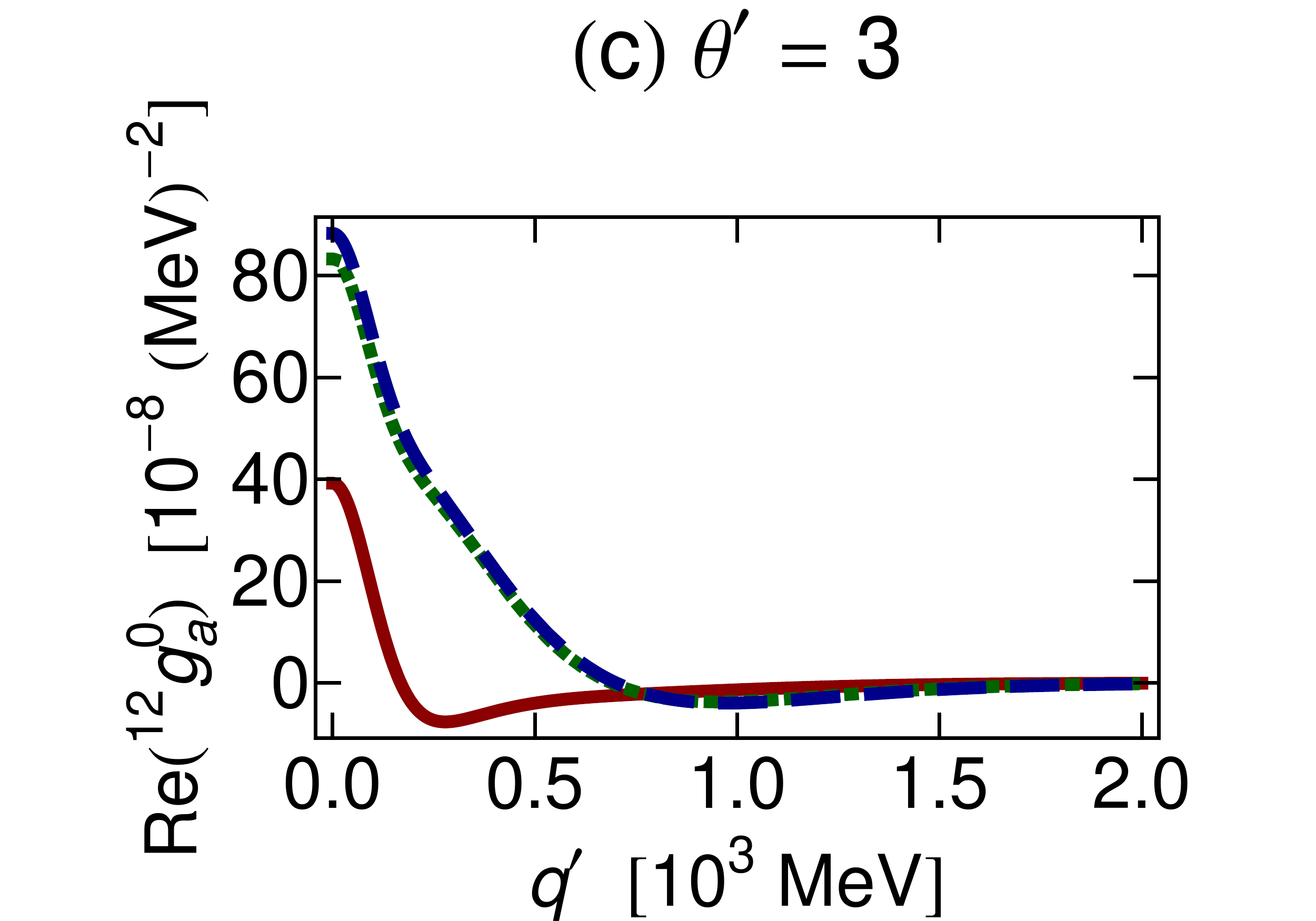}
}
\\
\subfloat{
\includegraphics[width=6cm]{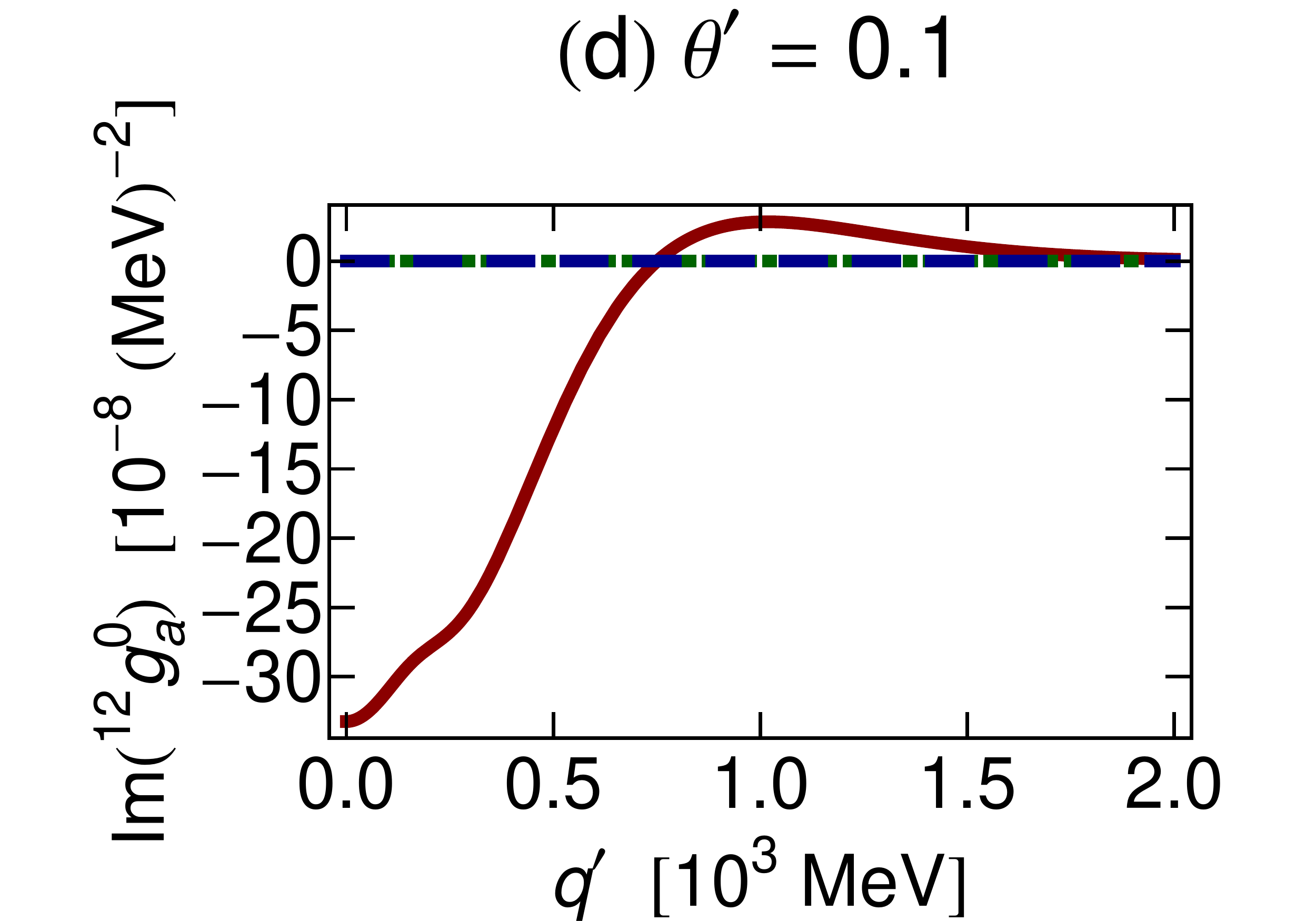}
}
\subfloat{
\includegraphics[width=6cm]{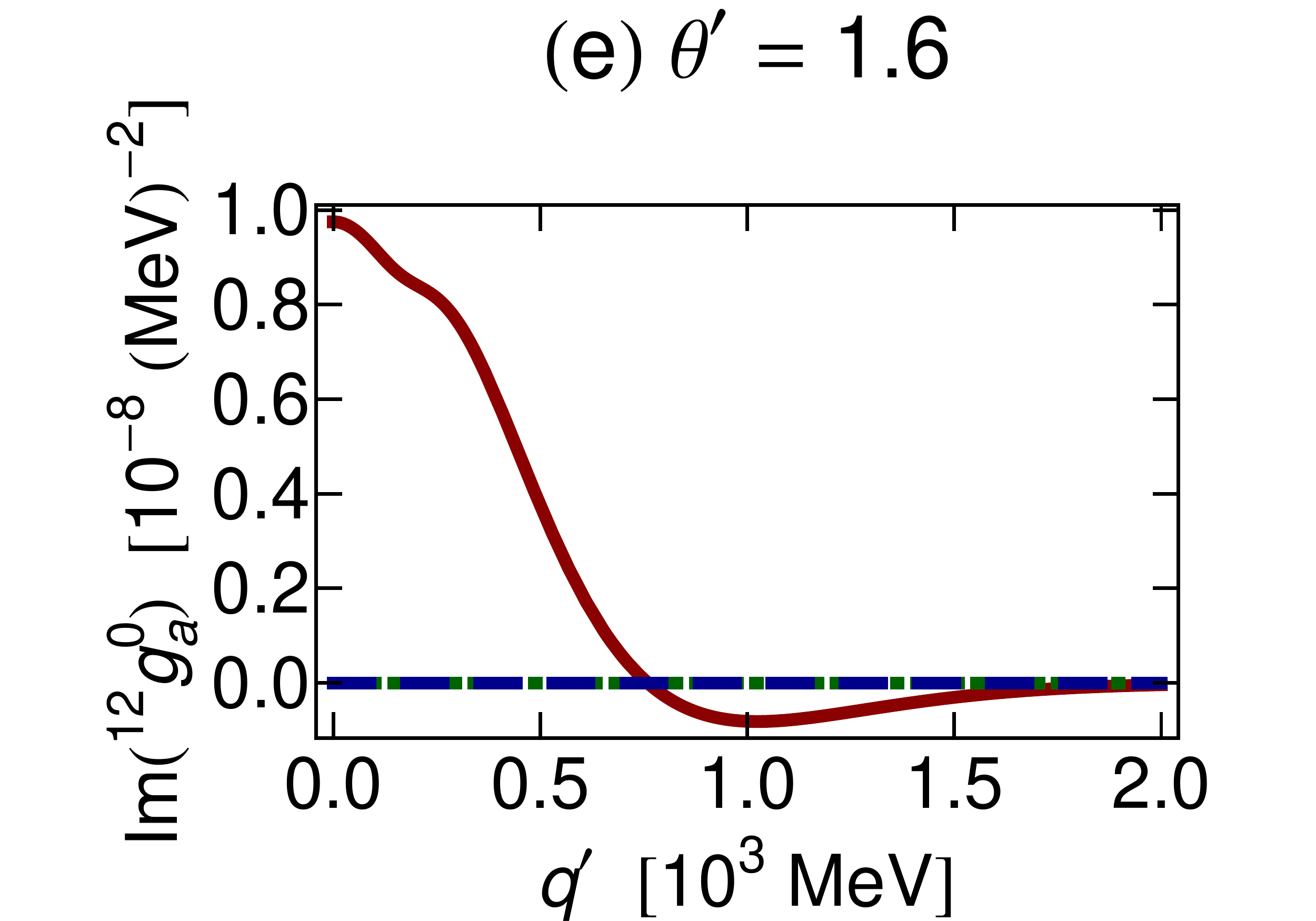}
}
\subfloat{
\includegraphics[width=6cm]{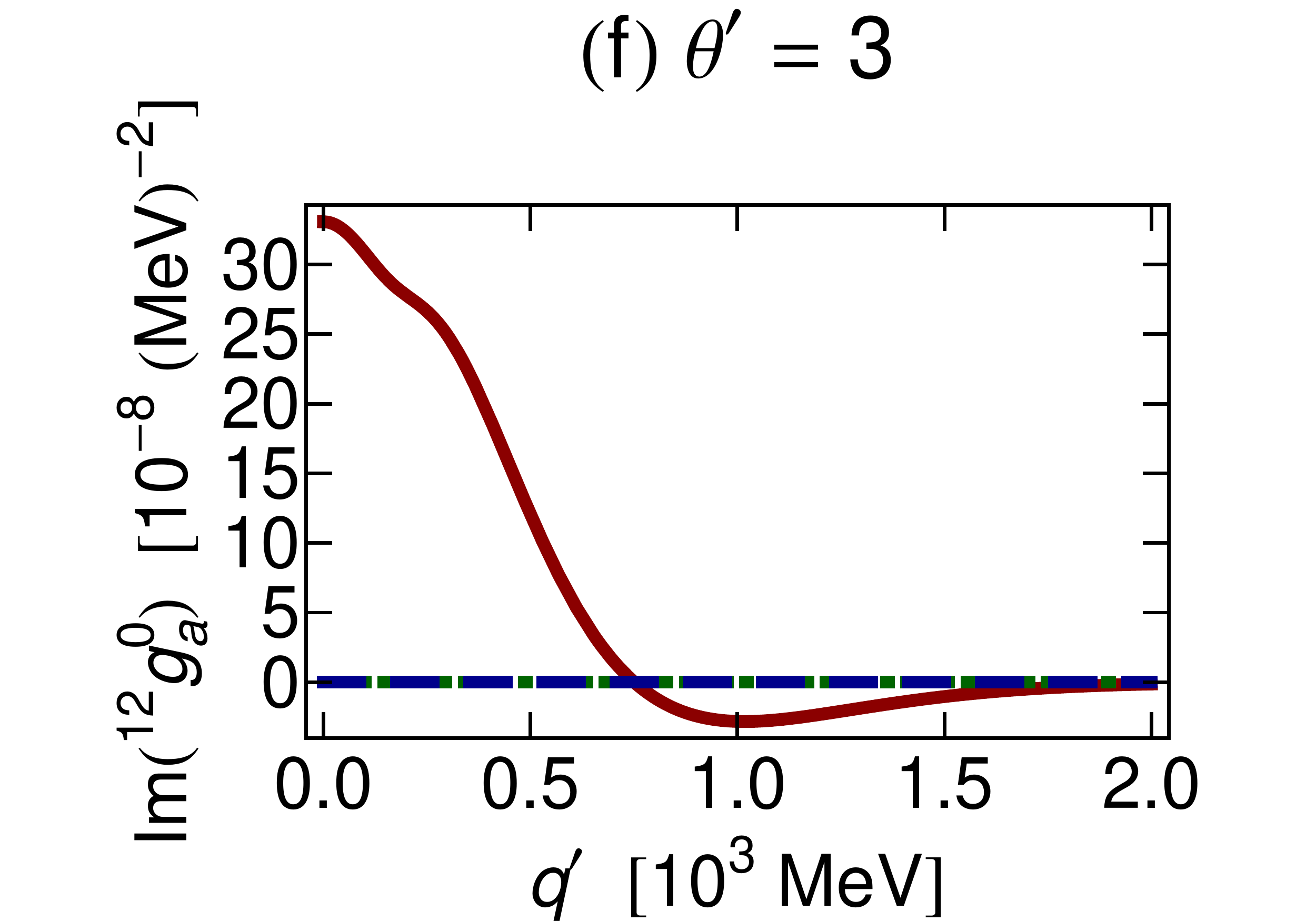}
}
\\
\subfloat{
\includegraphics[width=6cm]{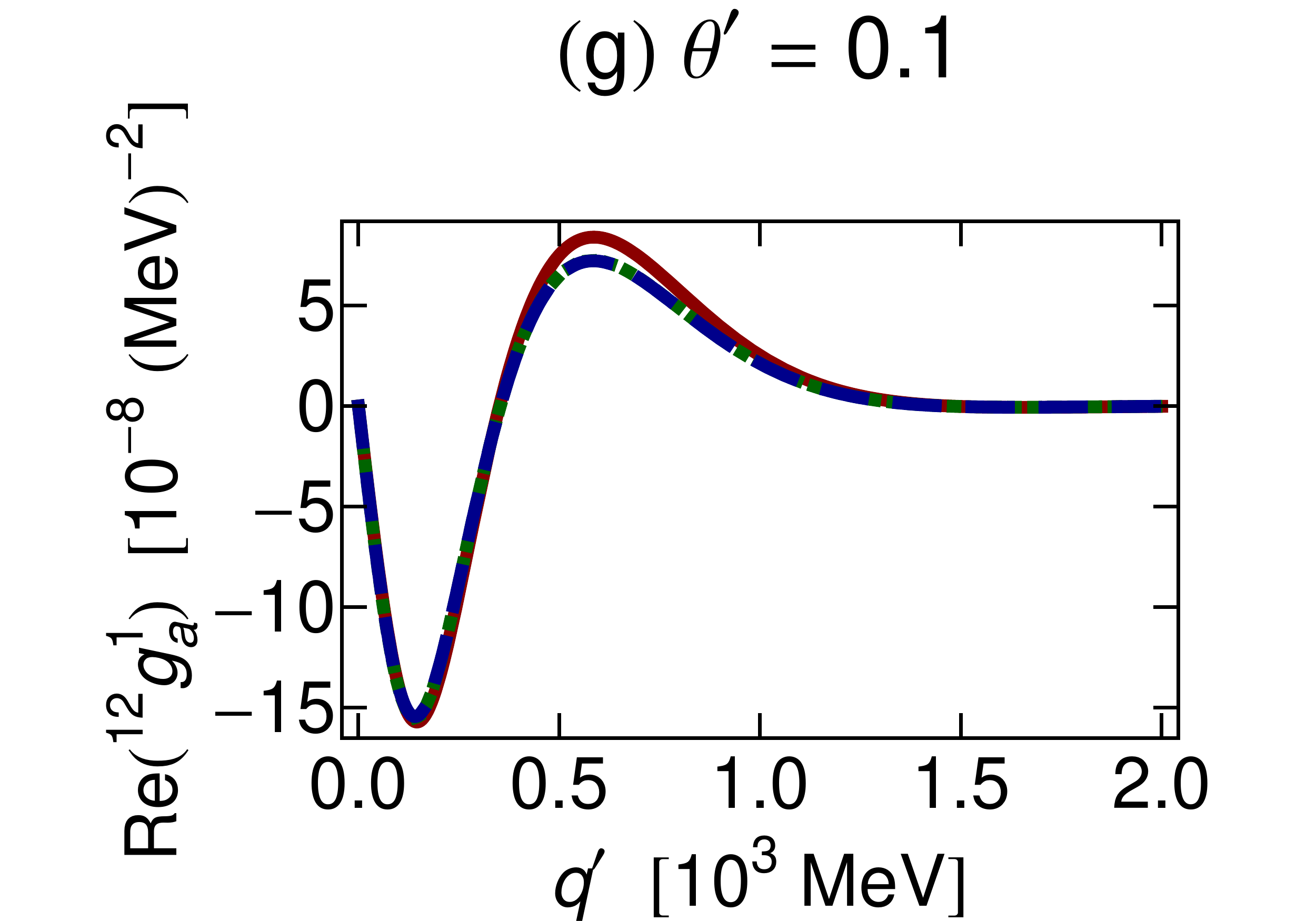}
}
\subfloat{
\includegraphics[width=6cm]{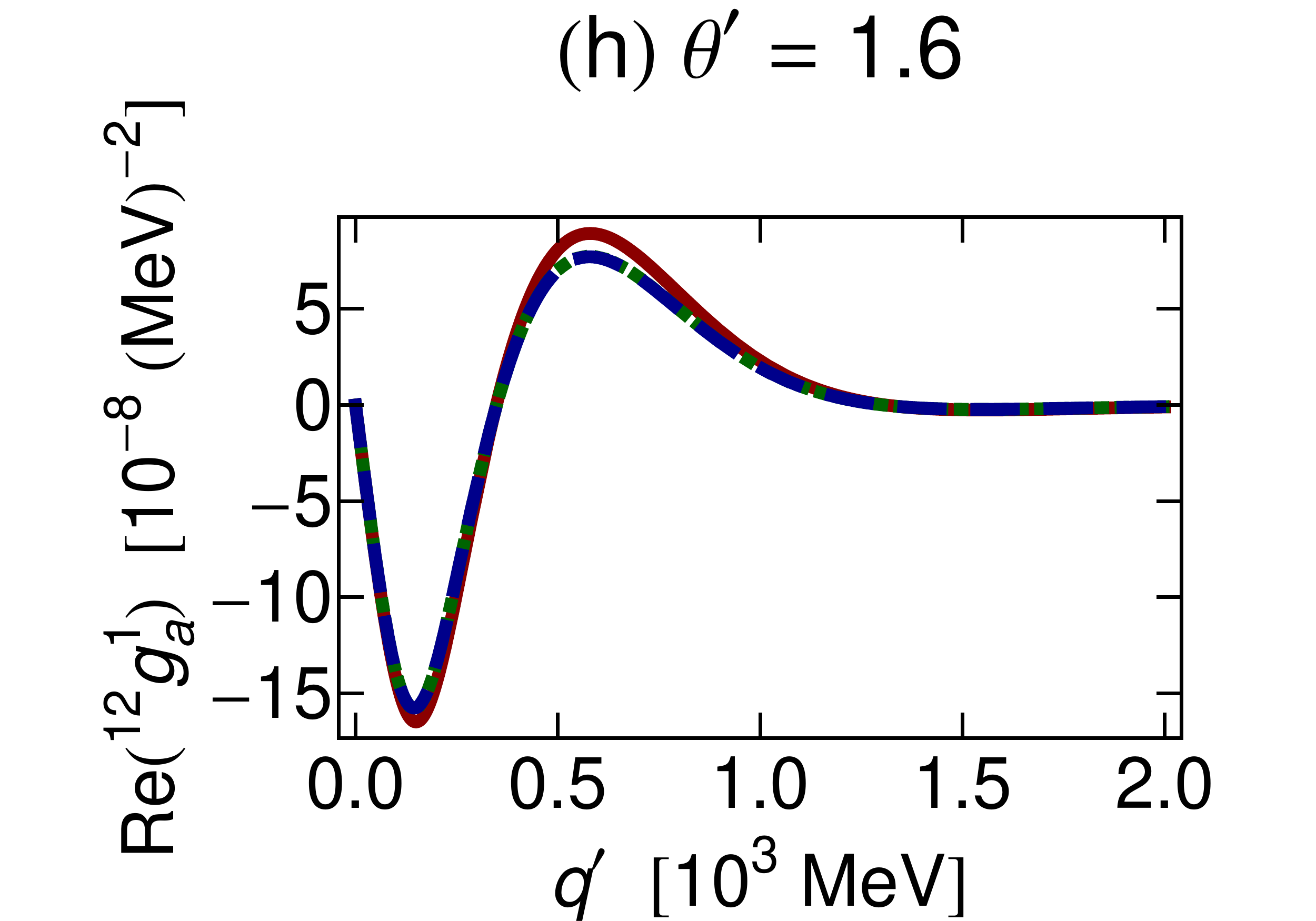}
}
\subfloat{
\includegraphics[width=6cm]{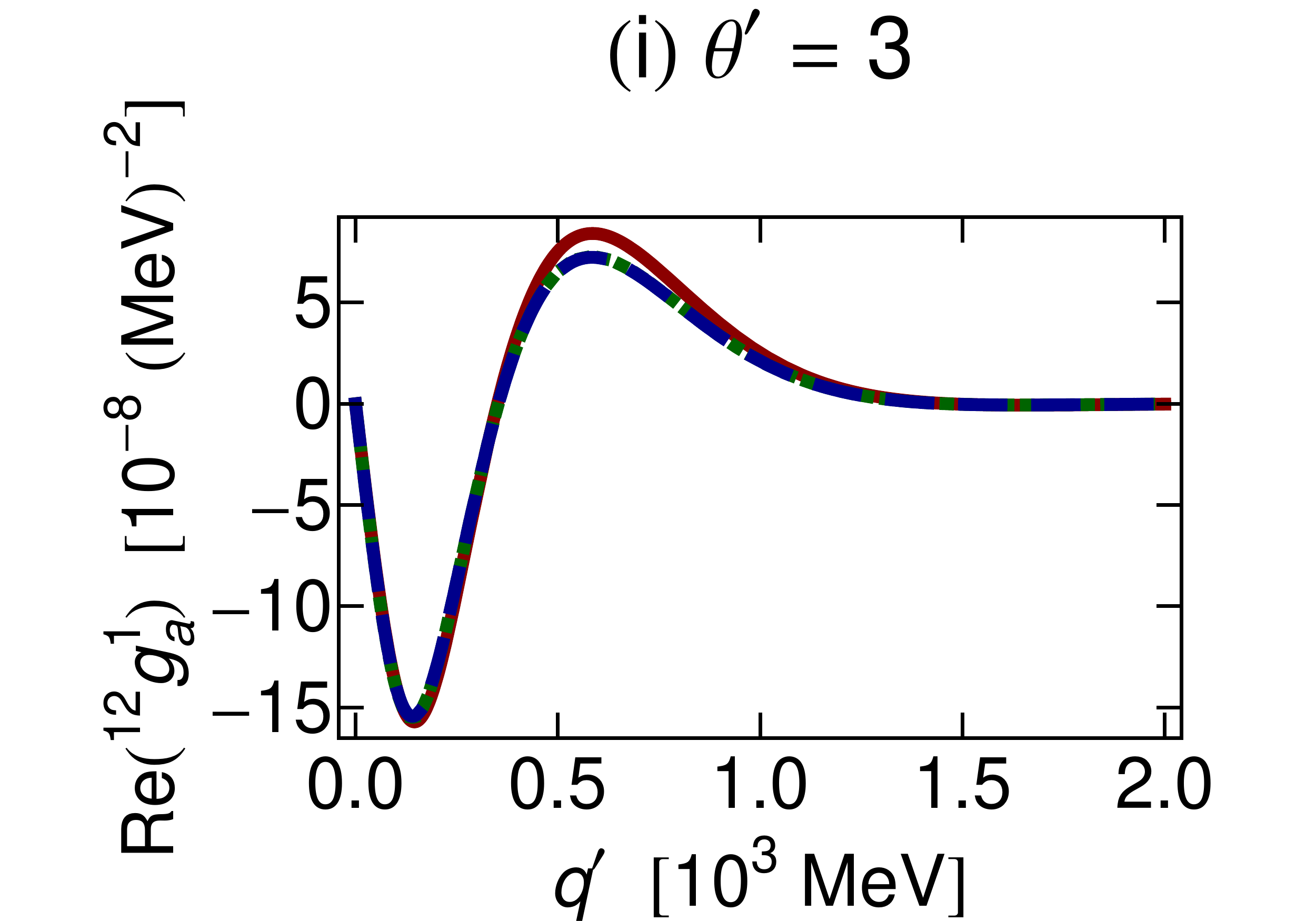}
}
\\
\subfloat{
\includegraphics[width=6cm]{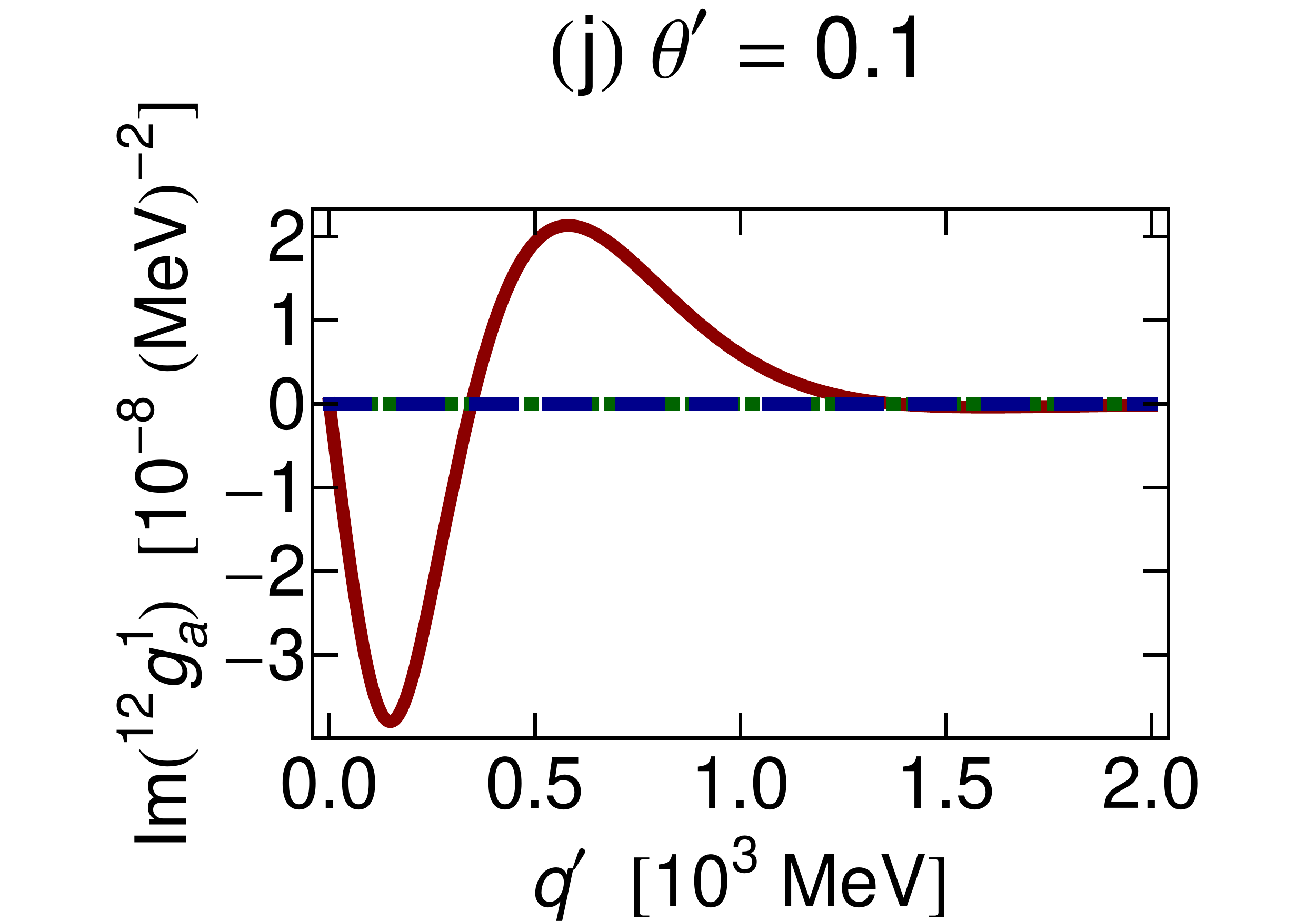}
}
\subfloat{
\includegraphics[width=6cm]{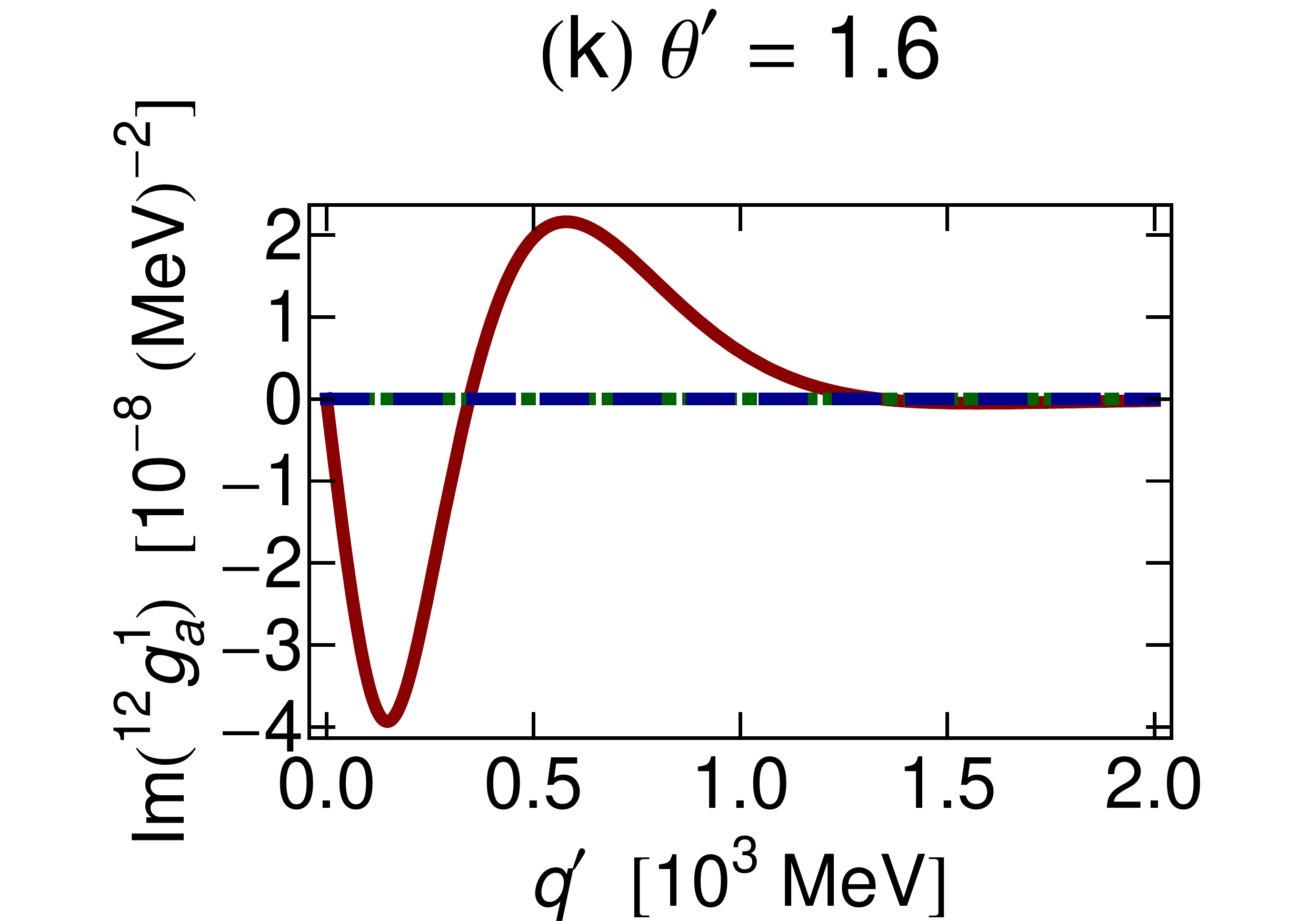}
}
\subfloat{
\includegraphics[width=6cm]{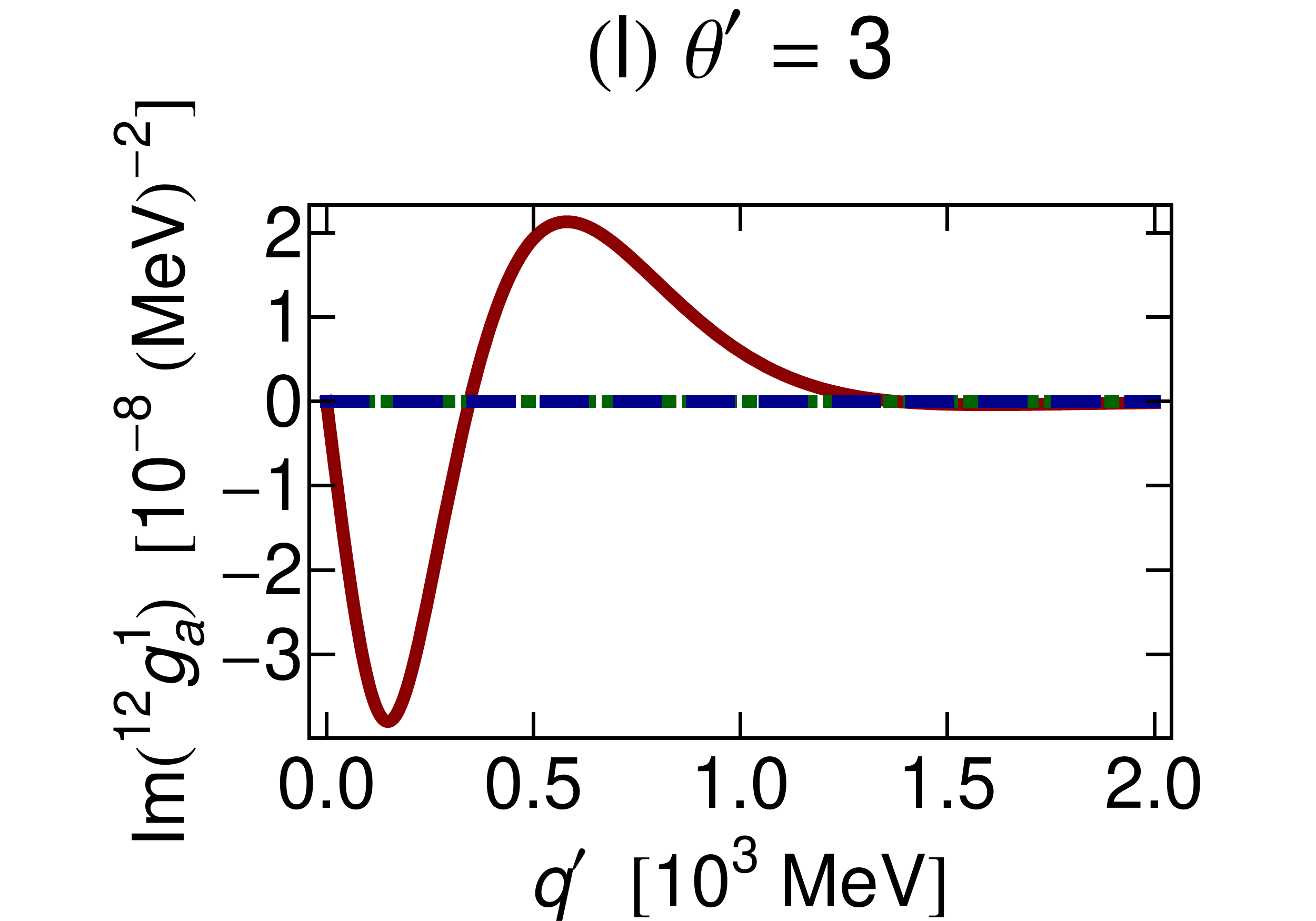}
}
\end{center}
\caption{(Color online) Same as Fig.~\ref{Fig:g_plots50_0} but for $\up{12}g^I_a$ and $\up{12}t^I_a$.}
\label{Fig:g_plots50_12}
\end{figure}

\begin{figure}[H]
\begin{center}
\subfloat{
\includegraphics[width=6cm]{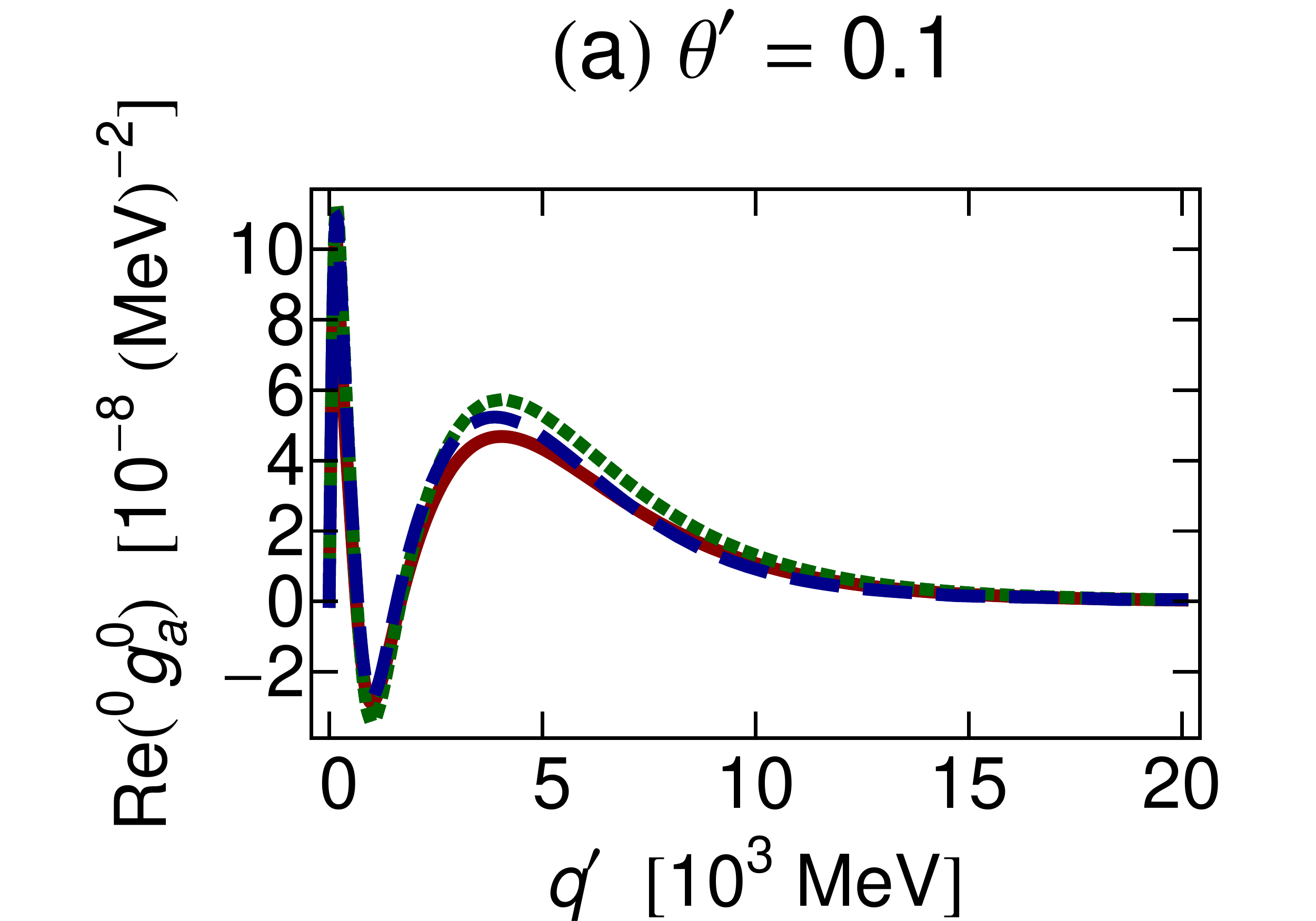}
}
\subfloat{
\includegraphics[width=6cm]{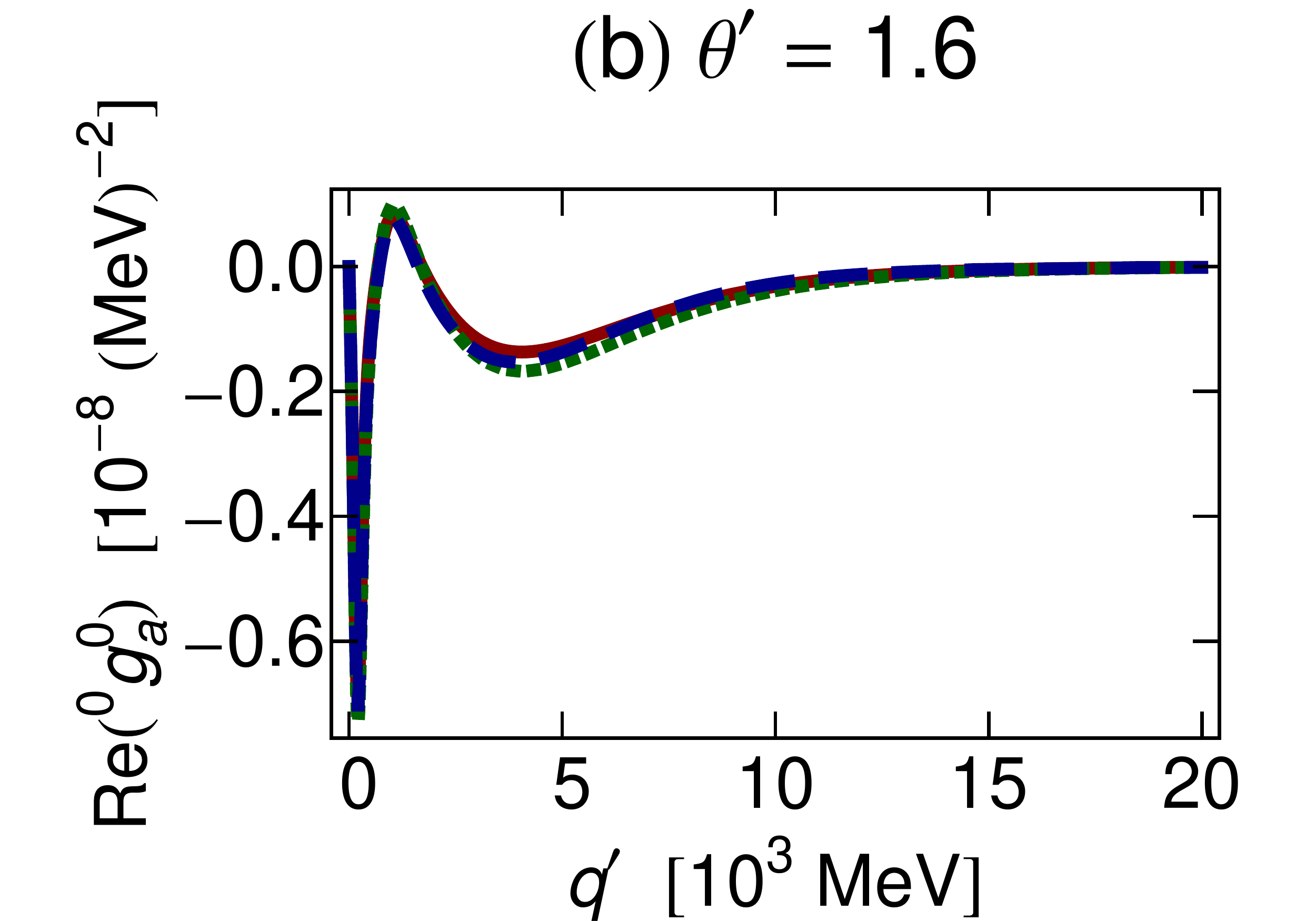}
}
\subfloat{
\includegraphics[width=6cm]{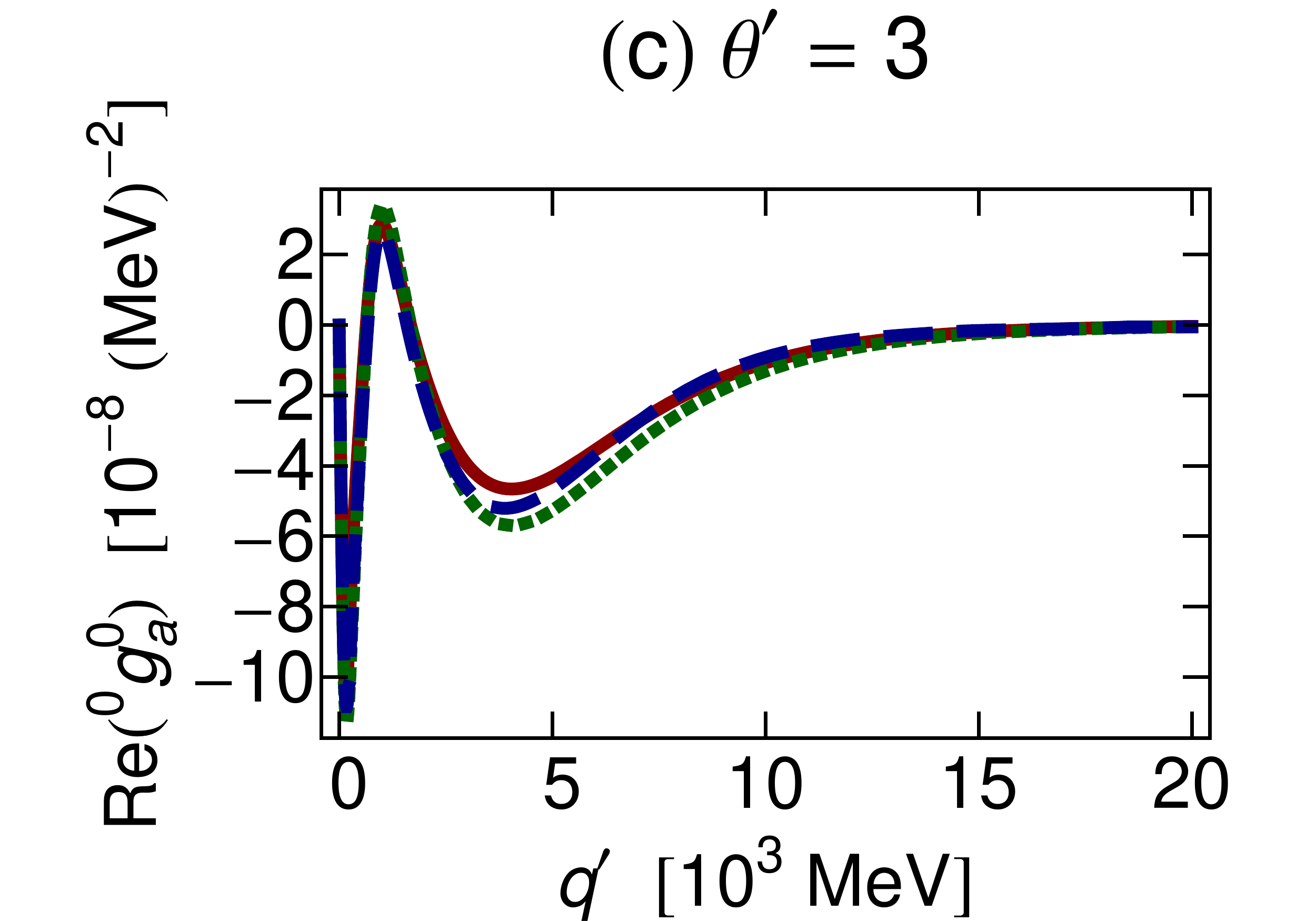}
}
\\
\subfloat{
\includegraphics[width=6cm]{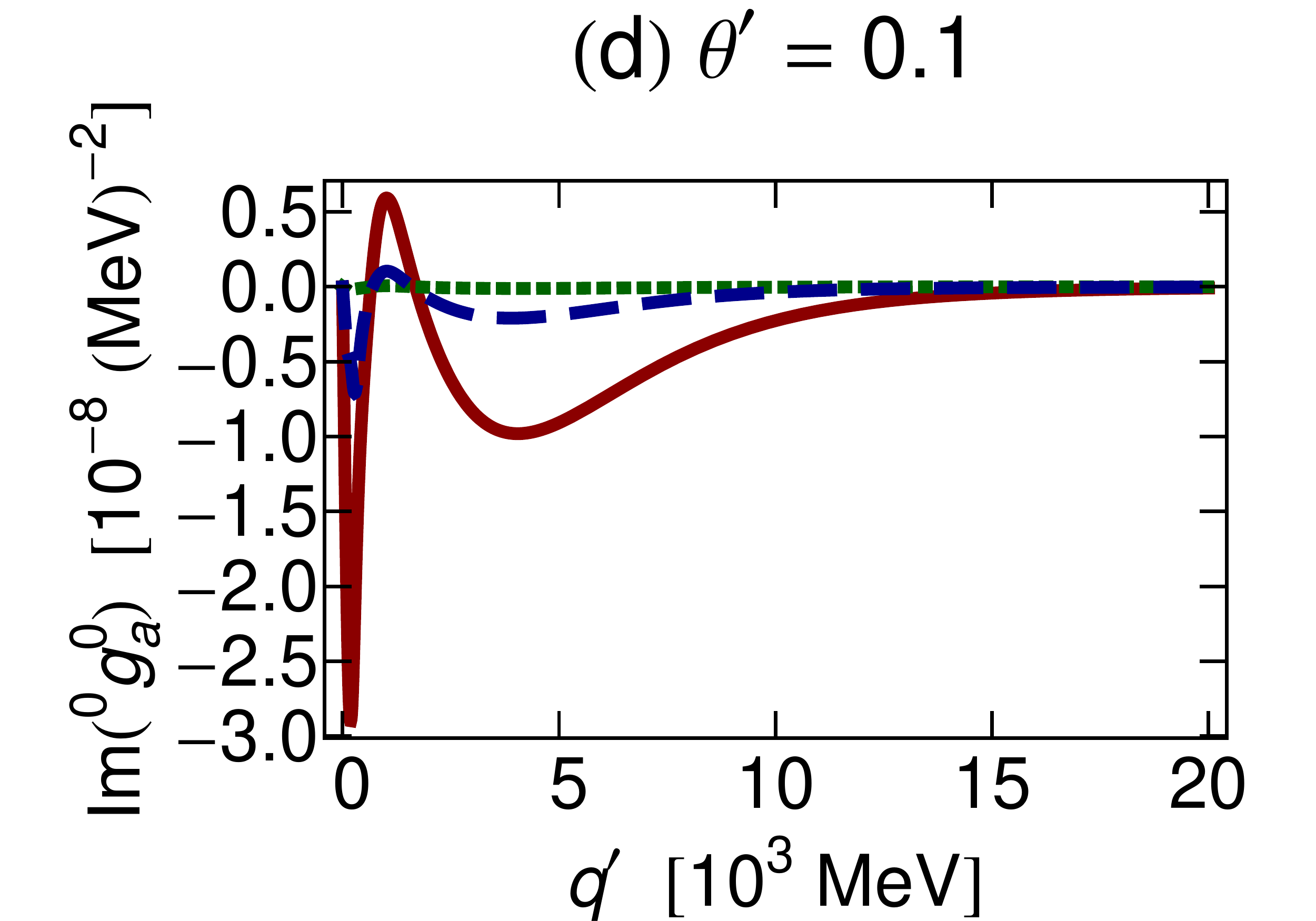}
}
\subfloat{
\includegraphics[width=6cm]{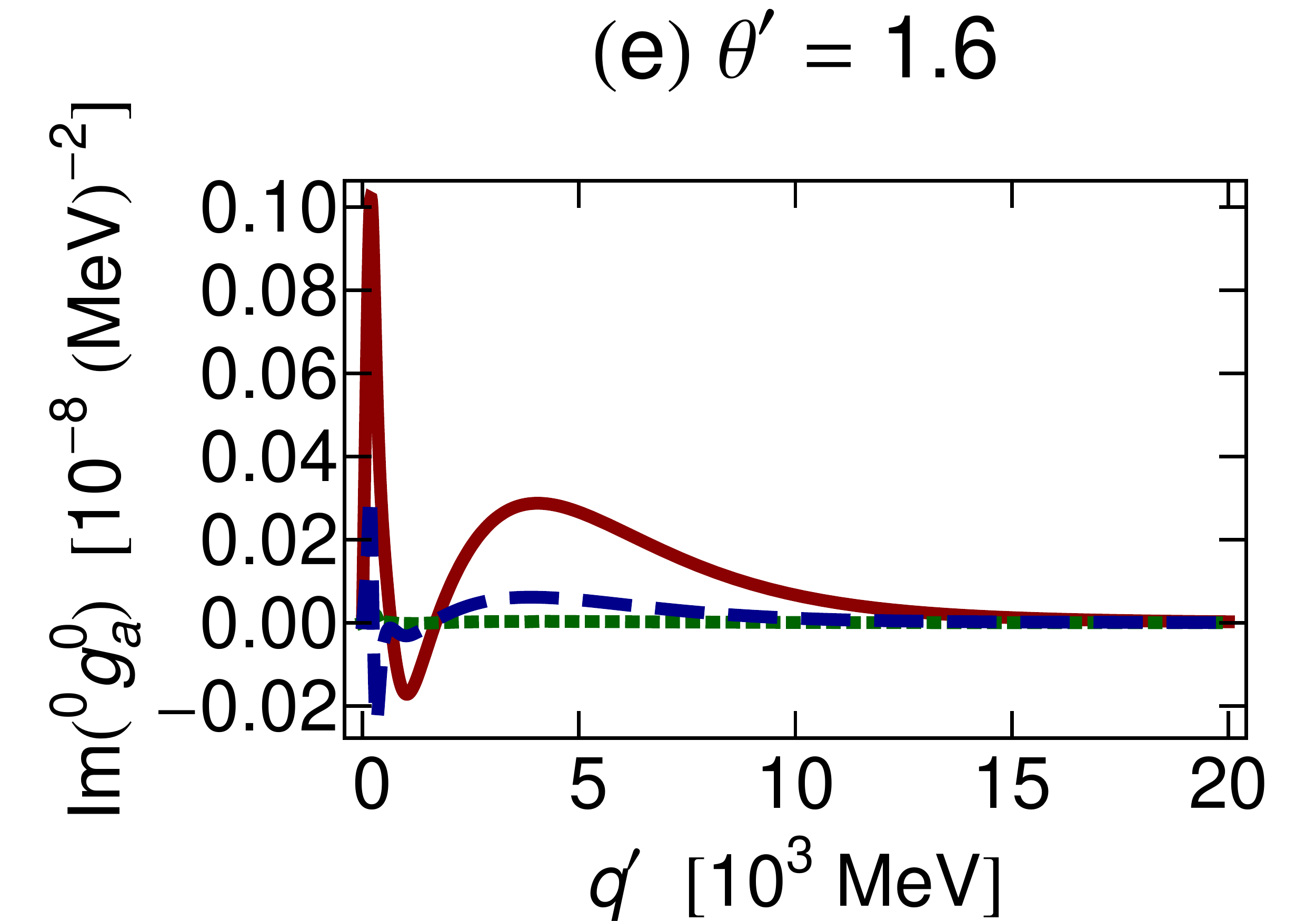}
}
\subfloat{
\includegraphics[width=6cm]{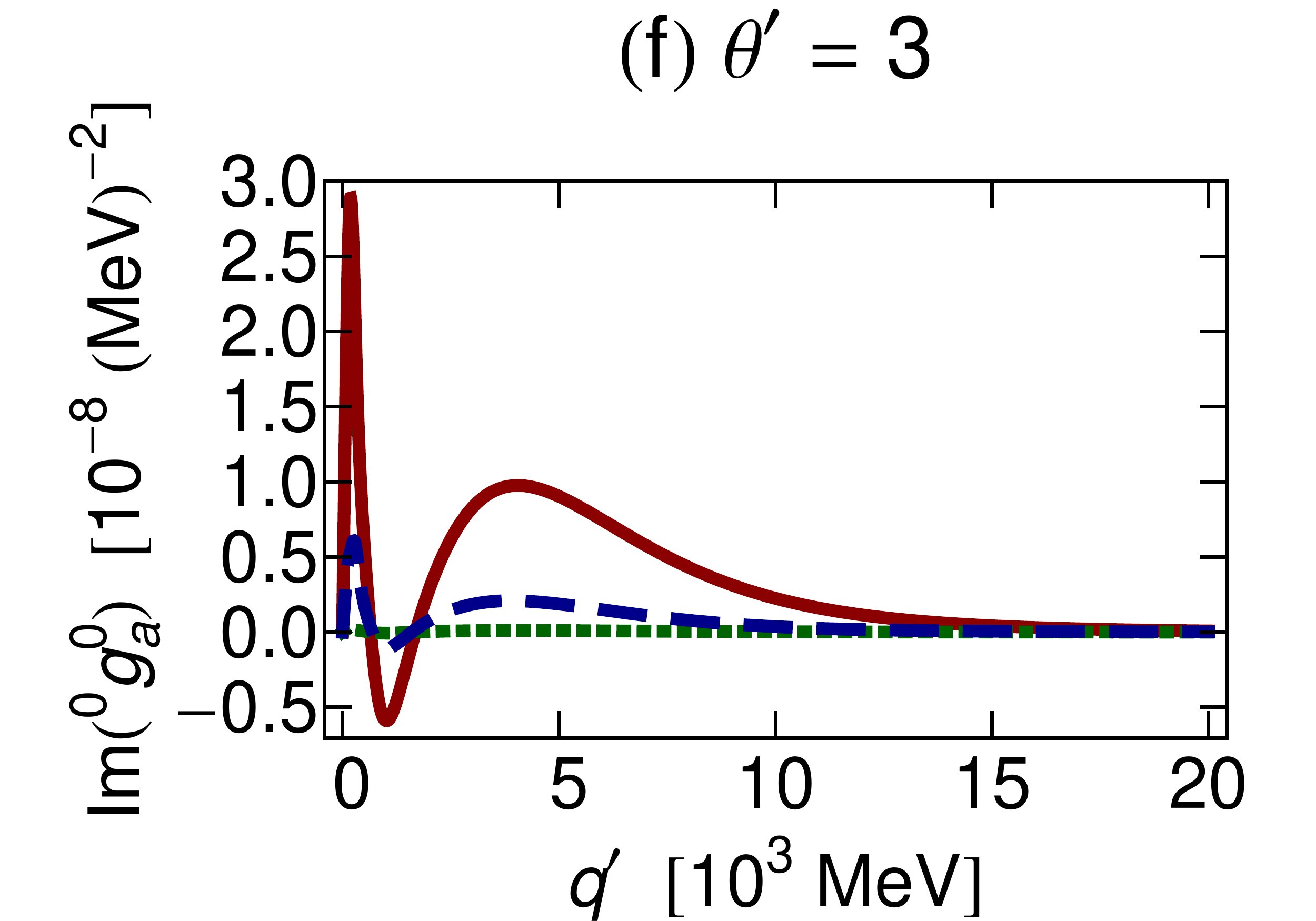}
}
\\
\subfloat{
\includegraphics[width=6cm]{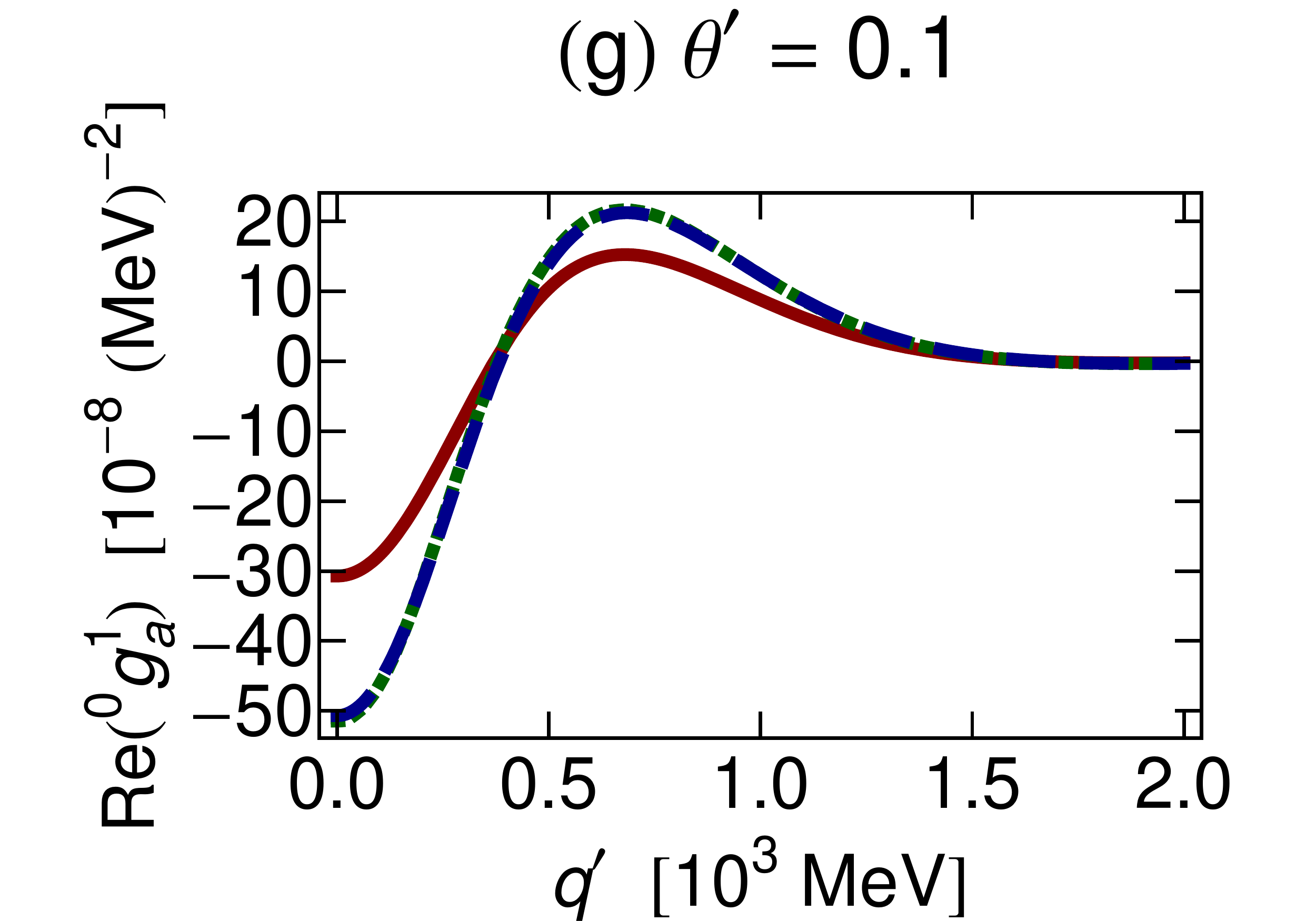}
}
\subfloat{
\includegraphics[width=6cm]{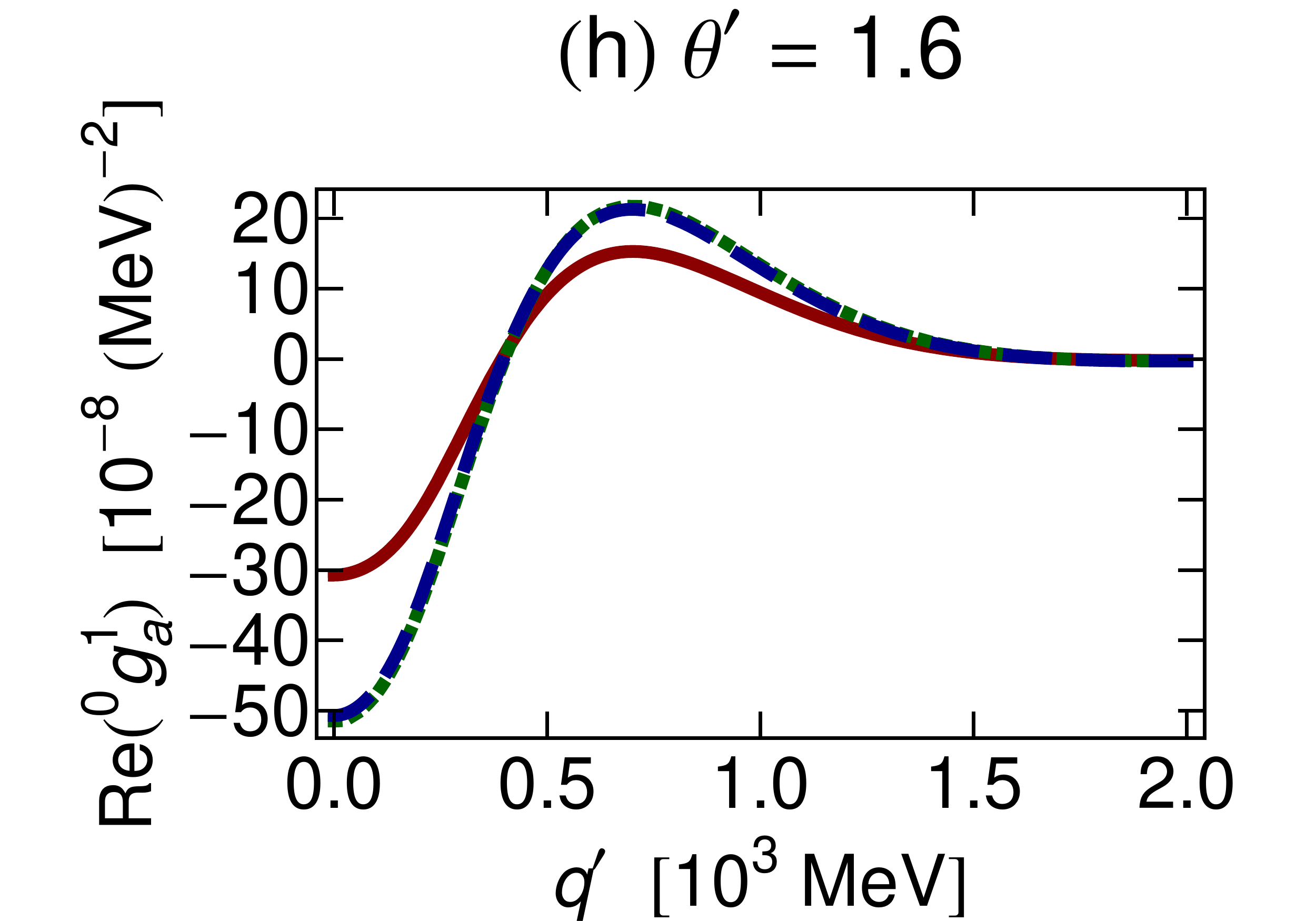}
}
\subfloat{
\includegraphics[width=6cm]{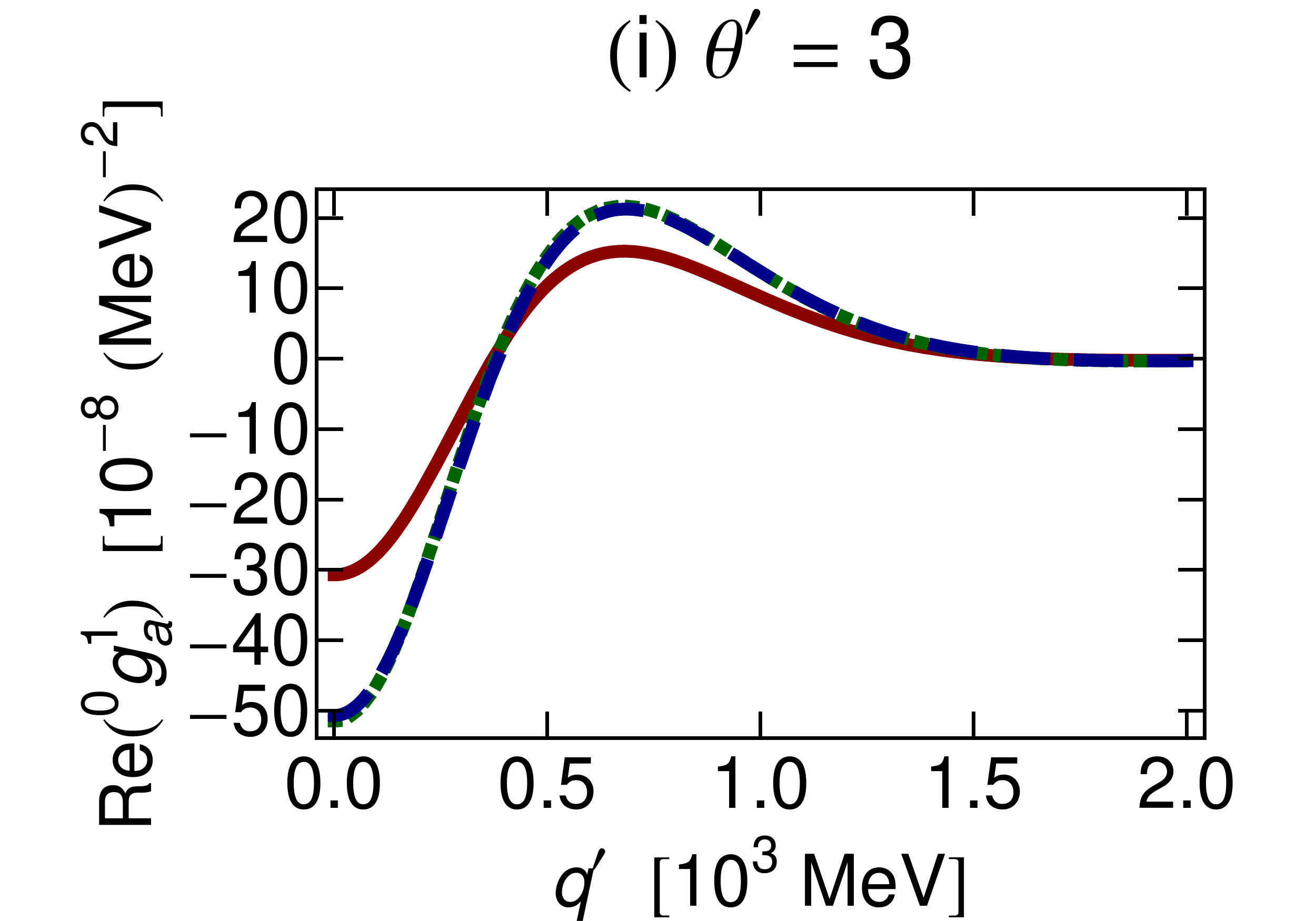}
}
\\
\subfloat{
\includegraphics[width=6cm]{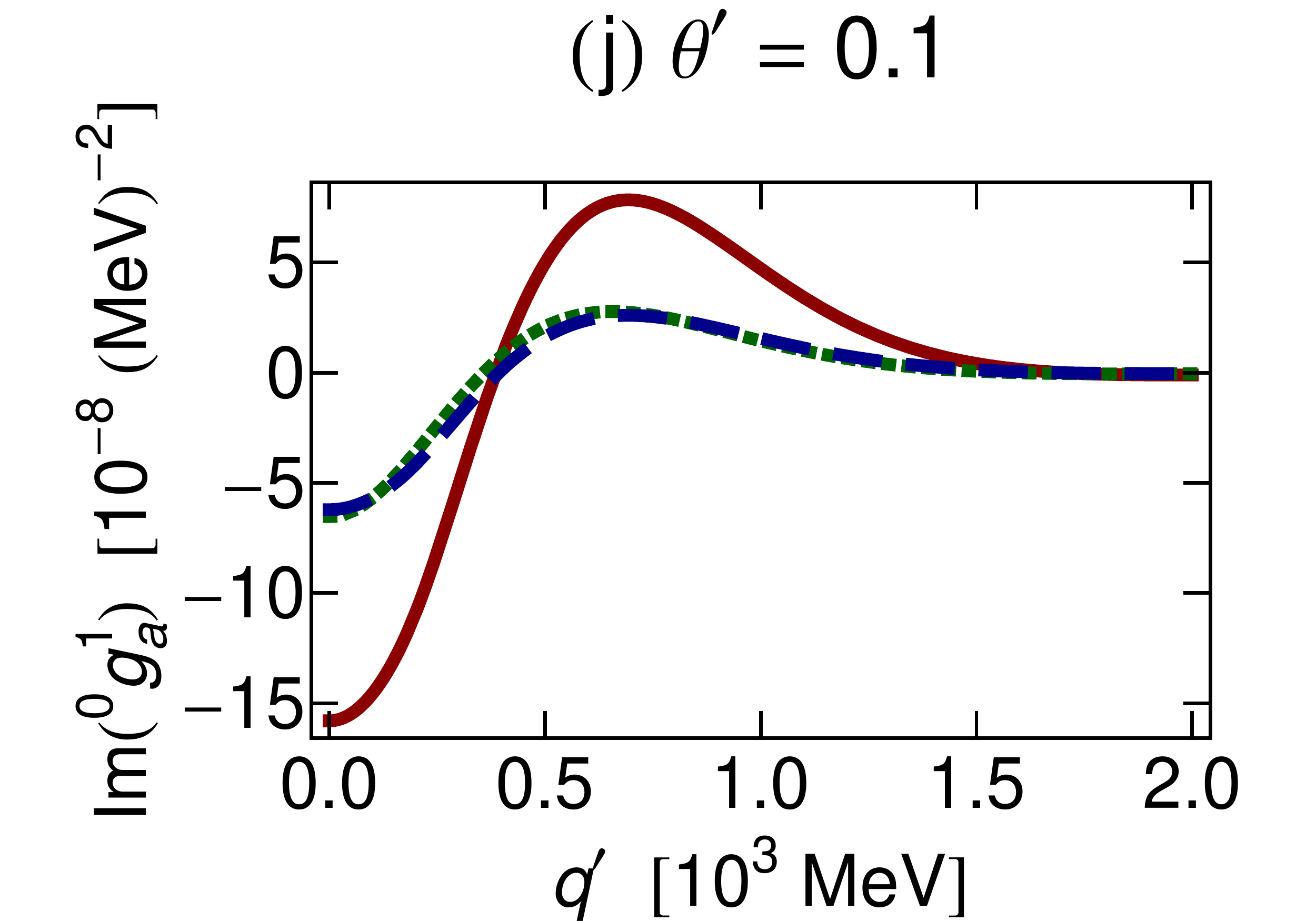}
}
\subfloat{
\includegraphics[width=6cm]{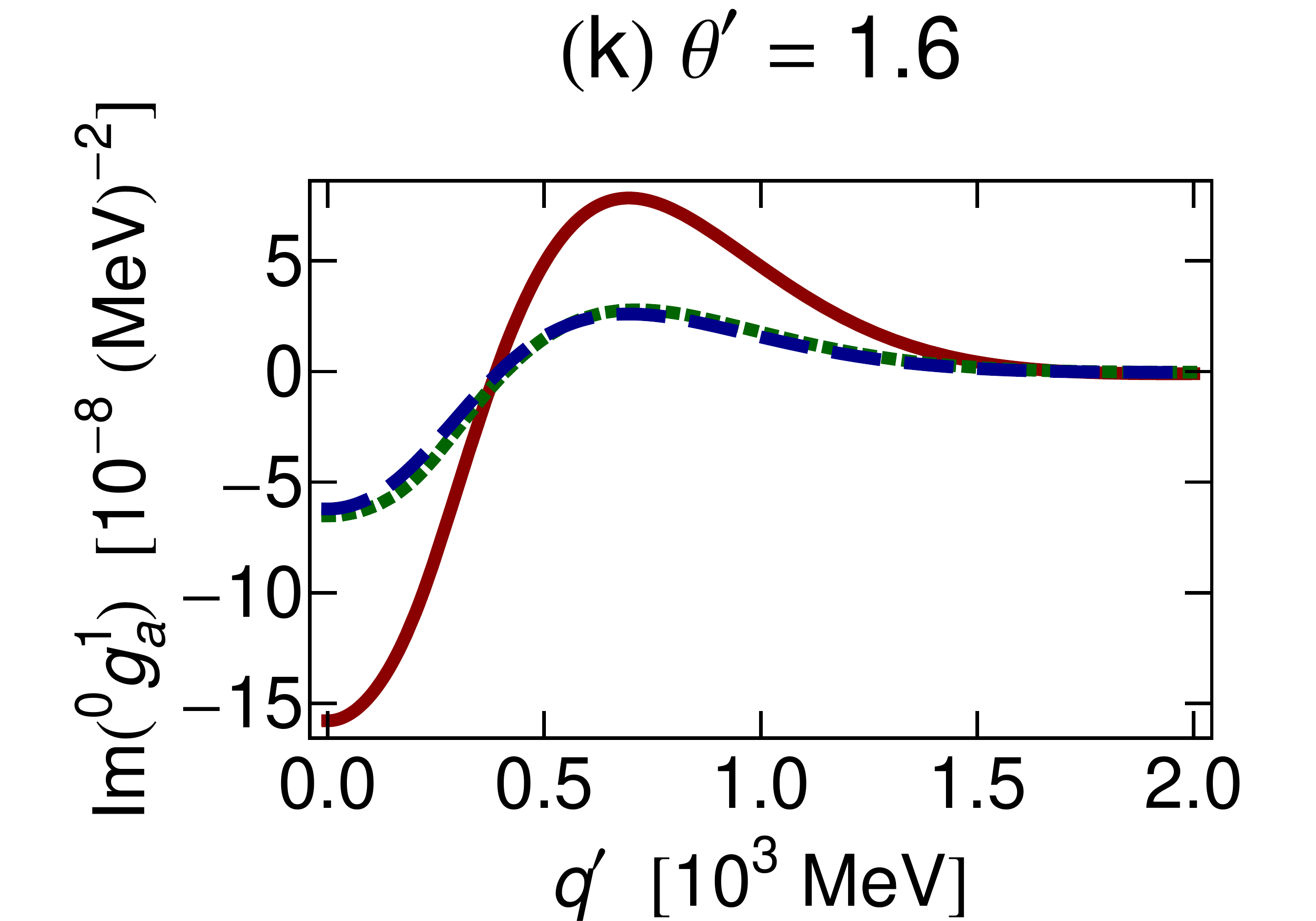}
}
\subfloat{
\includegraphics[width=6cm]{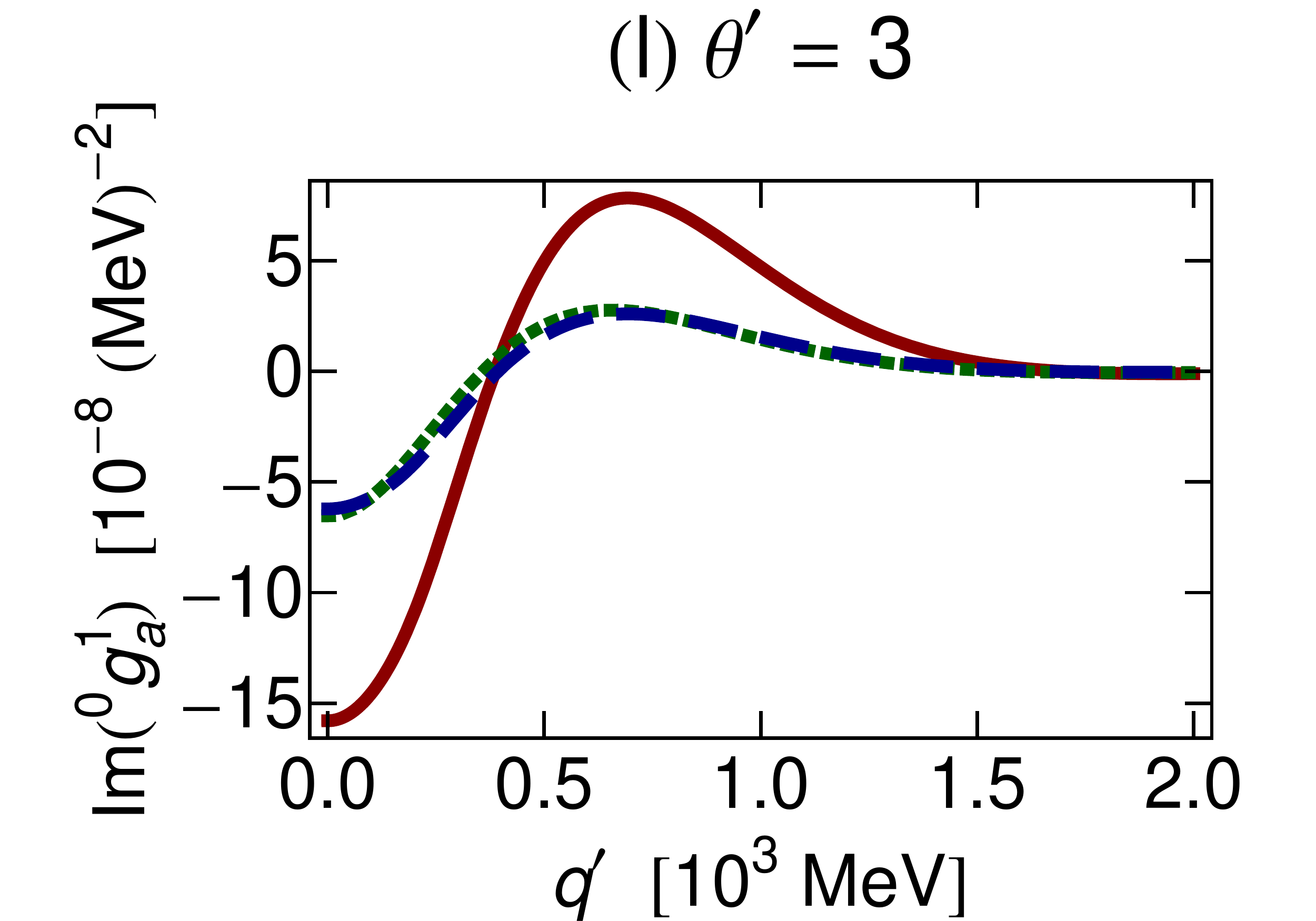}
}
\end{center}
\caption{(Color online) Same as Fig.~\ref{Fig:g_plots50_0} but at $q=216.67 \units{MeV}$.}
\label{Fig:g_plots100_0}
\end{figure}

\begin{figure}[H]
\begin{center}
\subfloat{
\includegraphics[width=6cm]{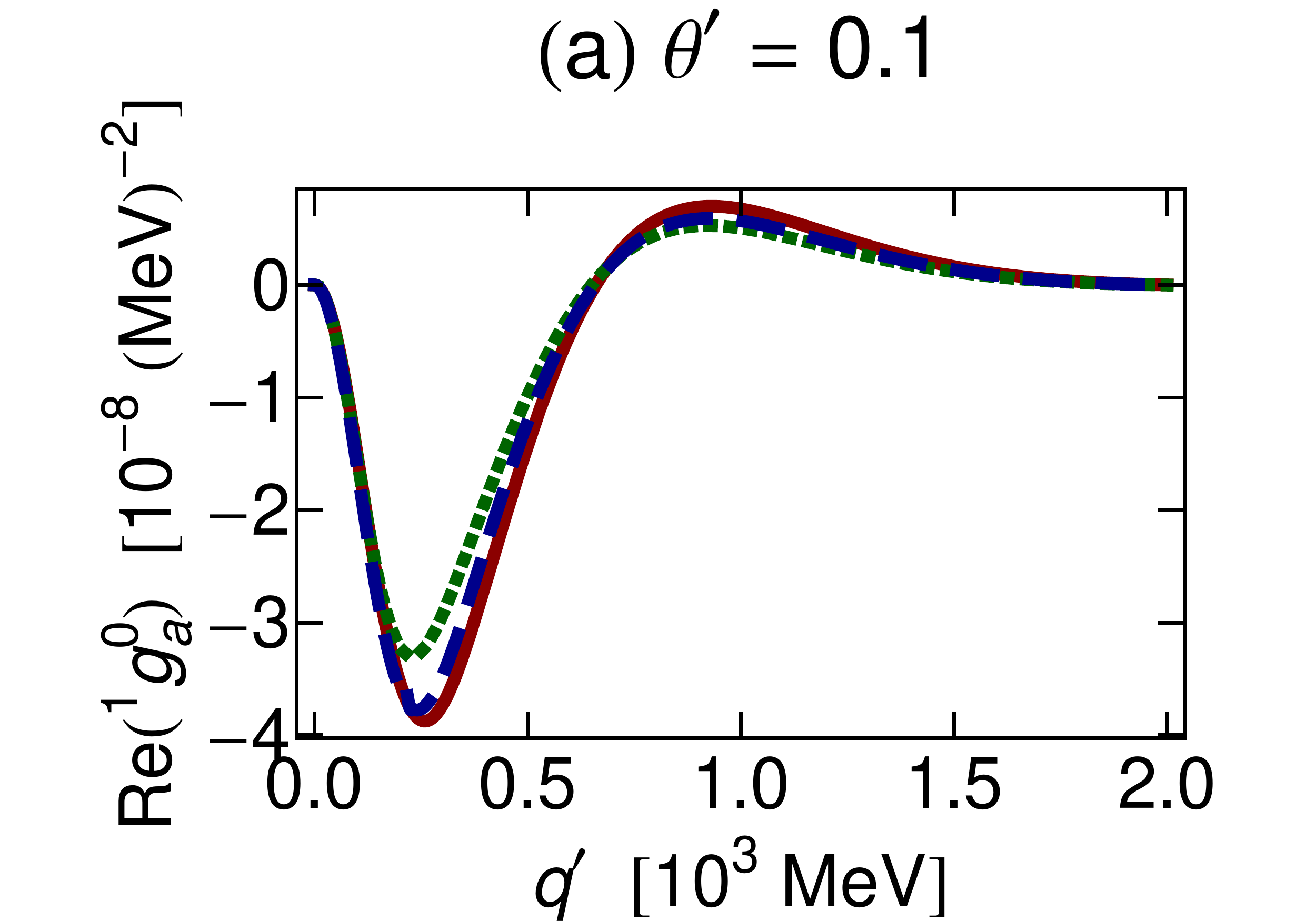}
}
\subfloat{
\includegraphics[width=6cm]{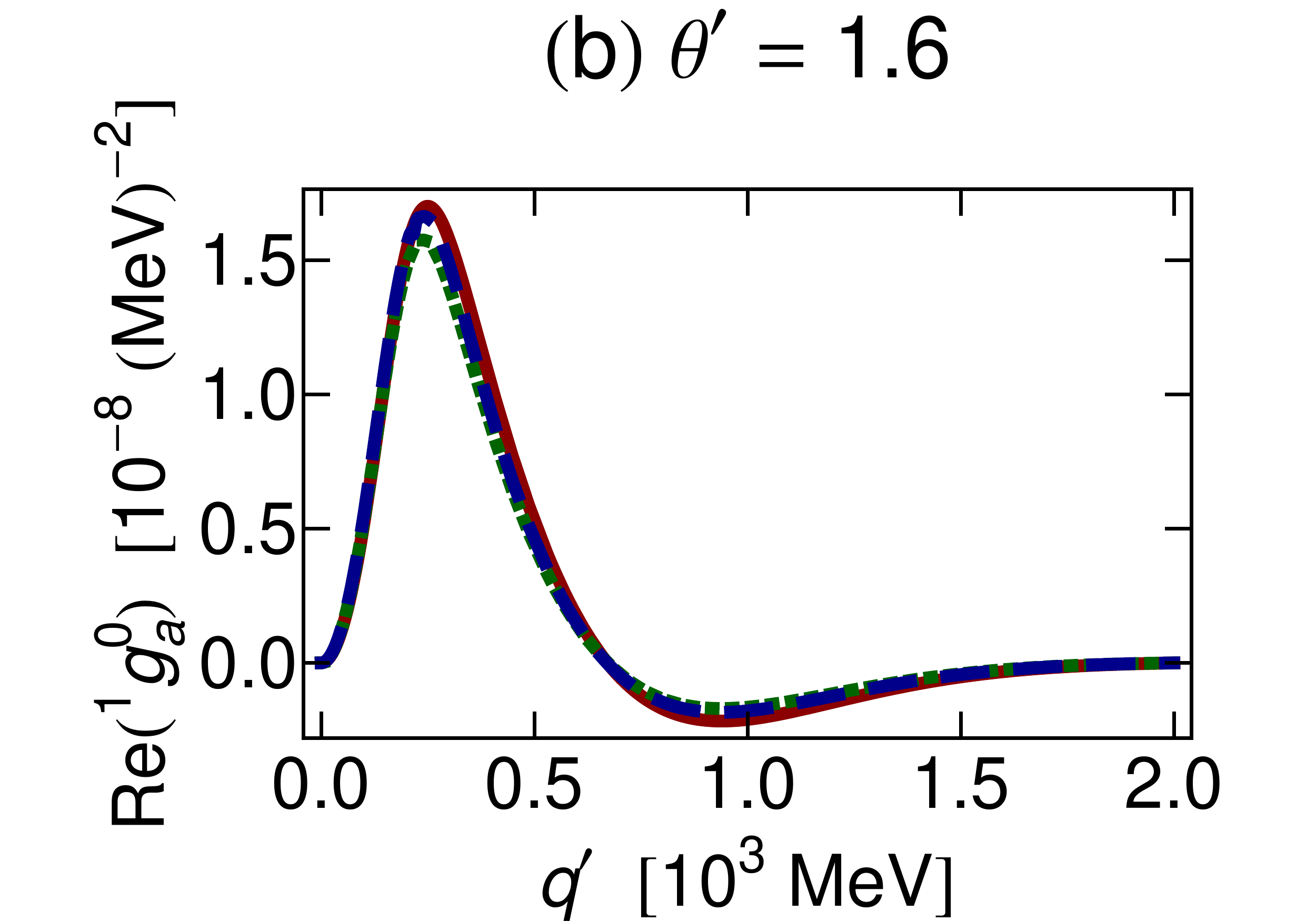}
}
\subfloat{
\includegraphics[width=6cm]{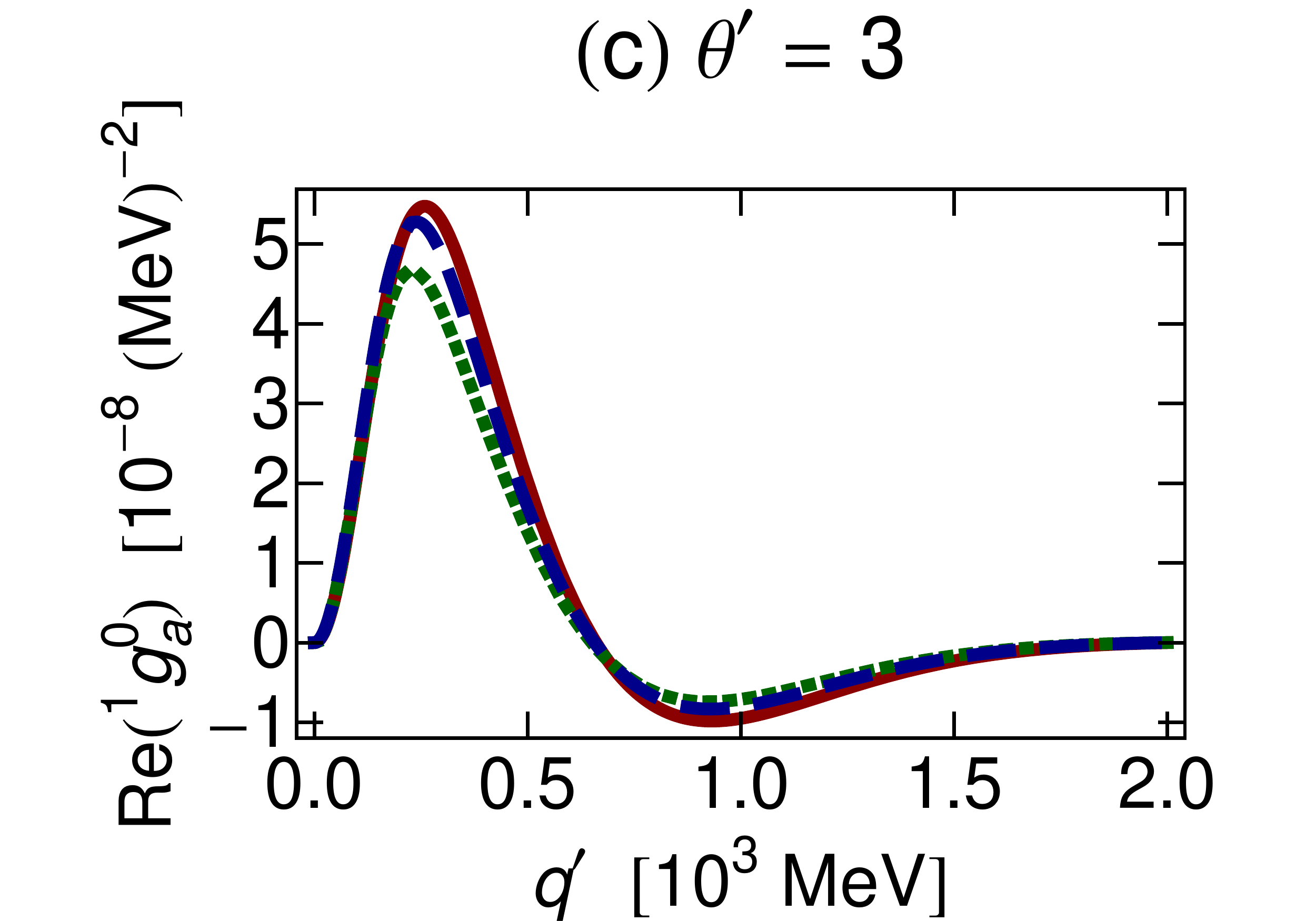}
}
\\
\subfloat{
\includegraphics[width=6cm]{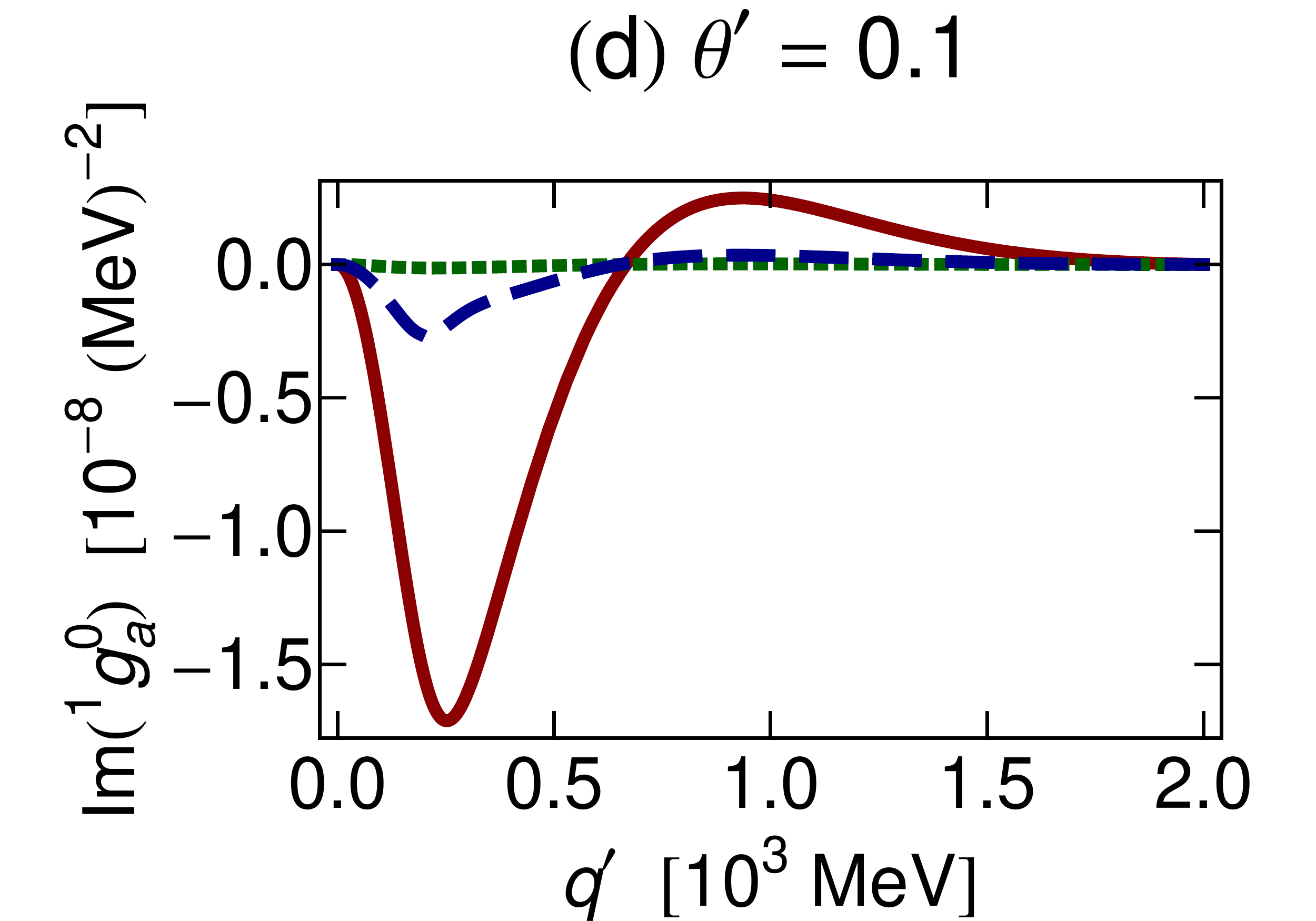}
}
\subfloat{
\includegraphics[width=6cm]{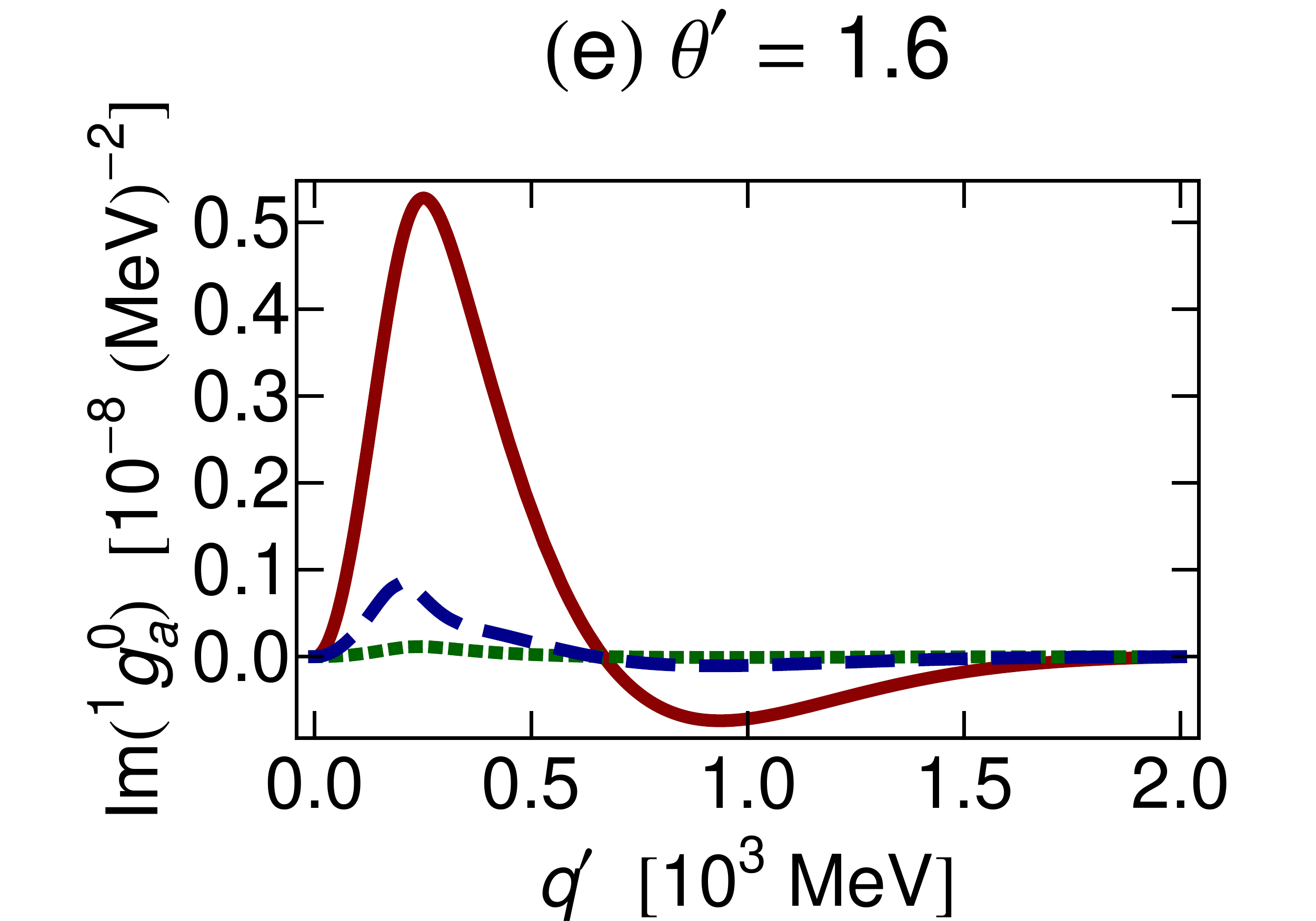}
}
\subfloat{
\includegraphics[width=6cm]{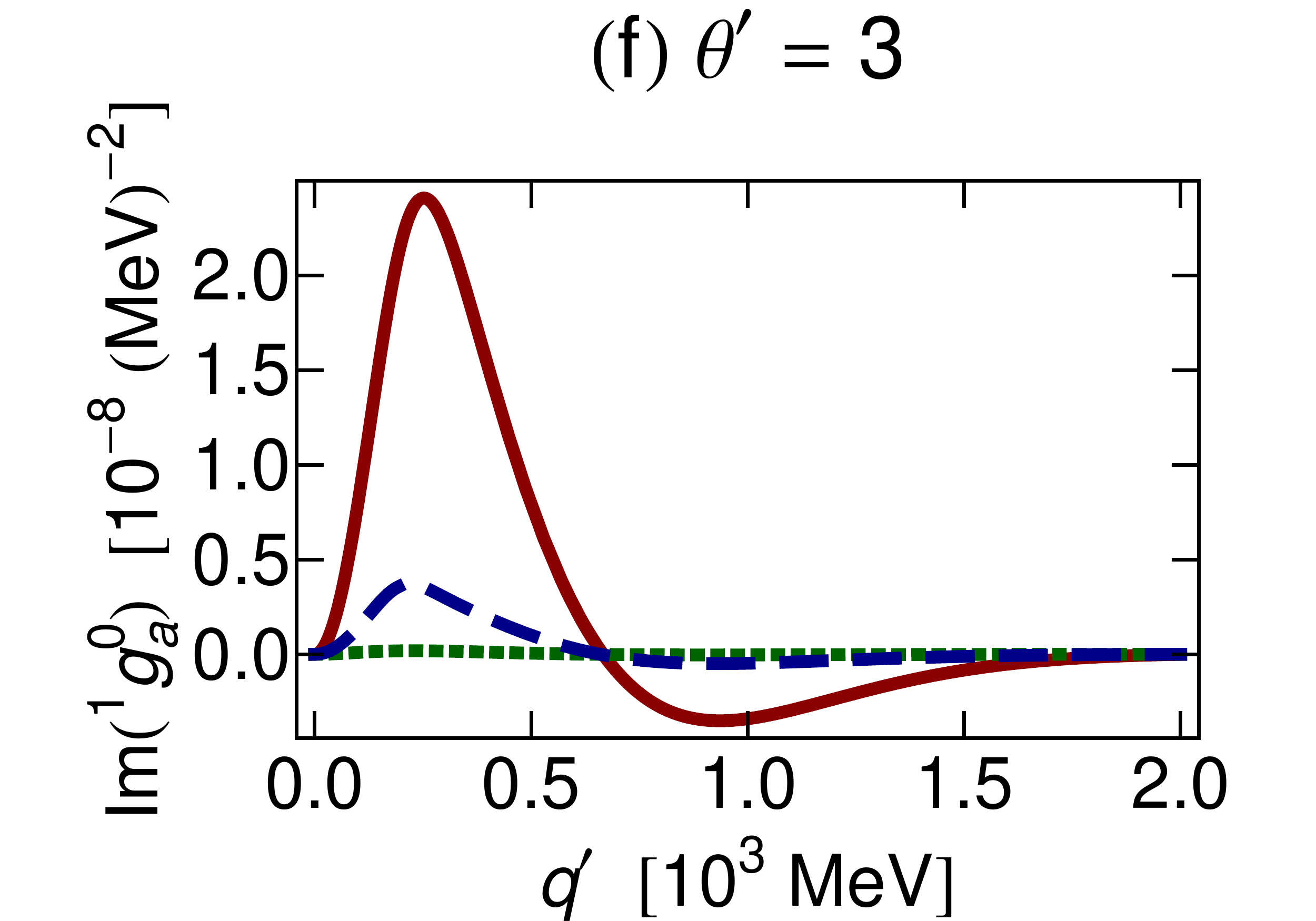}
}
\\
\subfloat{
\includegraphics[width=6cm]{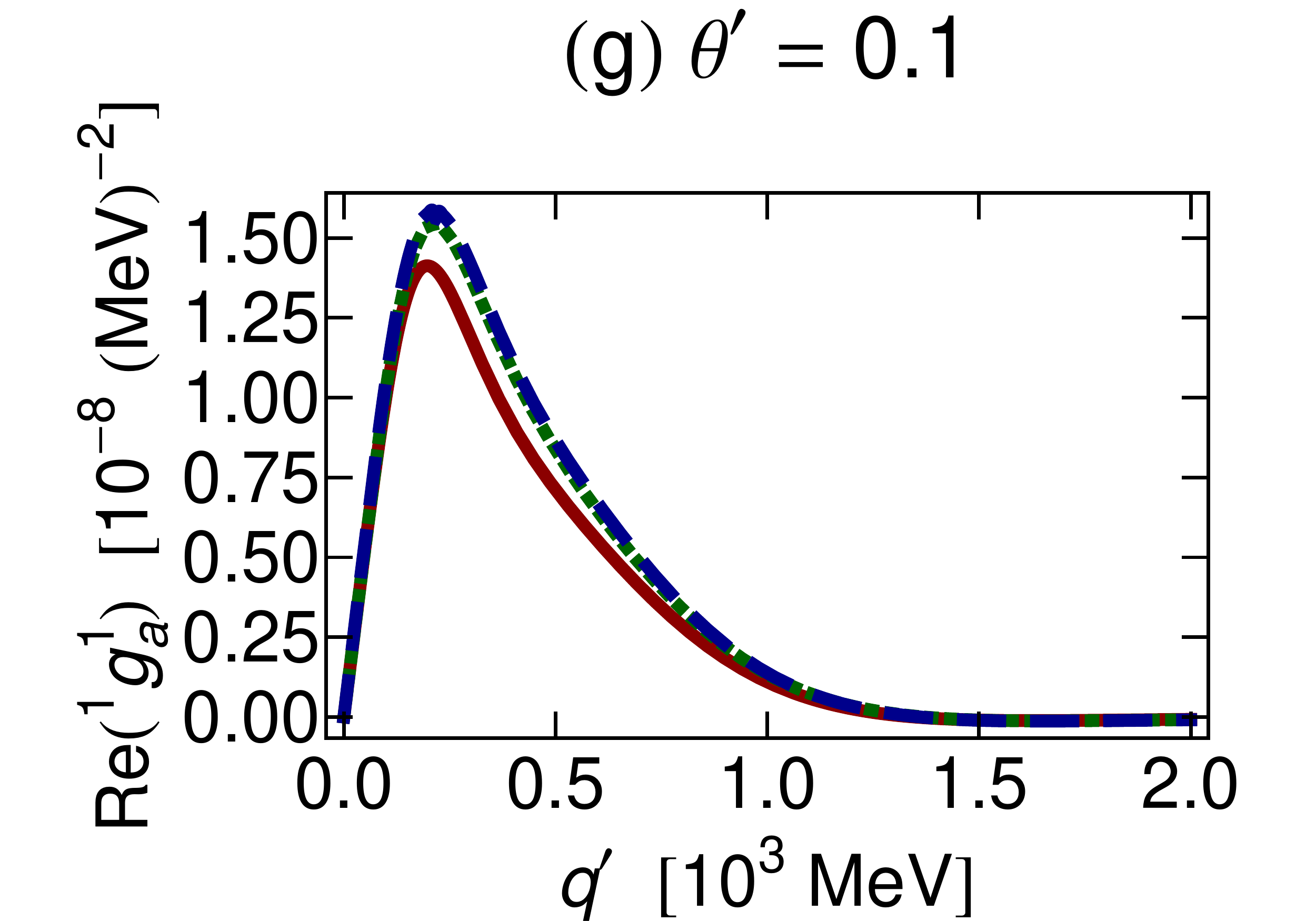}
}
\subfloat{
\includegraphics[width=6cm]{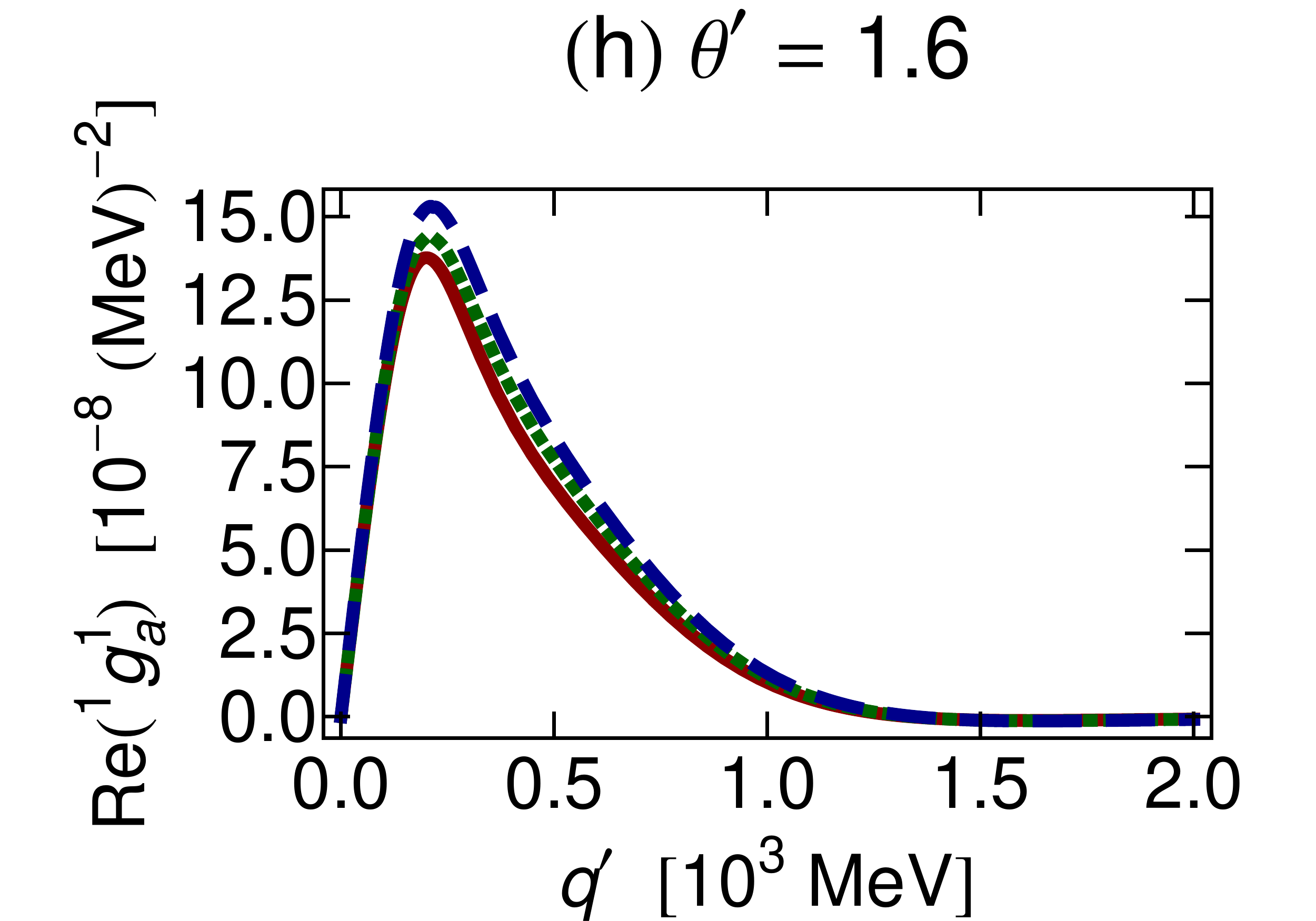}
}
\subfloat{
\includegraphics[width=6cm]{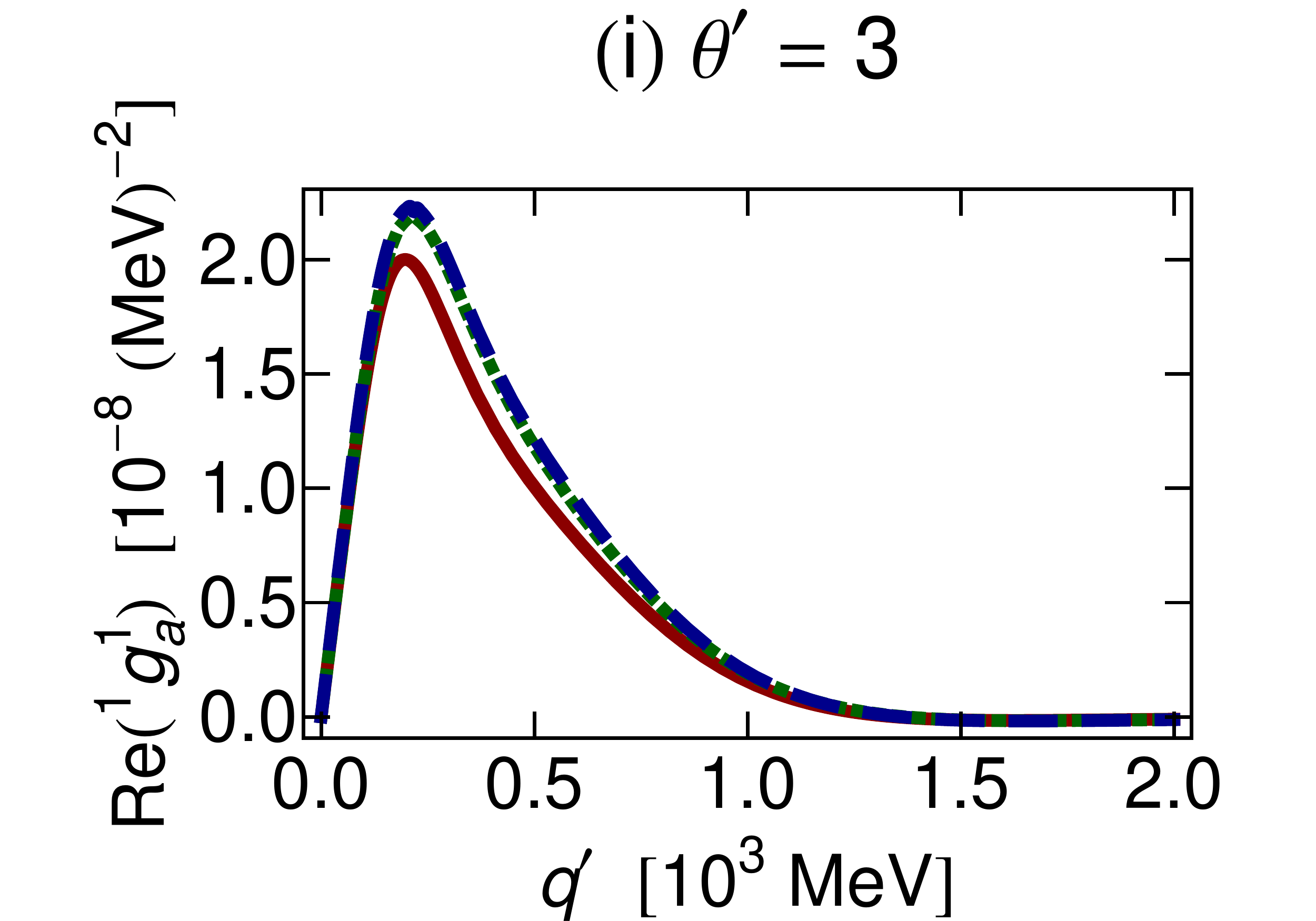}
}
\\
\subfloat{
\includegraphics[width=6cm]{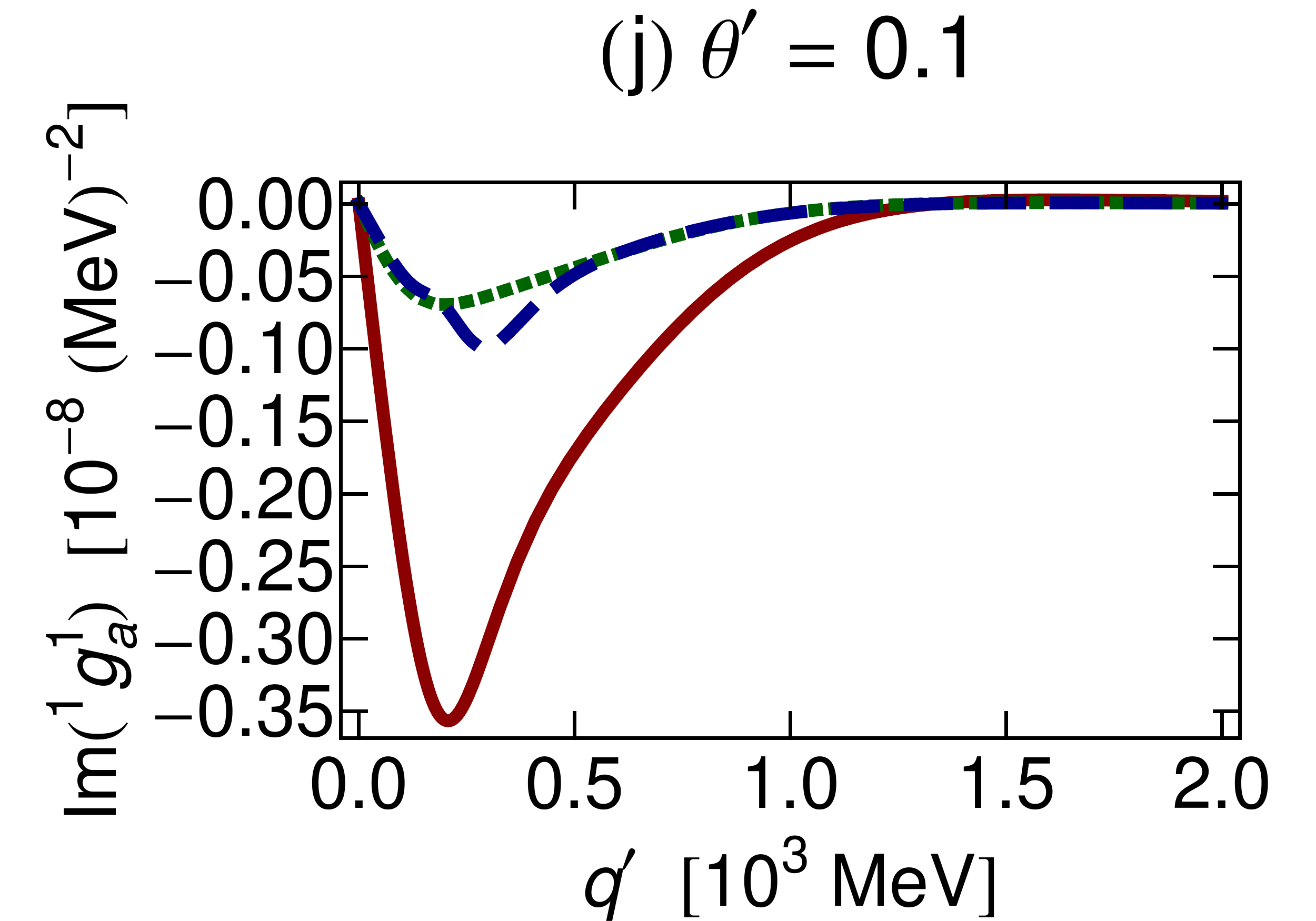}
}
\subfloat{
\includegraphics[width=6cm]{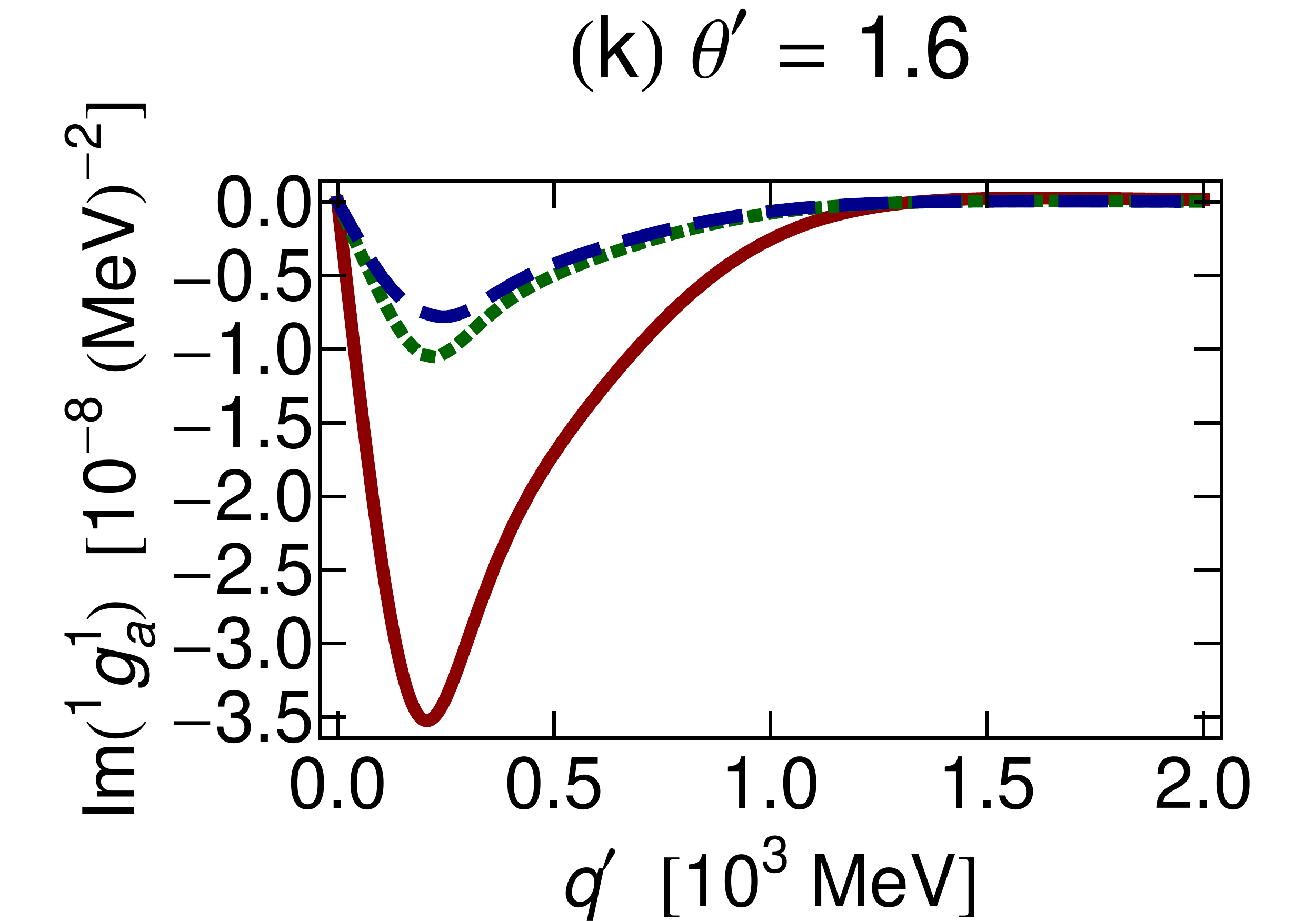}
}
\subfloat{
\includegraphics[width=6cm]{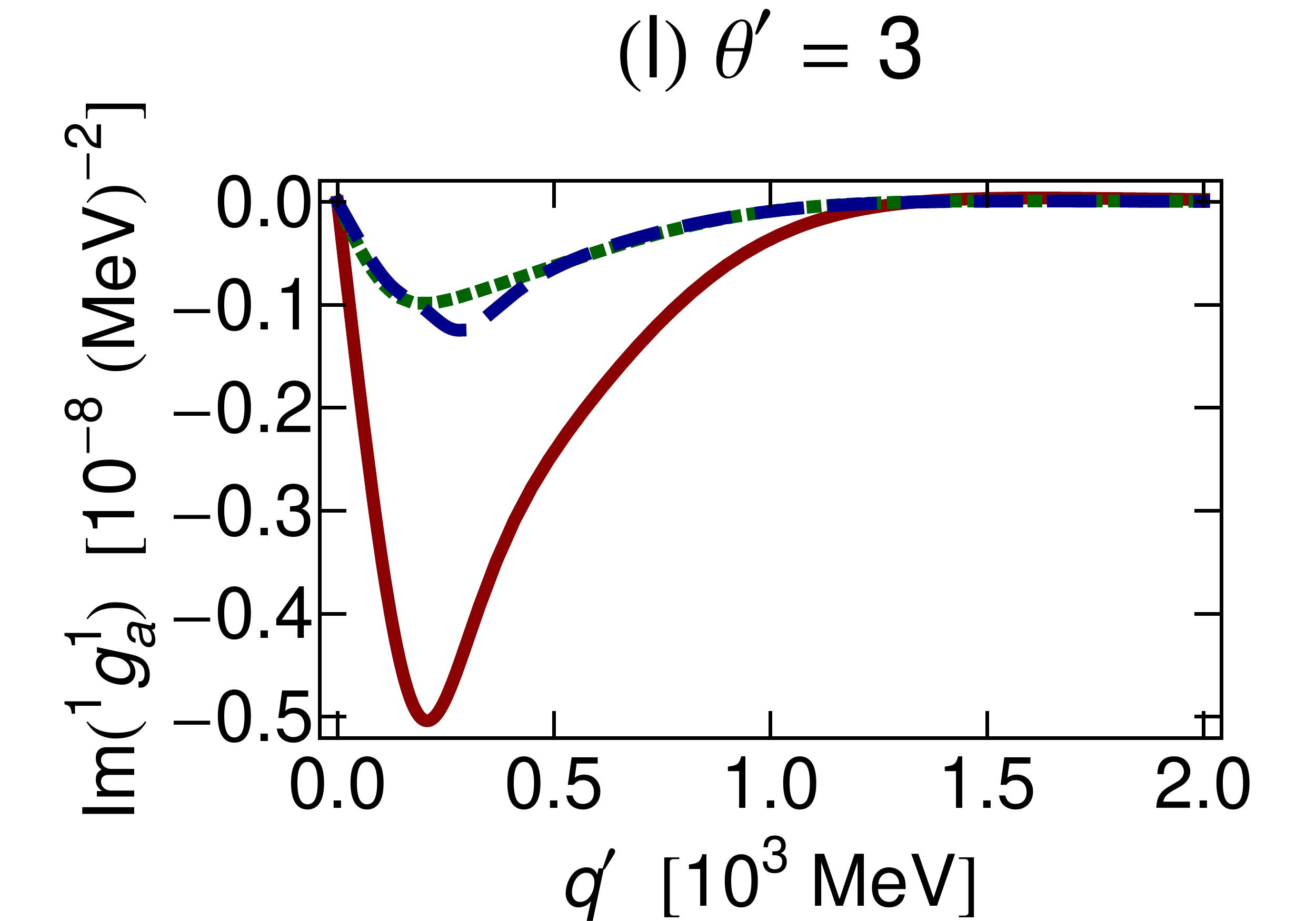}
}
\end{center}
\caption{(Color online) Same as Fig.~\ref{Fig:g_plots50_0} but for $\up{1}g^I_a$ and $\up{1}t^I_a$ at $q=216.67 \units{MeV}$.}
\label{Fig:g_plots100_1}
\end{figure}

\begin{figure}[H]
\begin{center}
\subfloat{
\includegraphics[width=6cm]{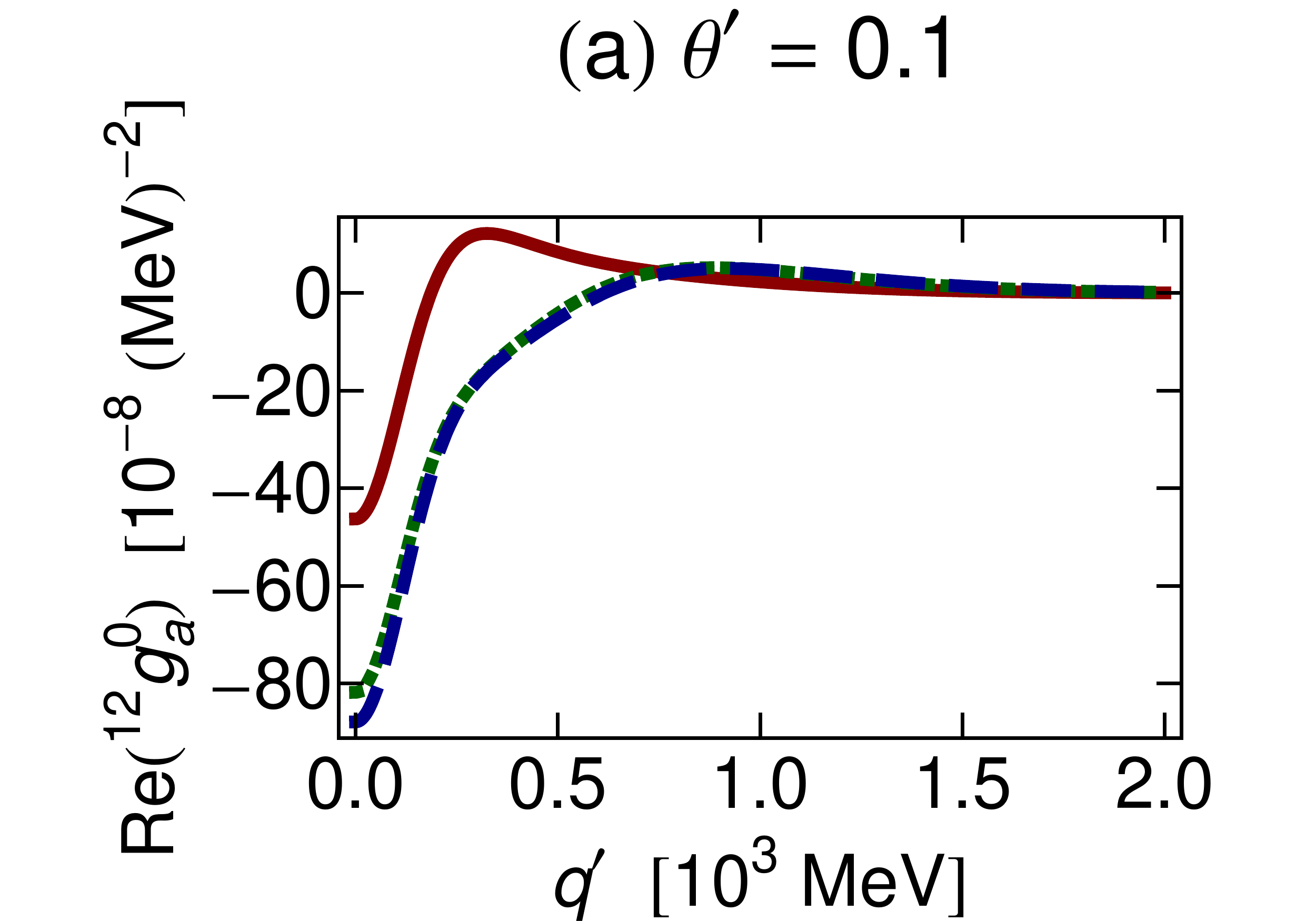}
}
\subfloat{
\includegraphics[width=6cm]{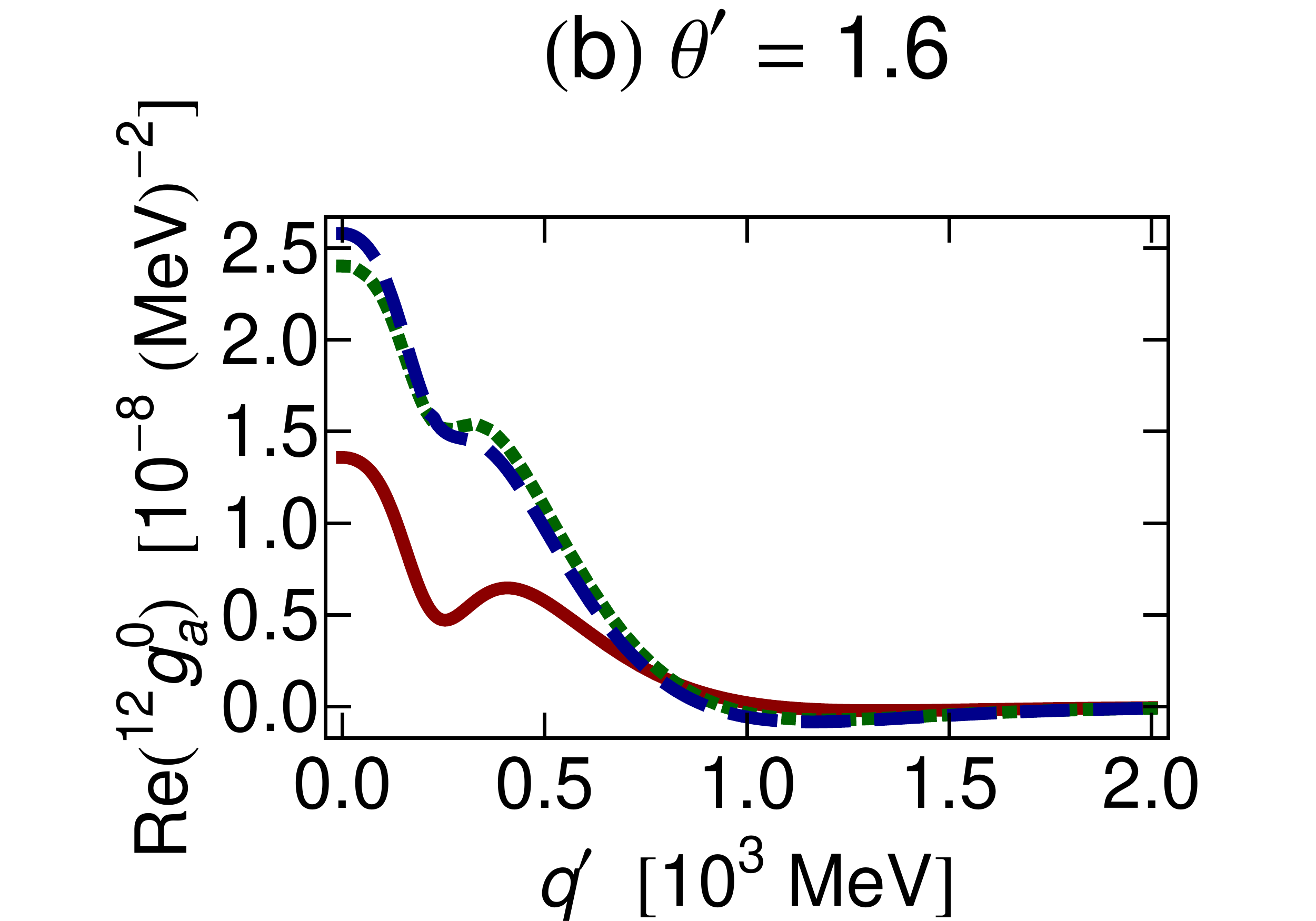}
}
\subfloat{
\includegraphics[width=6cm]{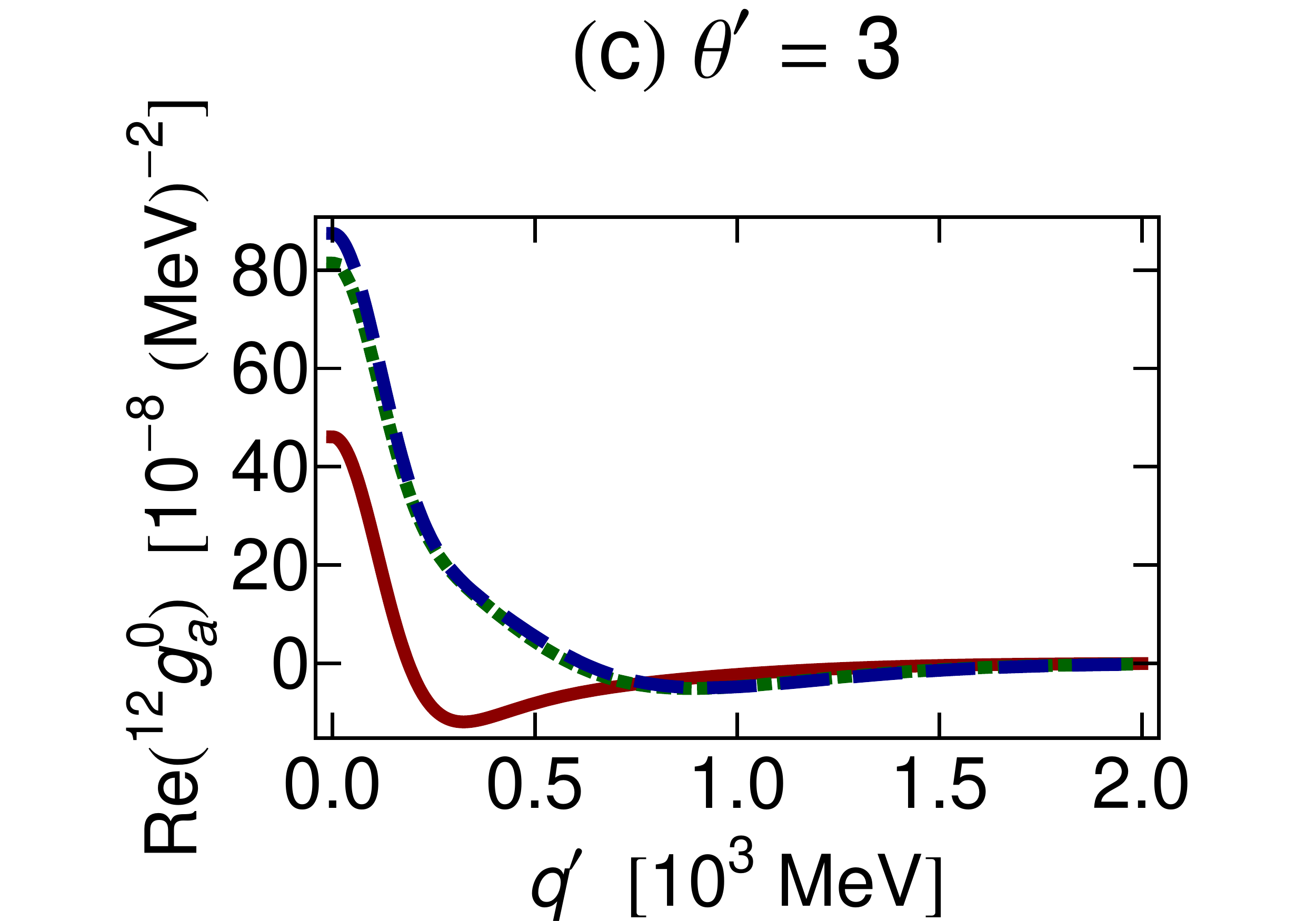}
}
\\
\subfloat{
\includegraphics[width=6cm]{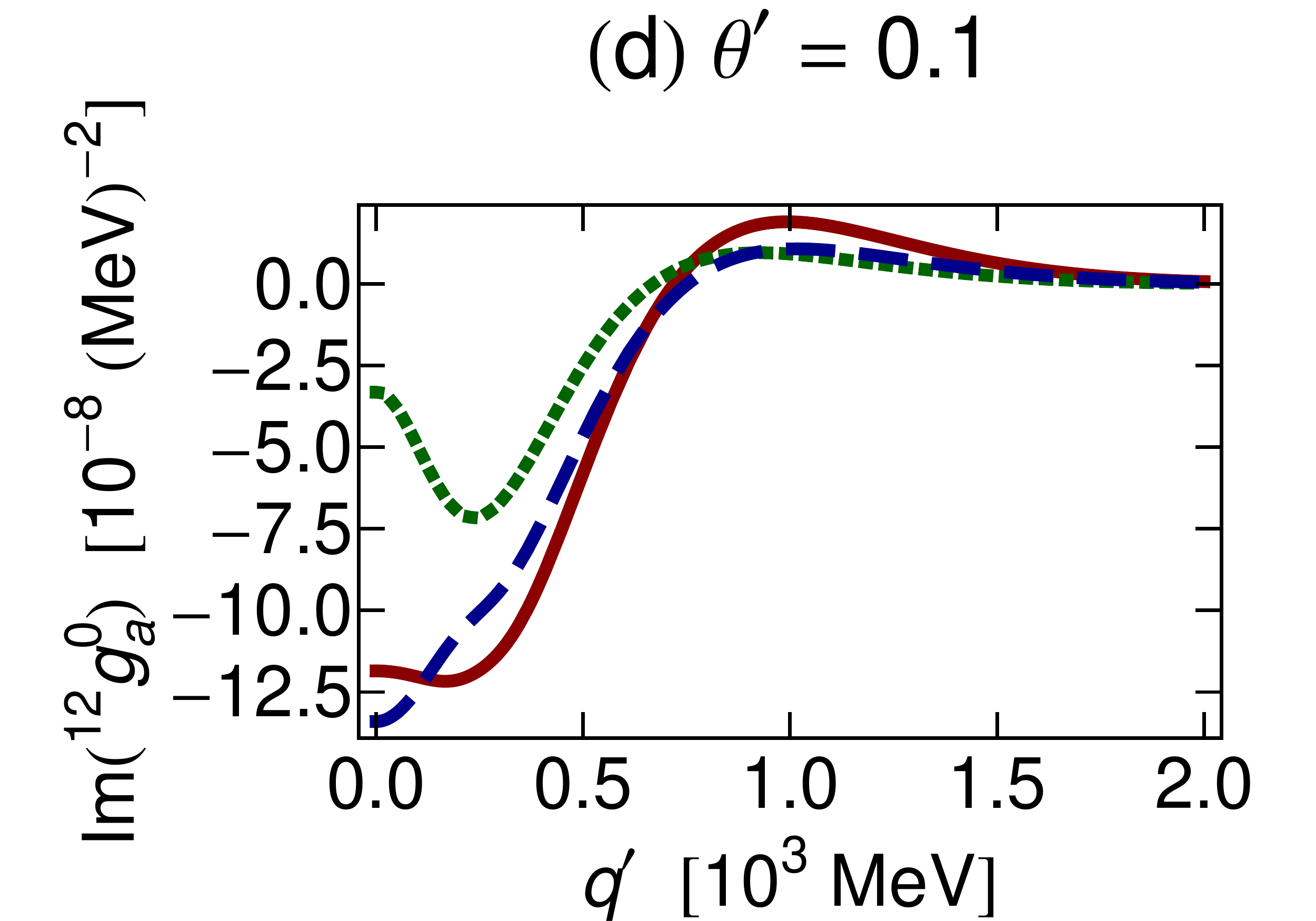}
}
\subfloat{
\includegraphics[width=6cm]{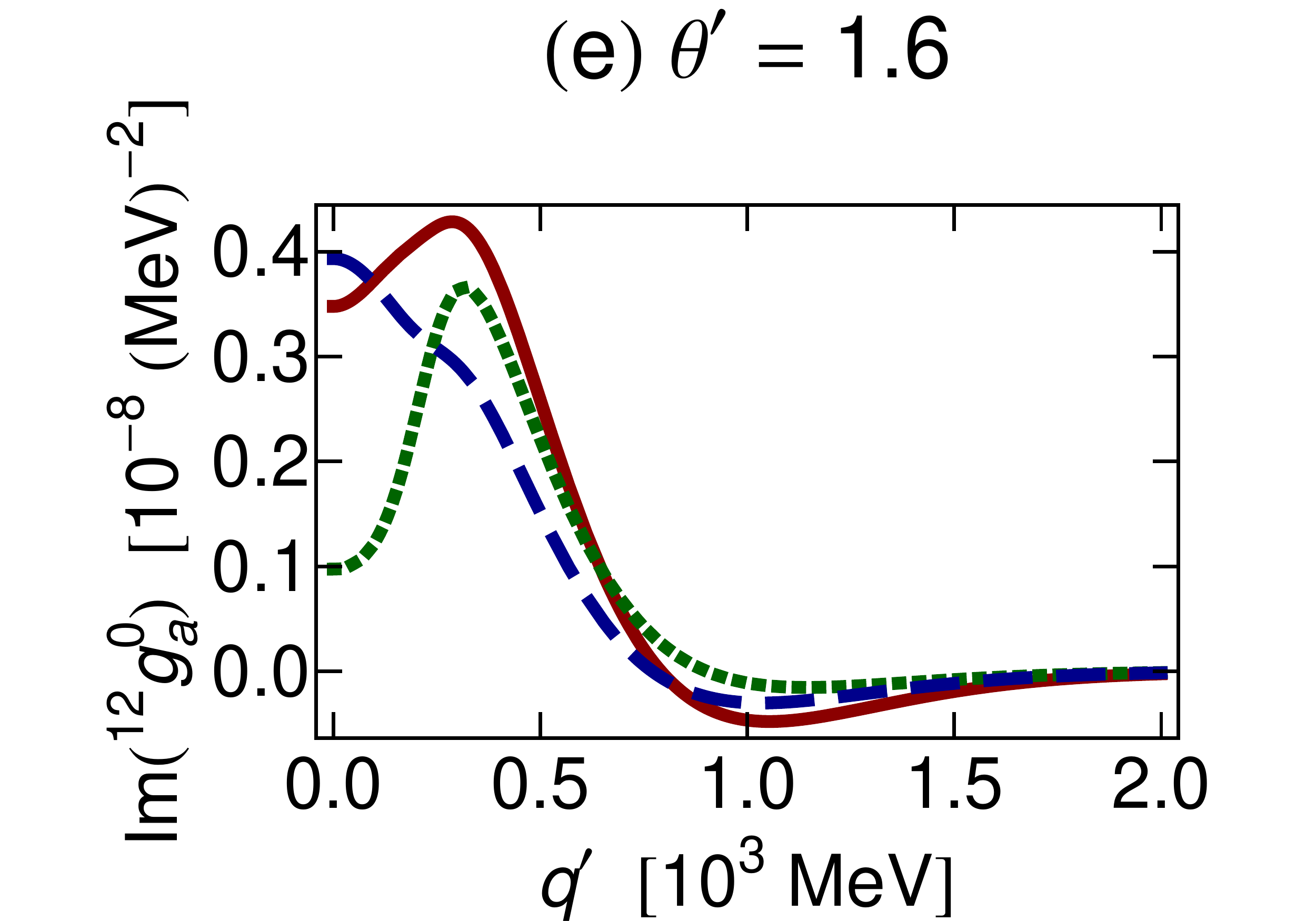}
}
\subfloat{
\includegraphics[width=6cm]{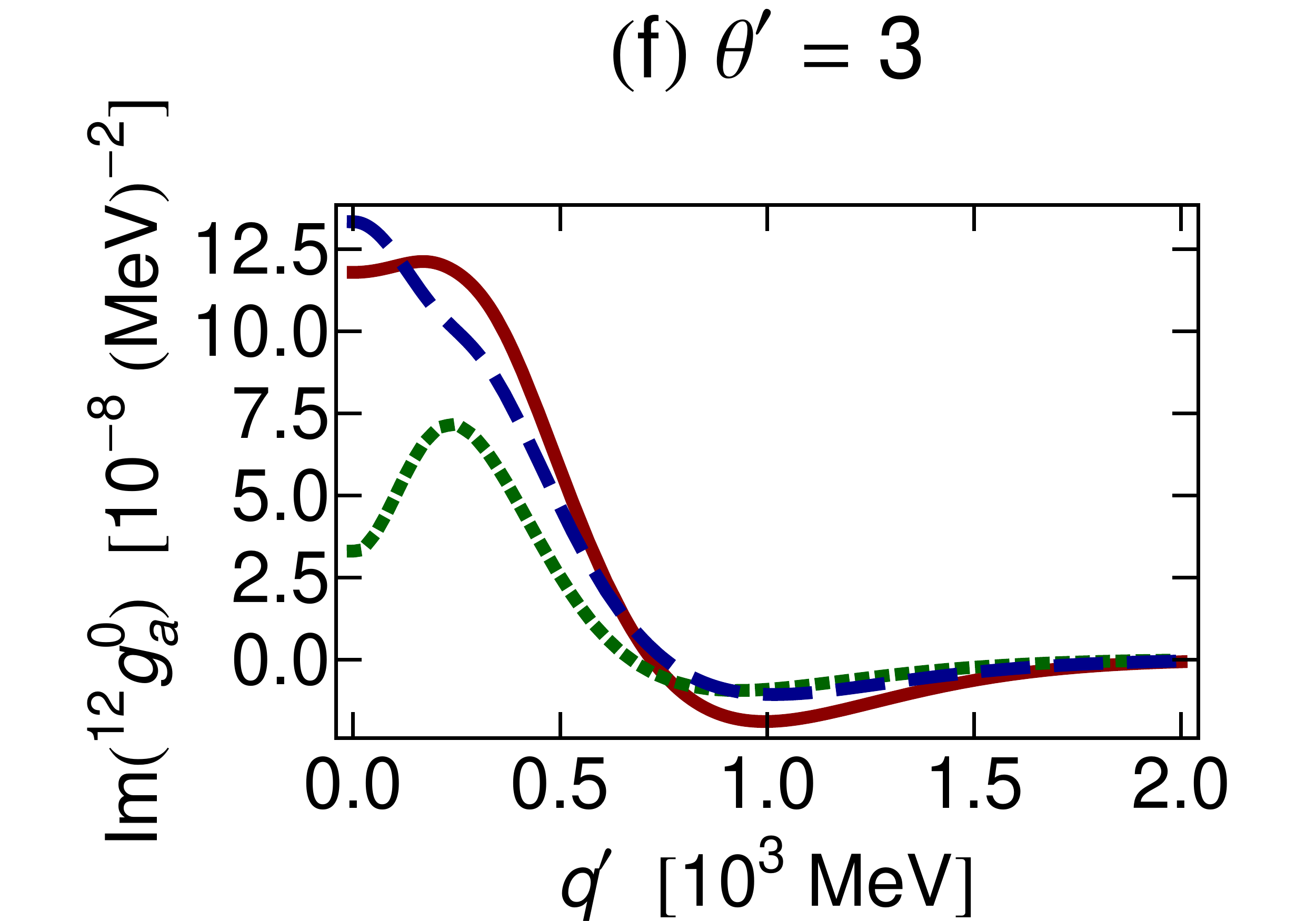}
}
\\
\subfloat{
\includegraphics[width=6cm]{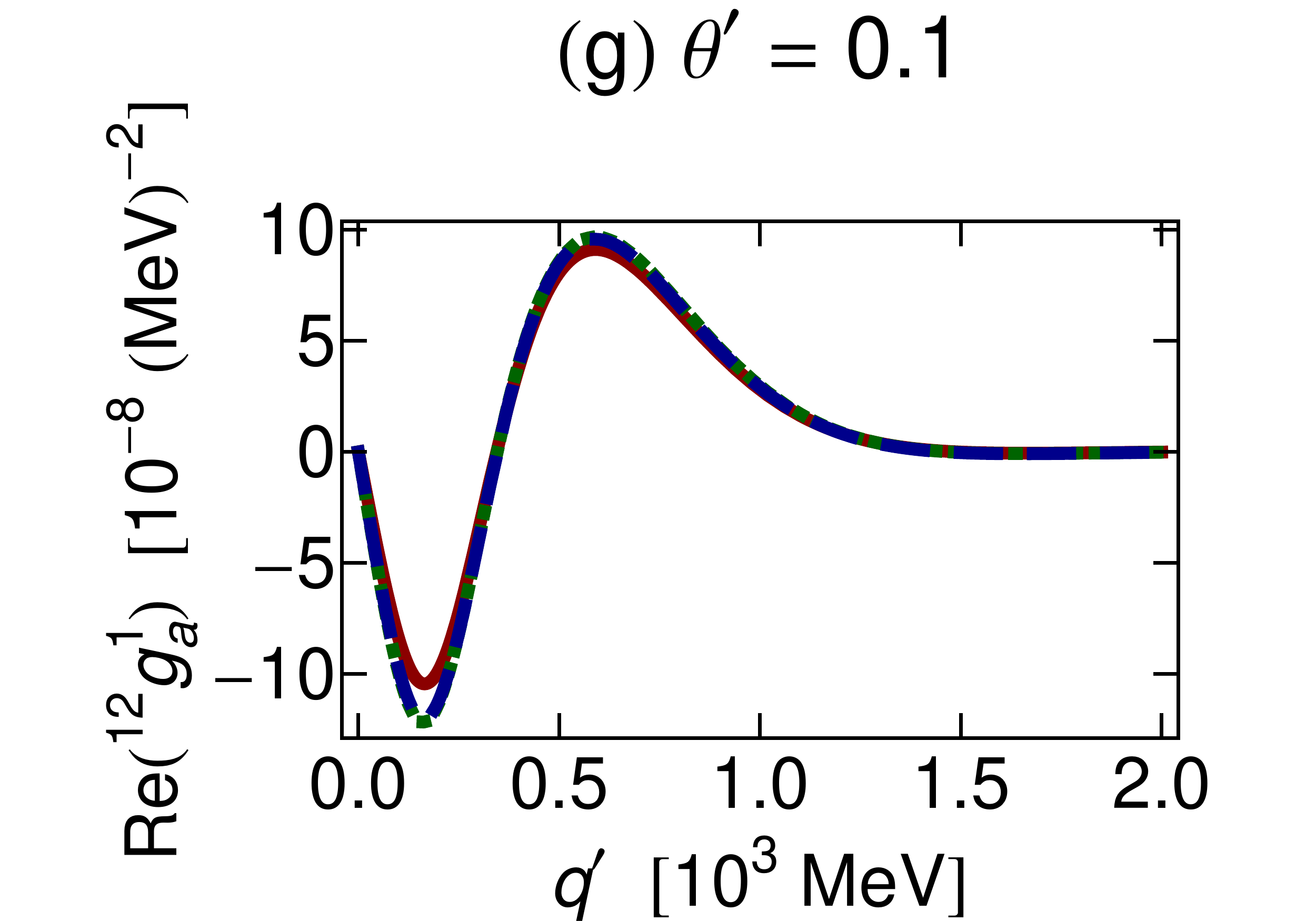}
}
\subfloat{
\includegraphics[width=6cm]{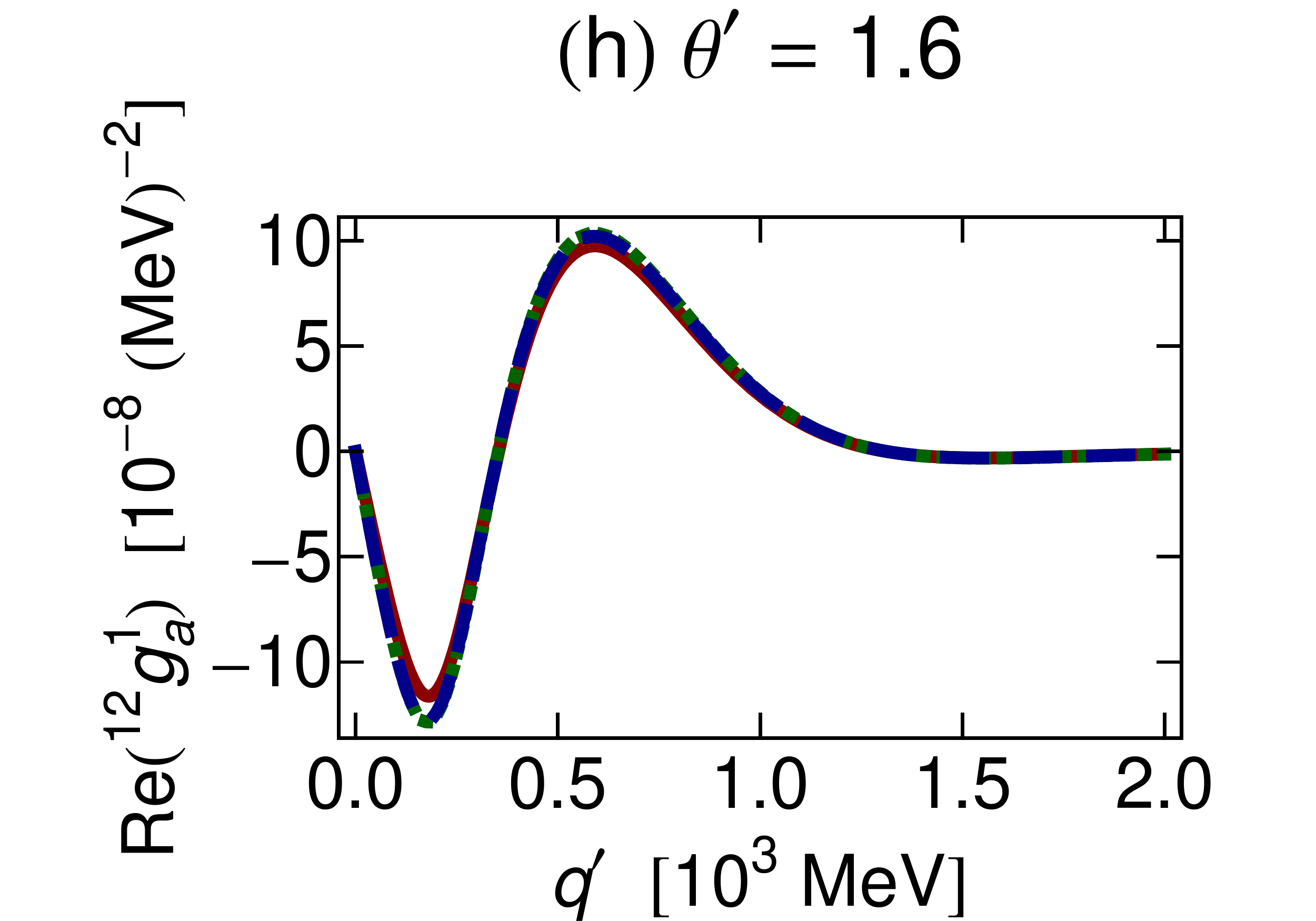}
}
\subfloat{
\includegraphics[width=6cm]{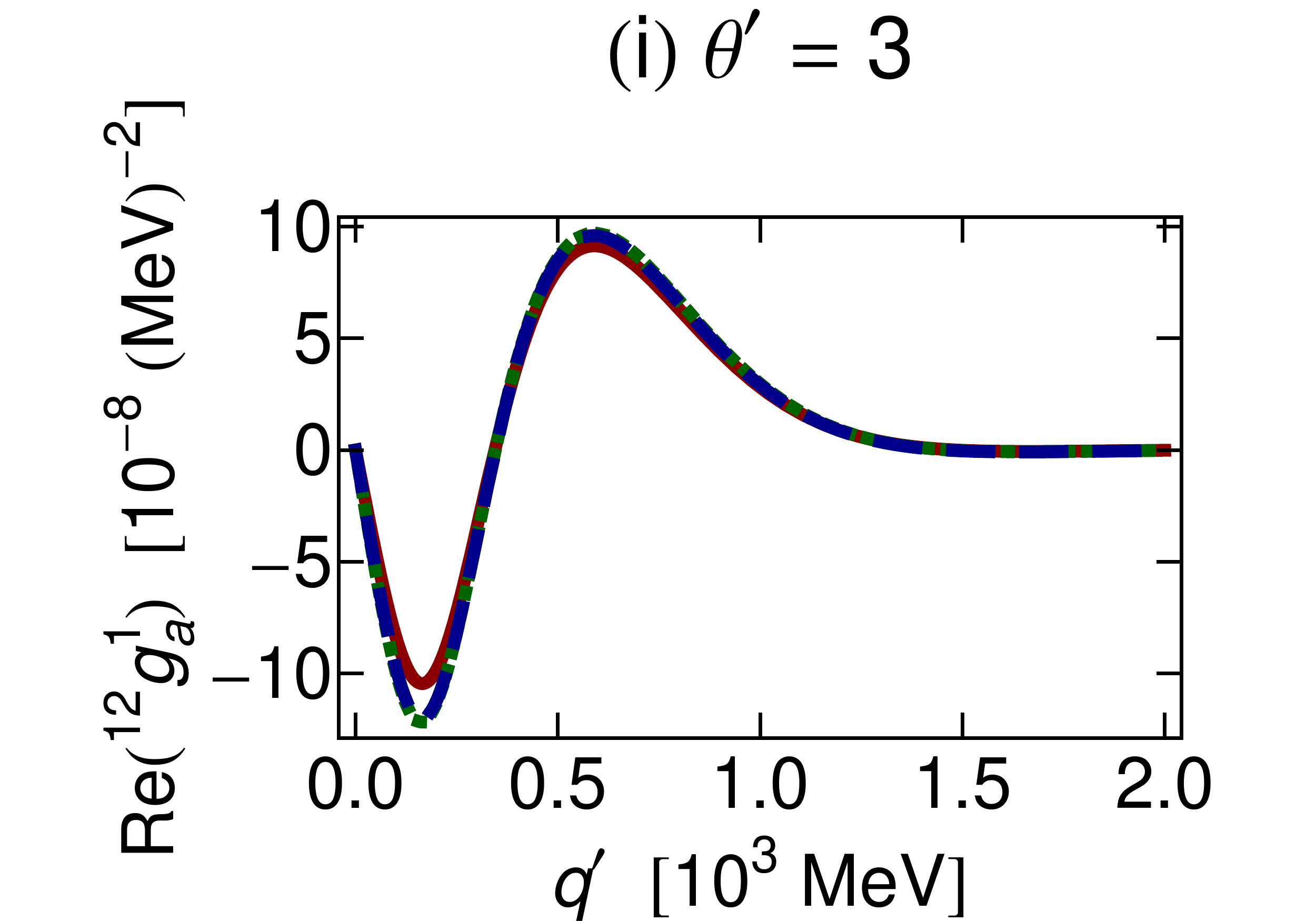}
}
\\
\subfloat{
\includegraphics[width=6cm]{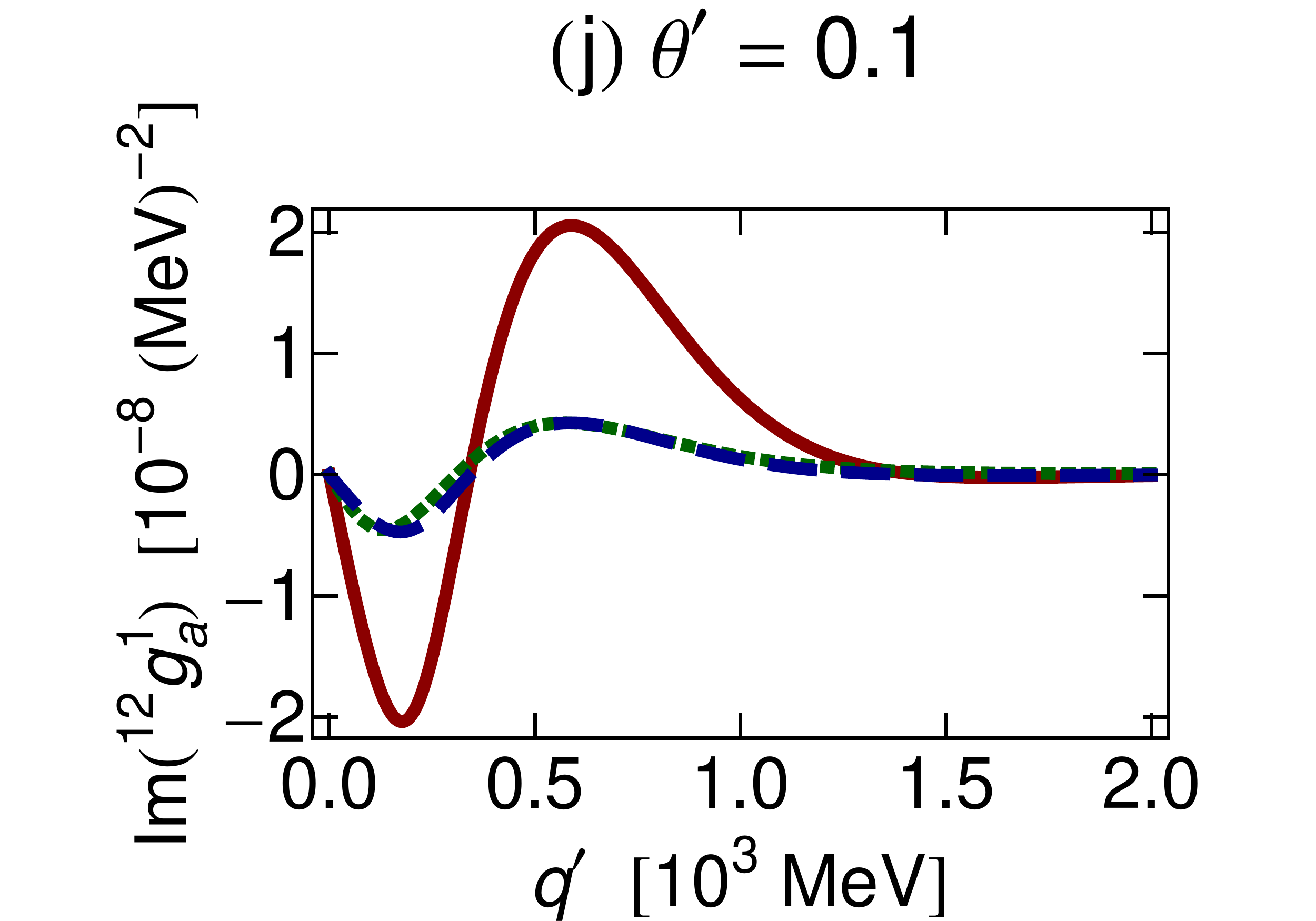}
}
\subfloat{
\includegraphics[width=6cm]{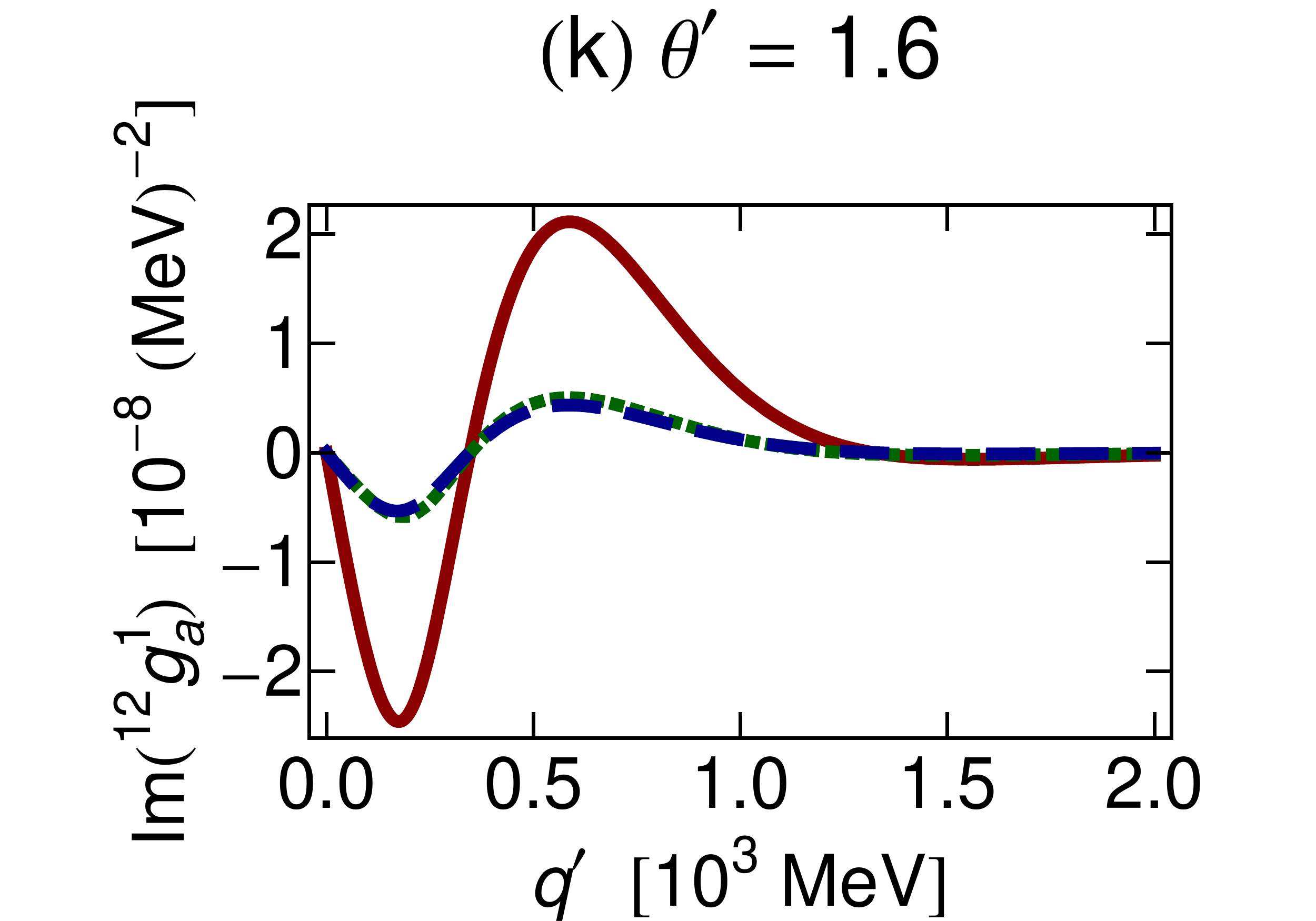}
}
\subfloat{
\includegraphics[width=6cm]{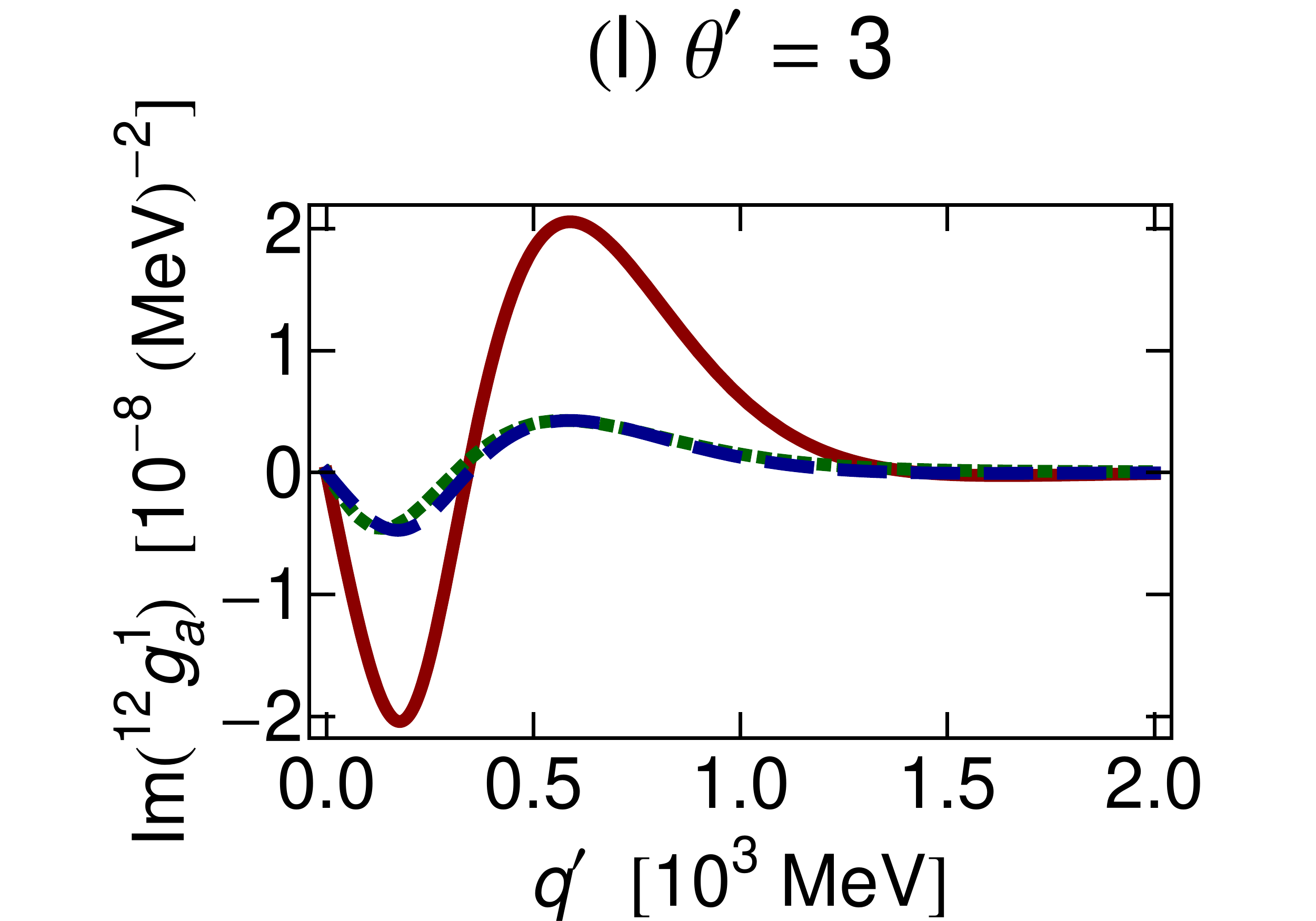}
}
\end{center}
\caption{(Color online) Same as Fig.~\ref{Fig:g_plots50_0} but for $\up{12}g^I_a$ and $\up{12}t^I_a$ at $q=216.67 \units{MeV}$.}
\label{Fig:g_plots100_12}
\end{figure}

\begin{figure}[H]
\begin{center}
\subfloat{
\includegraphics[width=6cm]{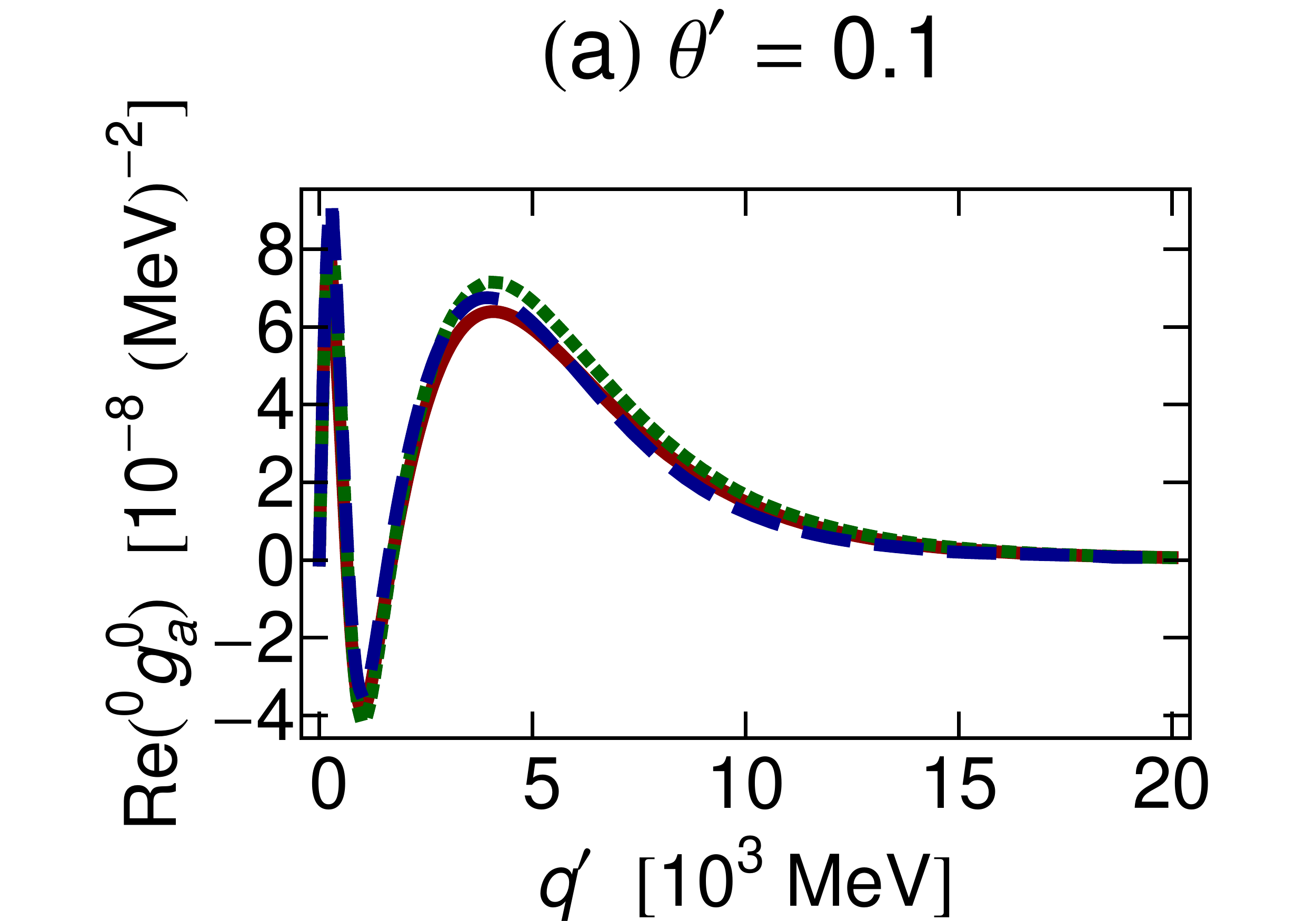}
}
\subfloat{
\includegraphics[width=6cm]{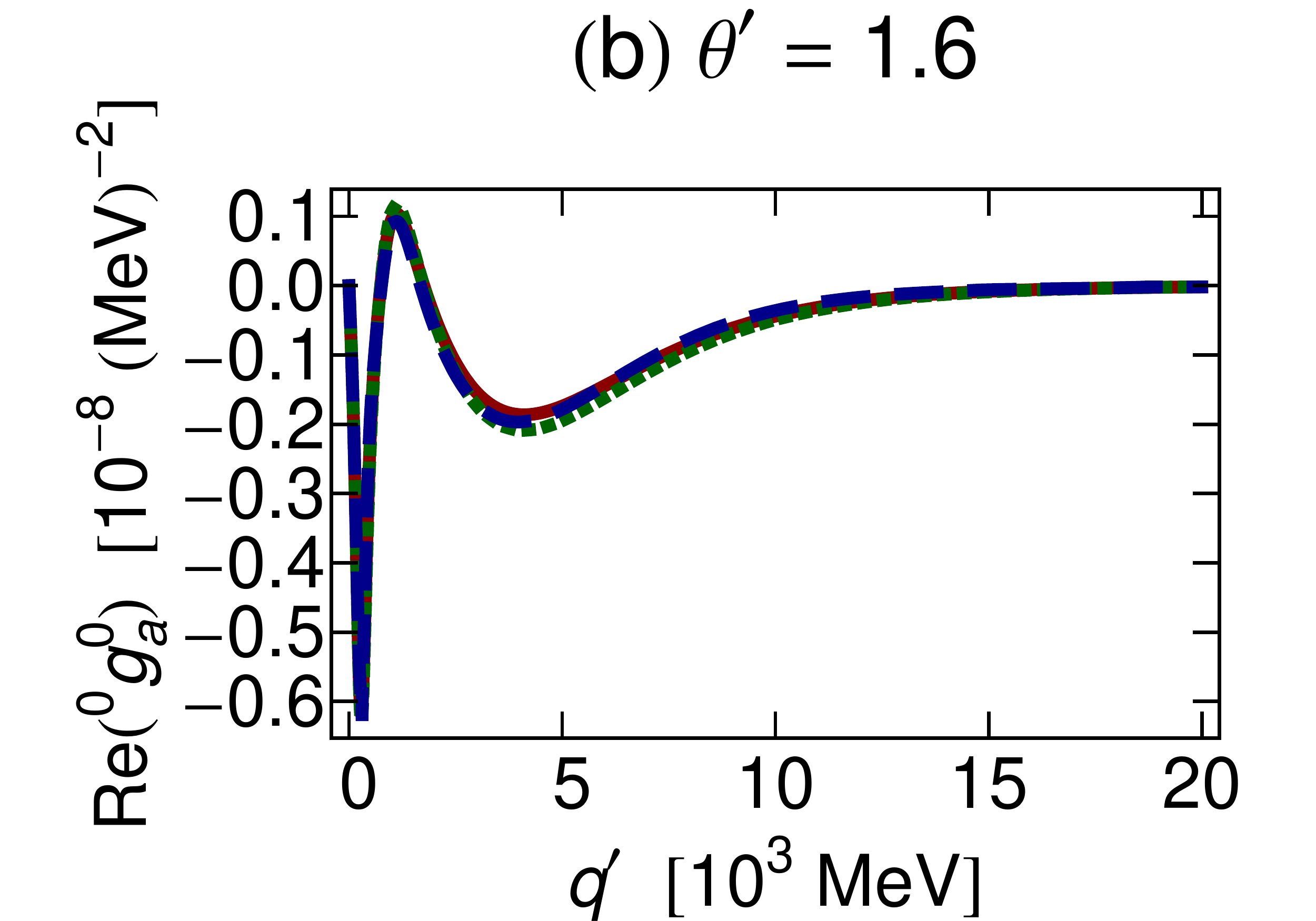}
}
\subfloat{
\includegraphics[width=6cm]{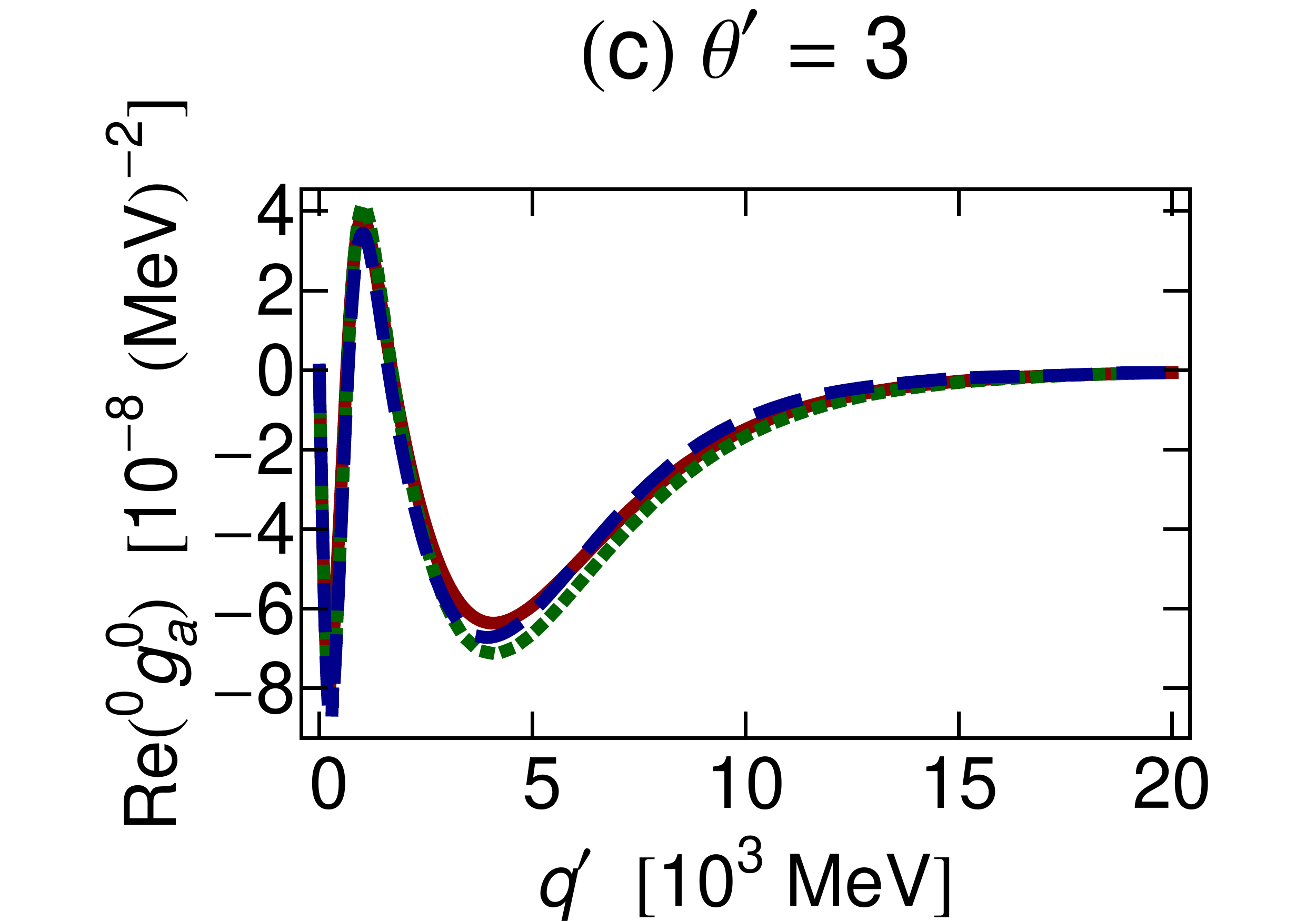}
}
\\
\subfloat{
\includegraphics[width=6cm]{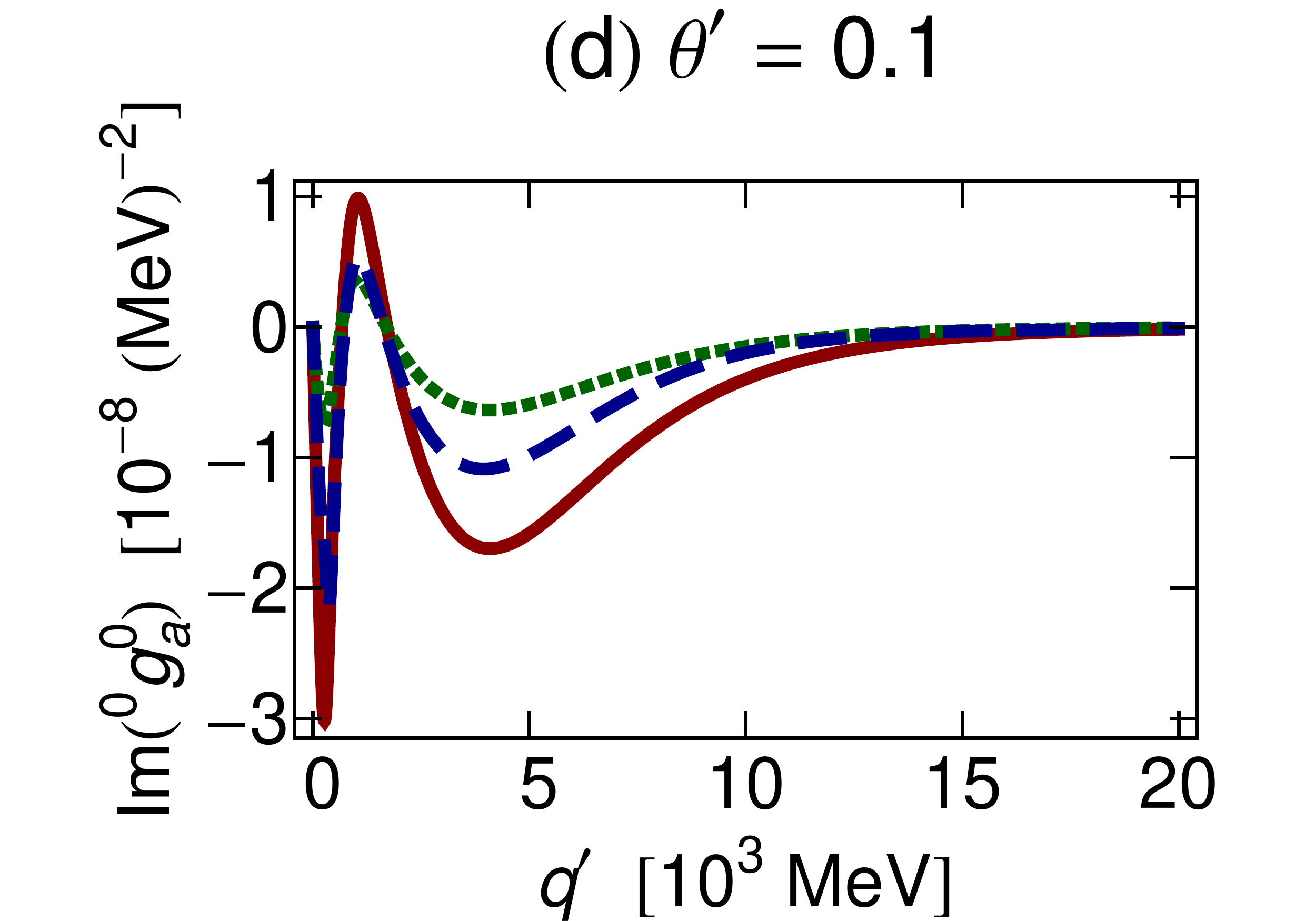}
}
\subfloat{
\includegraphics[width=6cm]{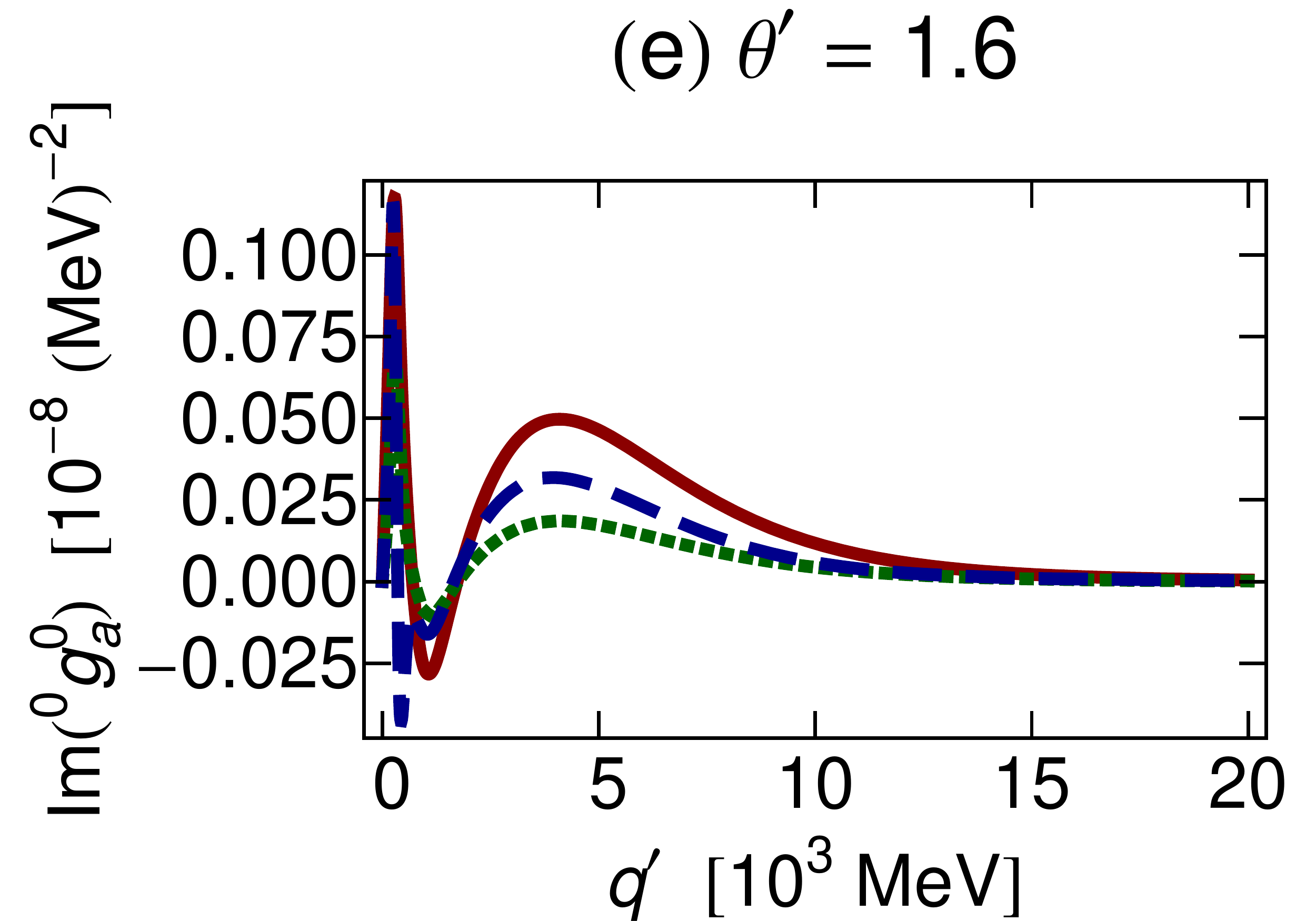}
}
\subfloat{
\includegraphics[width=6cm]{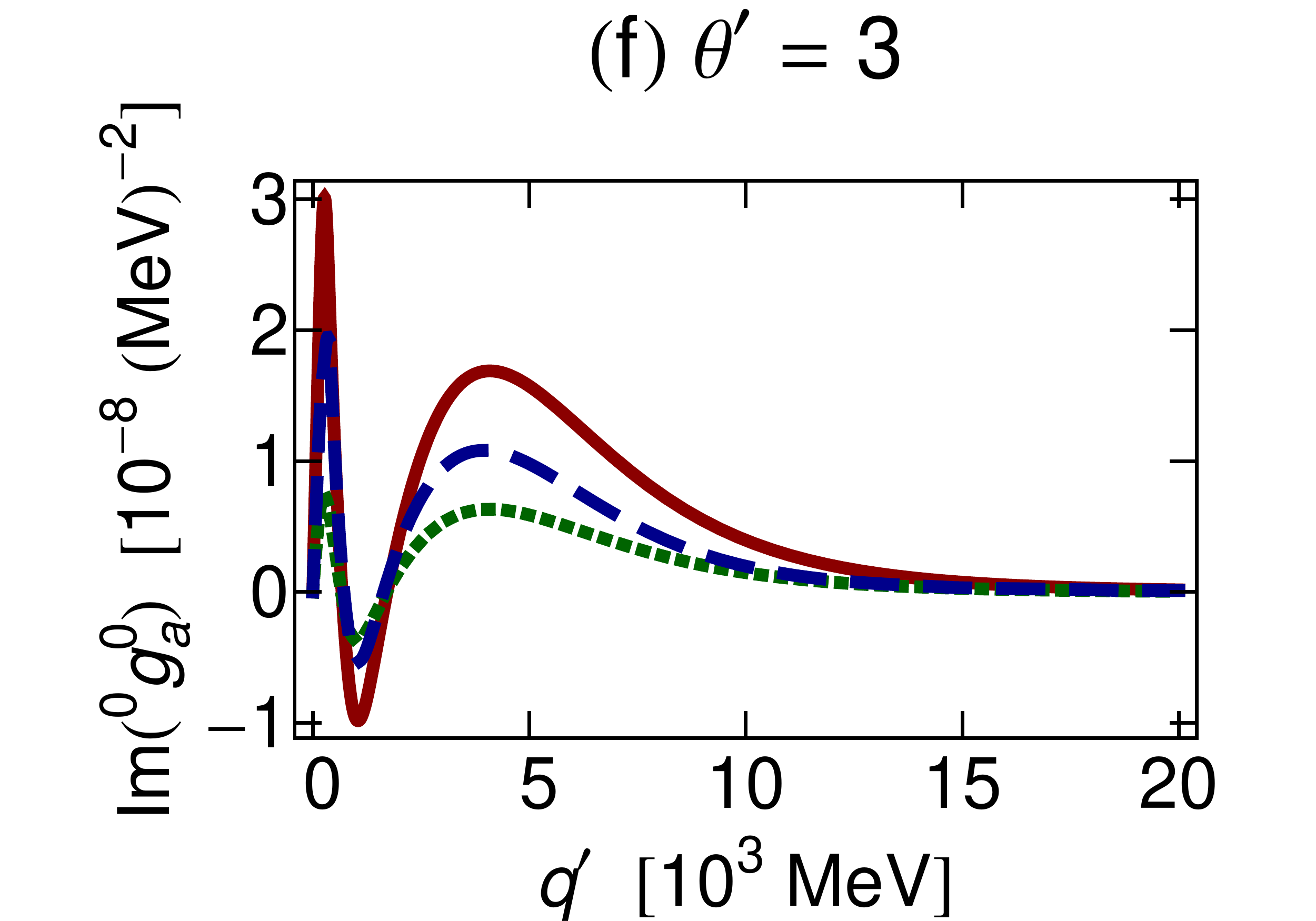}
}
\\
\subfloat{
\includegraphics[width=6cm]{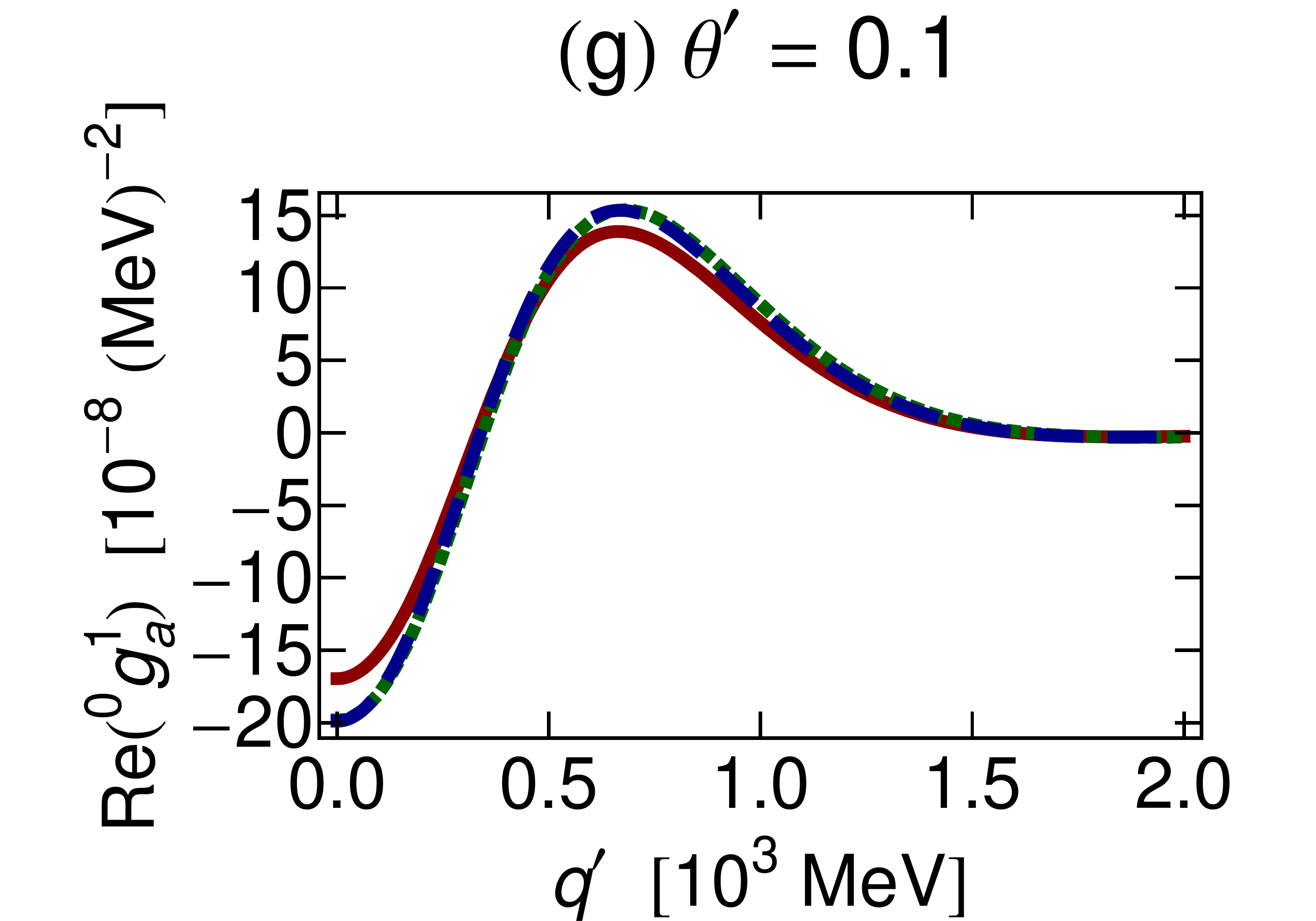}
}
\subfloat{
\includegraphics[width=6cm]{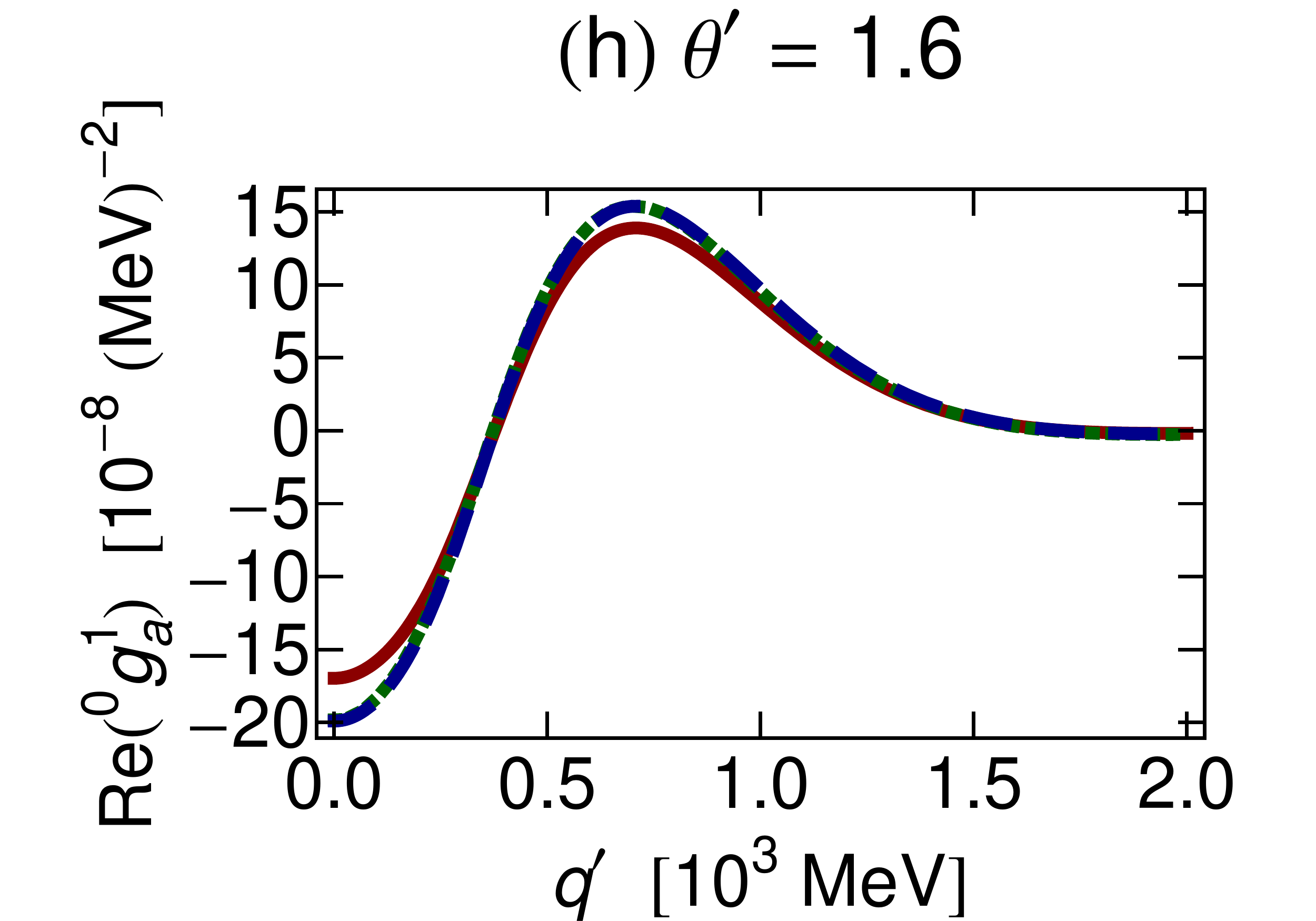}
}
\subfloat{
\includegraphics[width=6cm]{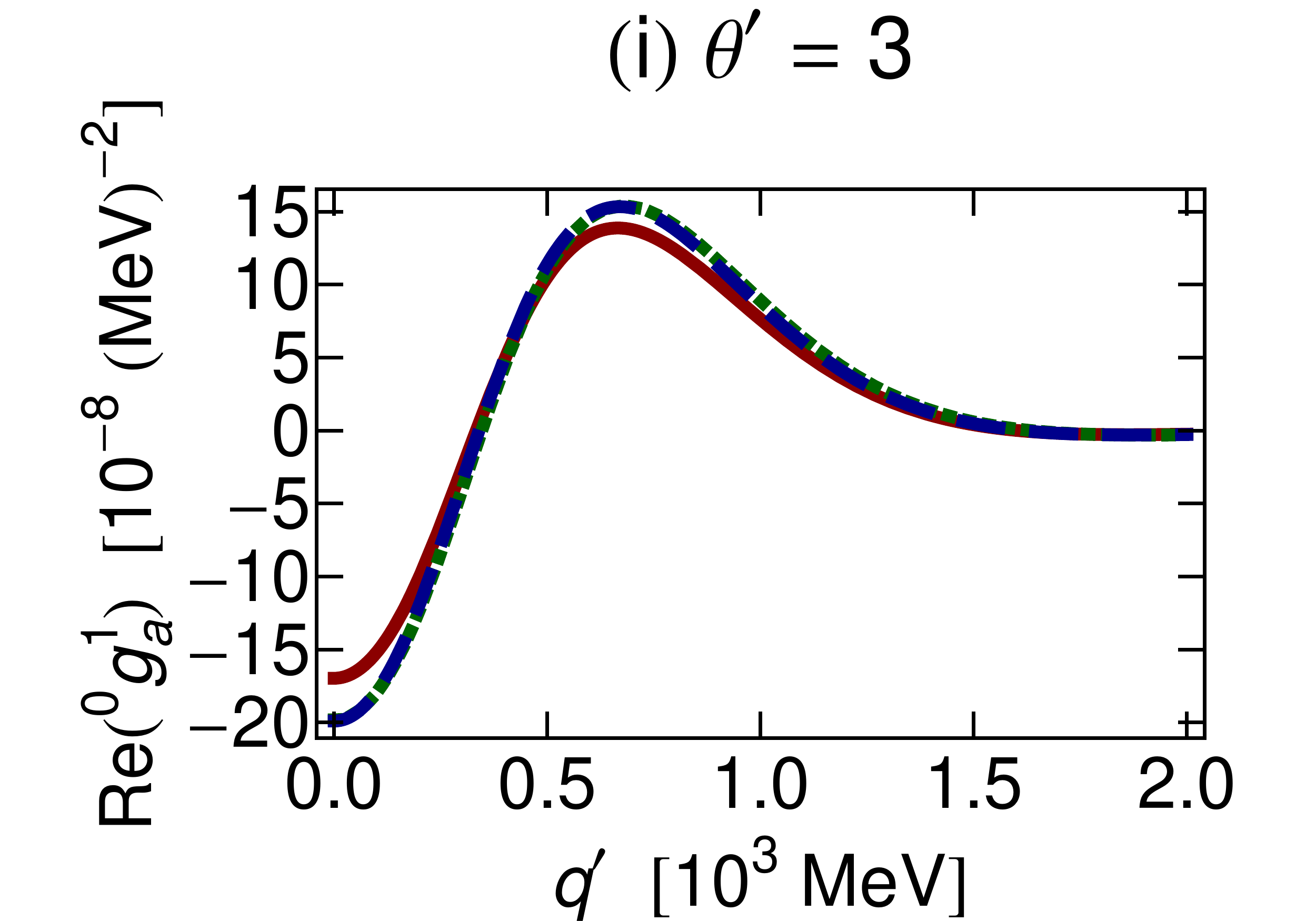}
}
\\
\subfloat{
\includegraphics[width=6cm]{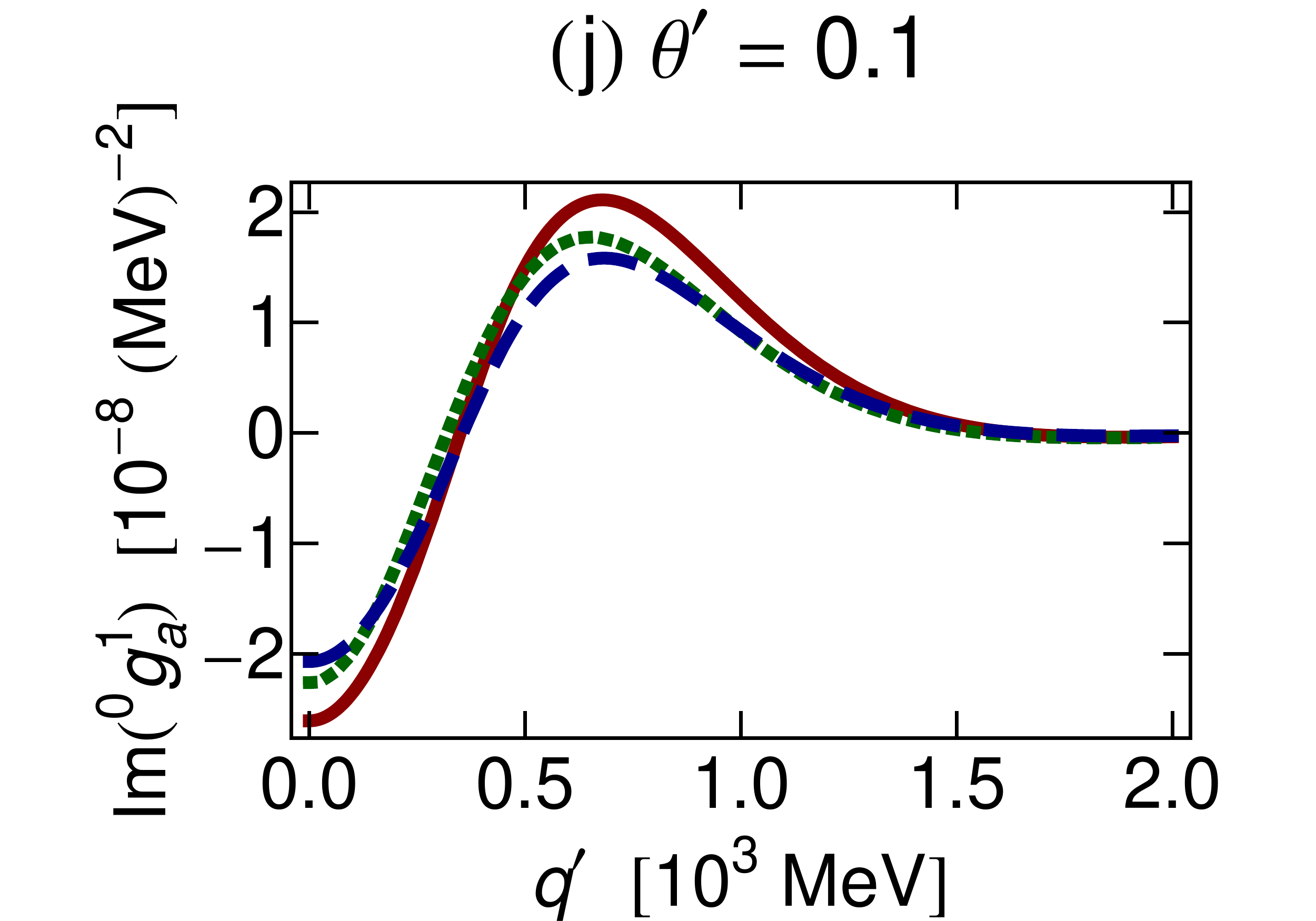}
}
\subfloat{
\includegraphics[width=6cm]{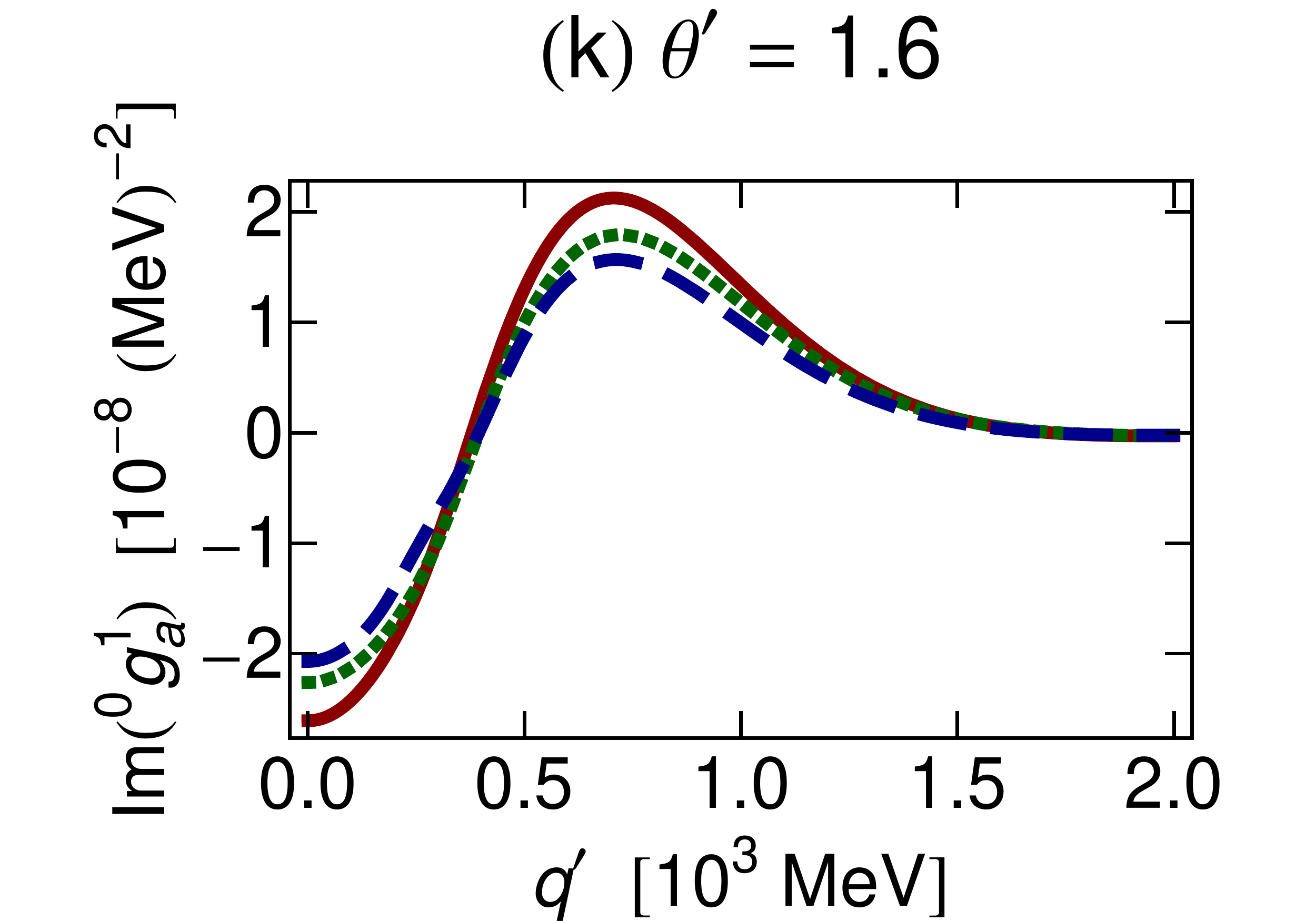}
}
\subfloat{
\includegraphics[width=6cm]{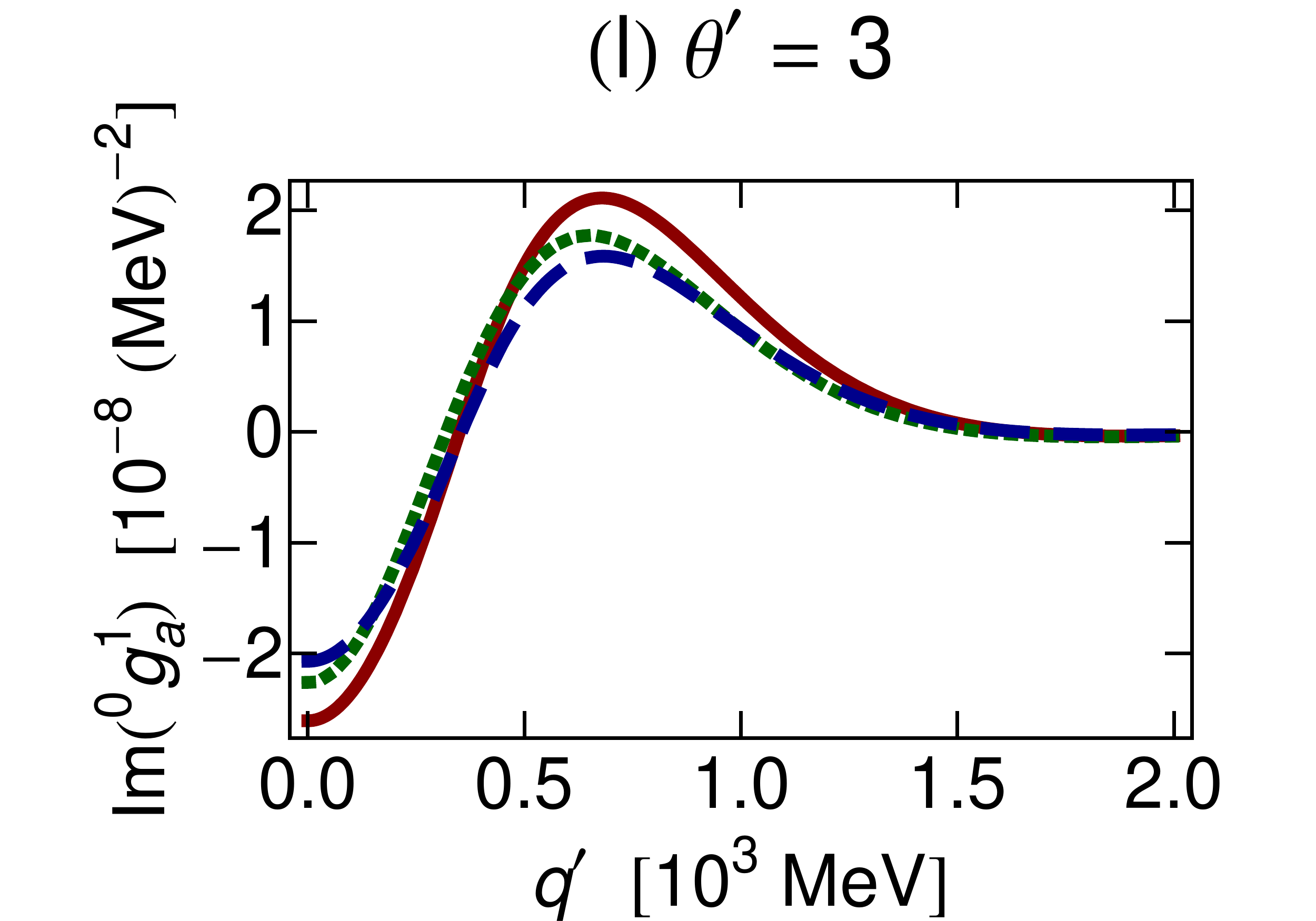}
}
\end{center}
\caption{(Color online) Same as Fig.~\ref{Fig:g_plots50_0} but at $q=306.42 \units{MeV}$.}
\label{Fig:g_plots200_0}
\end{figure}

\begin{figure}[H]
\begin{center}
\subfloat{
\includegraphics[width=6cm]{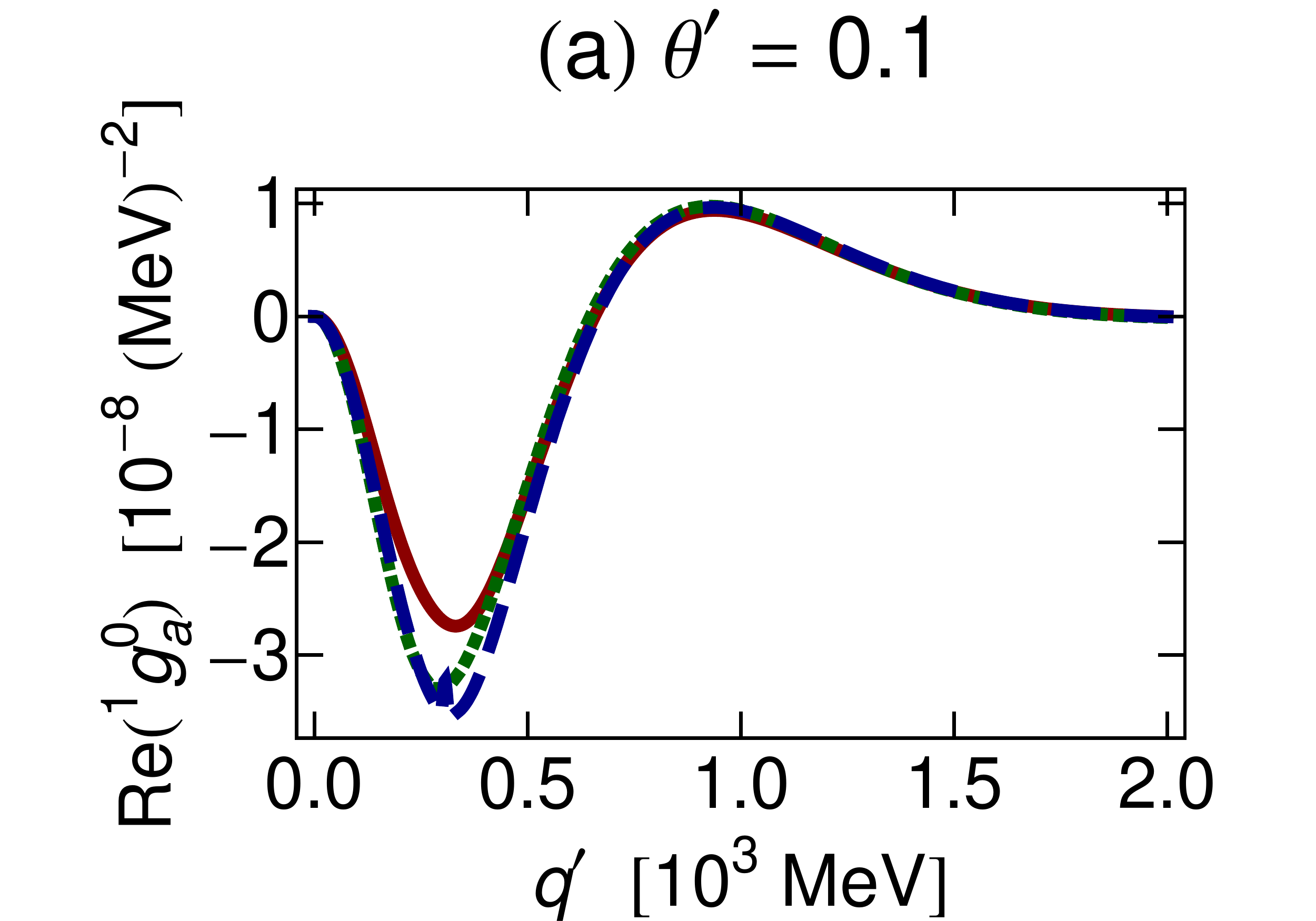}
}
\subfloat{
\includegraphics[width=6cm]{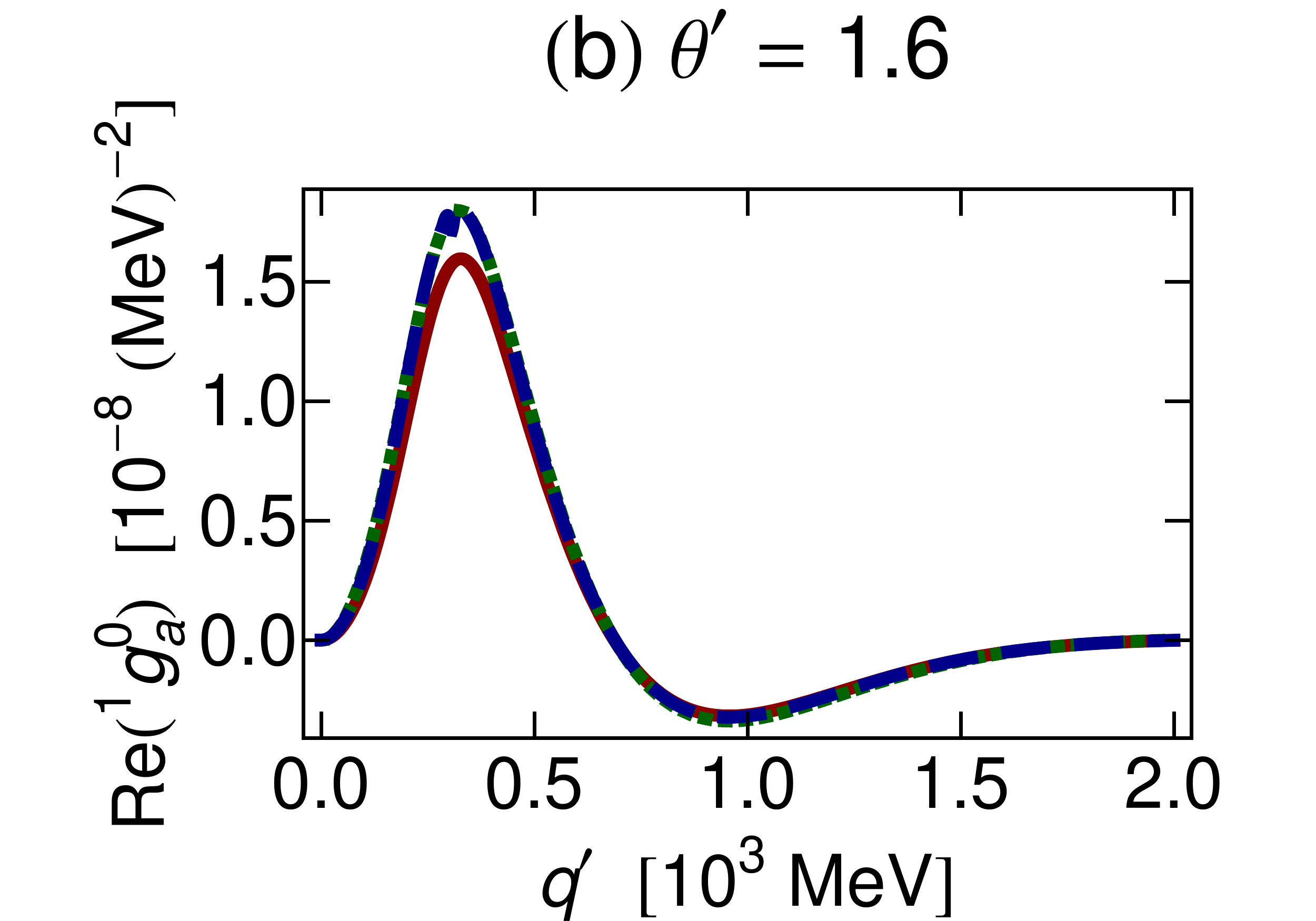}
}
\subfloat{
\includegraphics[width=6cm]{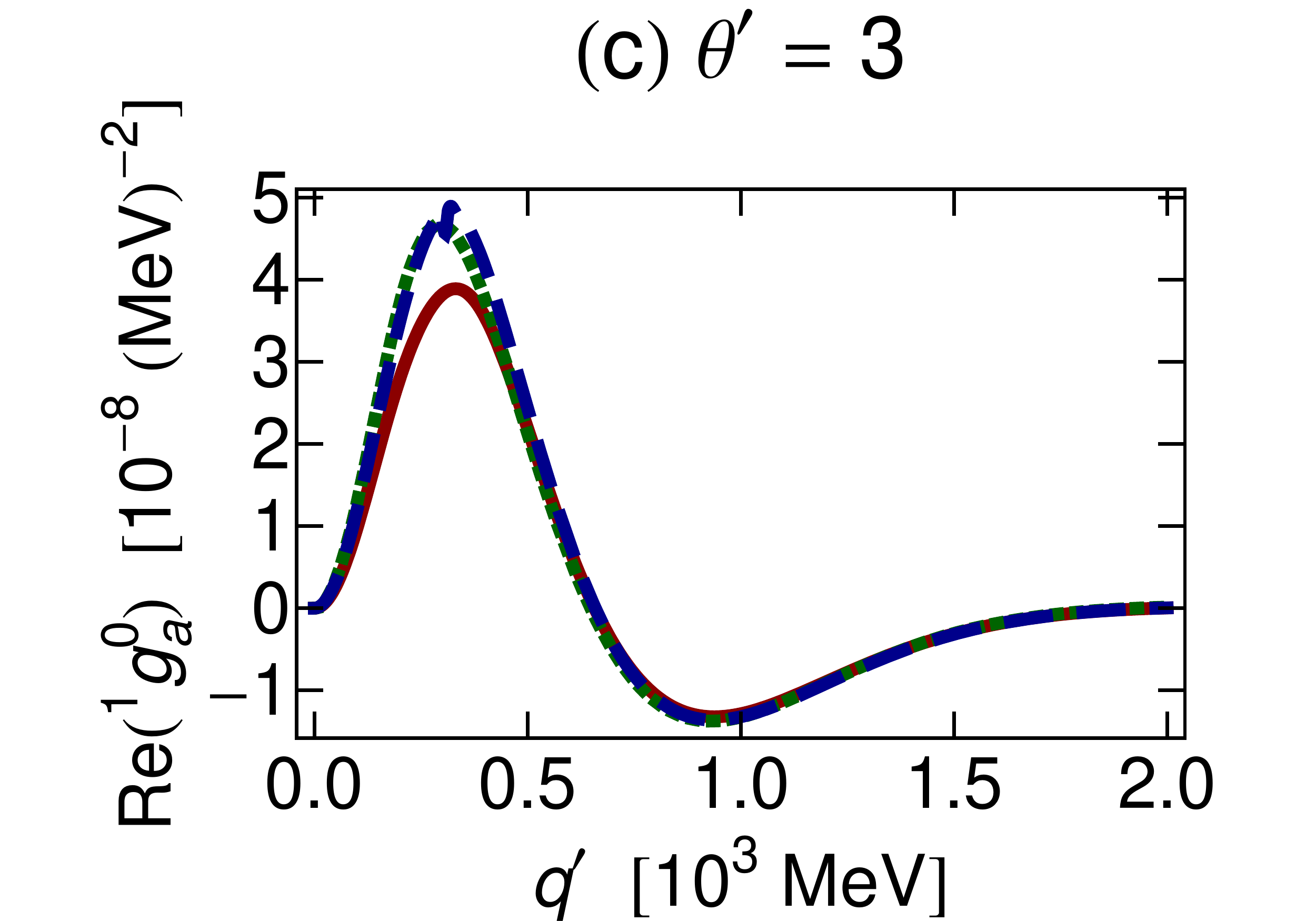}
}
\\
\subfloat{
\includegraphics[width=6cm]{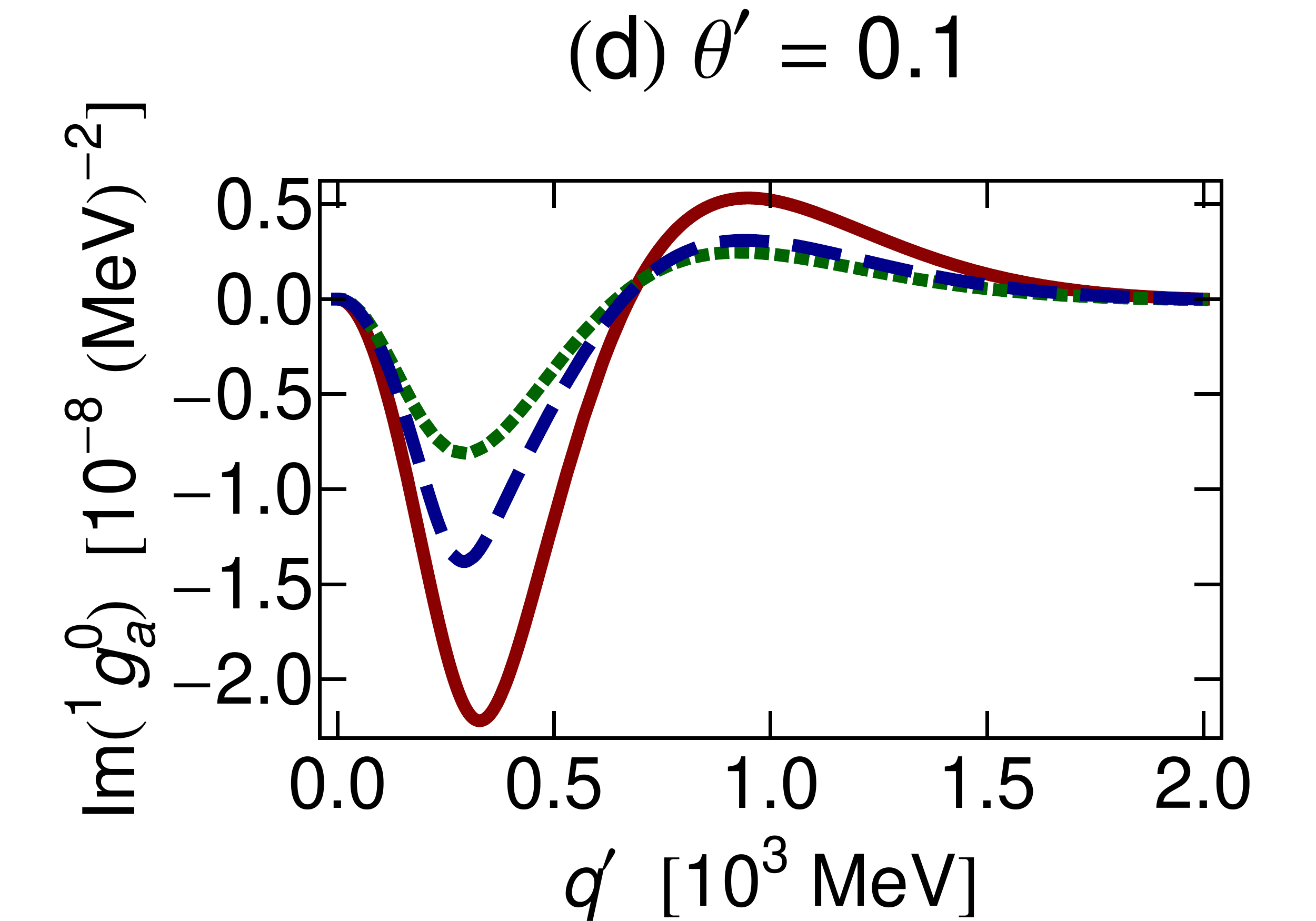}
}
\subfloat{
\includegraphics[width=6cm]{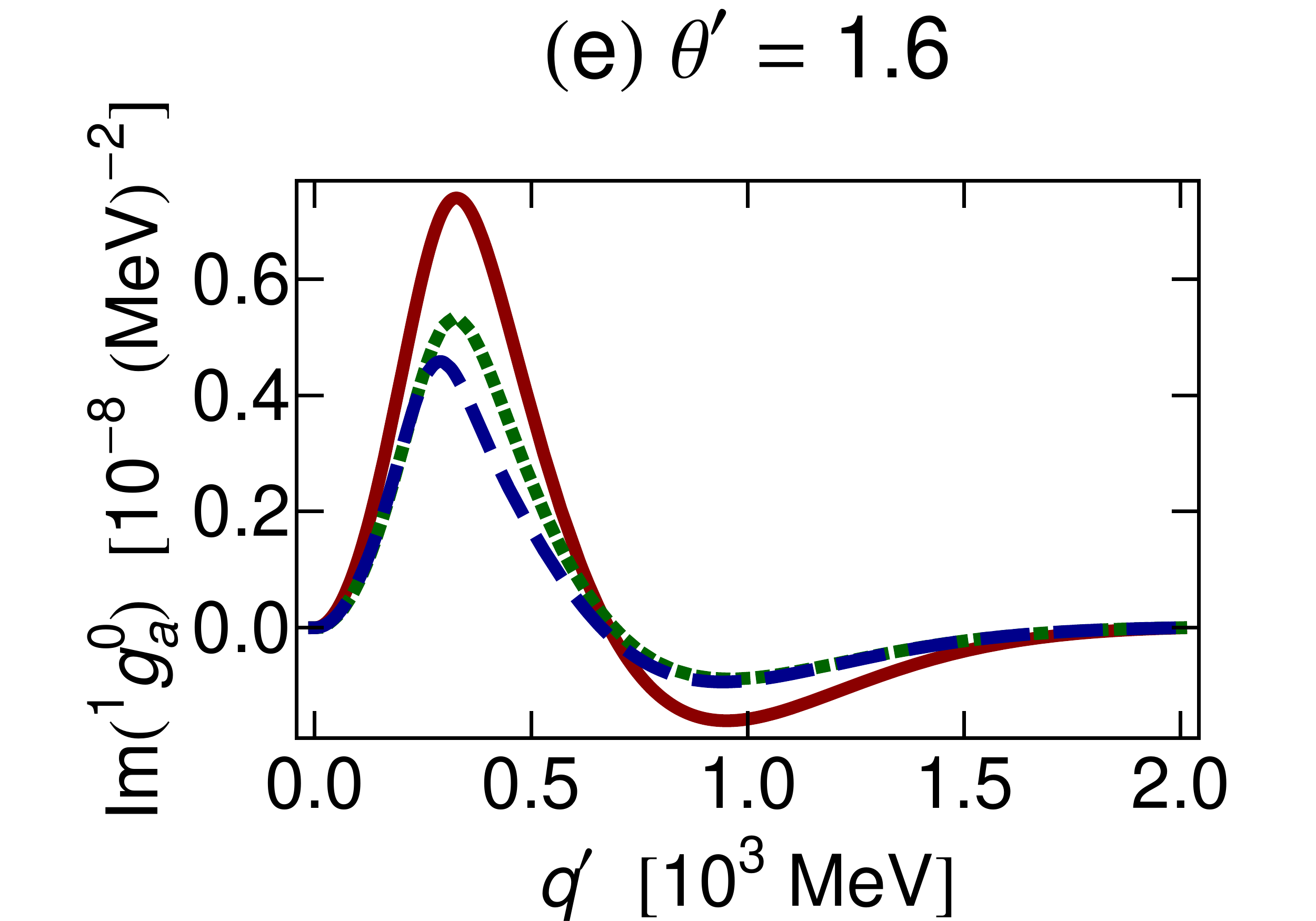}
}
\subfloat{
\includegraphics[width=6cm]{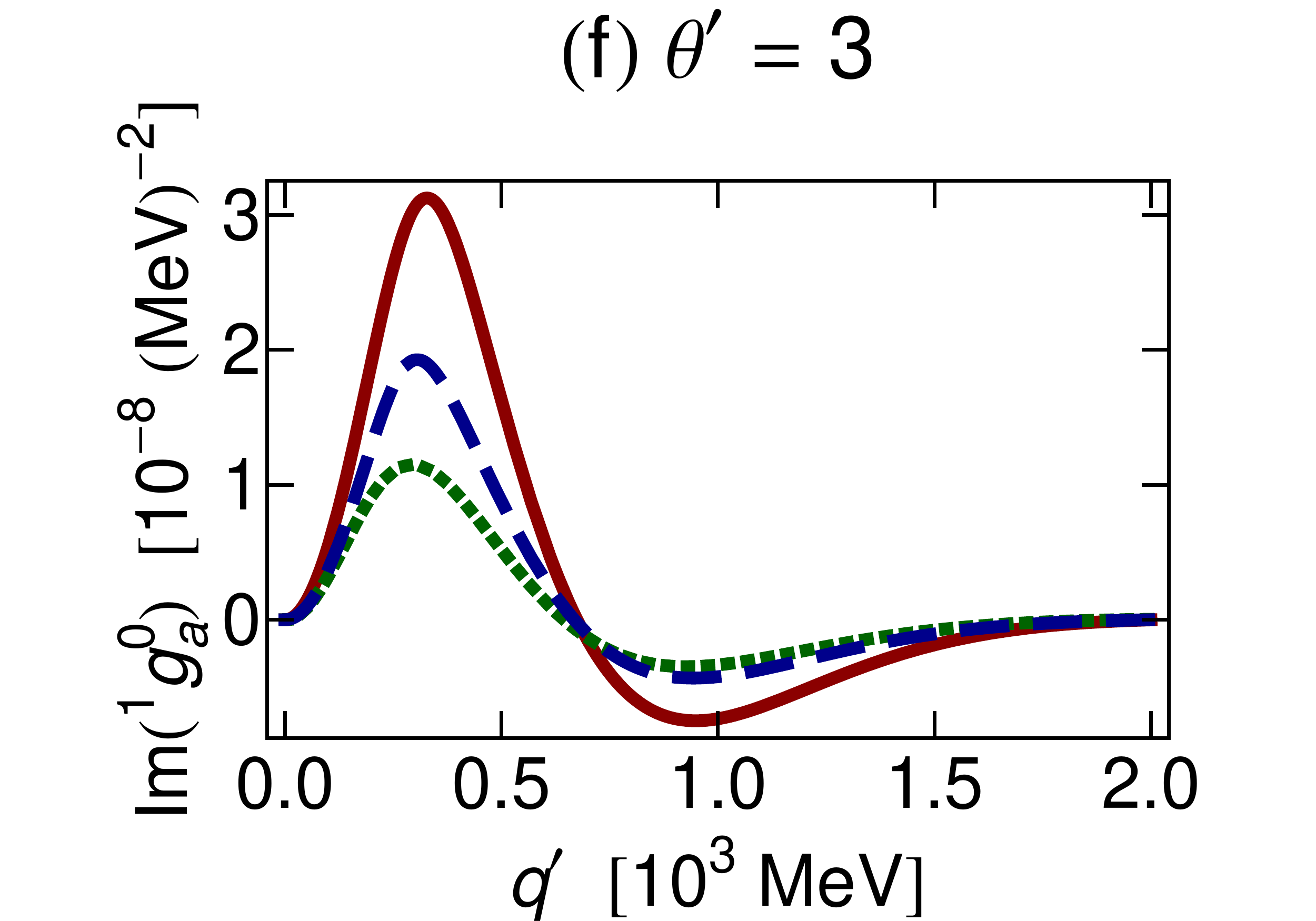}
}
\\
\subfloat{
\includegraphics[width=6cm]{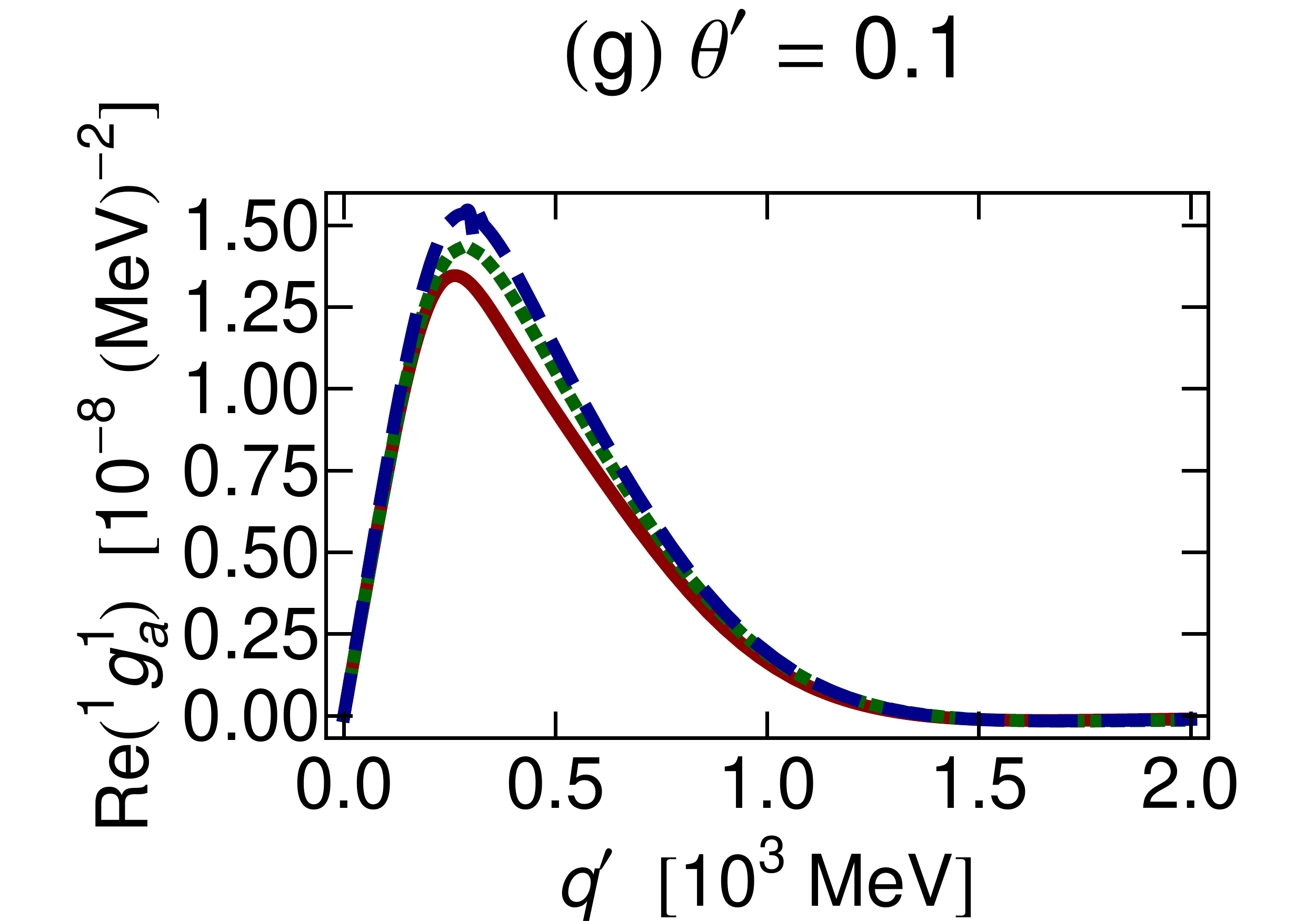}
}
\subfloat{
\includegraphics[width=6cm]{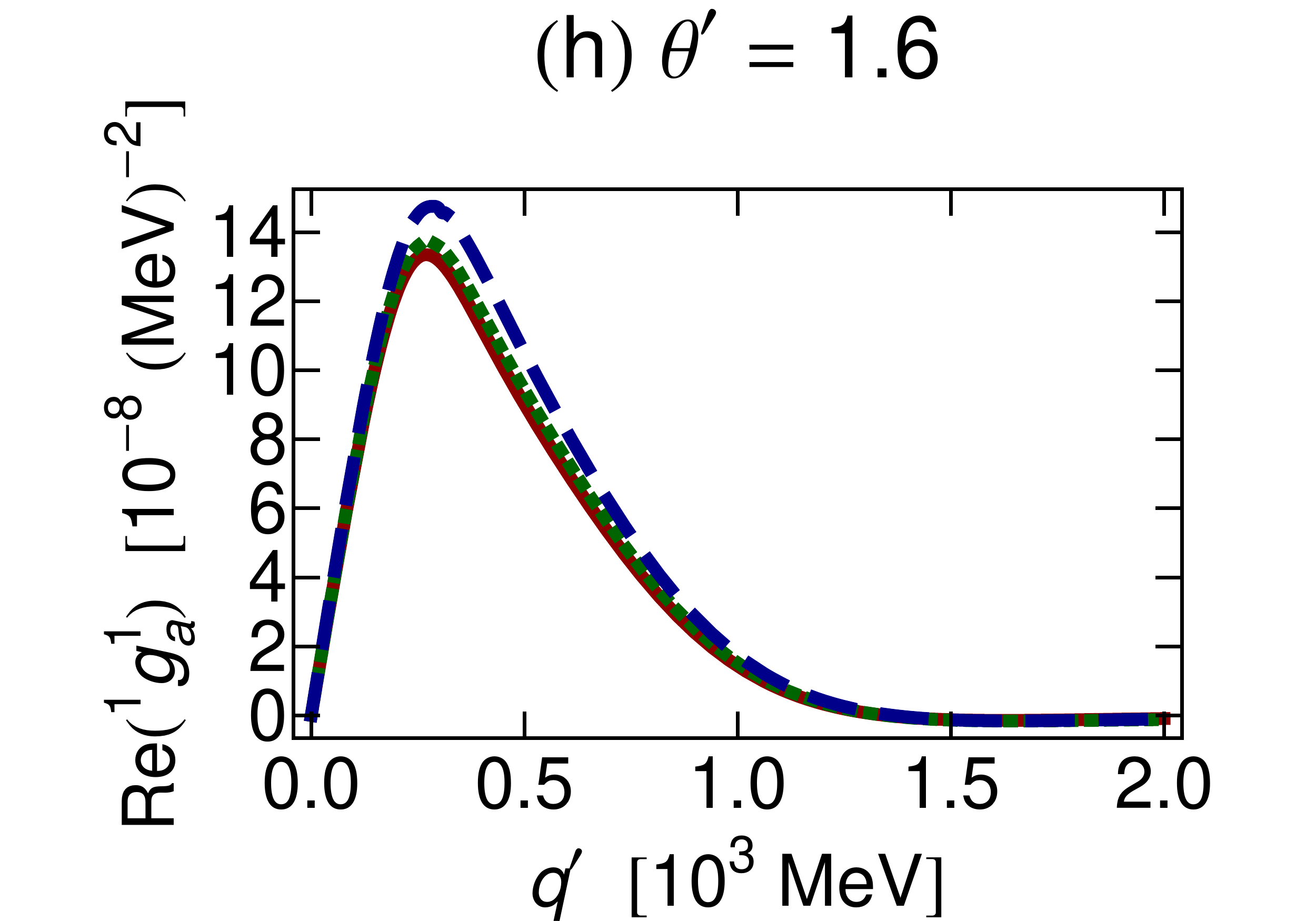}
}
\subfloat{
\includegraphics[width=6cm]{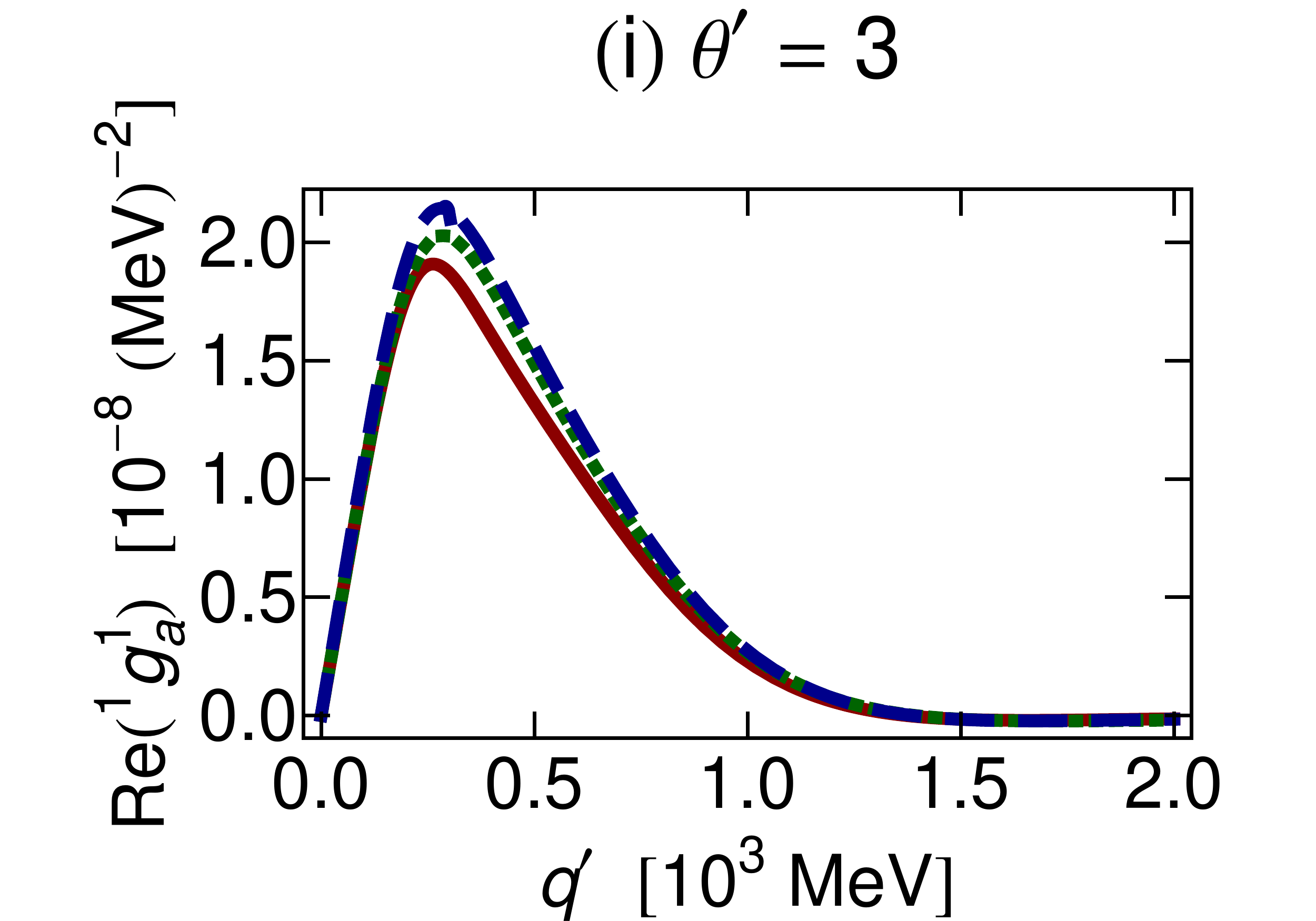}
}
\\
\subfloat{
\includegraphics[width=6cm]{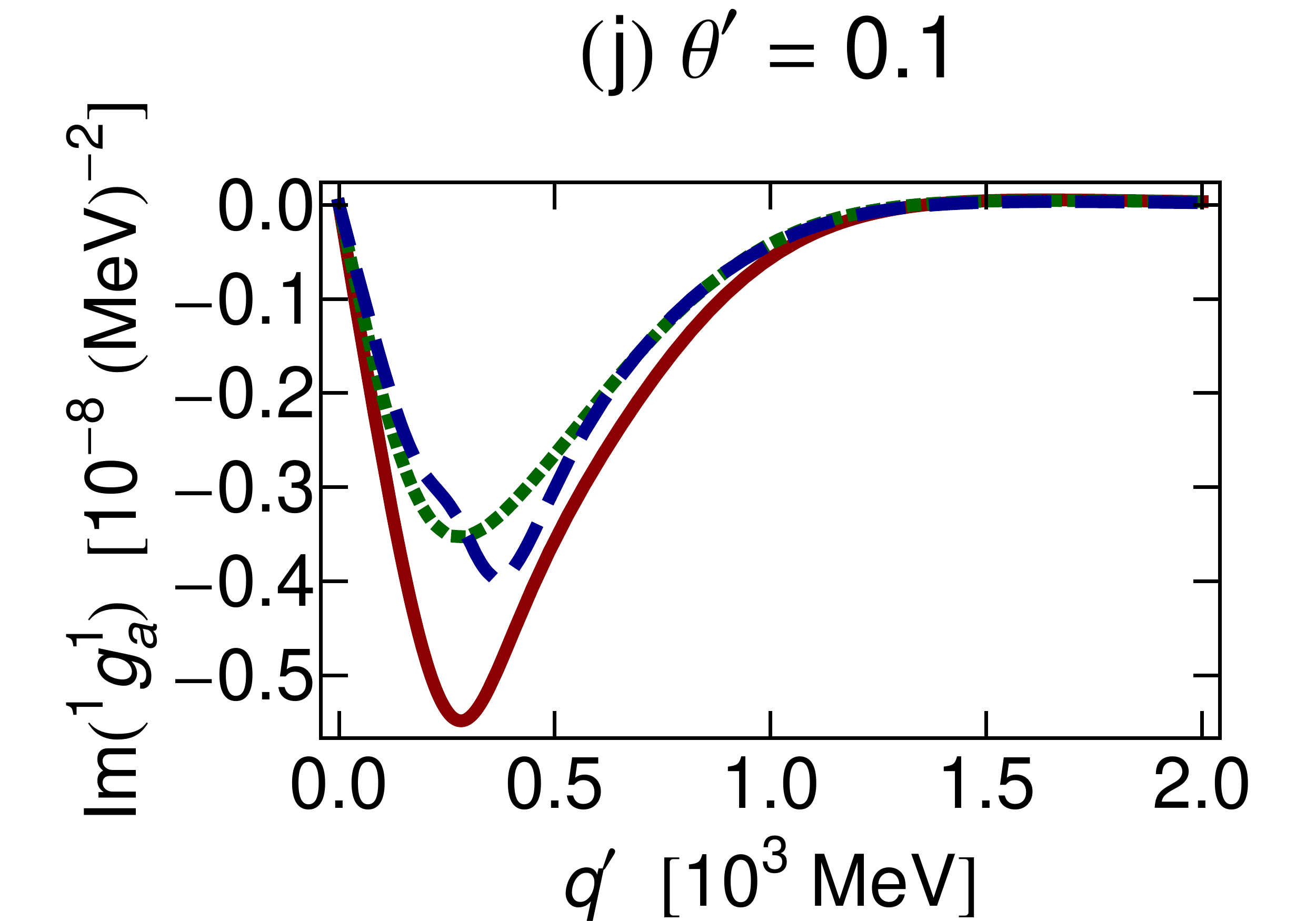}
}
\subfloat{
\includegraphics[width=6cm]{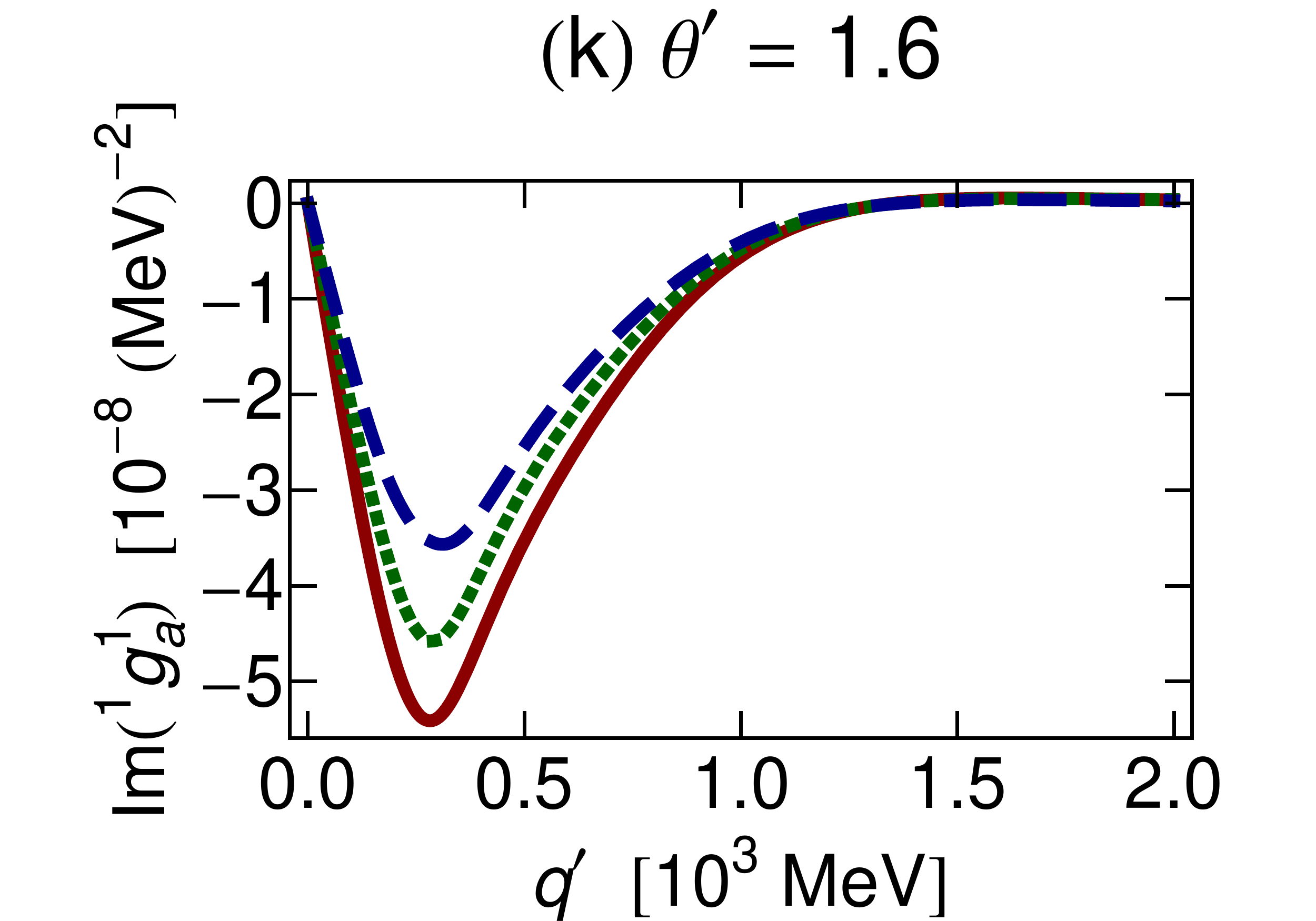}
}
\subfloat{
\includegraphics[width=6cm]{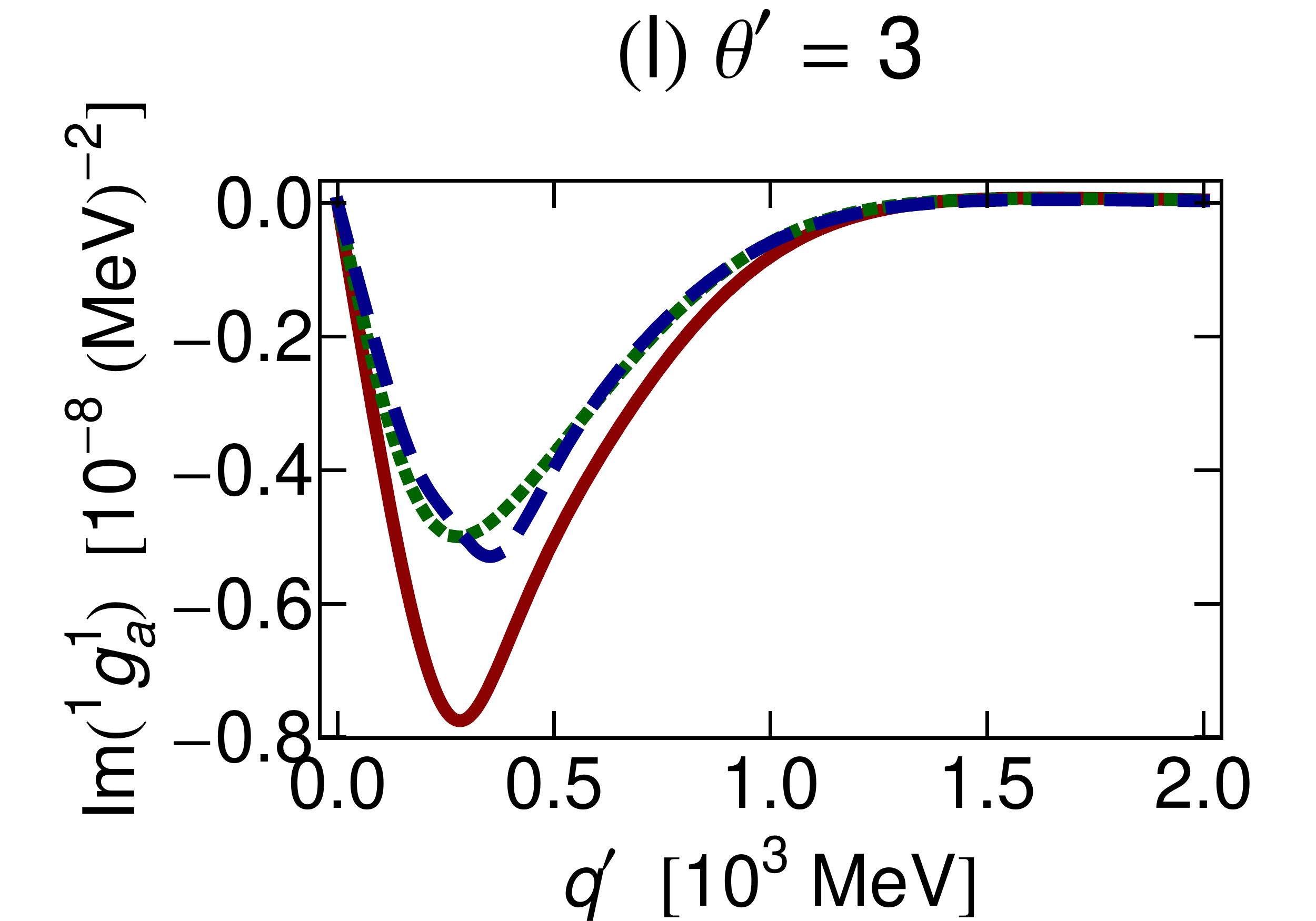}
}
\end{center}
\caption{(Color online) Same as Fig.~\ref{Fig:g_plots50_0} but for $\up{1}g^I_a$ and $\up{1}t^I_a$ at $q=306.42 \units{MeV}$.}
\label{Fig:g_plots200_1}
\end{figure}

\begin{figure}[H]
\begin{center}
\subfloat{
\includegraphics[width=6cm]{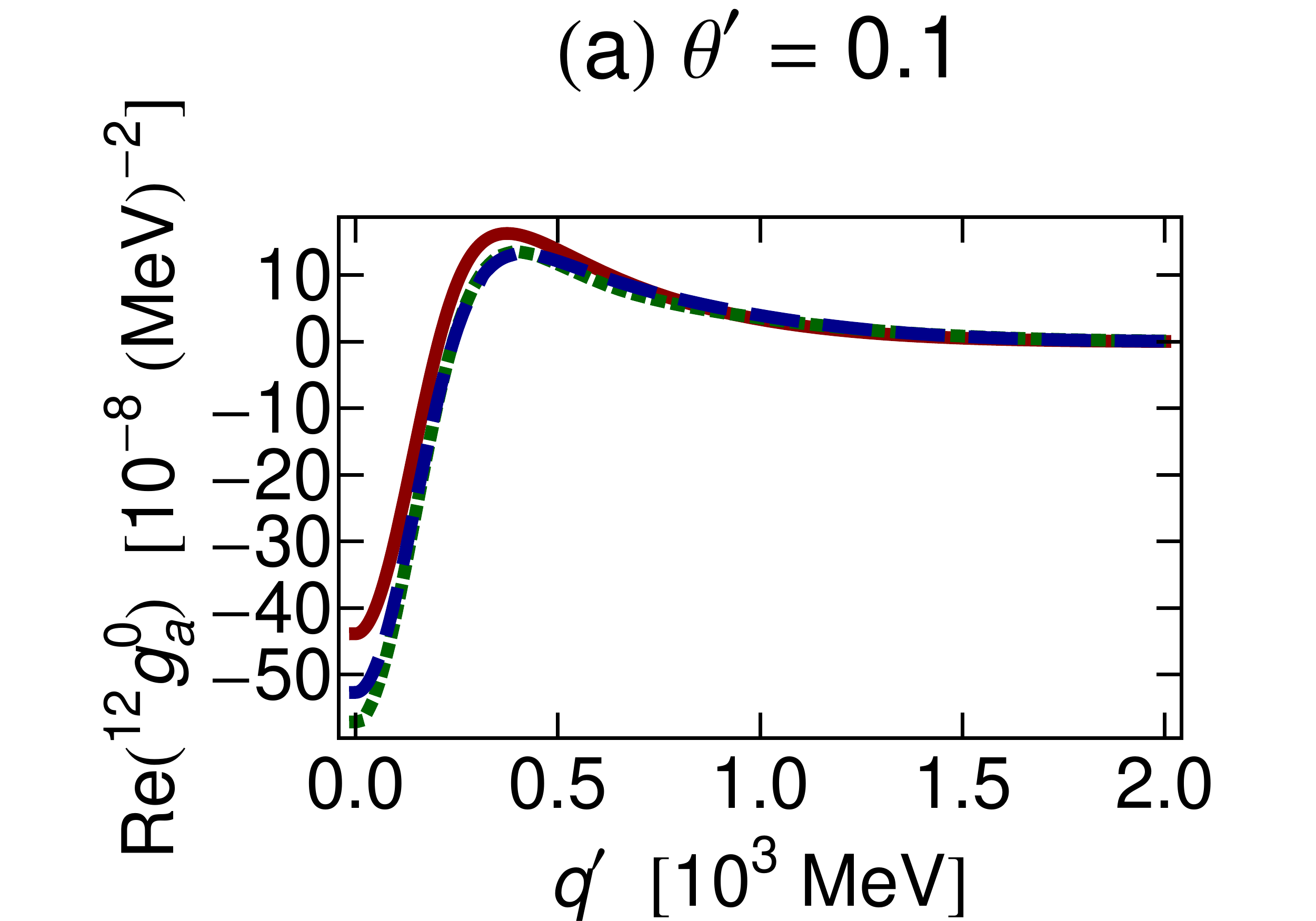}
}
\subfloat{
\includegraphics[width=6cm]{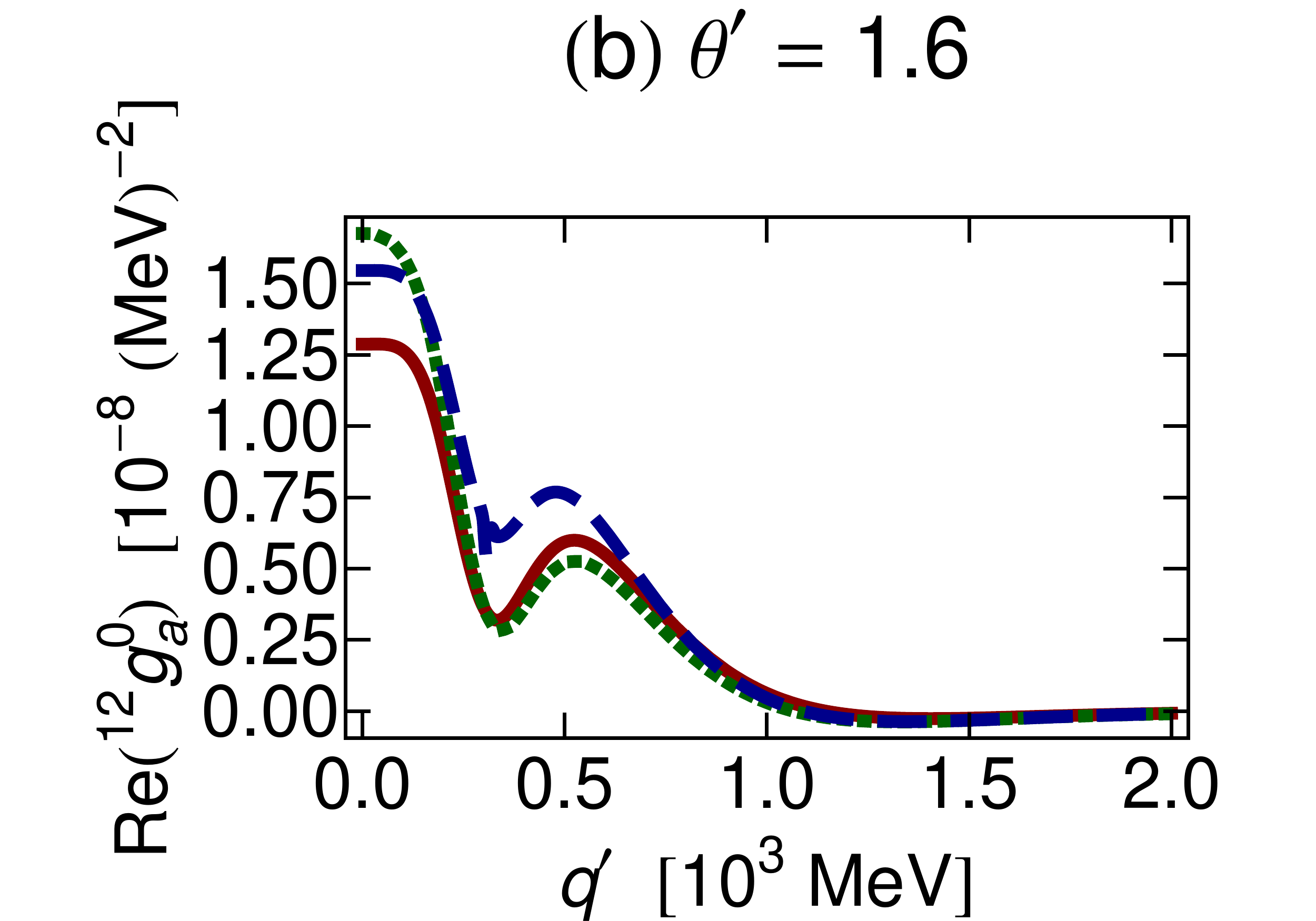}
}
\subfloat{
\includegraphics[width=6cm]{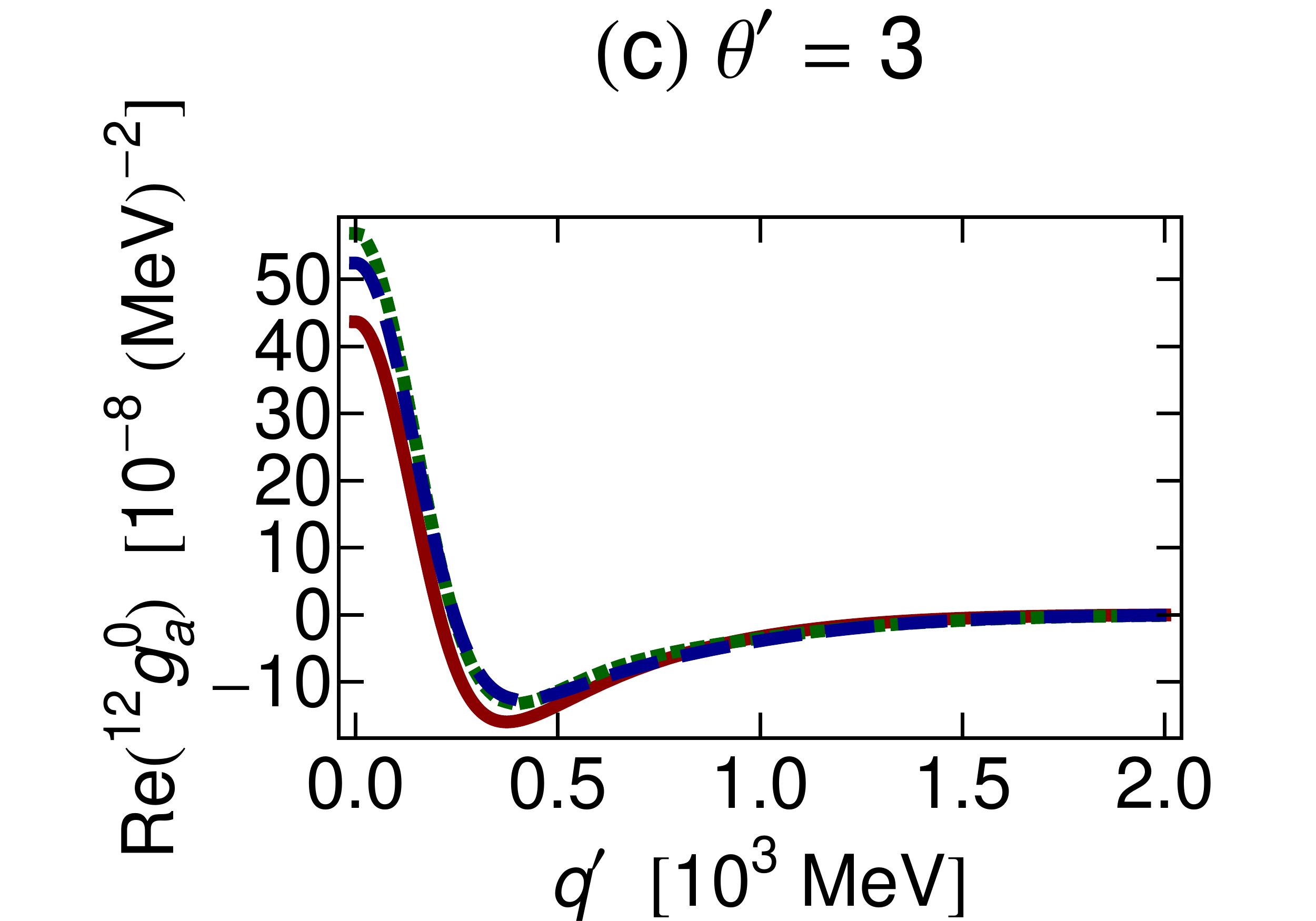}
}
\\
\subfloat{
\includegraphics[width=6cm]{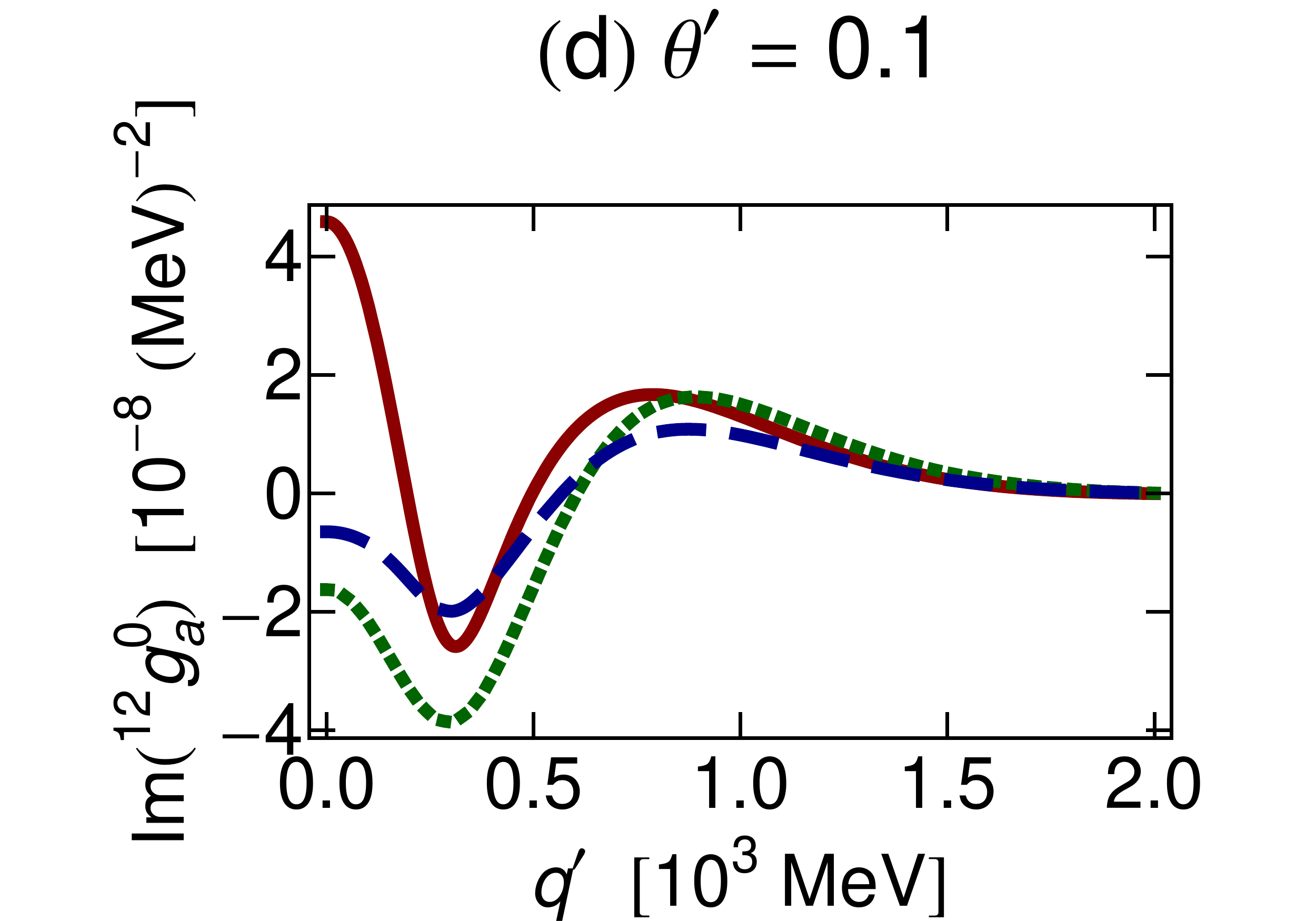}
}
\subfloat{
\includegraphics[width=6cm]{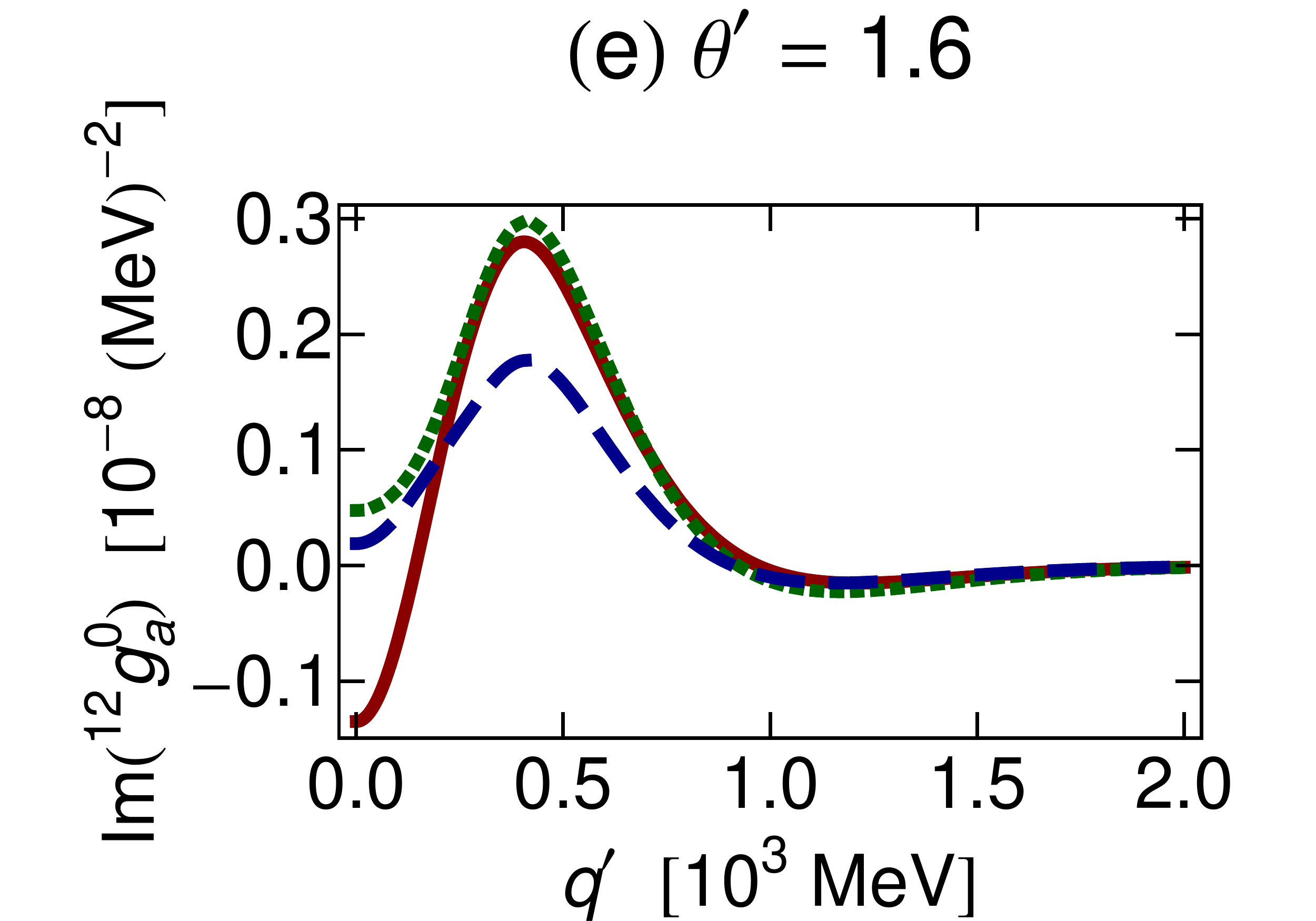}
}
\subfloat{
\includegraphics[width=6cm]{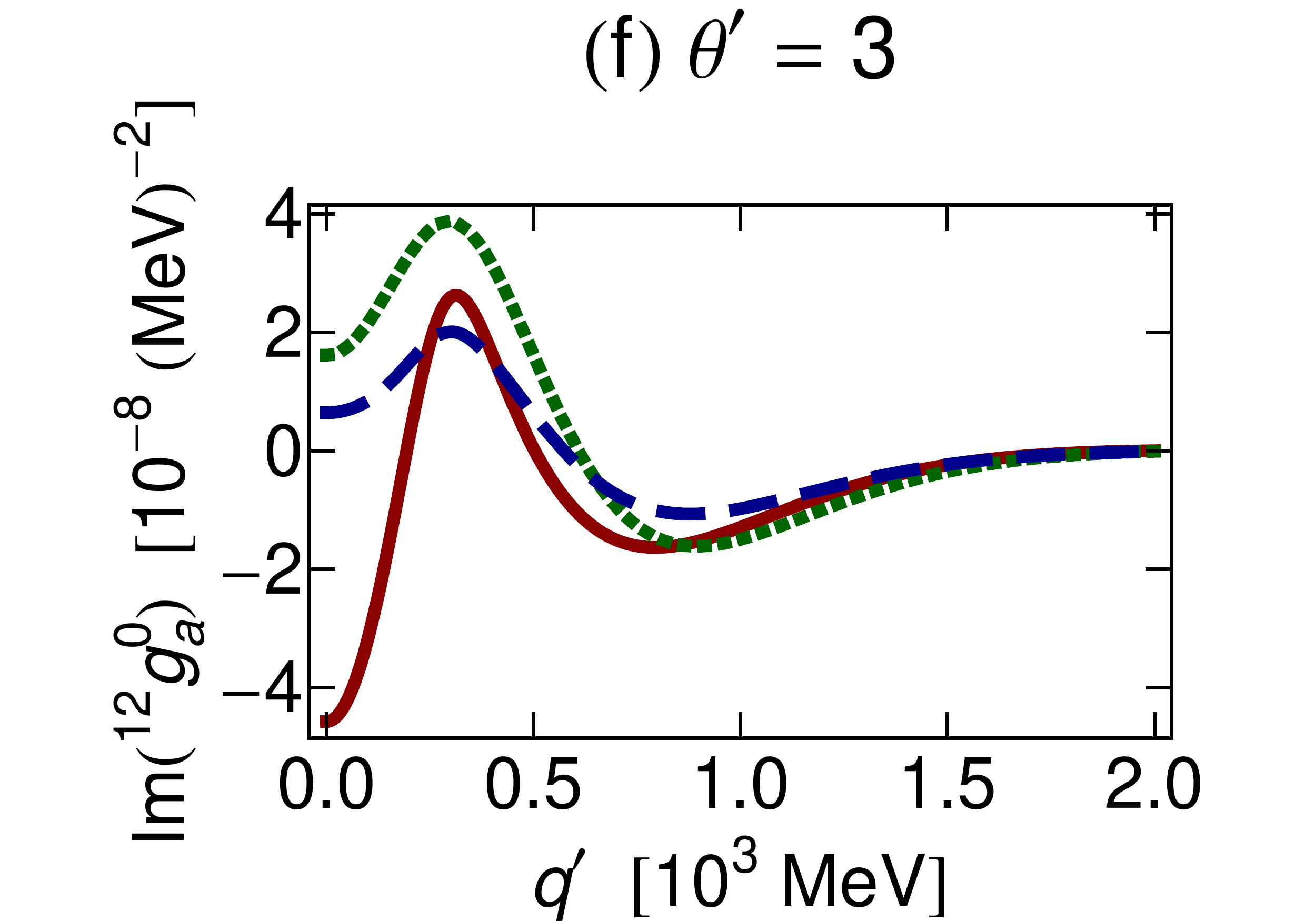}
}
\\
\subfloat{
\includegraphics[width=6cm]{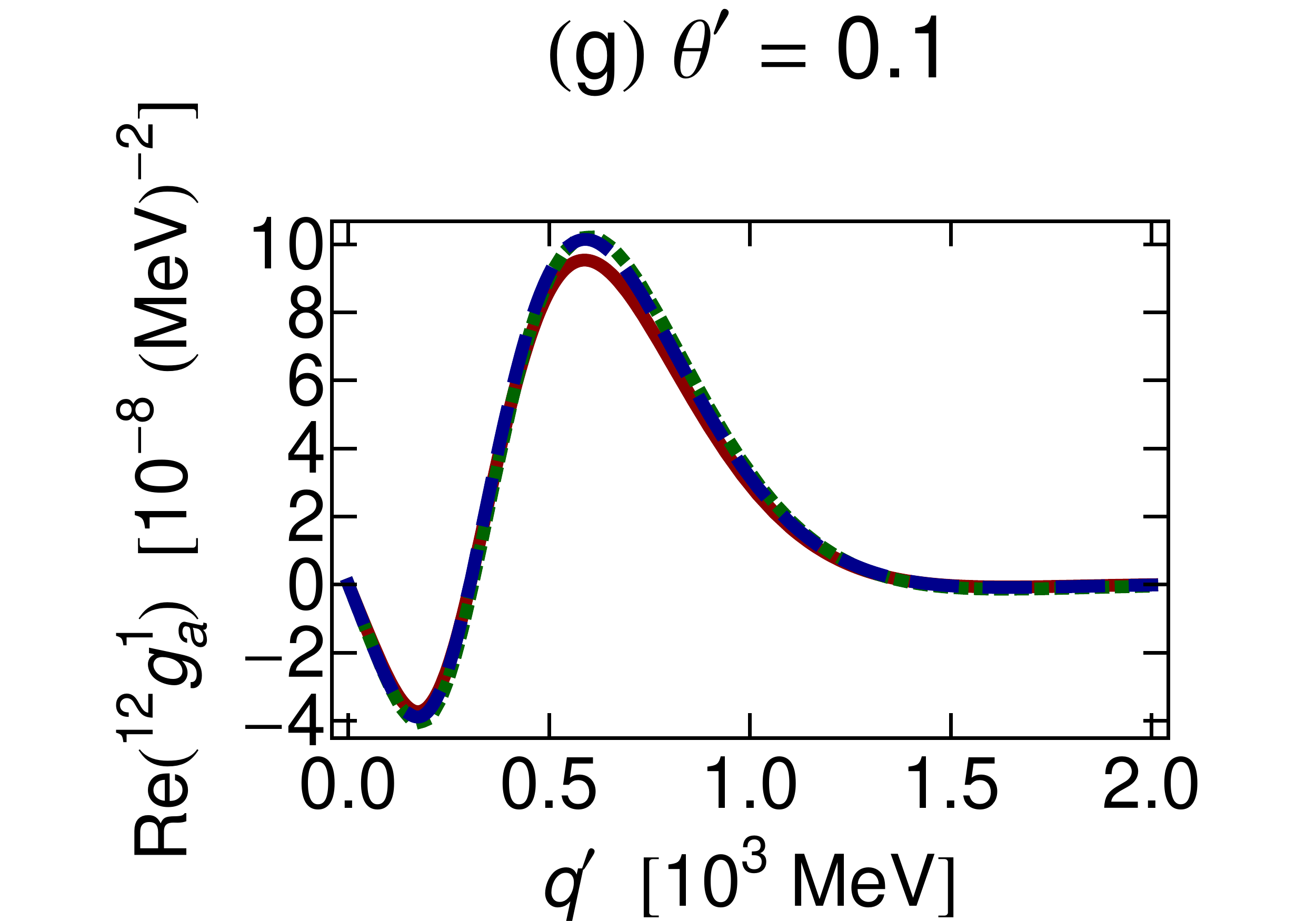}
}
\subfloat{
\includegraphics[width=6cm]{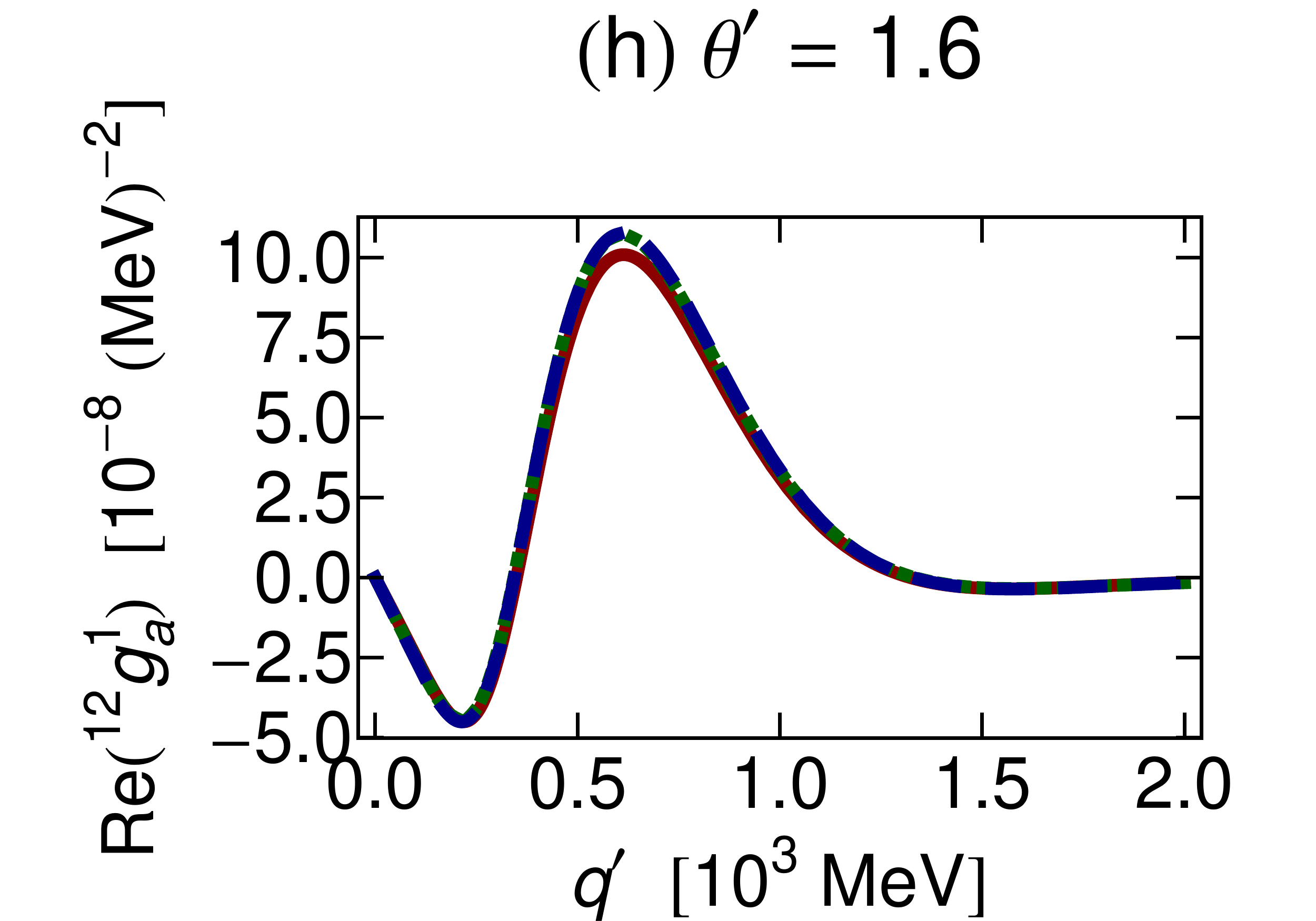}
}
\subfloat{
\includegraphics[width=6cm]{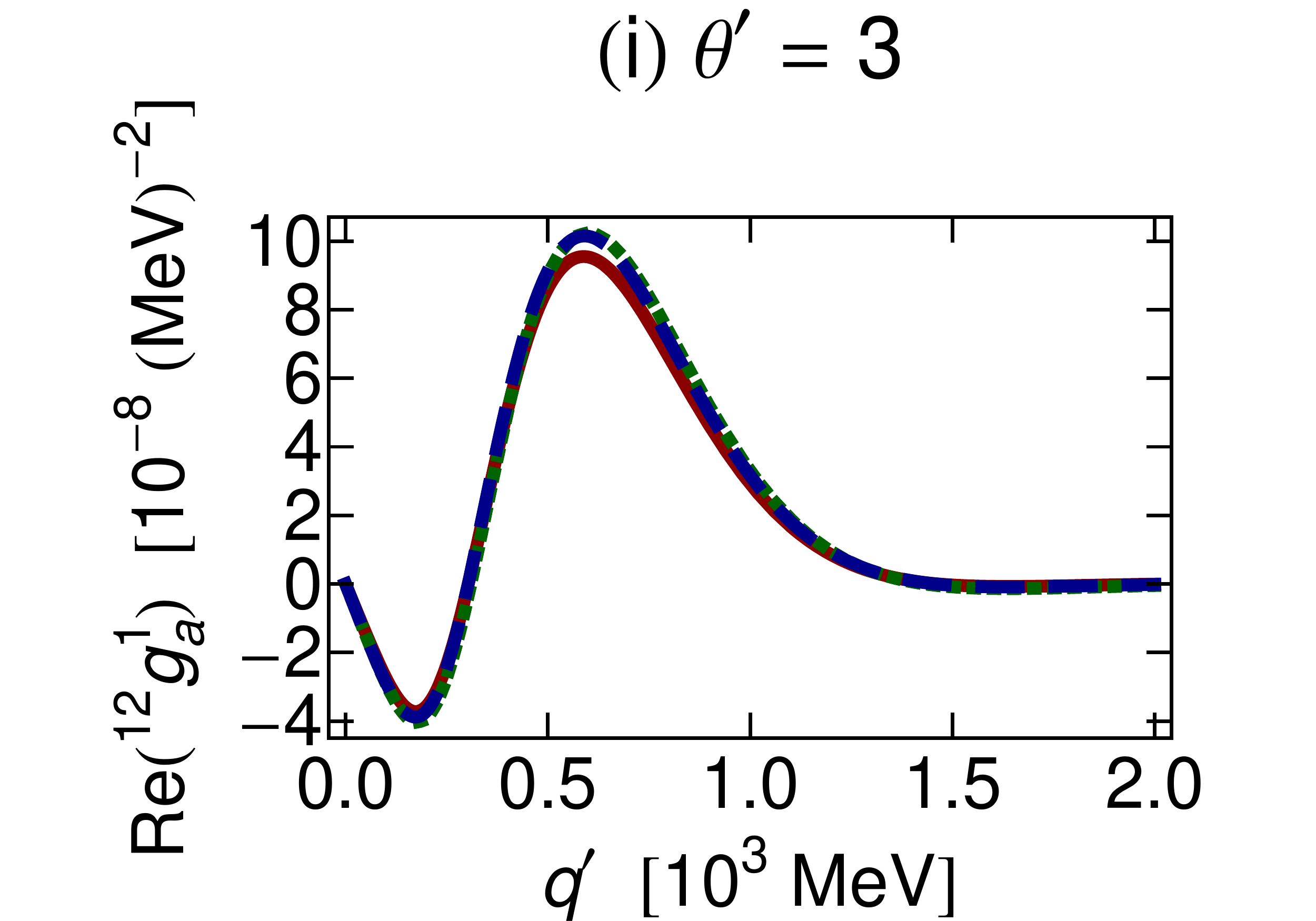}
}
\\
\subfloat{
\includegraphics[width=6cm]{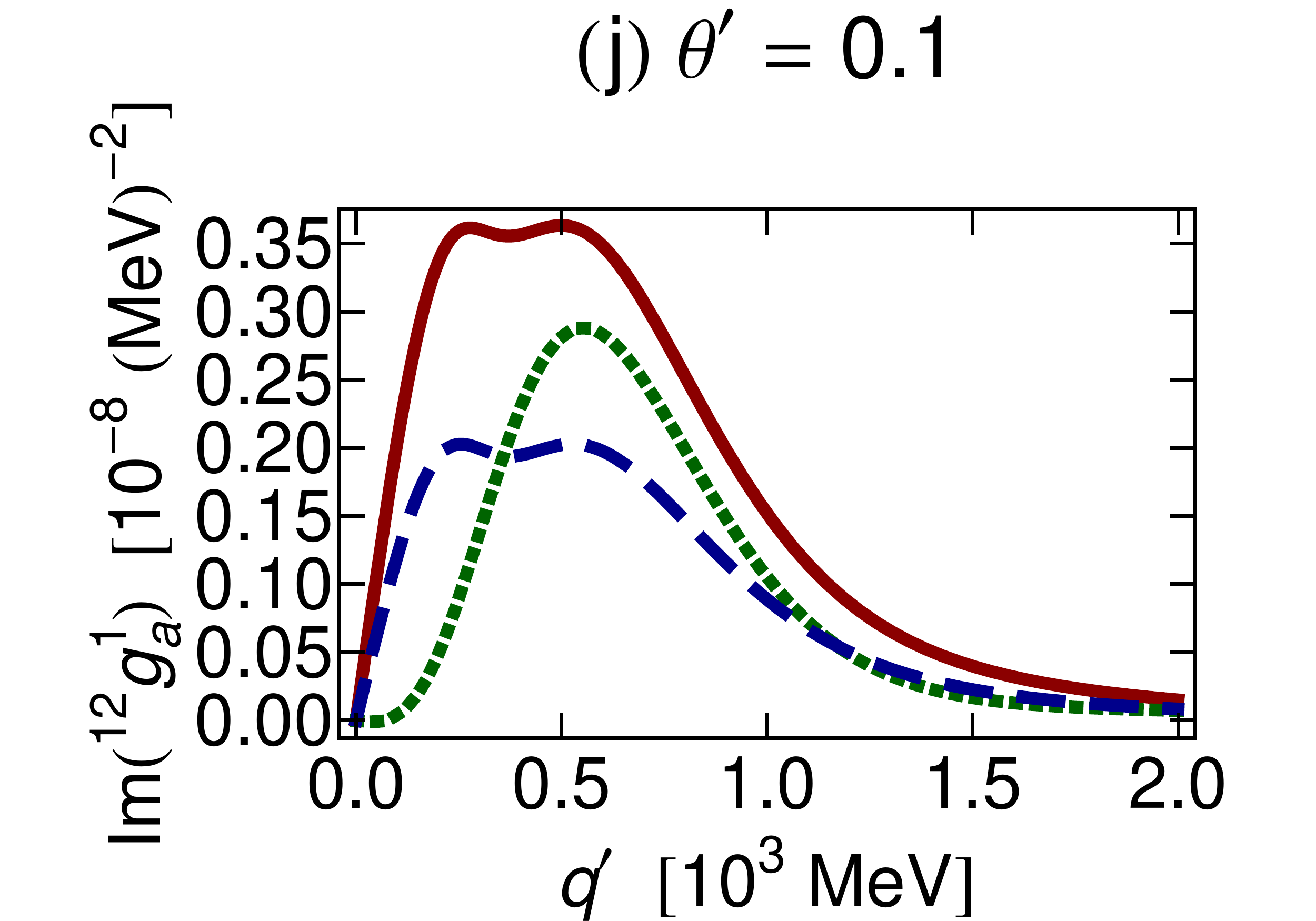}
}
\subfloat{
\includegraphics[width=6cm]{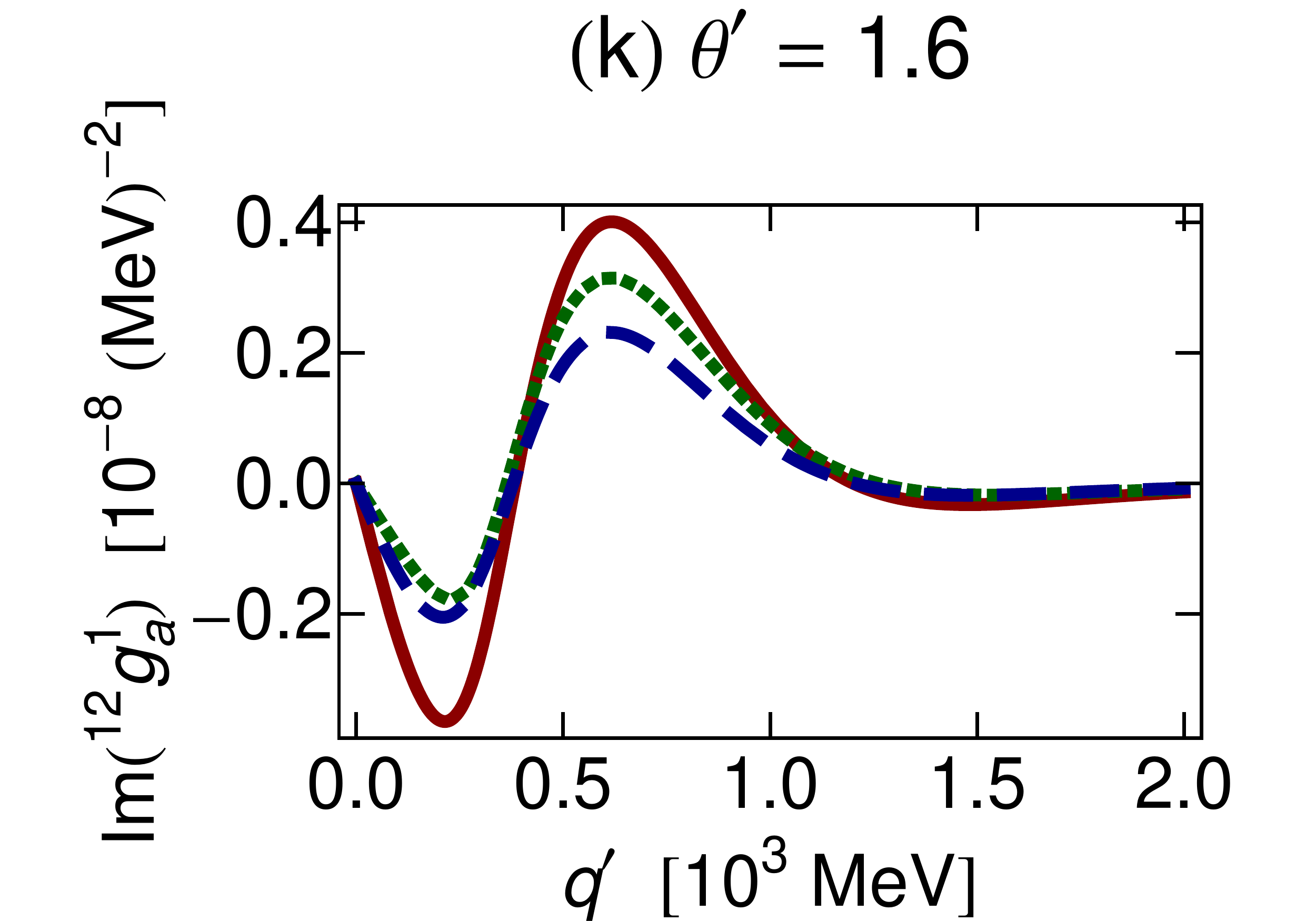}
}
\subfloat{
\includegraphics[width=6cm]{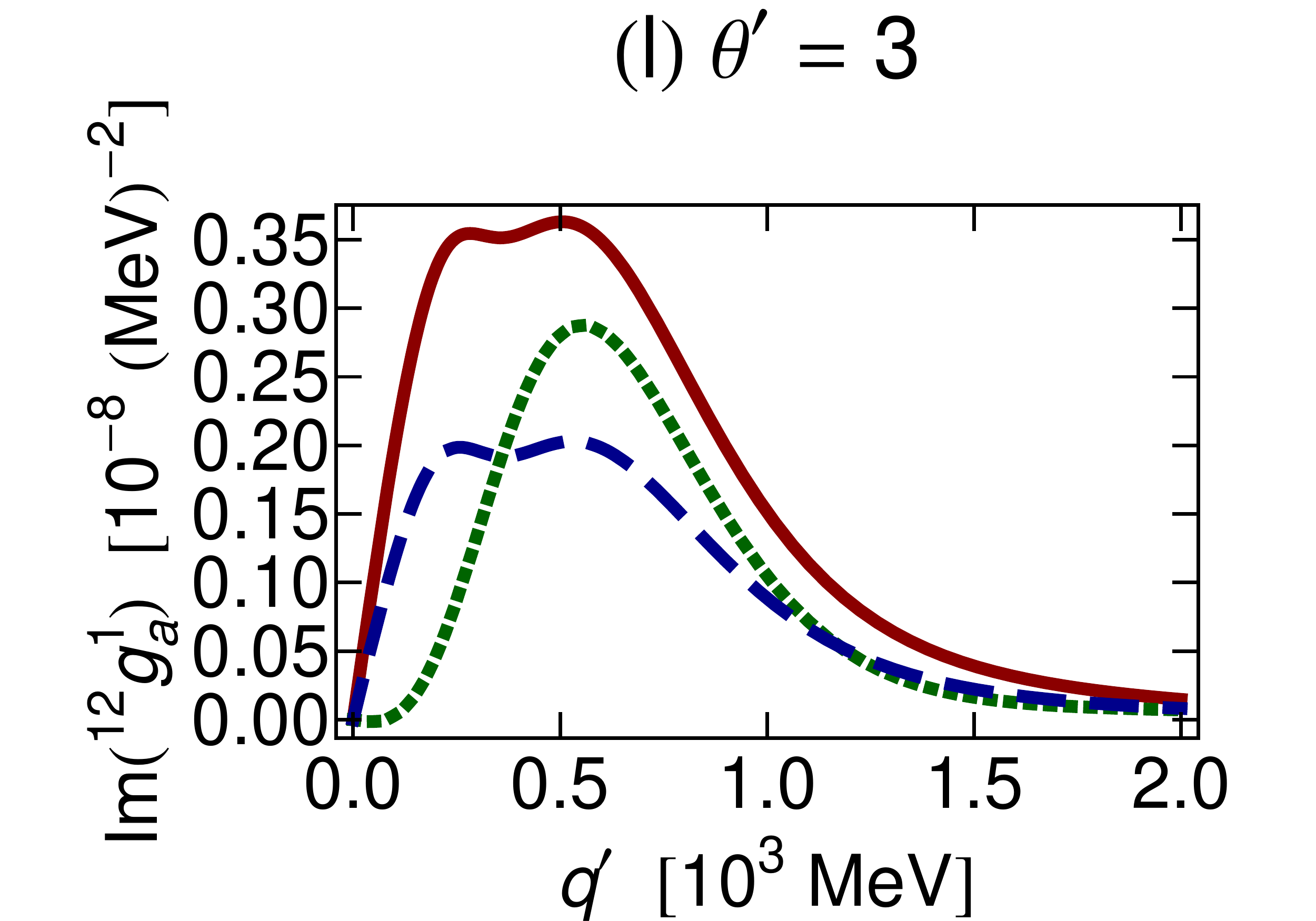}
}
\end{center}
\caption{(Color online) Same as Fig.~\ref{Fig:g_plots50_0} but for $\up{12}g^I_a$ and $\up{12}t^I_a$ at $q=306.42 \units{MeV}$.}
\label{Fig:g_plots200_12}
\end{figure}

\begin{figure}[H]
\begin{center}
\subfloat{
\includegraphics[width=6cm]{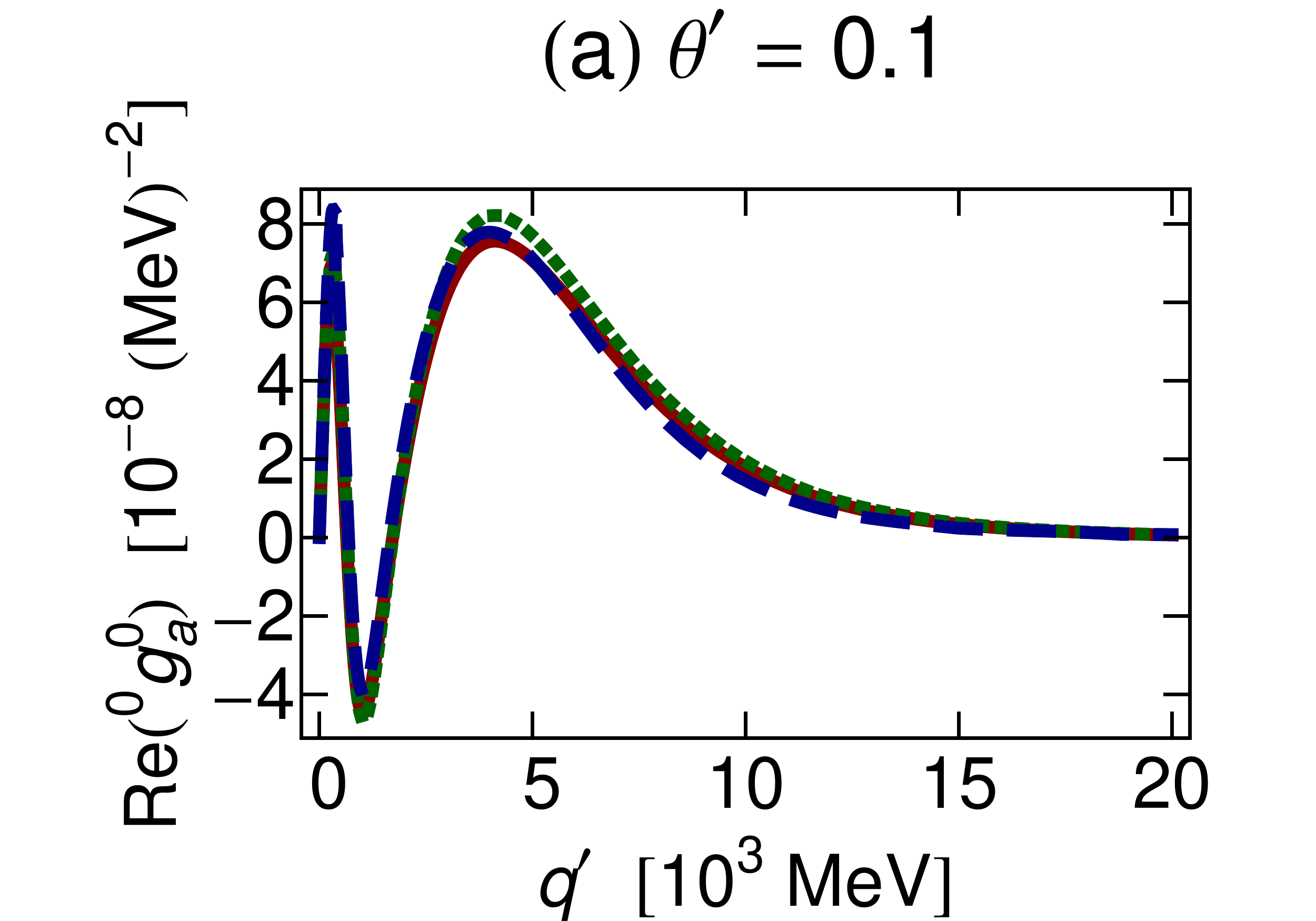}
}
\subfloat{
\includegraphics[width=6cm]{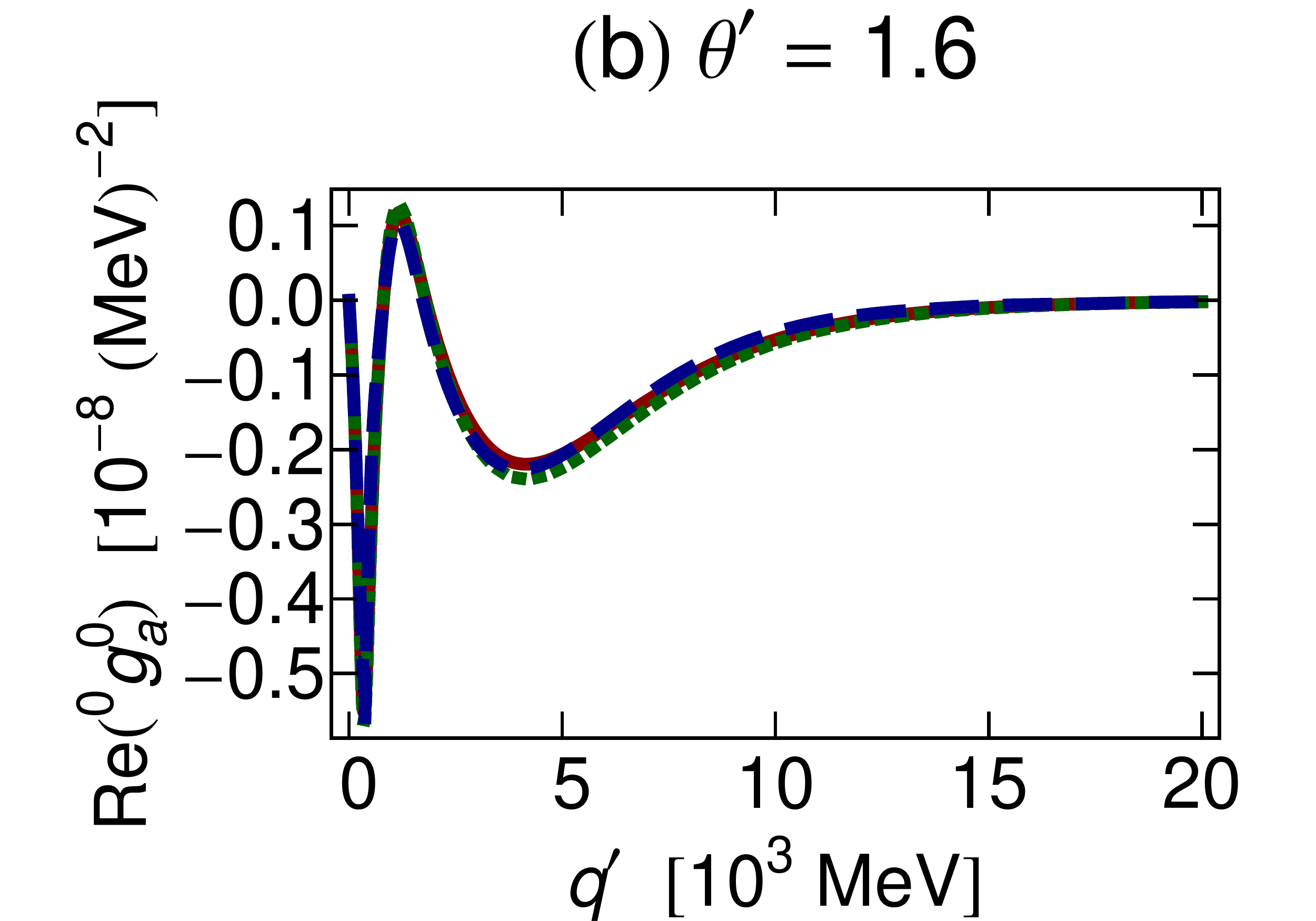}
}
\subfloat{
\includegraphics[width=6cm]{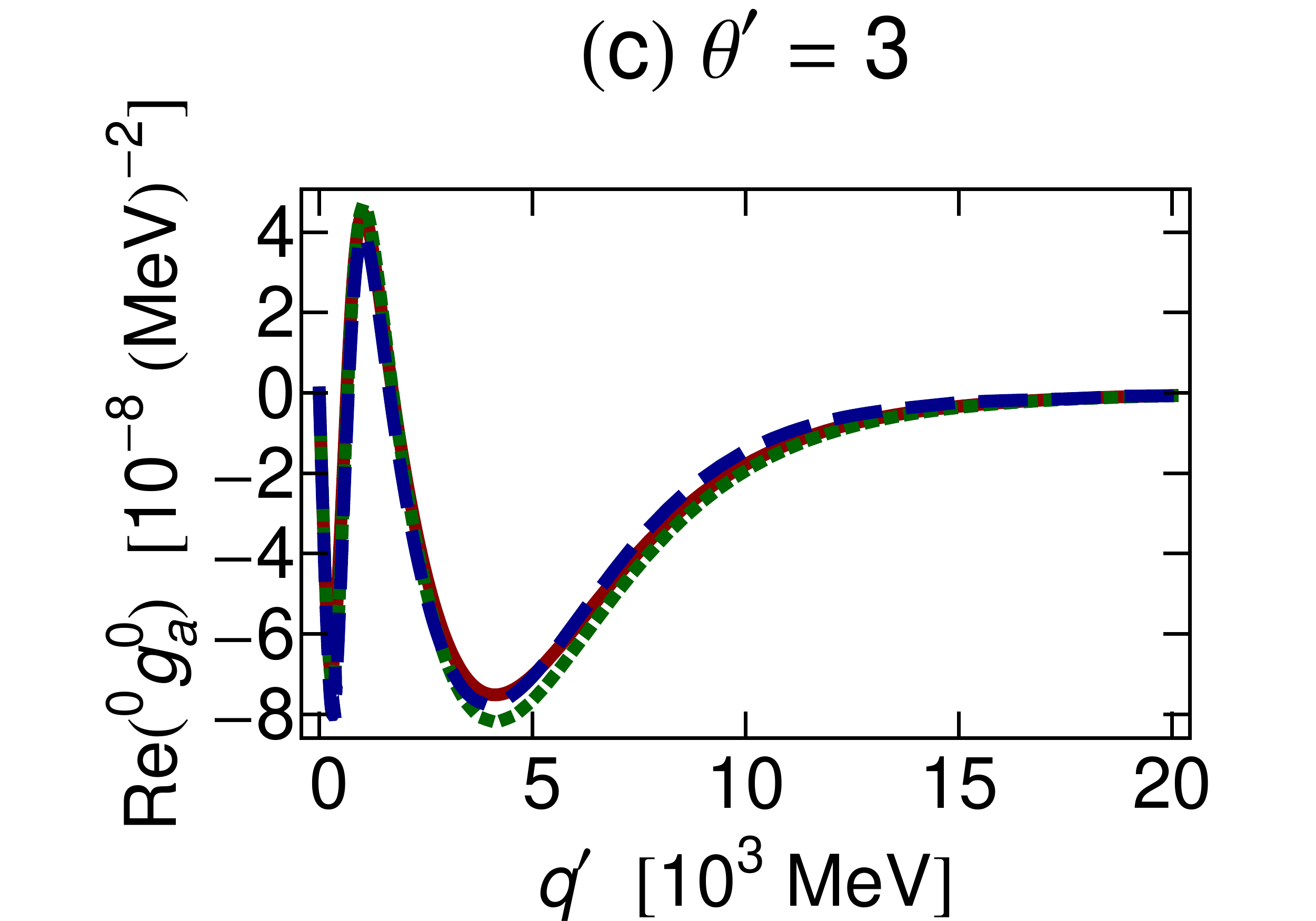}
}
\\
\subfloat{
\includegraphics[width=6cm]{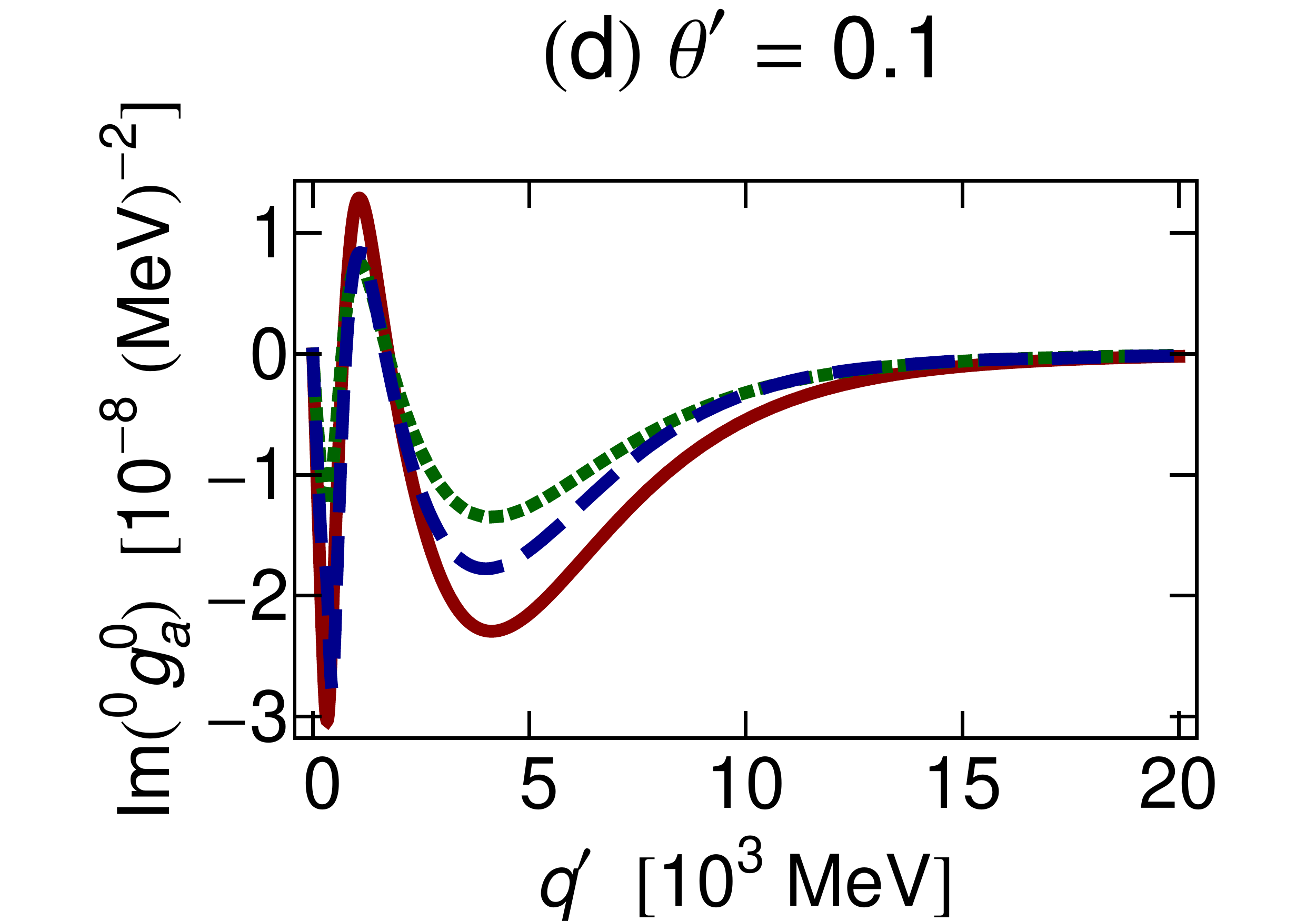}
}
\subfloat{
\includegraphics[width=6cm]{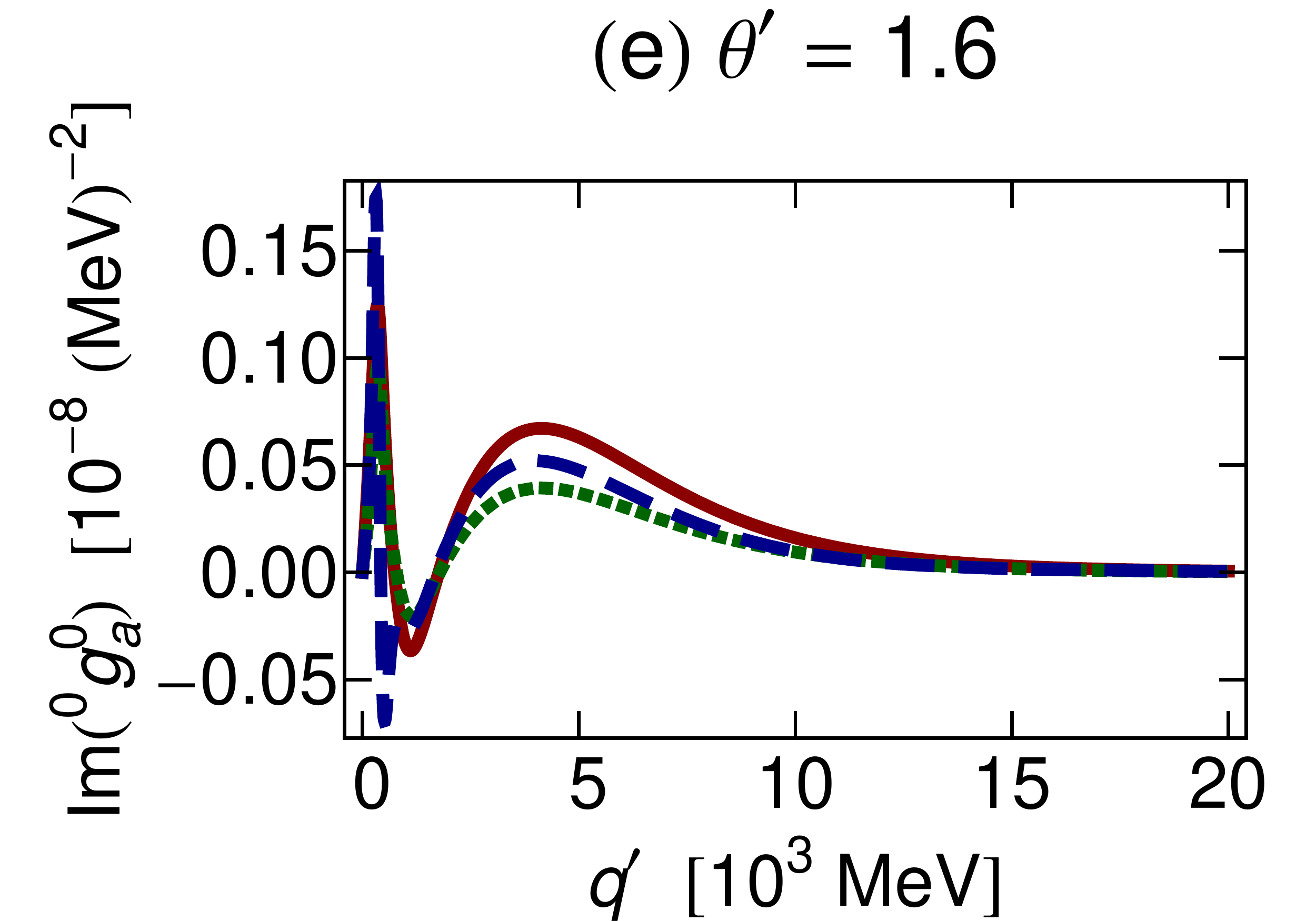}
}
\subfloat{
\includegraphics[width=6cm]{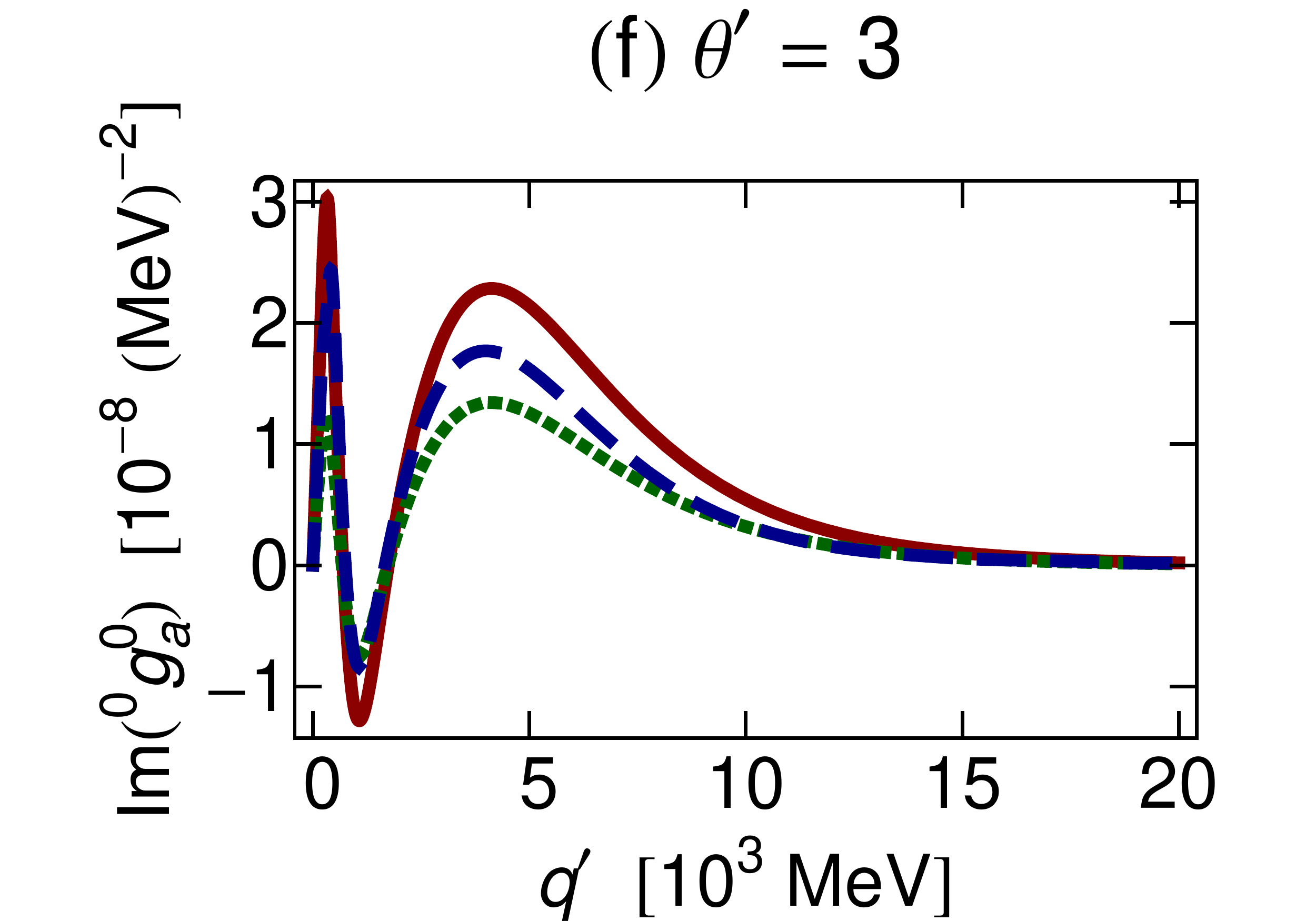}
}
\\
\subfloat{
\includegraphics[width=6cm]{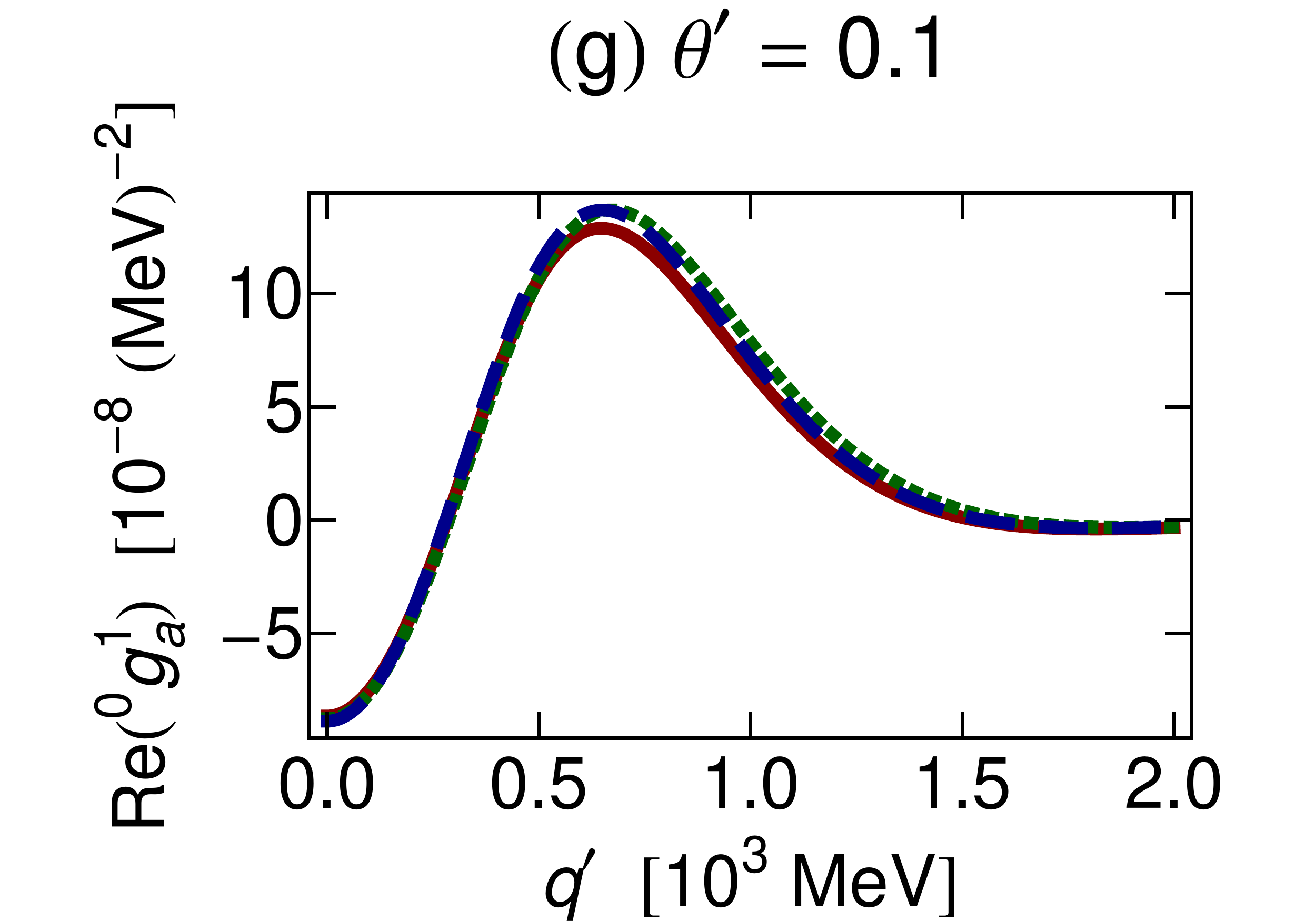}
}
\subfloat{
\includegraphics[width=6cm]{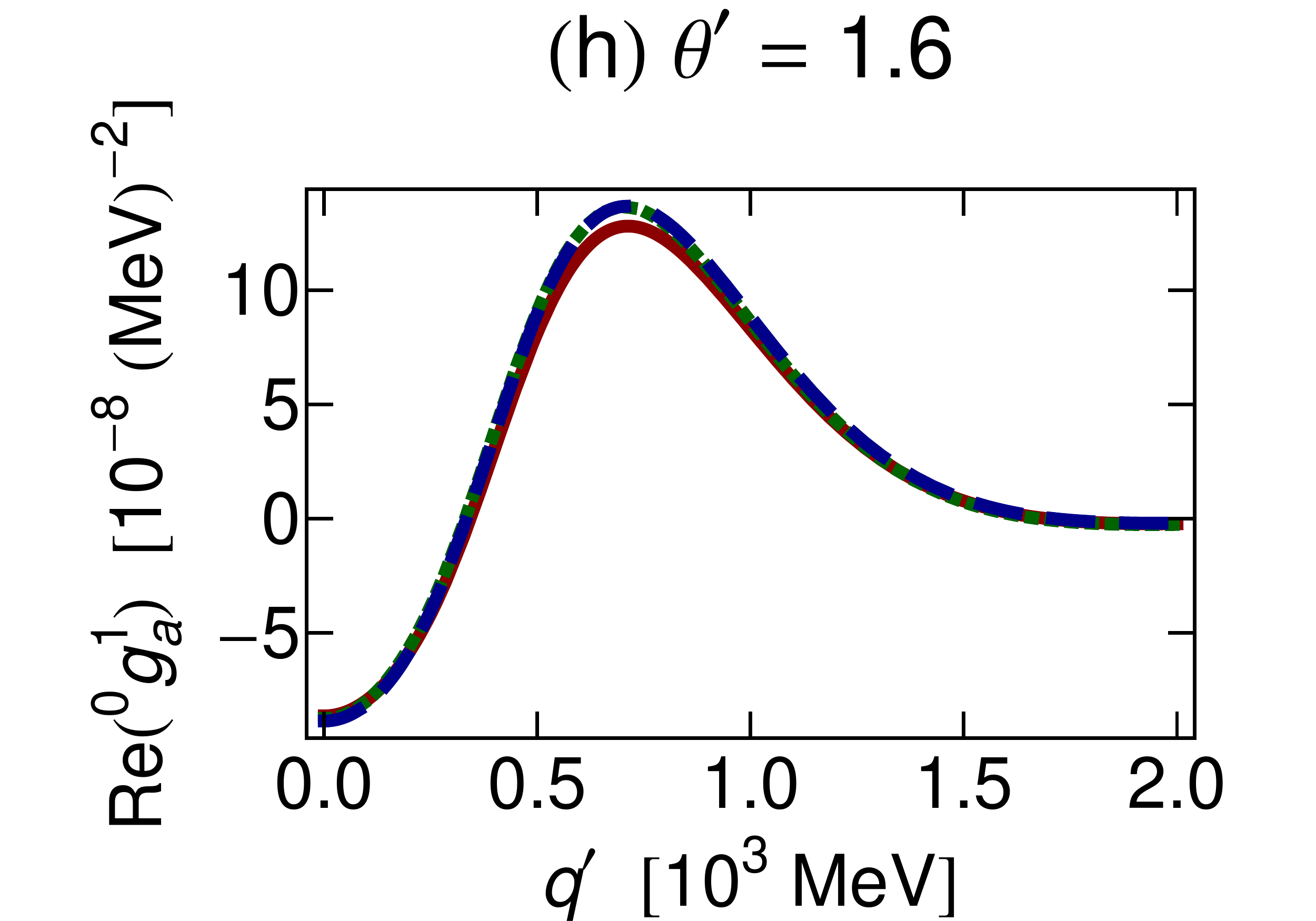}
}
\subfloat{
\includegraphics[width=6cm]{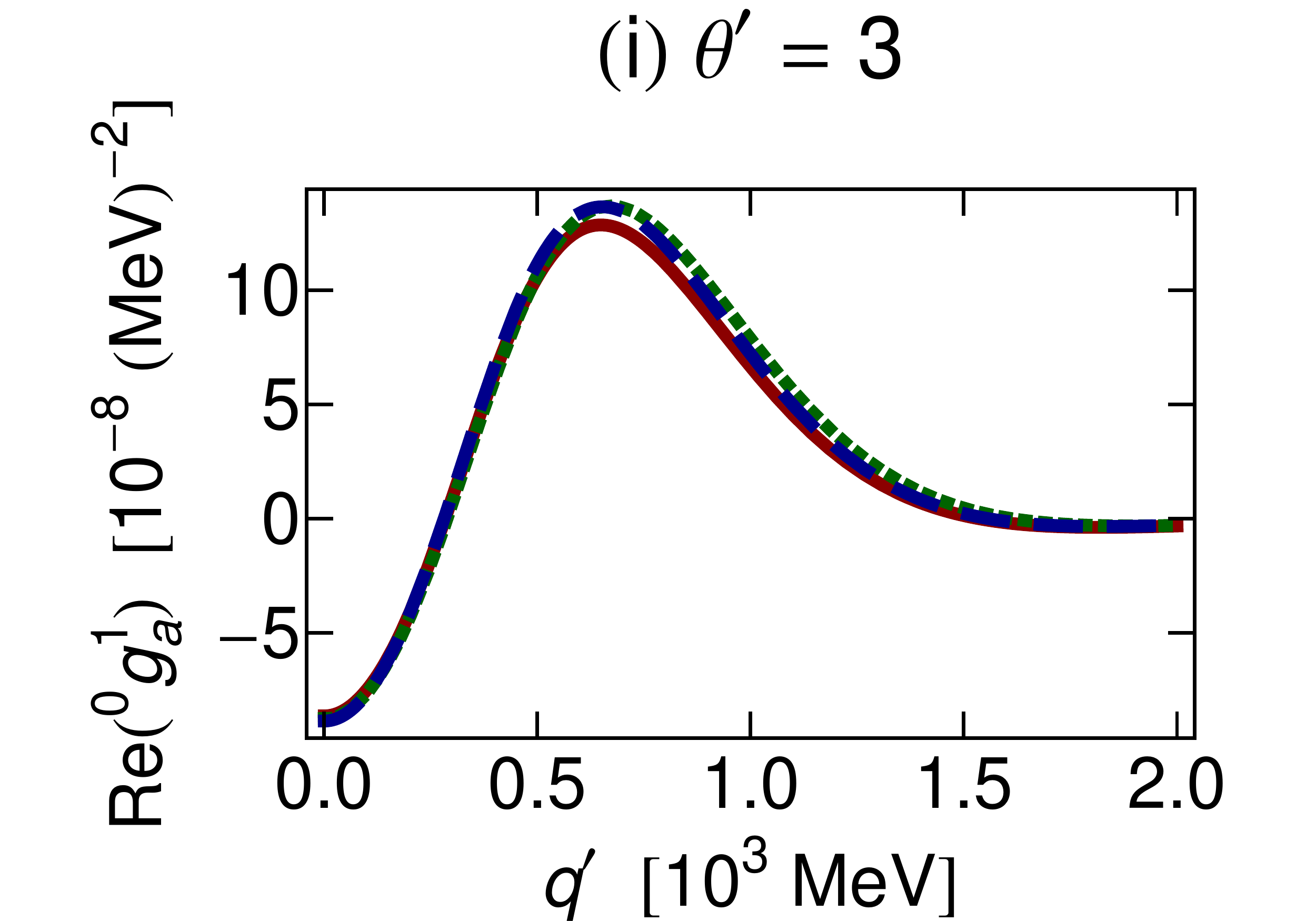}
}
\\
\subfloat{
\includegraphics[width=6cm]{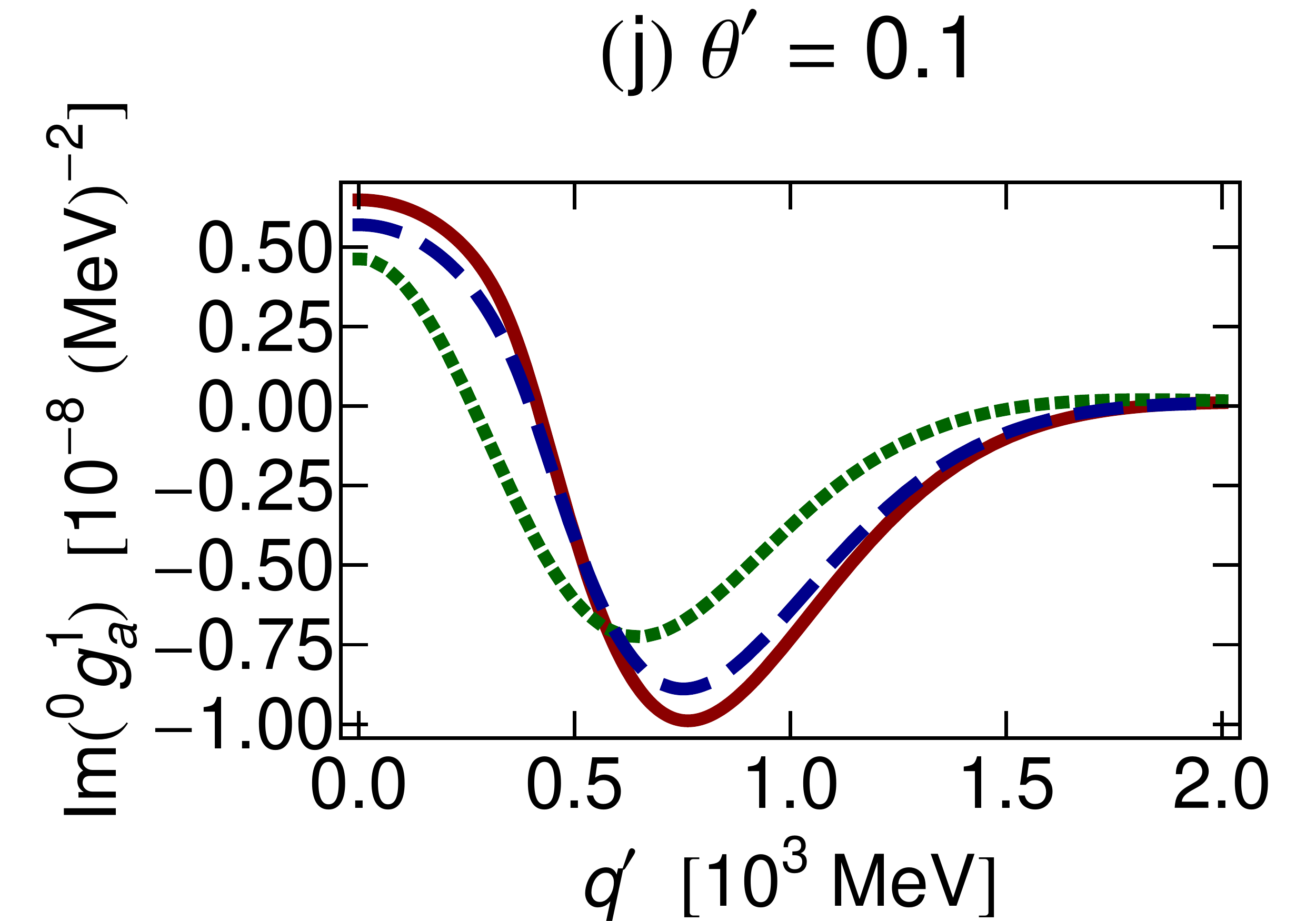}
}
\subfloat{
\includegraphics[width=6cm]{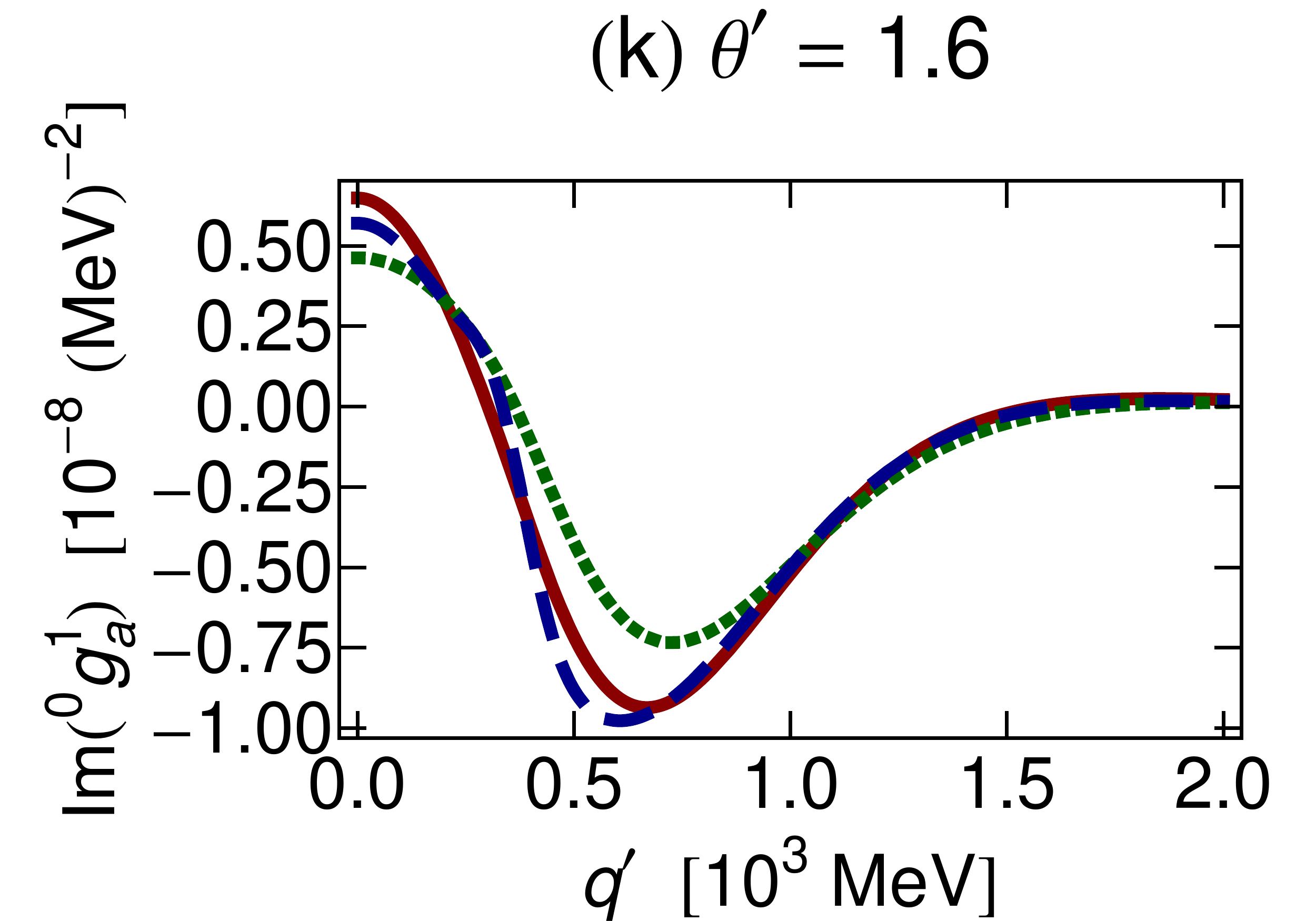}
}
\subfloat{
\includegraphics[width=6cm]{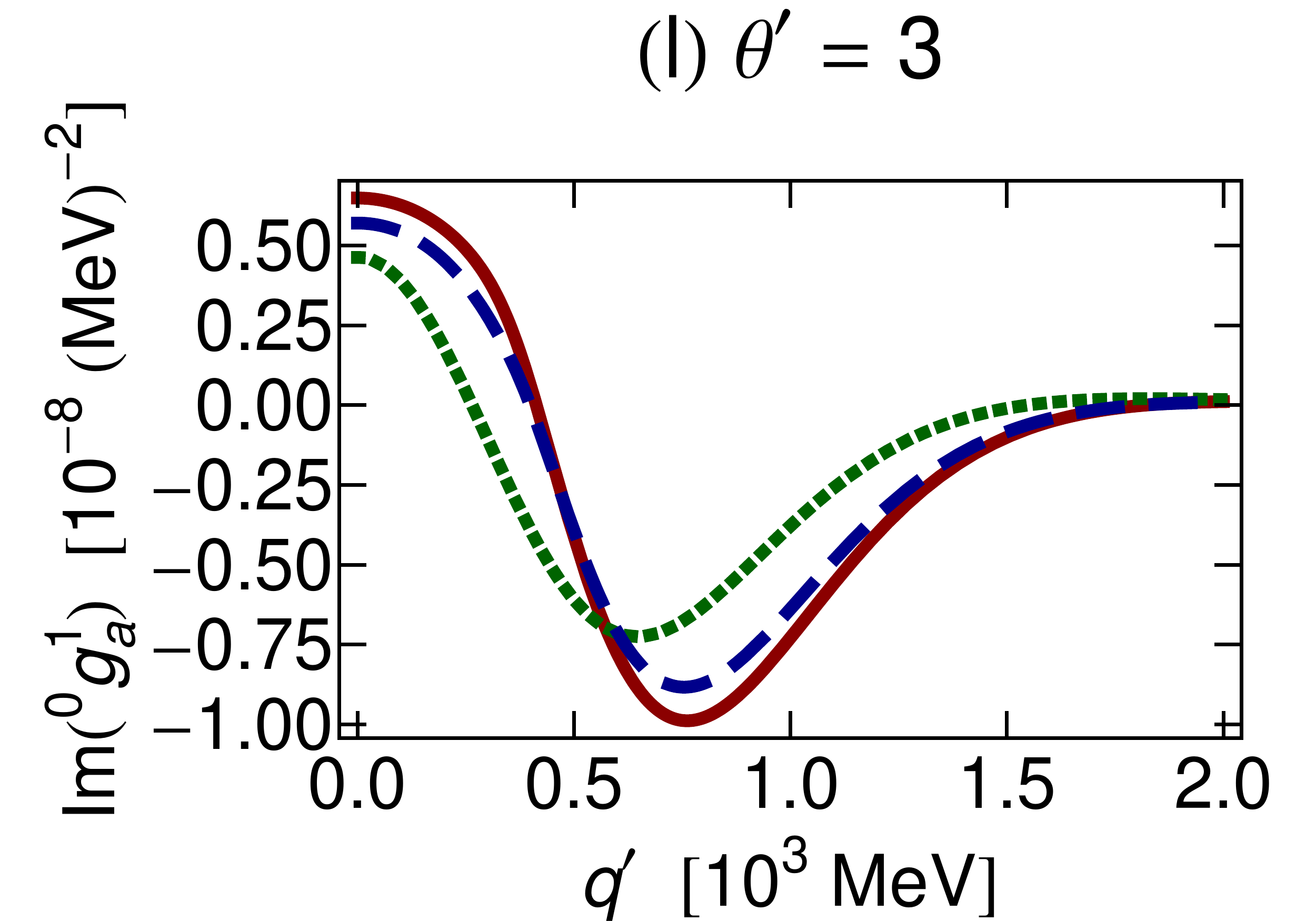}
}
\end{center}
\caption{(Color online) Same as Fig.~\ref{Fig:g_plots50_0} but at $q=375.29 \units{MeV}$.}
\label{Fig:g_plots300_0}
\end{figure}

\begin{figure}[H]
\begin{center}
\subfloat{
\includegraphics[width=6cm]{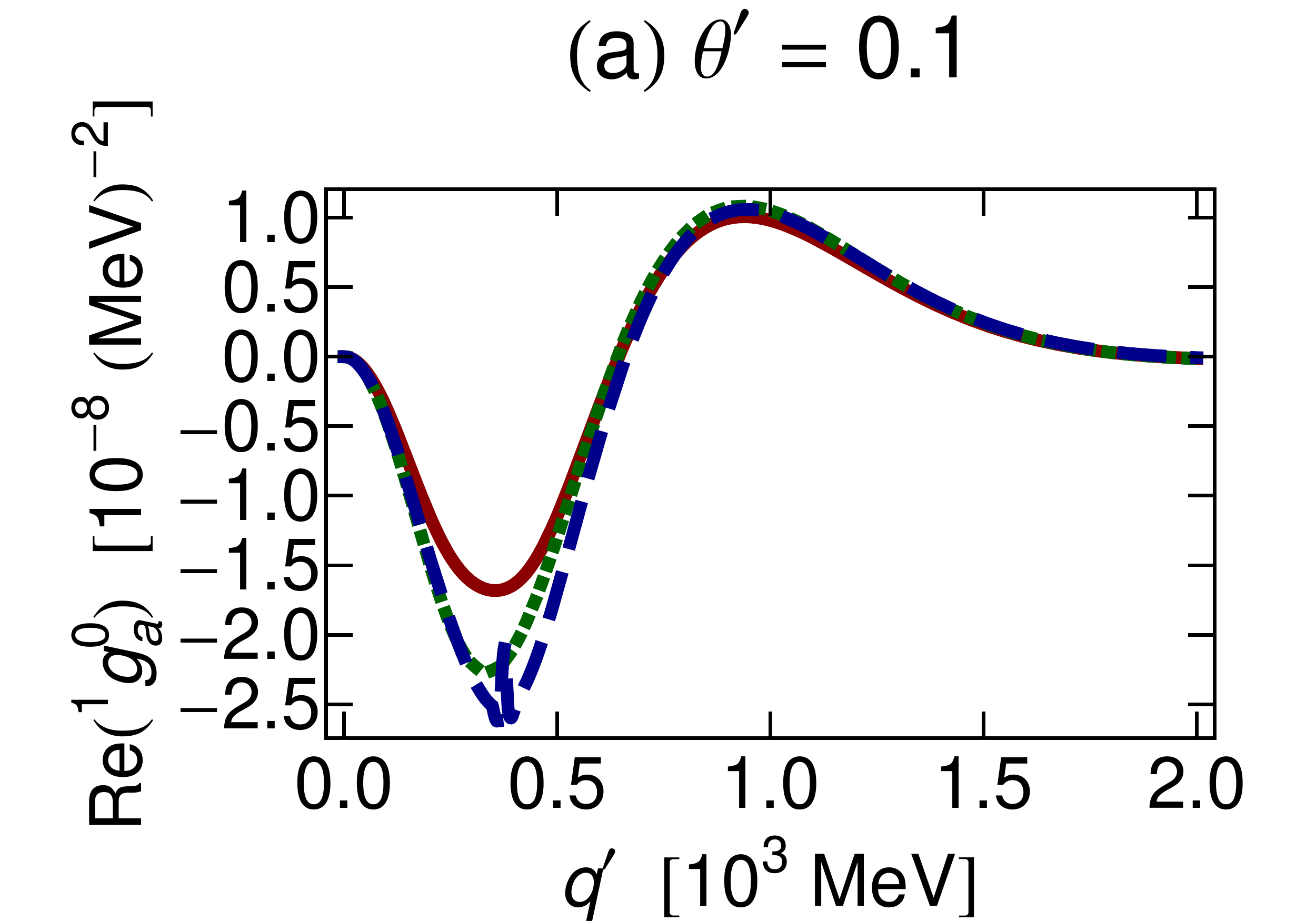}
}
\subfloat{
\includegraphics[width=6cm]{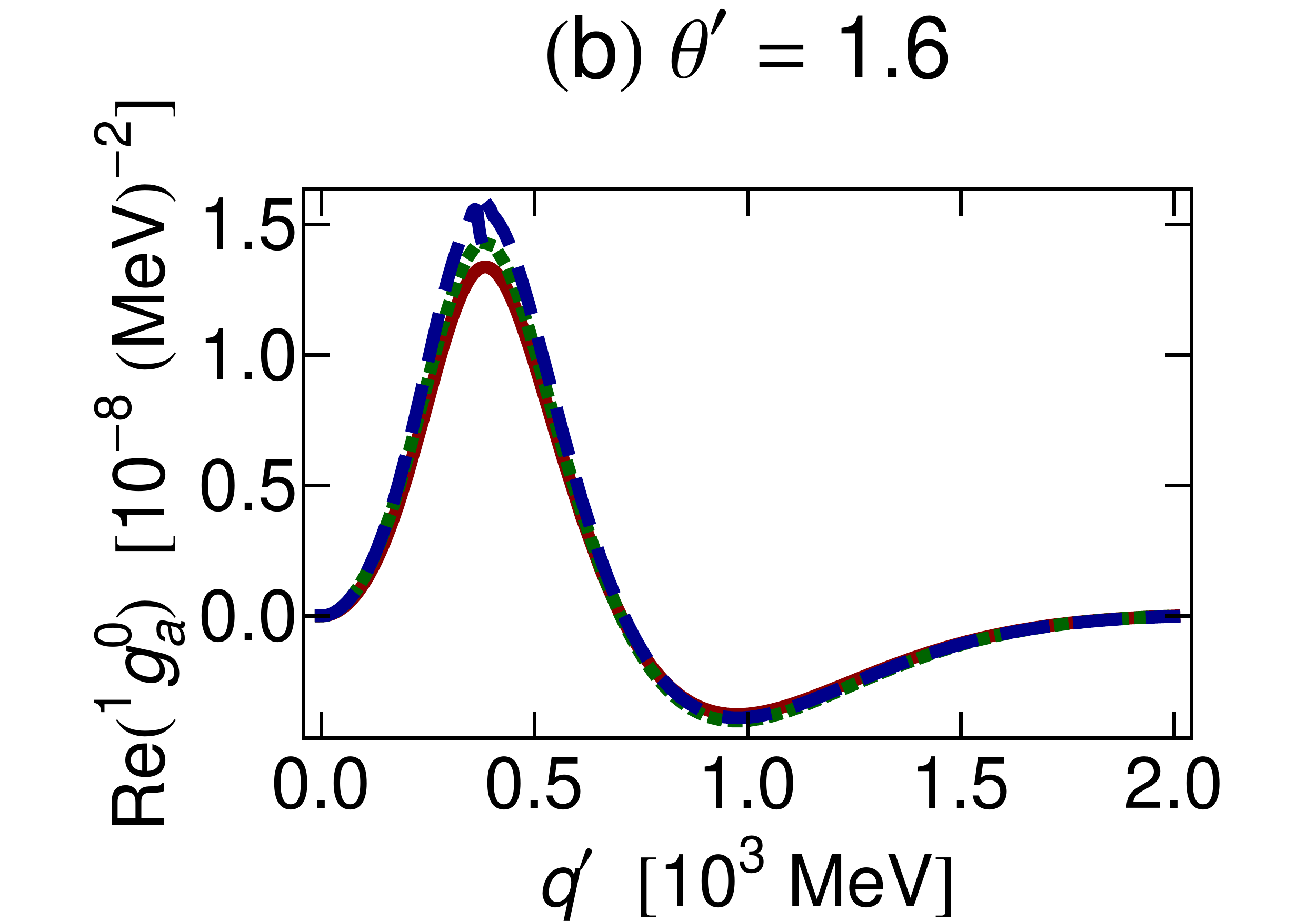}
}
\subfloat{
\includegraphics[width=6cm]{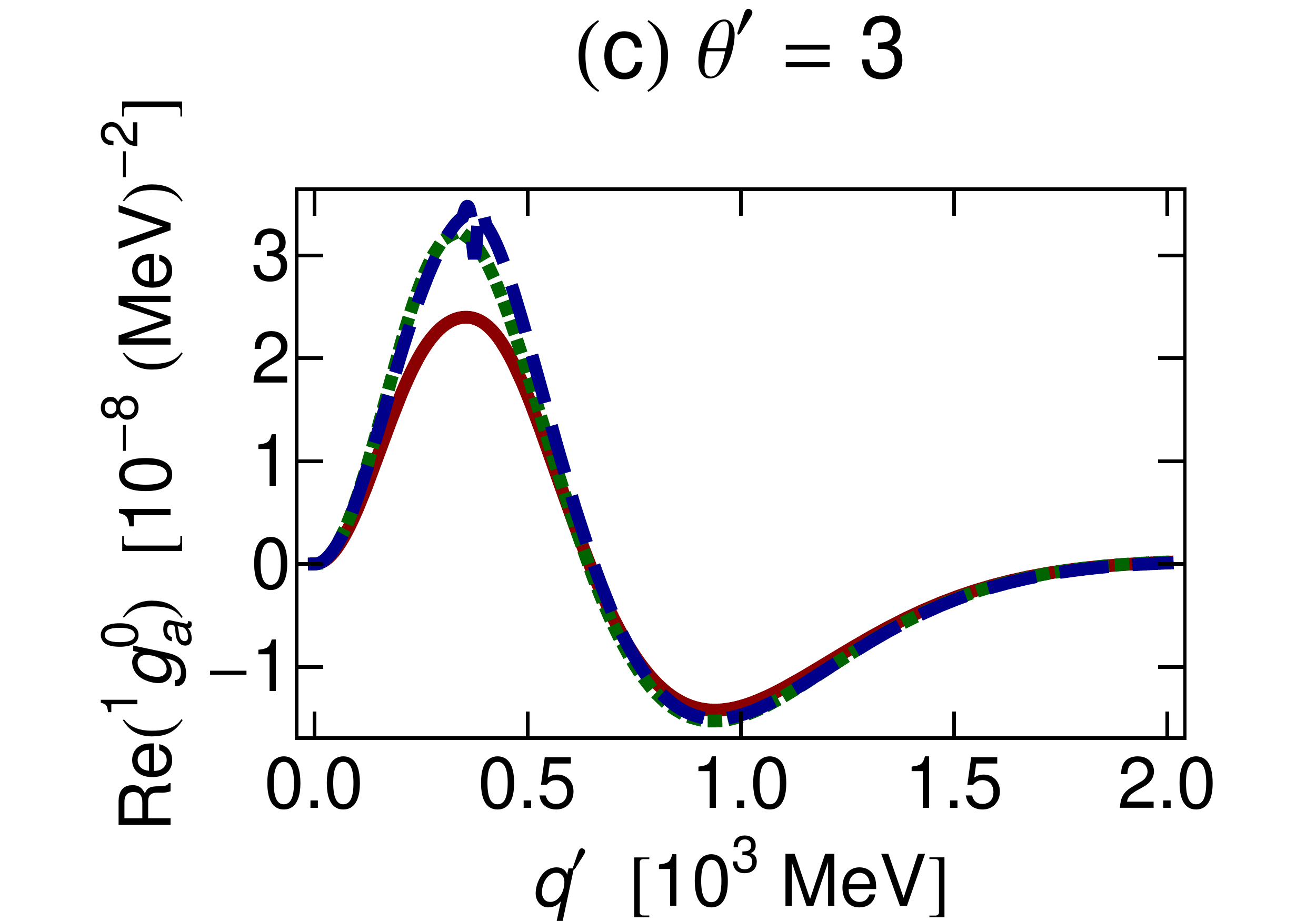}
}
\\
\subfloat{
\includegraphics[width=6cm]{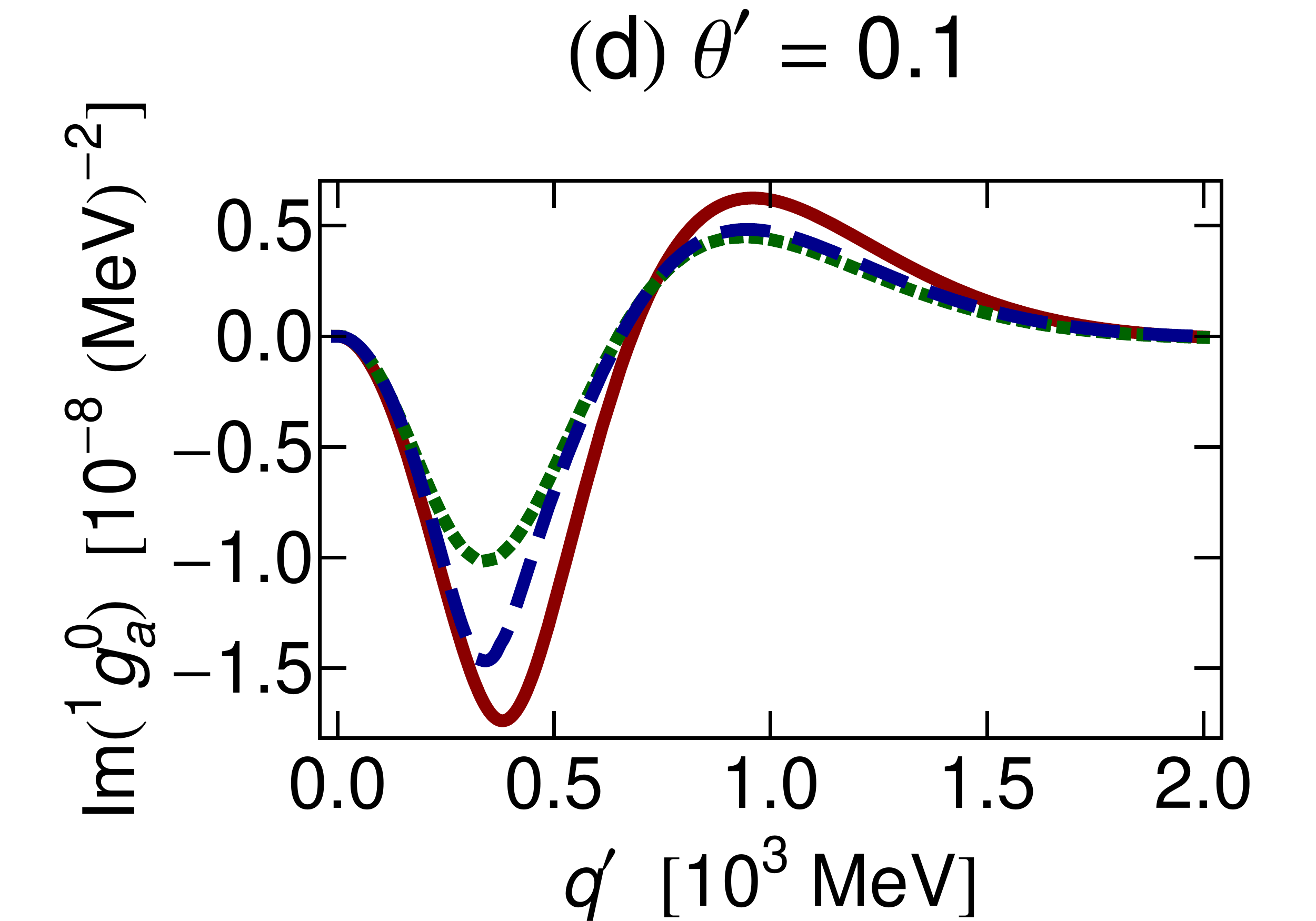}
}
\subfloat{
\includegraphics[width=6cm]{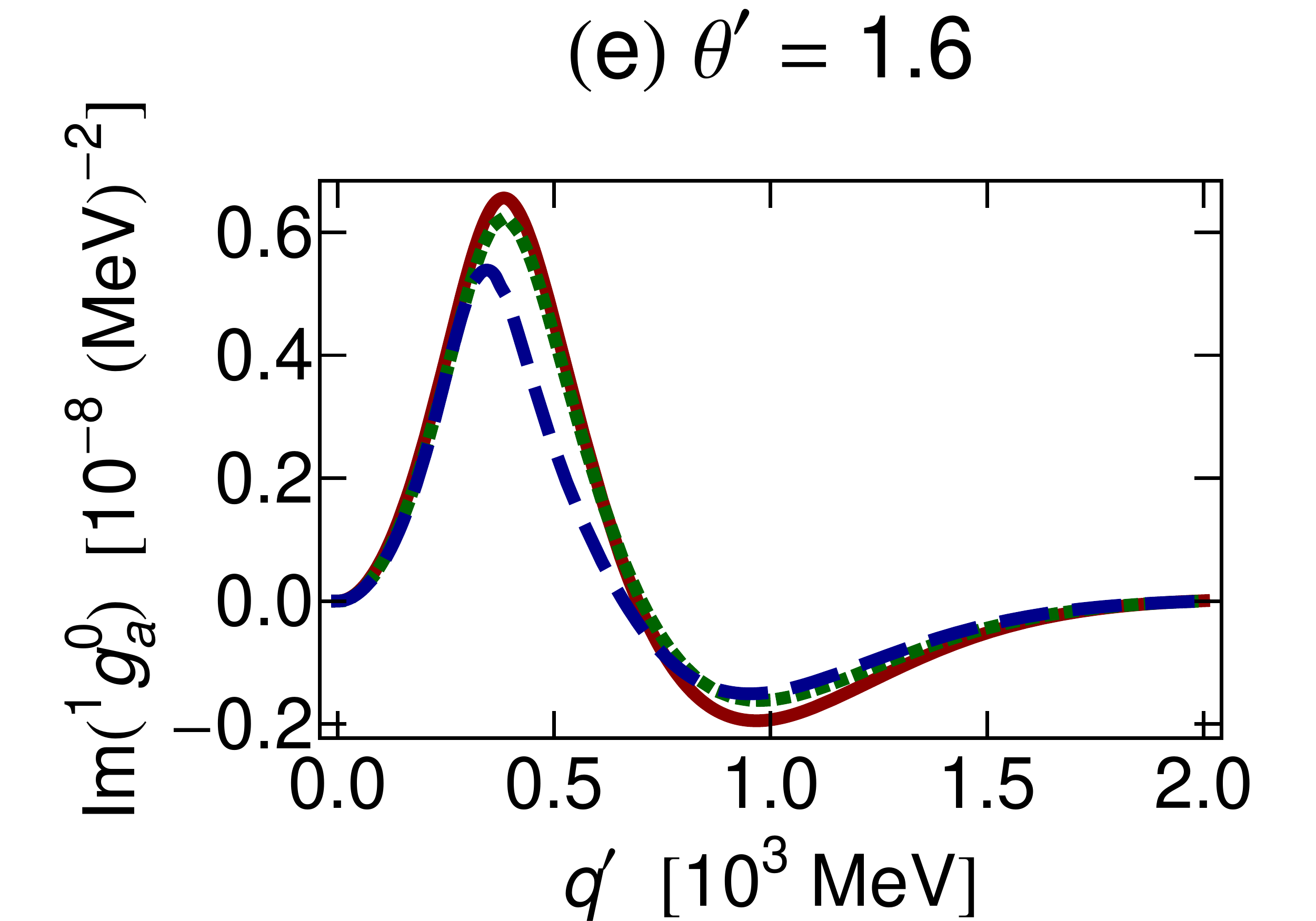}
}
\subfloat{
\includegraphics[width=6cm]{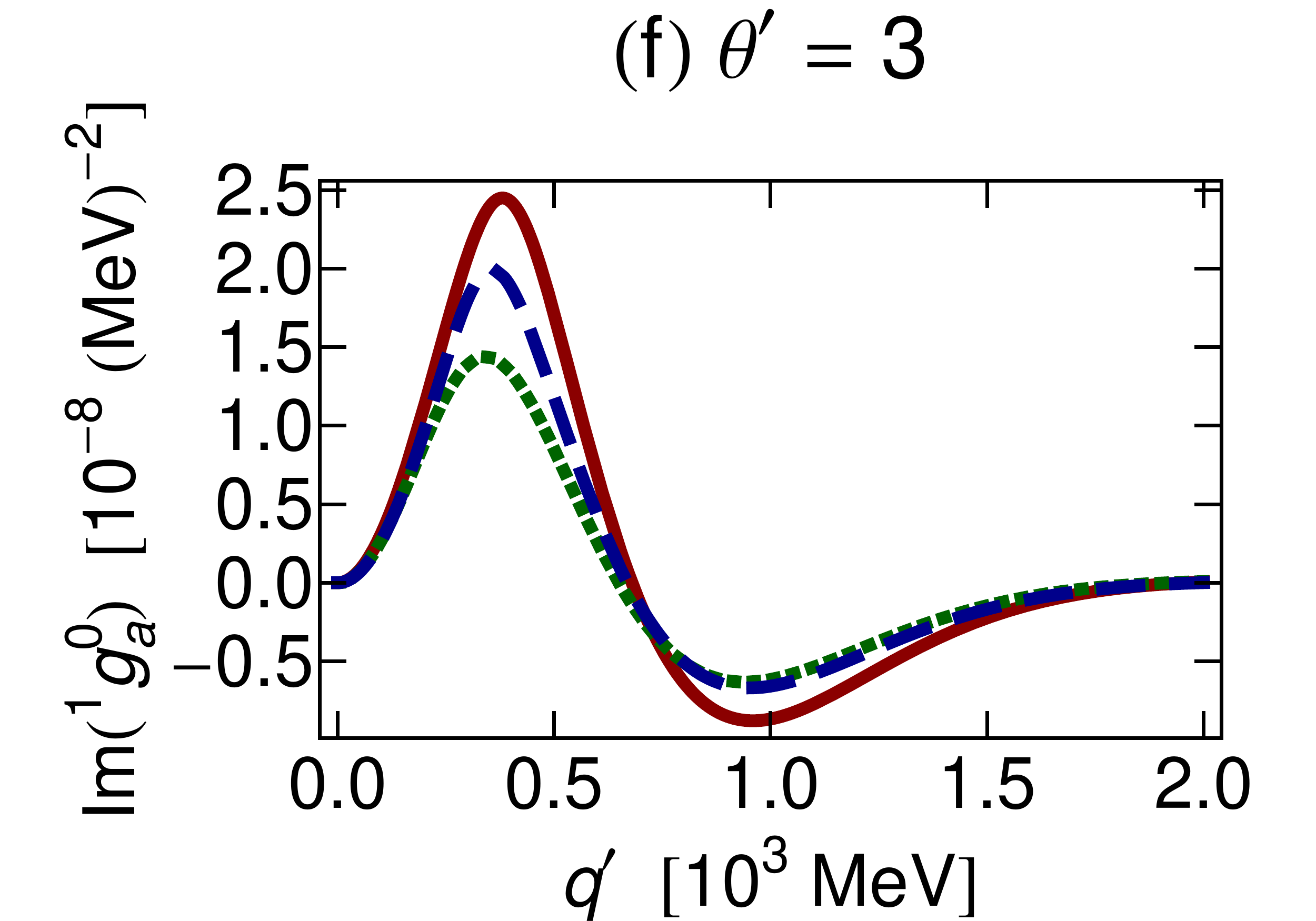}
}
\\
\subfloat{
\includegraphics[width=6cm]{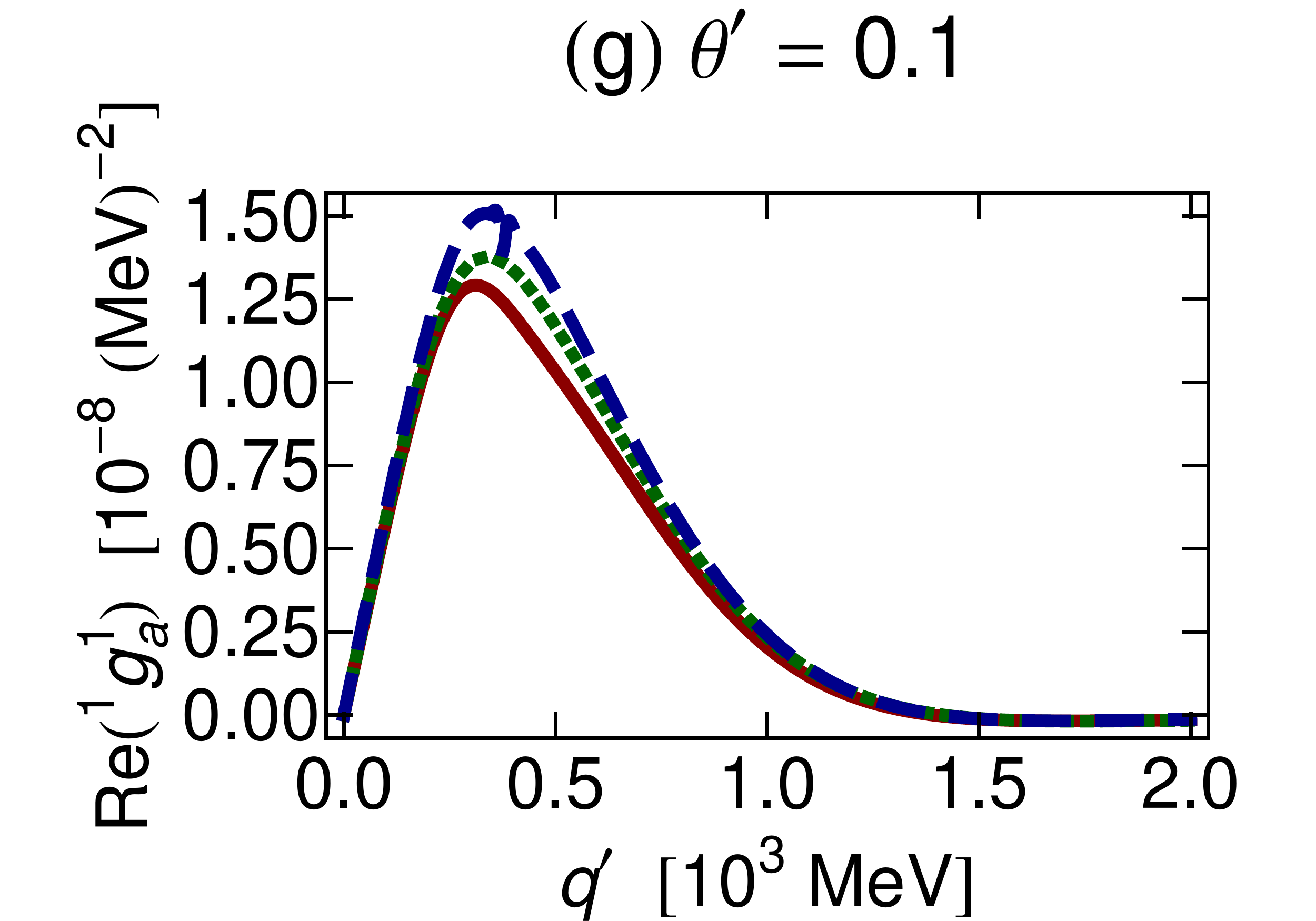}
}
\subfloat{
\includegraphics[width=6cm]{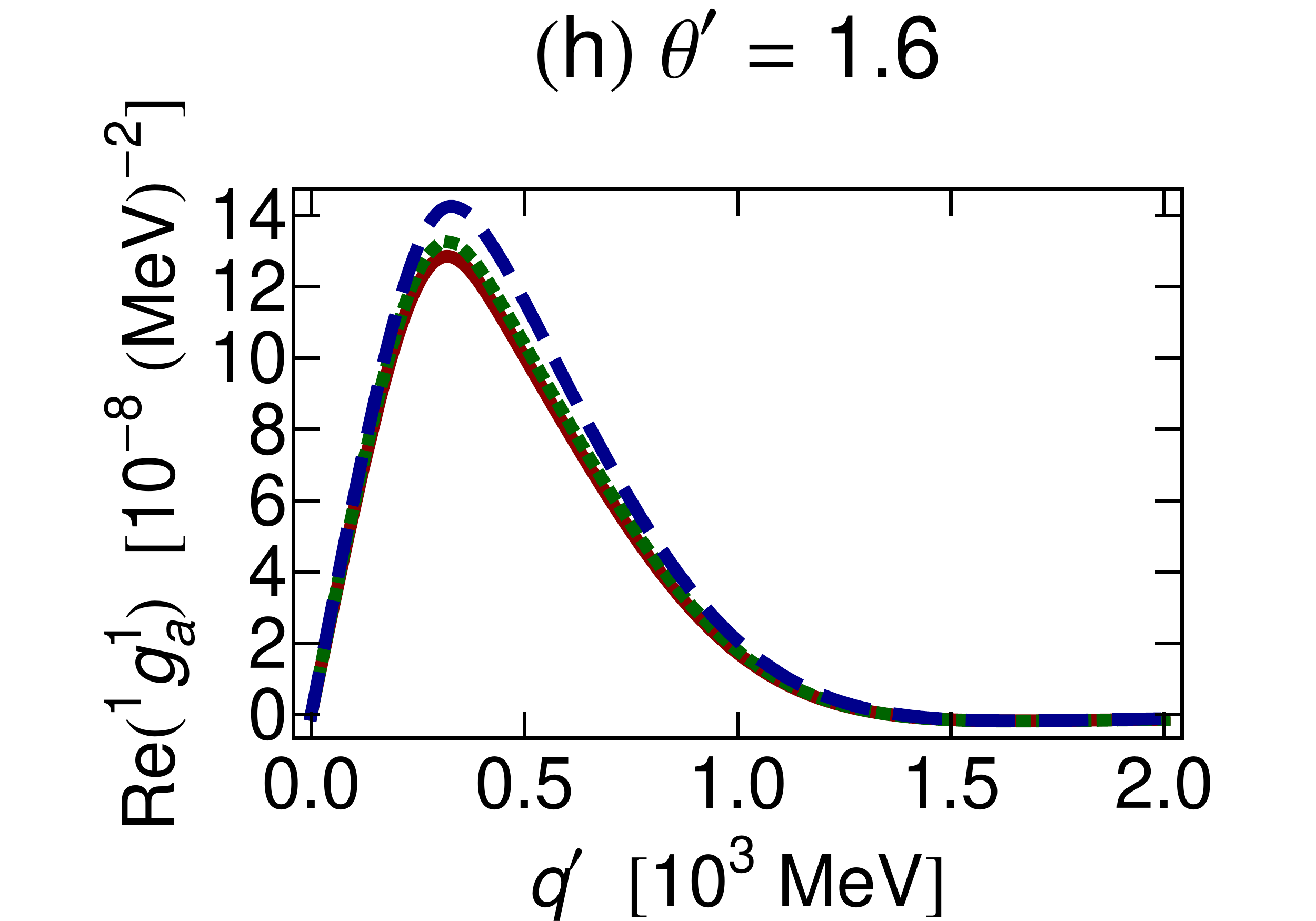}
}
\subfloat{
\includegraphics[width=6cm]{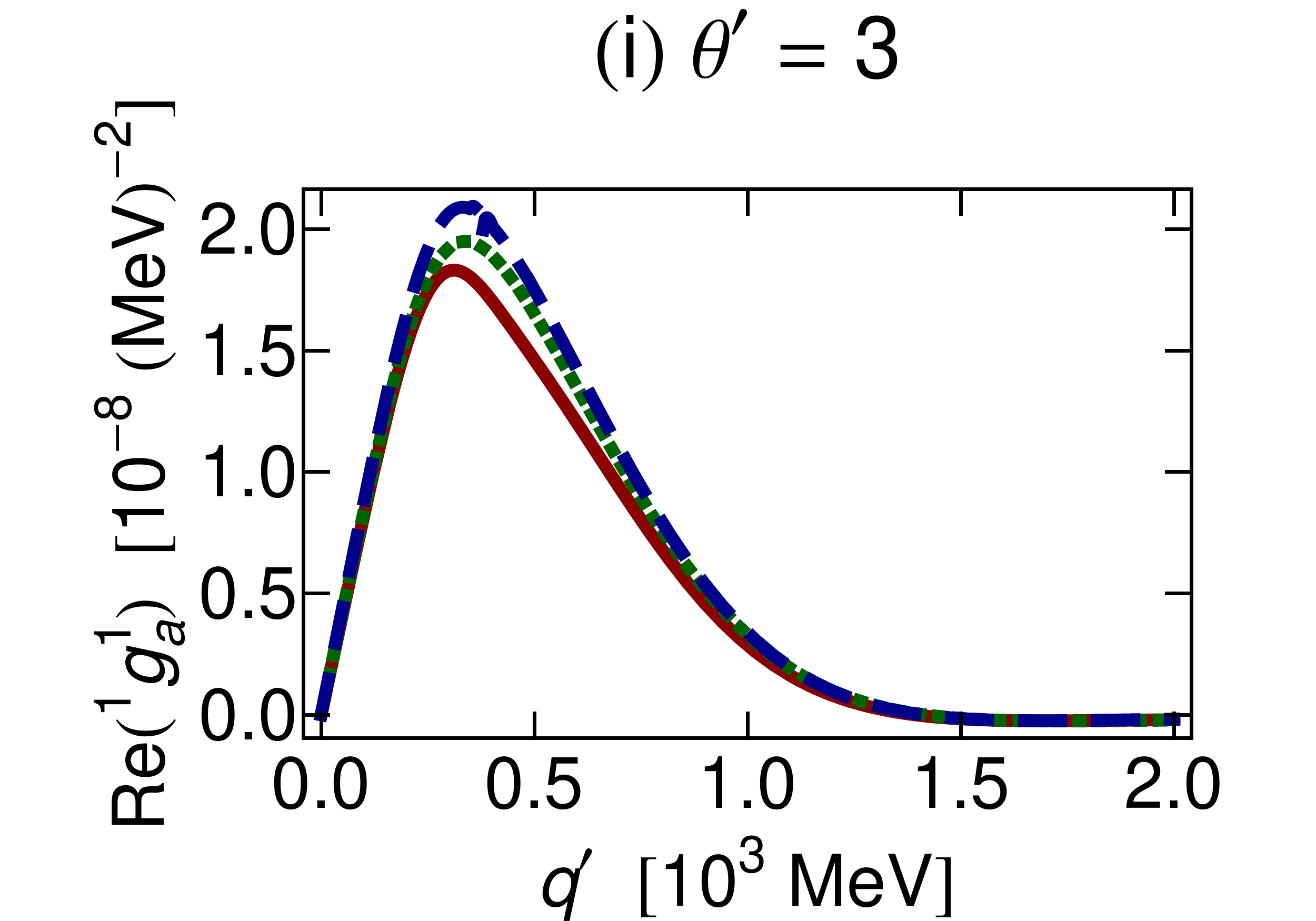}
}
\\
\subfloat{
\includegraphics[width=6cm]{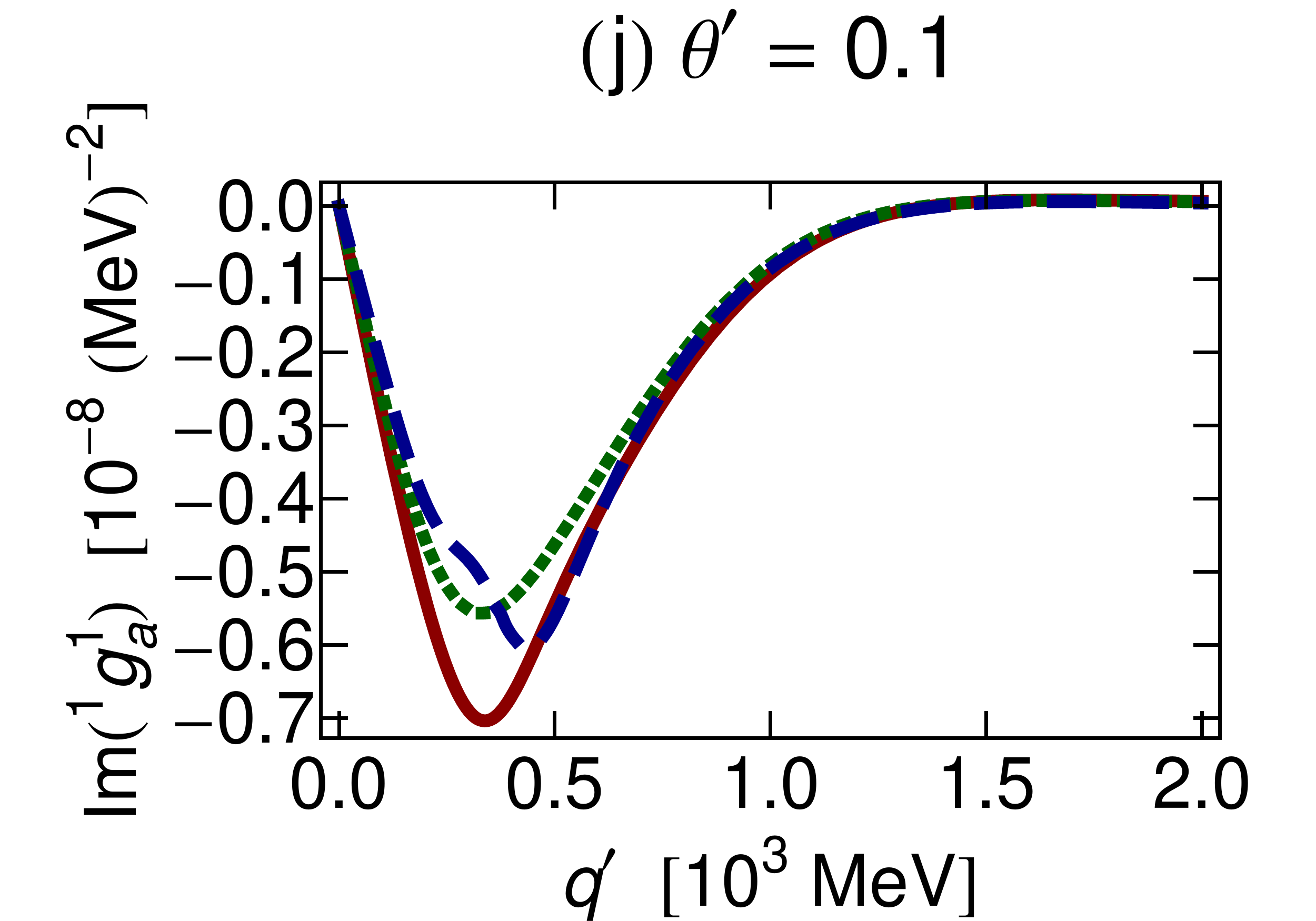}
}
\subfloat{
\includegraphics[width=6cm]{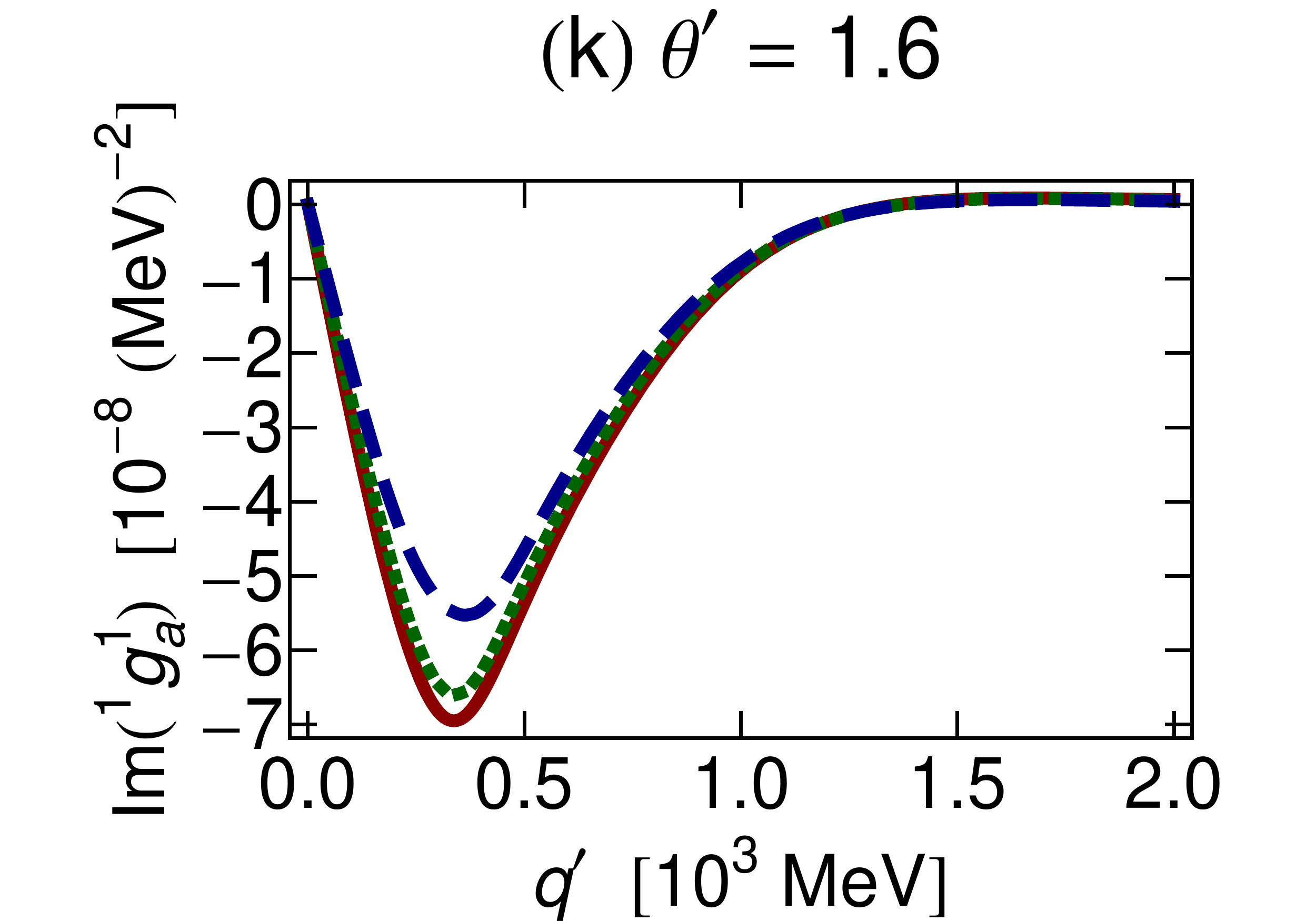}
}
\subfloat{
\includegraphics[width=6cm]{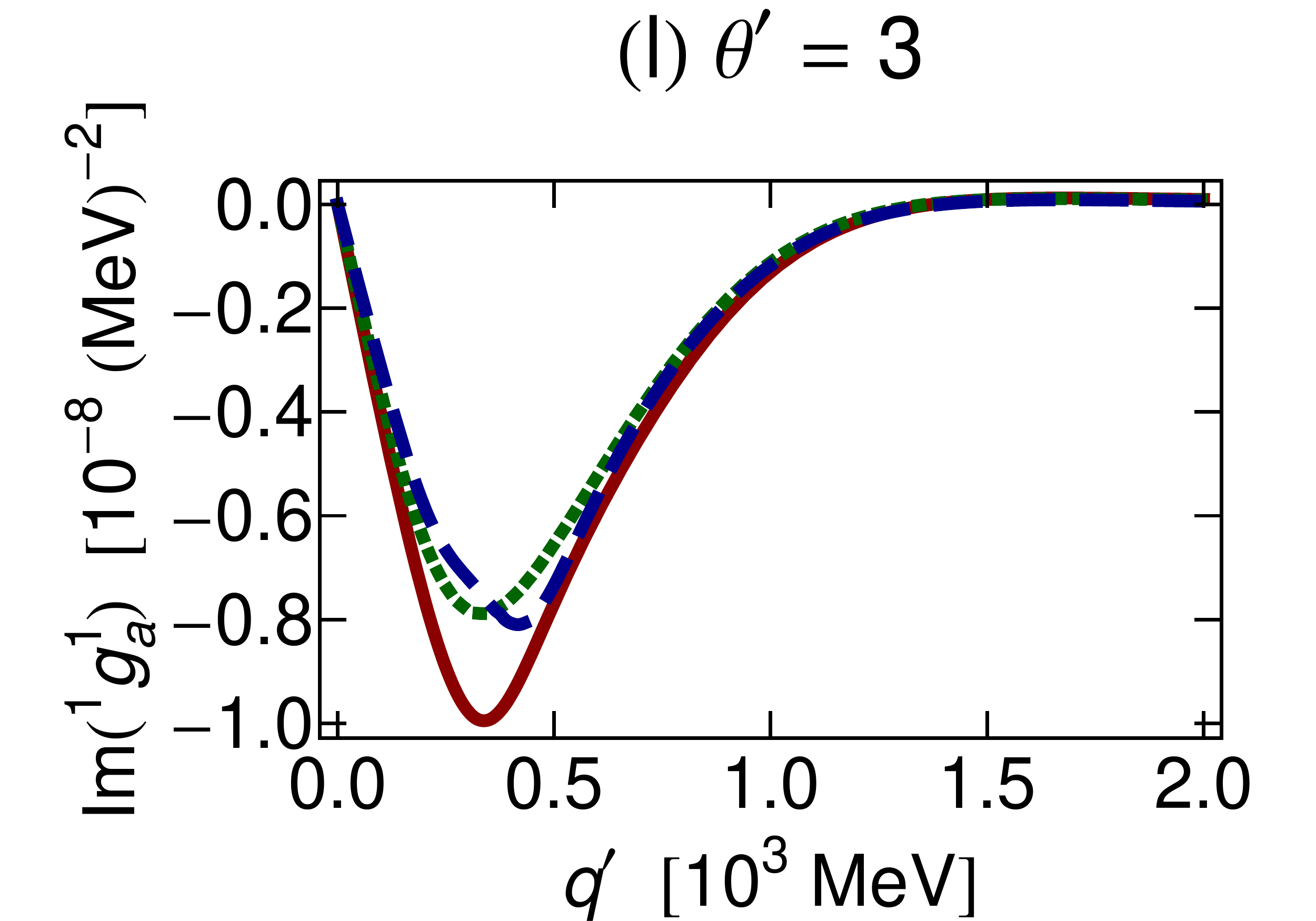}
}
\end{center}
\caption{(Color online) Same as Fig.~\ref{Fig:g_plots50_0} but for $\up{1}g^I_a$ and $\up{1}t^I_a$ at $q=375.29 \units{MeV}$.}
\label{Fig:g_plots300_1}
\end{figure}

\begin{figure}[H]
\begin{center}
\subfloat{
\includegraphics[width=6cm]{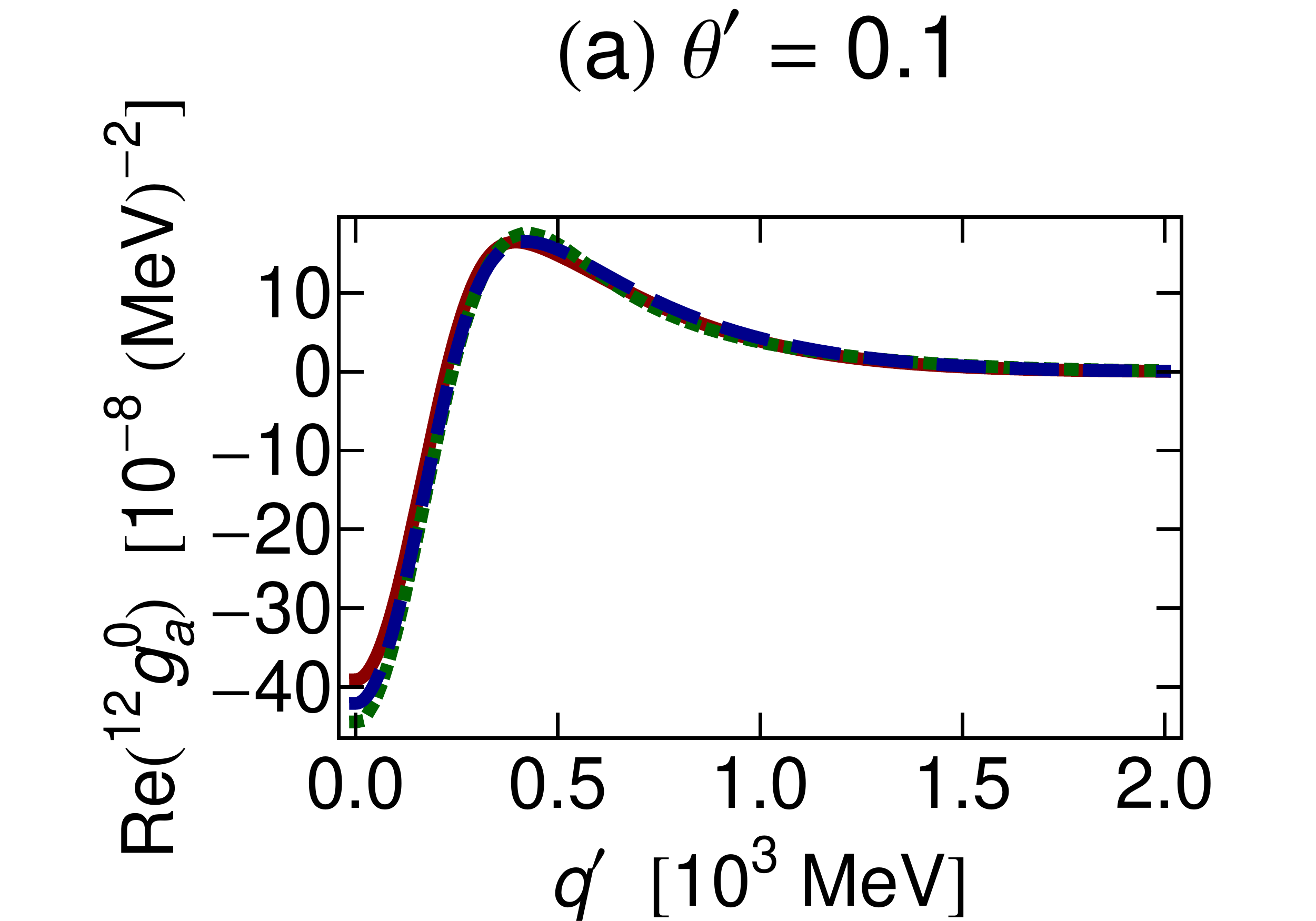}
}
\subfloat{
\includegraphics[width=6cm]{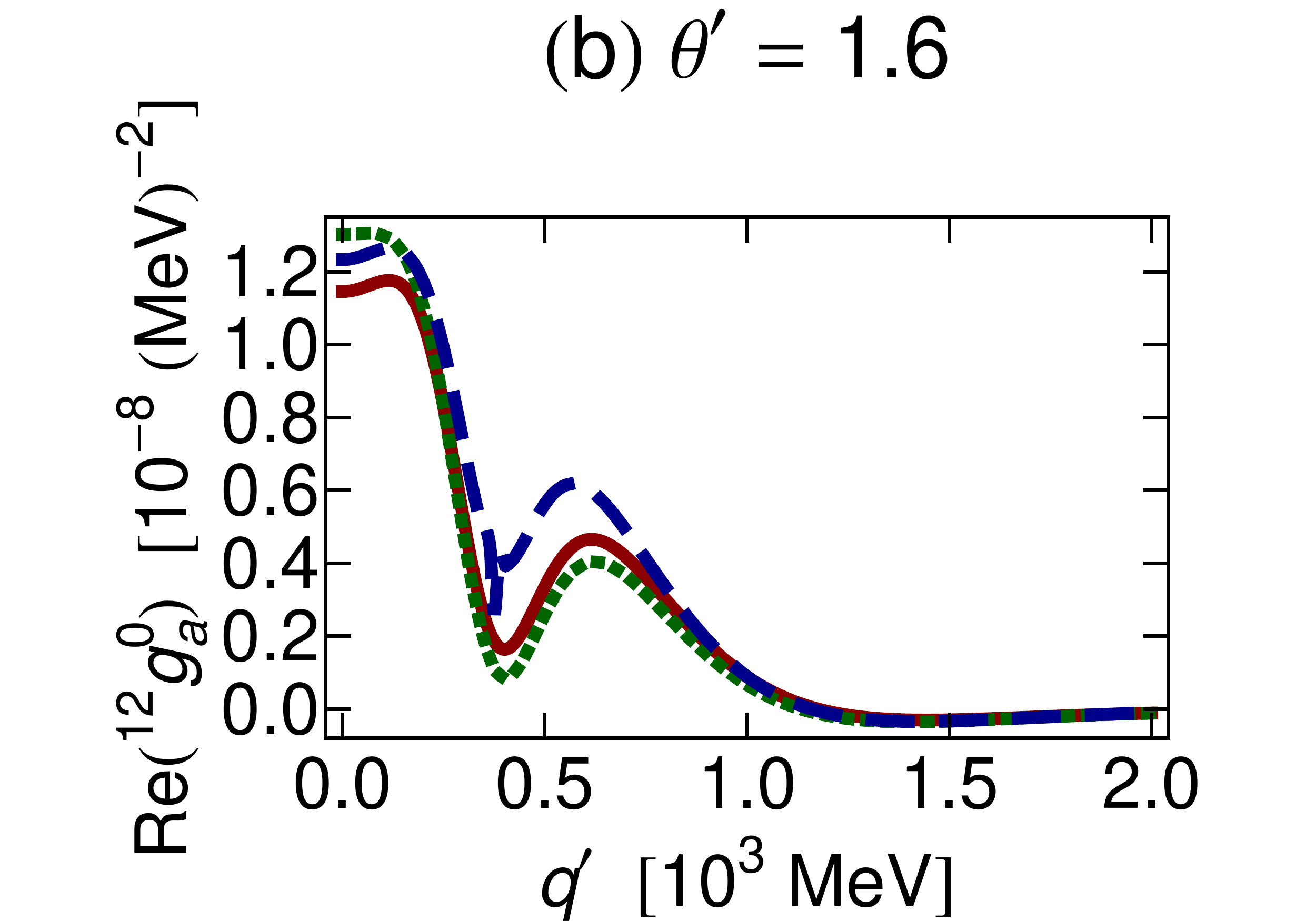}
}
\subfloat{
\includegraphics[width=6cm]{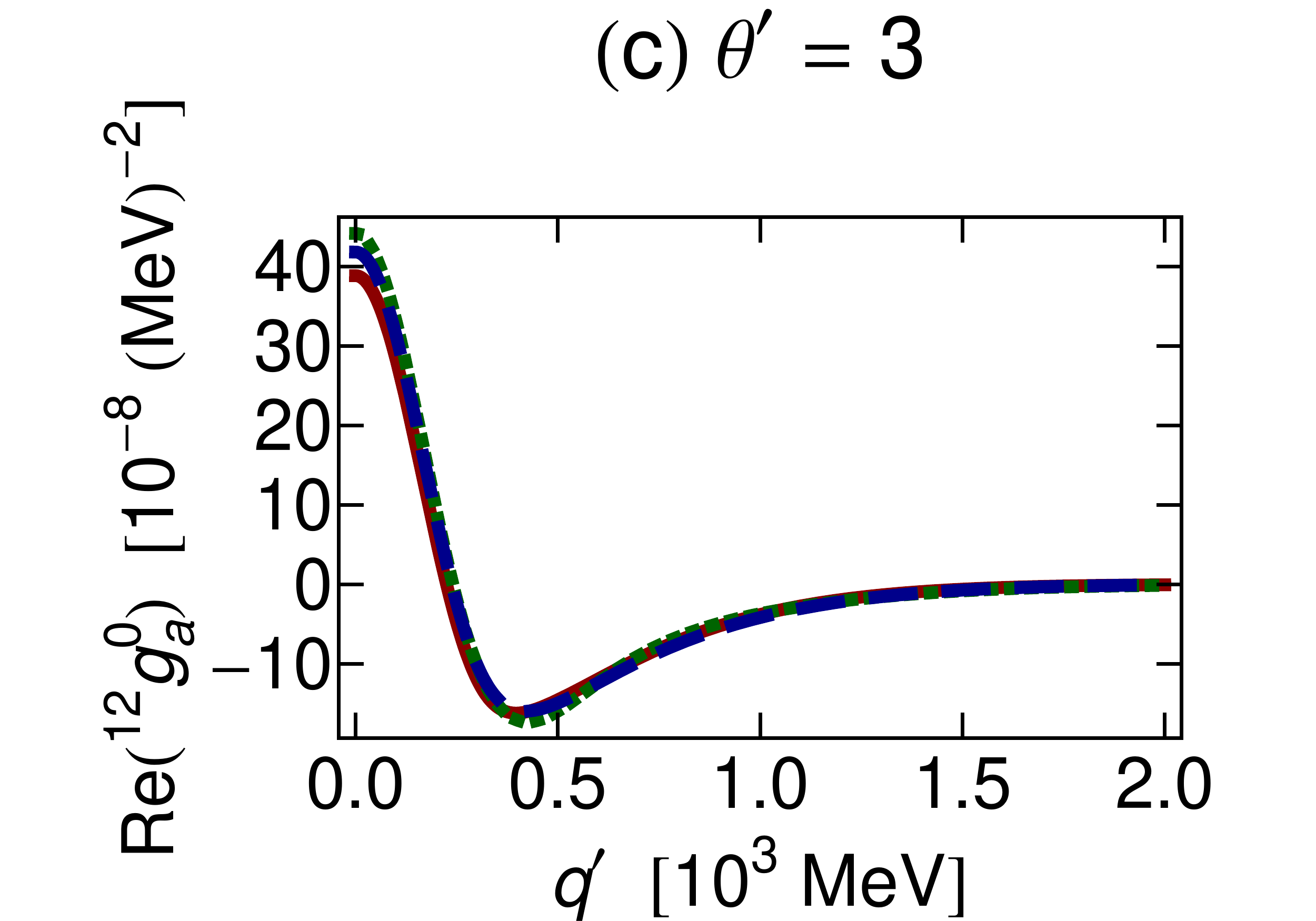}
}
\\
\subfloat{
\includegraphics[width=6cm]{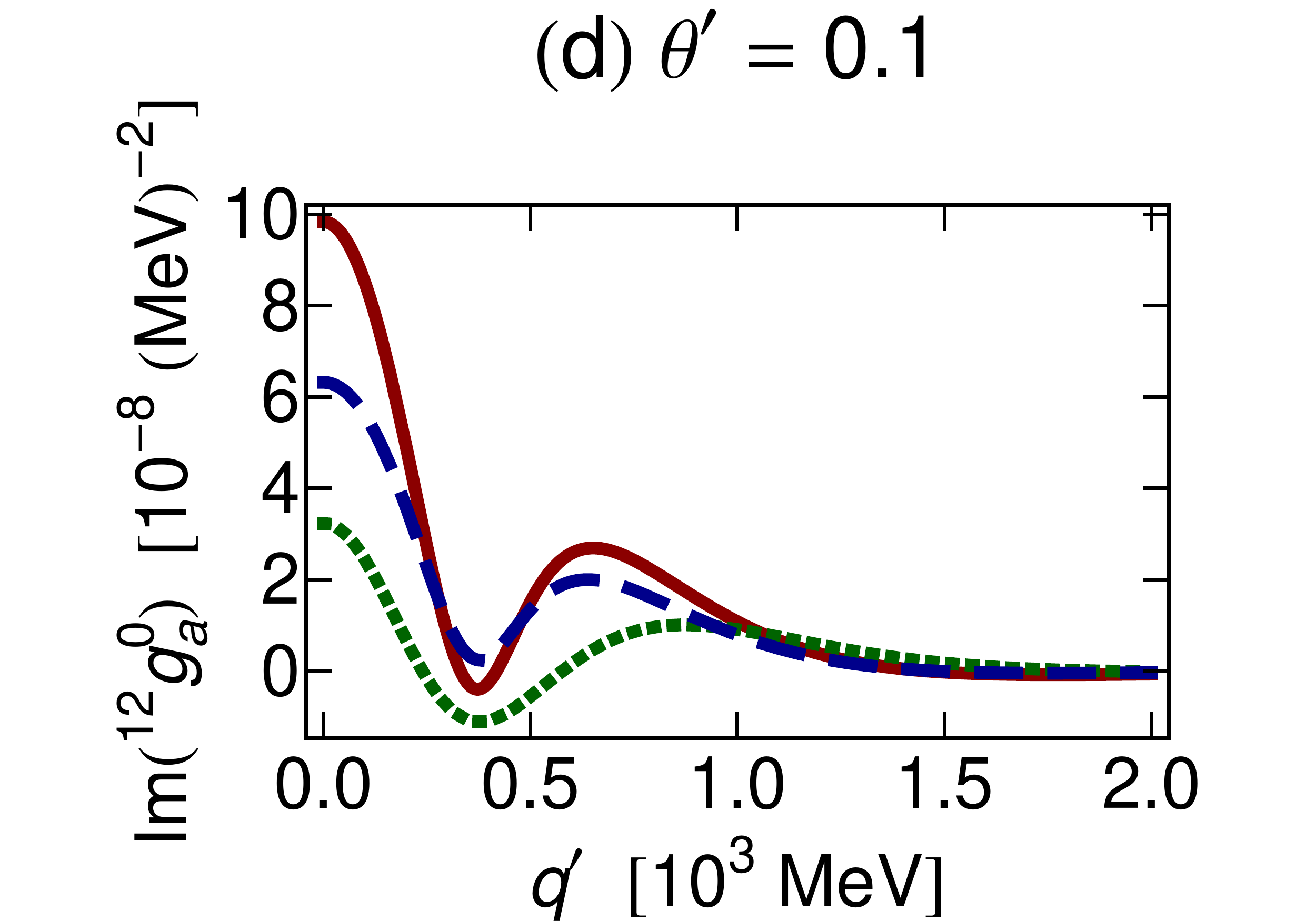}
}
\subfloat{
\includegraphics[width=6cm]{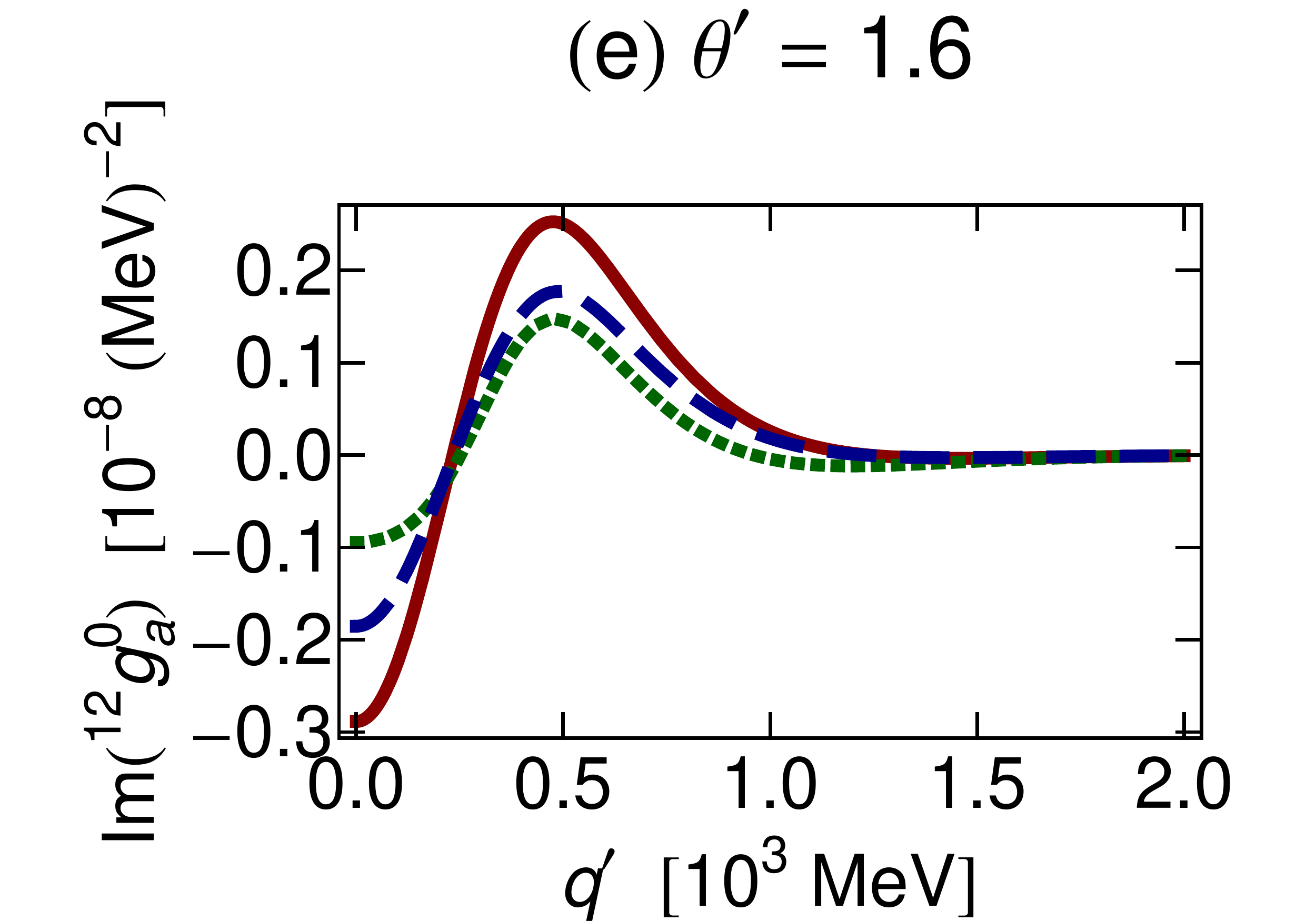}
}
\subfloat{
\includegraphics[width=6cm]{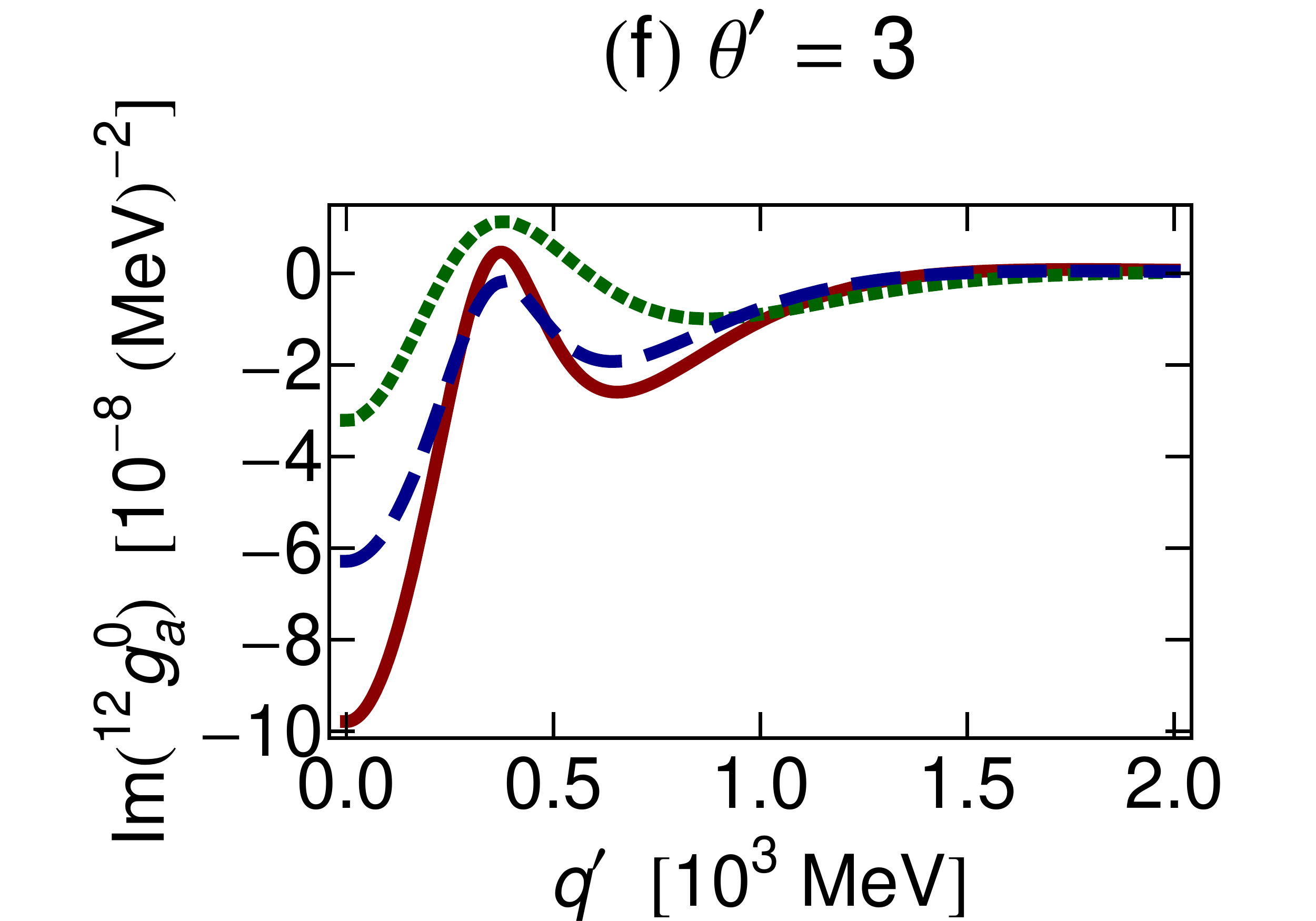}
}
\\
\subfloat{
\includegraphics[width=6cm]{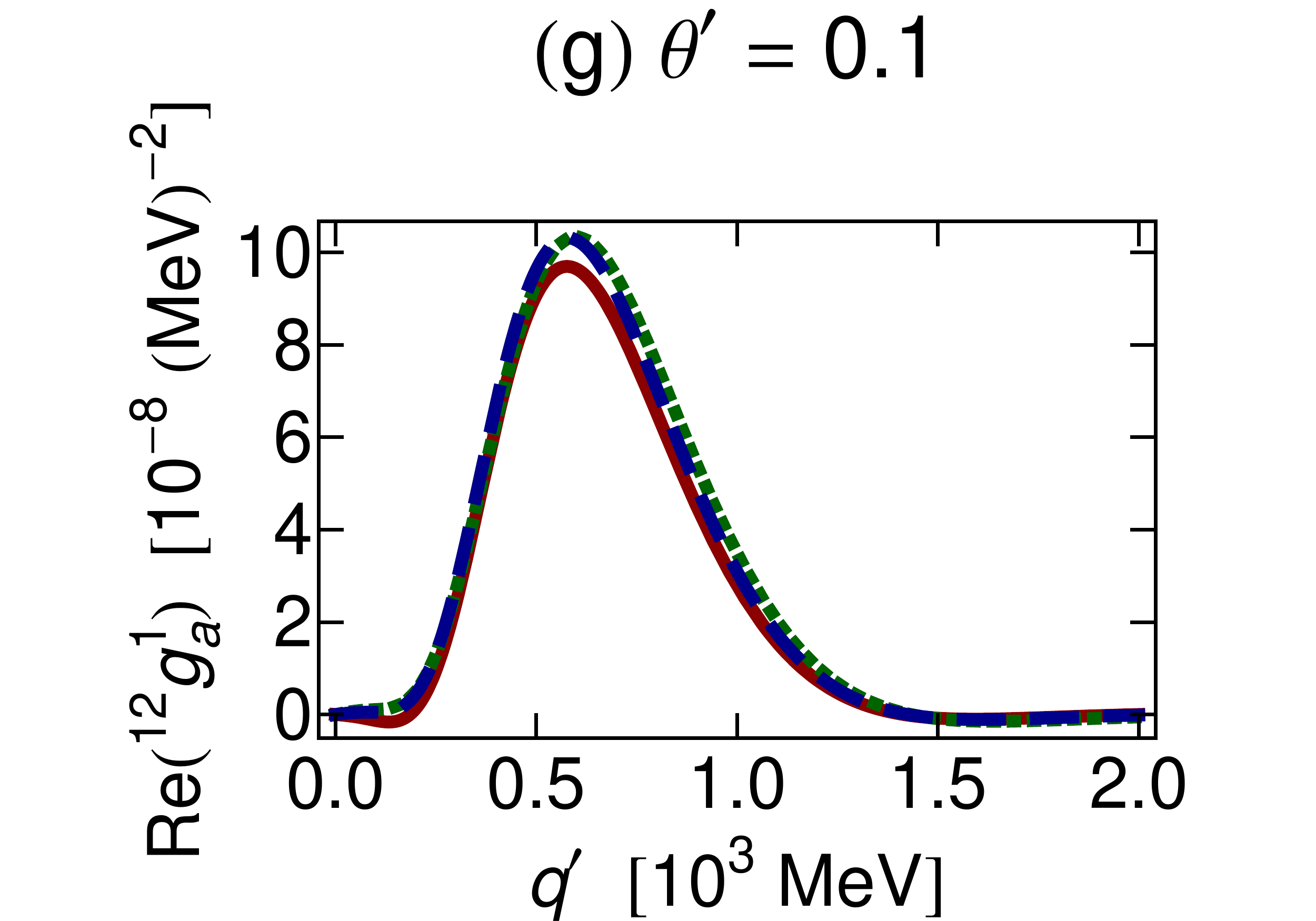}
}
\subfloat{
\includegraphics[width=6cm]{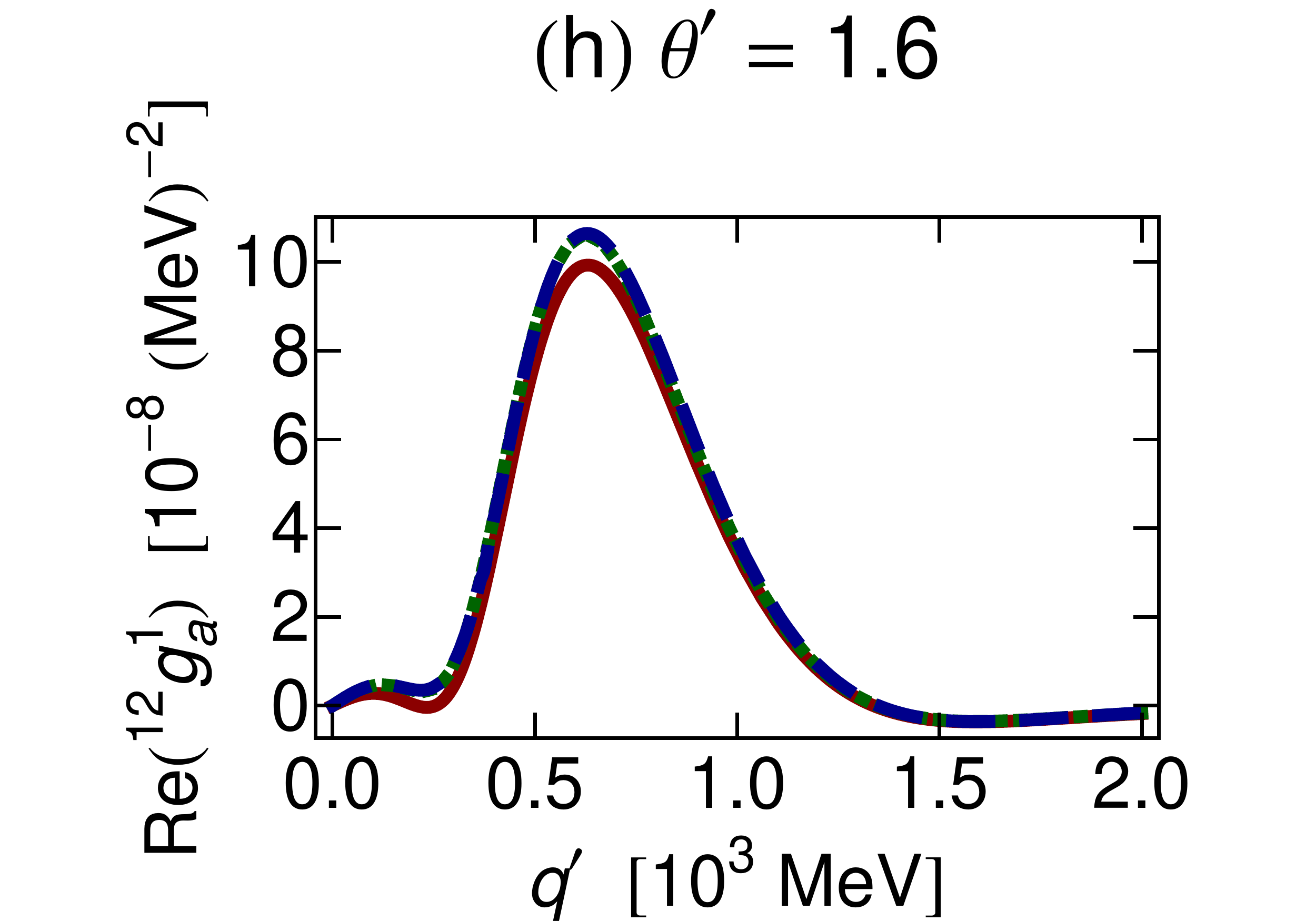}
}
\subfloat{
\includegraphics[width=6cm]{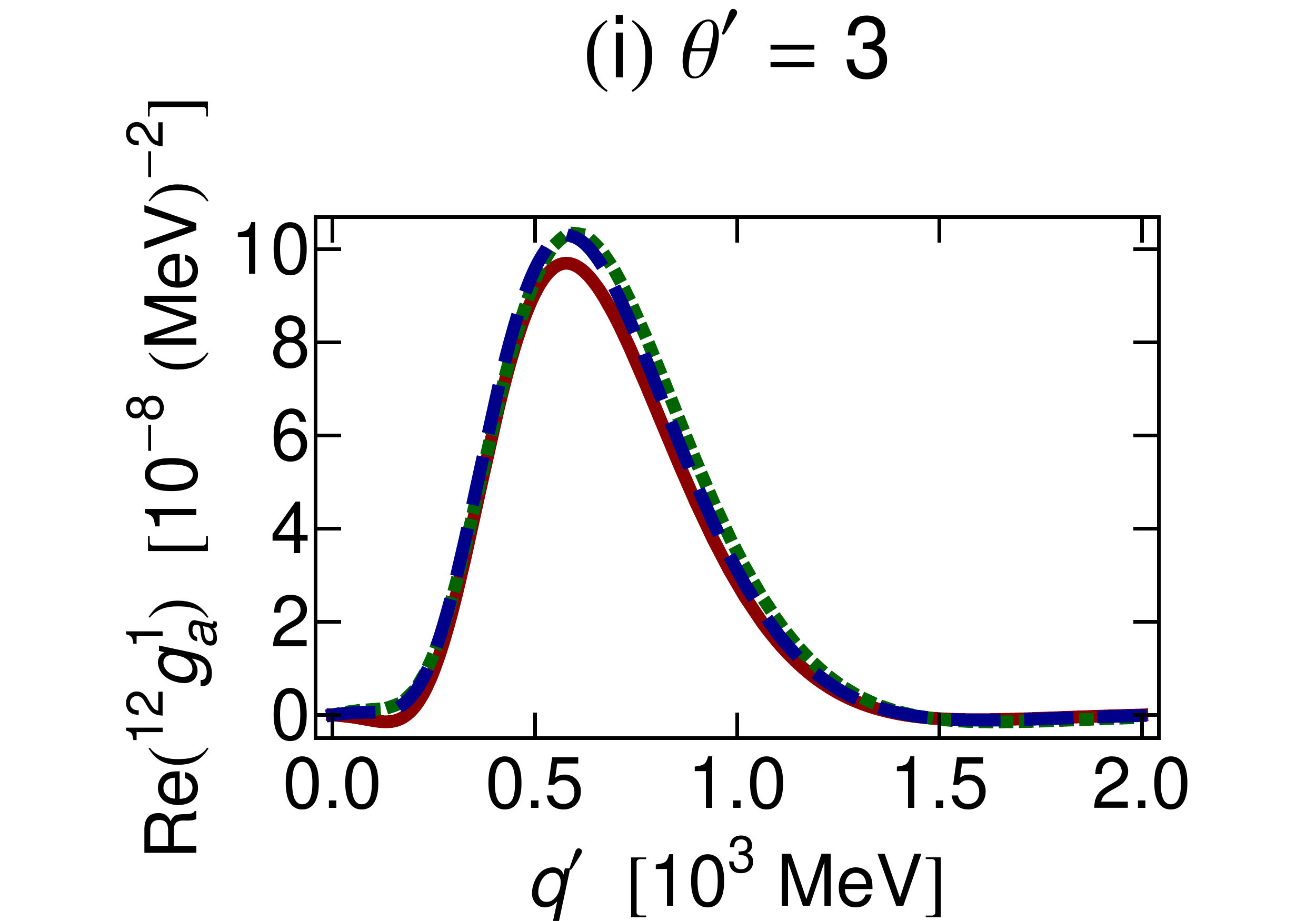}
}
\\
\subfloat{
\includegraphics[width=6cm]{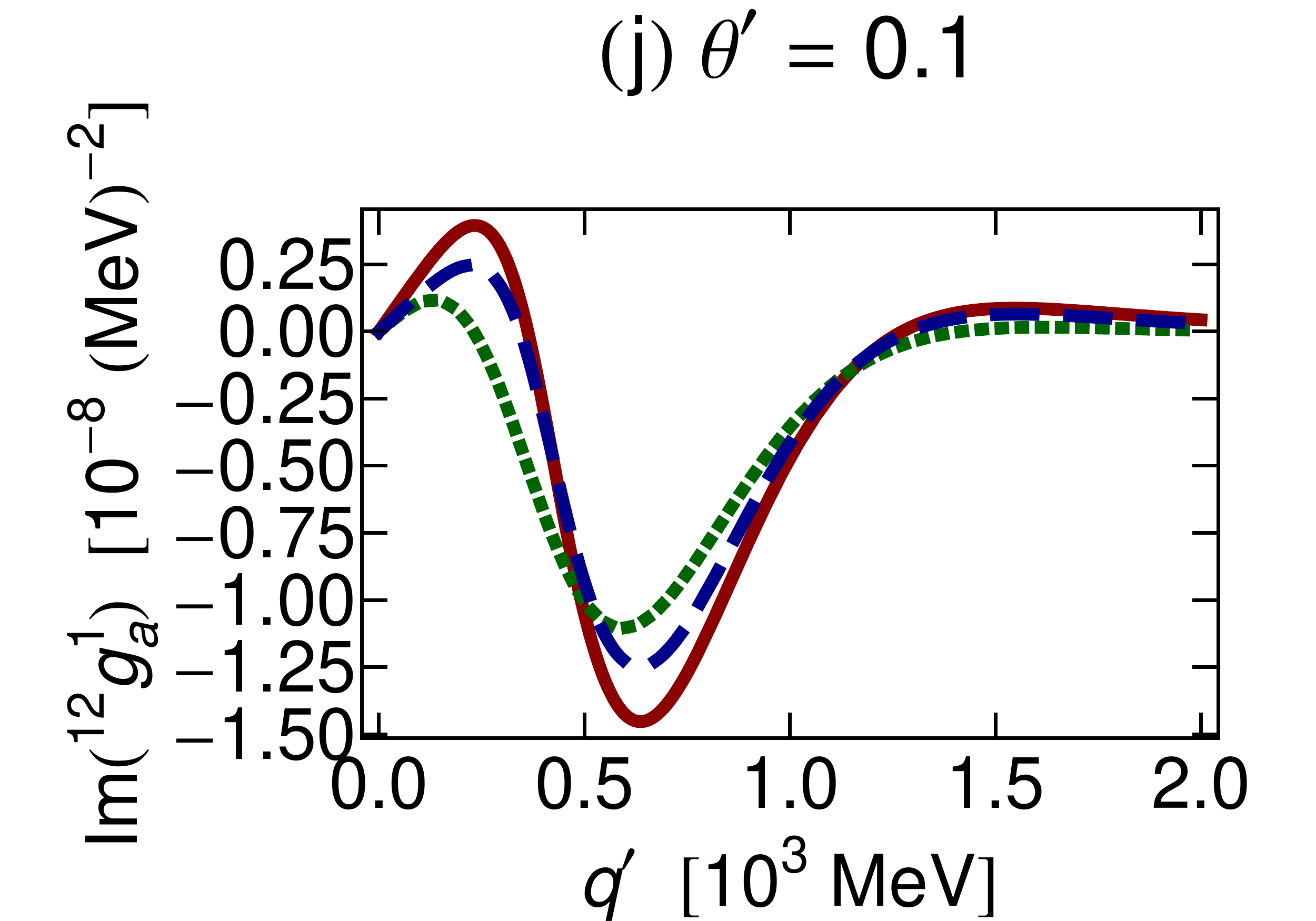}
}
\subfloat{
\includegraphics[width=6cm]{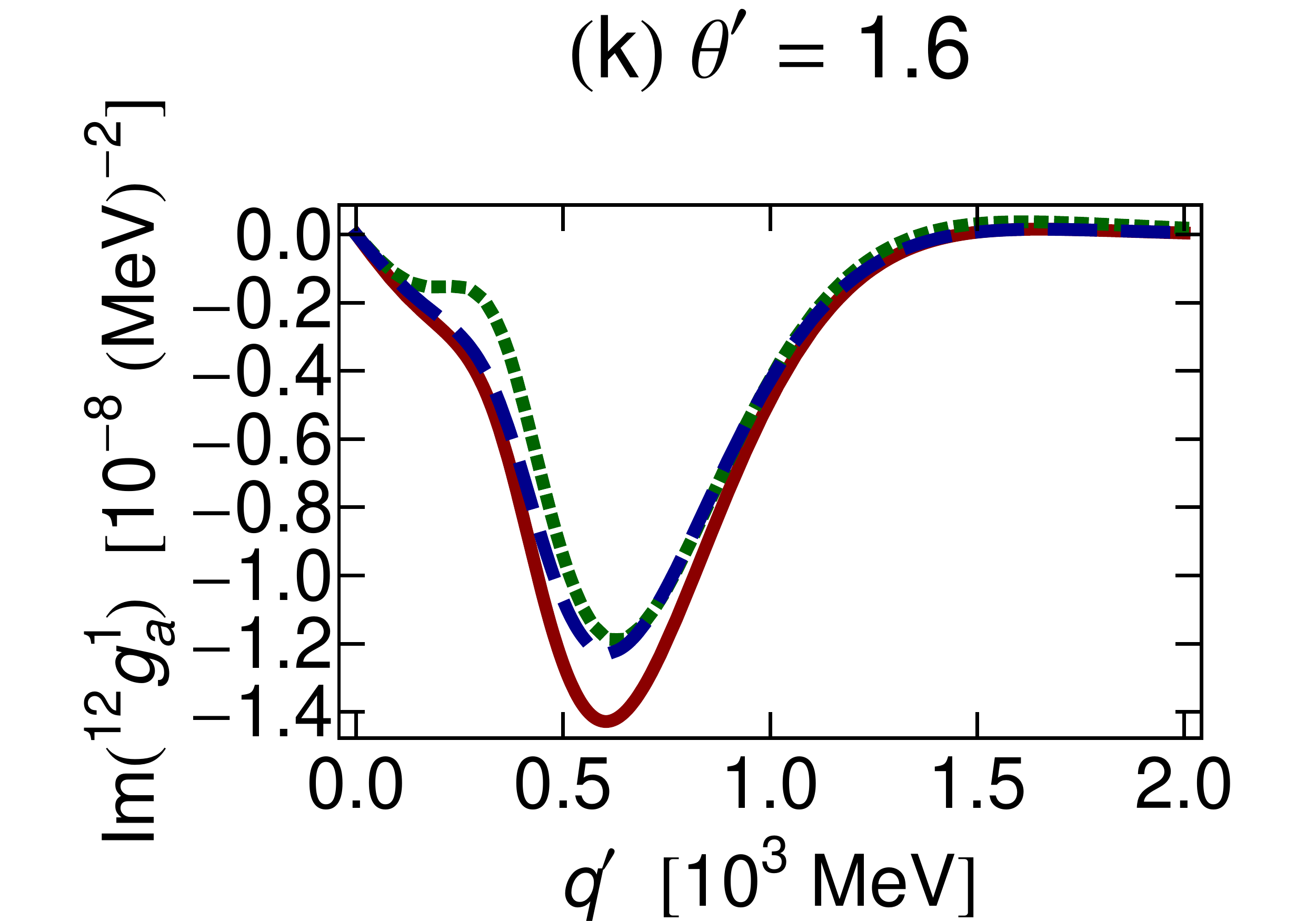}
}
\subfloat{
\includegraphics[width=6cm]{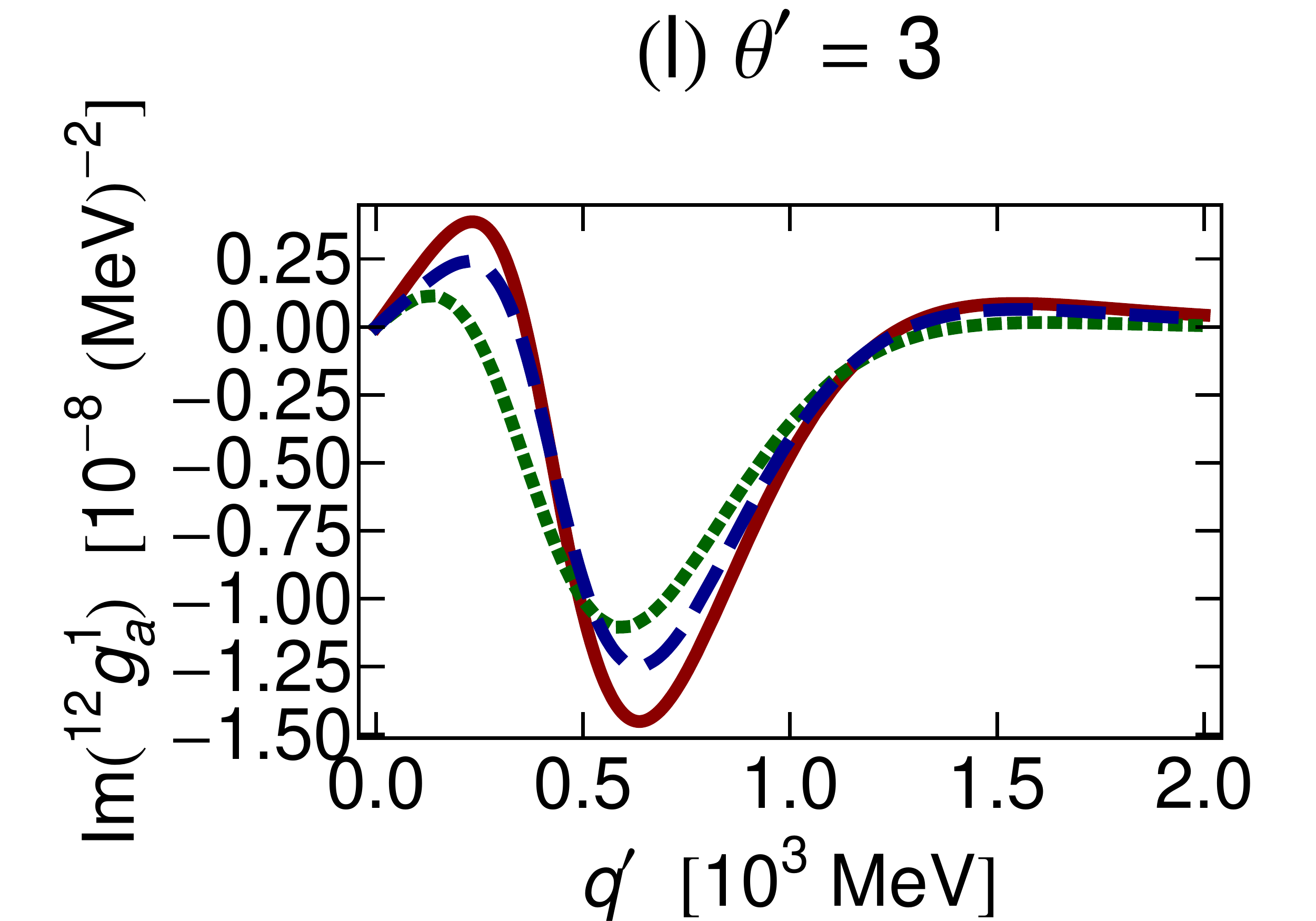}
}
\end{center}
\caption{(Color online) Same as Fig.~\ref{Fig:g_plots50_0} but for $\up{12}g^I_a$ and $\up{12}t^I_a$ at $q=375.29 \units{MeV}$.}
\label{Fig:g_plots2}
\label{Fig:g_plots300_12}
\end{figure}

In Figs.~\ref{Fig:g_plots50_0}-\ref{Fig:g_plots300_12} the real and imaginary parts of amplitudes $\up{0}g_a^I$, $\up{1}g_a^I$, and $\up{12}g_a^I$
are displayed as a function of the off-shell momentum $q^\prime$. In each figure, the on-shell momentum $q$ and initial polar angle $\theta$ are held fixed. Furthermore, each frame corresponds to a selected value of $\theta^\prime$. Both isospin states are displayed.   
The four on-shell momenta selected for Figs.~\ref{Fig:g_plots50_0}-\ref{Fig:g_plots50_12}, \ref{Fig:g_plots100_0}-\ref{Fig:g_plots100_12}, \ref{Fig:g_plots200_0}-\ref{Fig:g_plots200_12}, and \ref{Fig:g_plots300_0}-\ref{Fig:g_plots300_12} correspond to (in-vacuum) laboratory energies equal to $50$, $100$, $200$, and $300 \units{MeV}$, respectively. In all frames, the solid (red) curve shows the predictions in free space, while the dashed (blue) and the
dotted (green) curves show the predictions obtained with the angle-averaged and exact Pauli operator, respectively, close to nuclear matter density. 

As a general pattern, the imaginary part is considerably more sensitive to the
handling of Pauli blocking. This is not surprising, as the absence (or presence) of an imaginary part
arising from the residue in Eq.~(\ref{Eq:BG_final}) depends on whether $Q$ vanishes (or not) for a particular
combination of $q$, $P$, and $k_F$; thus, it should be sensitive to how $Q$ is defined and treated.

Concerning energy dependence, the impact of Pauli blocking i.e. the
differences between the solid (red) curve and either of the other two, is larger at lower on-shell
momentum, as expected. However, differences between the two sets of Pauli-modified calculations
tend to be more noticeable at those on-shell momenta where the $g$-matrix is complex.

For a given on-shell momentum (or in-vacuum energy), model dependence is largest at smaller values of $q^\prime$, but
comparable at all angles considered in the figures.

Interesting observations can be made with regard to how the different types of physical amplitudes respond to the improved description
of Pauli blocking. The least sensitive is the uncoupled singlet $\up{0}g_a^I$, shown in Figs.~\ref{Fig:g_plots50_0}, \ref{Fig:g_plots100_0}, \ref{Fig:g_plots200_0}, \ref{Fig:g_plots300_0}. Its real part shows hardly any sensitivity
to the removal of the spherical approximation, whereas
the imaginary part reveals some small to moderate sensitivity at $q=216.67 \units{MeV}$ and $q=306.42 \units{MeV}$.
This can be understood. Although the connection to the conventional description in terms of $LSJ$ states must be taken with caution (for the reasons
explained earlier), such connection is not entirely lost. Thus, we recall that a major 
singlet state is the $\up{1}S_0$ partial wave, which is not expected
to be sensitive to the introduction of non-spherical components in the Pauli operator.

The uncoupled triplet states $\up{1}g_a^I$ show moderate sensitivity, mostly in their imaginary parts,
see Figs.~\ref{Fig:g_plots50_1}, \ref{Fig:g_plots100_1}, \ref{Fig:g_plots200_1}, \ref{Fig:g_plots300_1}.
On the other hand, the coupled triplet states $\up{n}g_a^I$ for $n=12,34,55,66$,
show some remarkable differences between the two sets of predictions.
As a member of the four coupled states defined in Eqs.~(\ref{Eq:partialwave2d12})-(\ref{Eq:partialwave2d66}), we selected $\up{12}g_a^I$, shown in Figs.~\ref{Fig:g_plots50_12}, \ref{Fig:g_plots100_12}, \ref{Fig:g_plots200_12}, \ref{Fig:g_plots300_12}. Differences between the dotted (blue) and dashed (green) curves at low $q^\prime$ can be substantial in all cases where the imaginary part is non-zero.

Concerning isospin dependence, generally the pattern is similar for $I=0$ and $I=1$, with
slightly more sensitivity in $I=0$ states. In terms of $LSJ$ states, the $\up{3}S_1$ wave, which receives large contribution from the tensor force, is likely to be sensitive to a non-spherical treatment of Pauli blocking.

At this point it's appropriate to elaborate further on the fact that
the largest differences between predictions originating from different handling's of such an important medium effect
as Pauli blocking occur in the imaginary part of the $g$-matrix. We note that
such differences would be entirely suppressed if for instance {\it in-medium} differential cross sections, which are an
important ingredient of transport models in heavy-ion collisions, were
calculated using the real $R$-matrix (also known as the $K$-matrix).
In-medium equivalence of the $R$-matrix and $T$-matrix formalisms (an assumption which is correct in-vacuum provided there are no open inelastic channels),
implies the validity of free-space unitarity. However, the latter is violated in the medium due to the presence of
Pauli-blocked (but otherwise energetically open) channels. We believe this renders the use of the $R$-matrix unsuitable in the medium, even in the absence of inelasticities in the potential. The present observation of the imaginary part being the most sensitive to modifications in the Pauli operator appears to strengthen this point.

\begin{figure}[H]
\begin{center}
\subfloat{
\includegraphics[width=6cm]{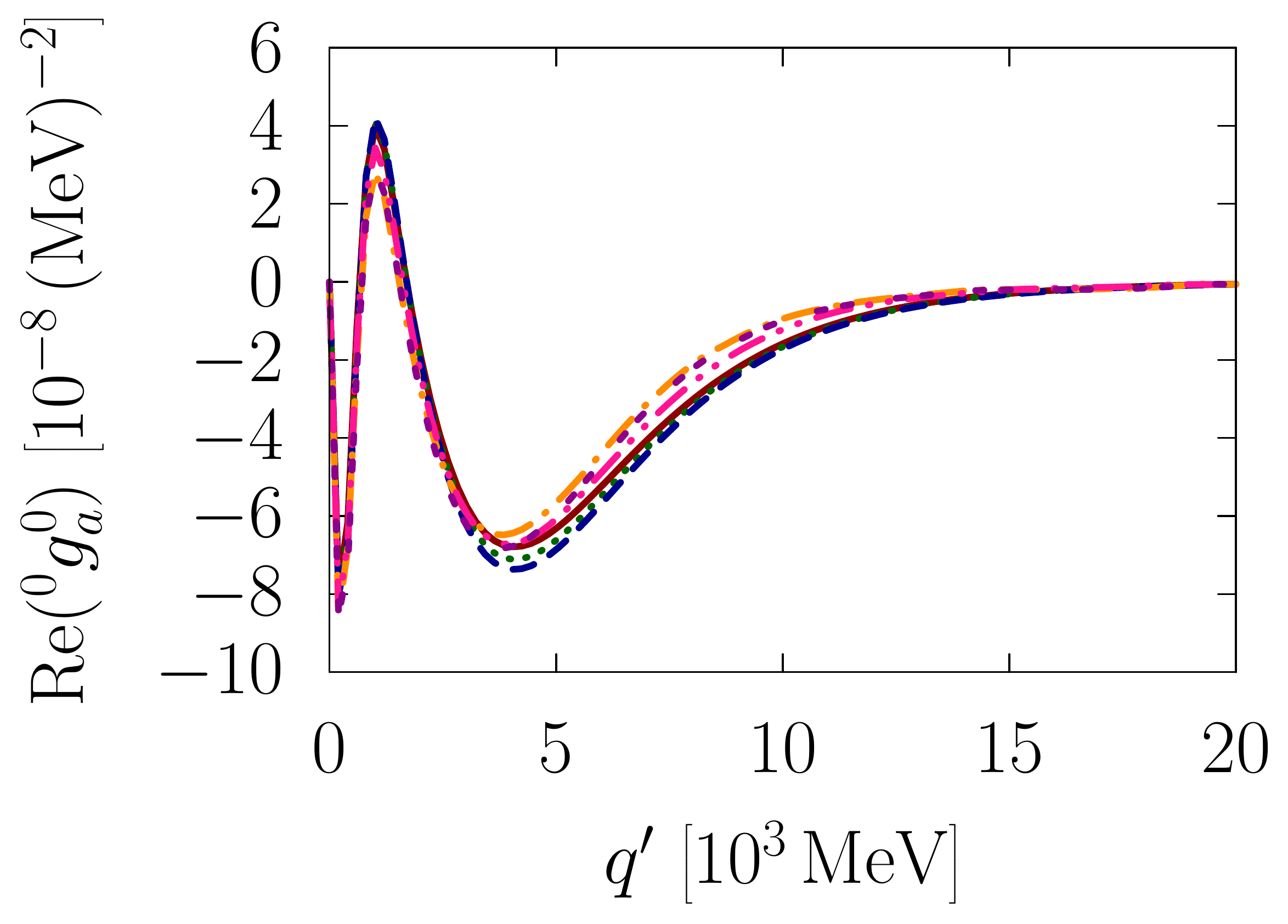}
}
\subfloat{
\includegraphics[width=6cm]{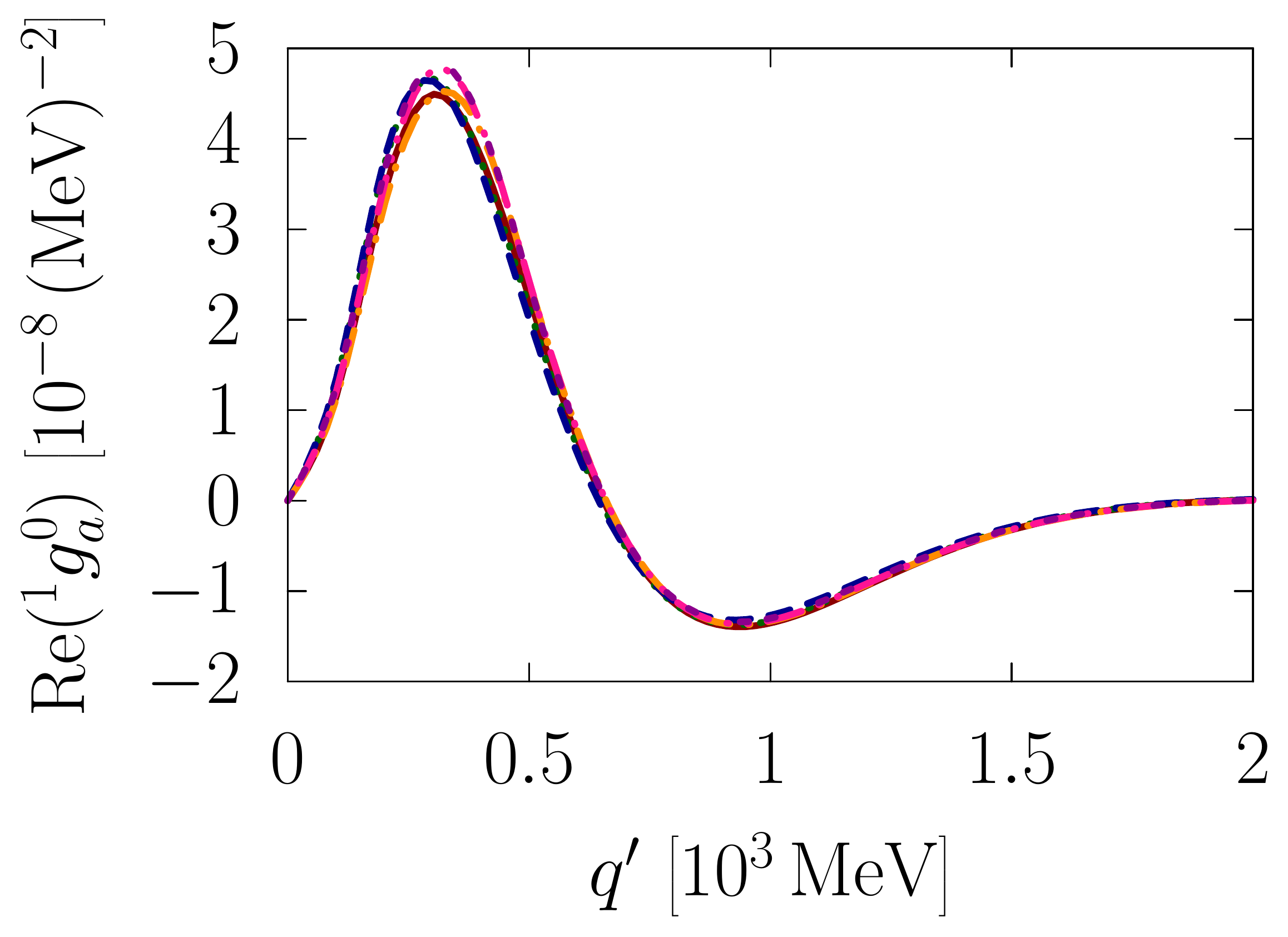}
}
\subfloat{
\includegraphics[width=6cm]{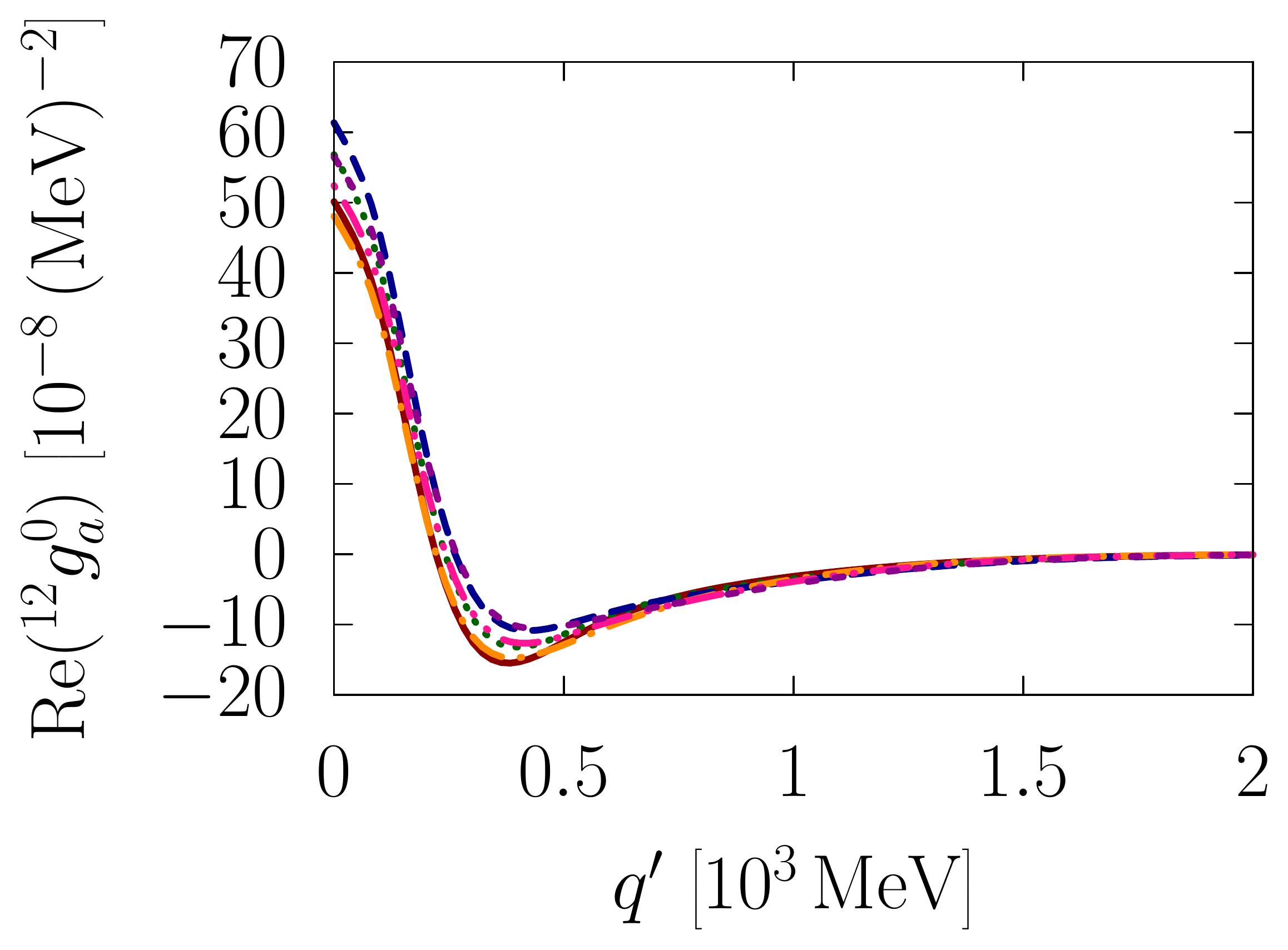}
}
\\
\subfloat{
\includegraphics[width=6cm]{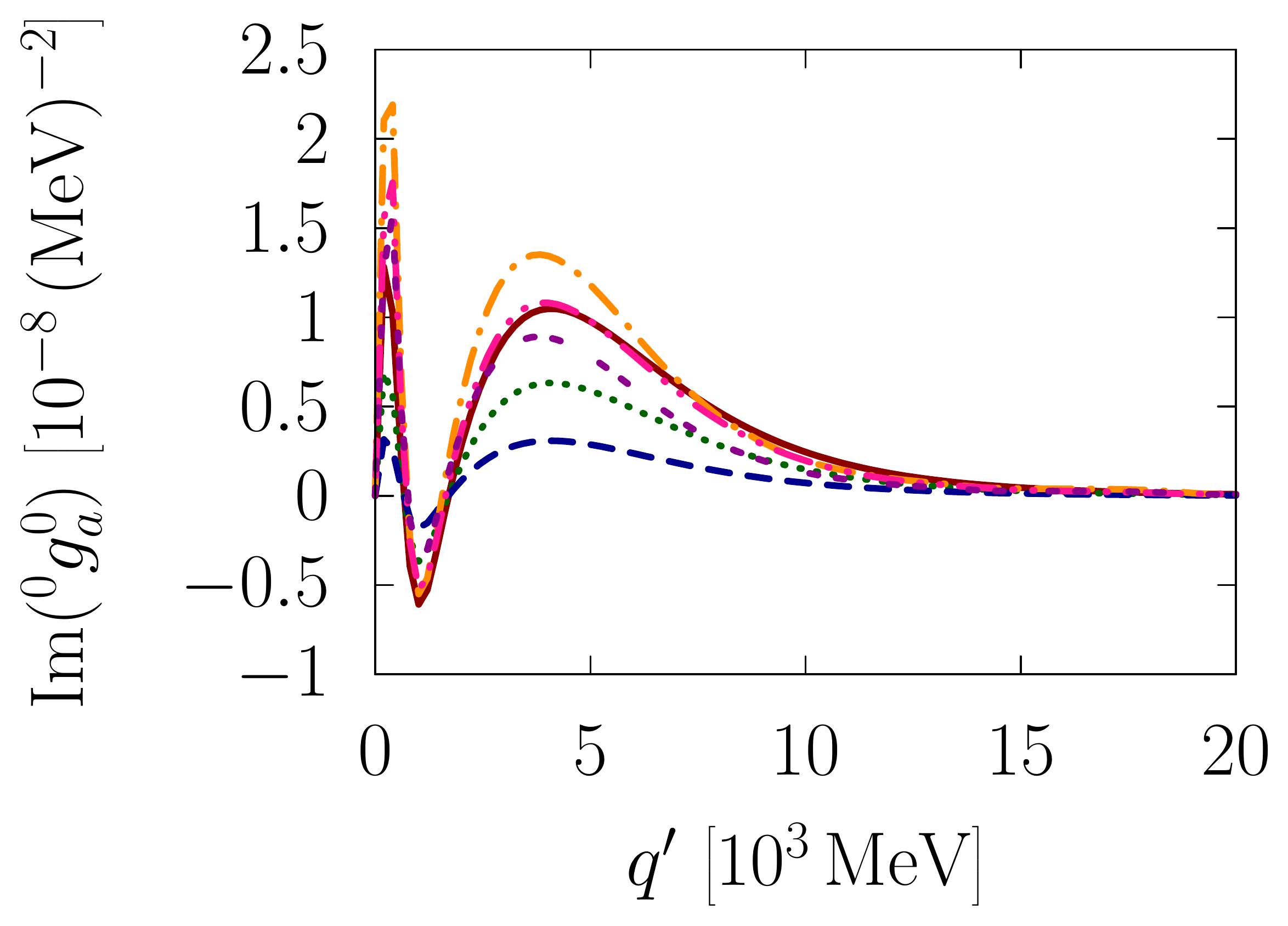}
}
\subfloat{
\includegraphics[width=6cm]{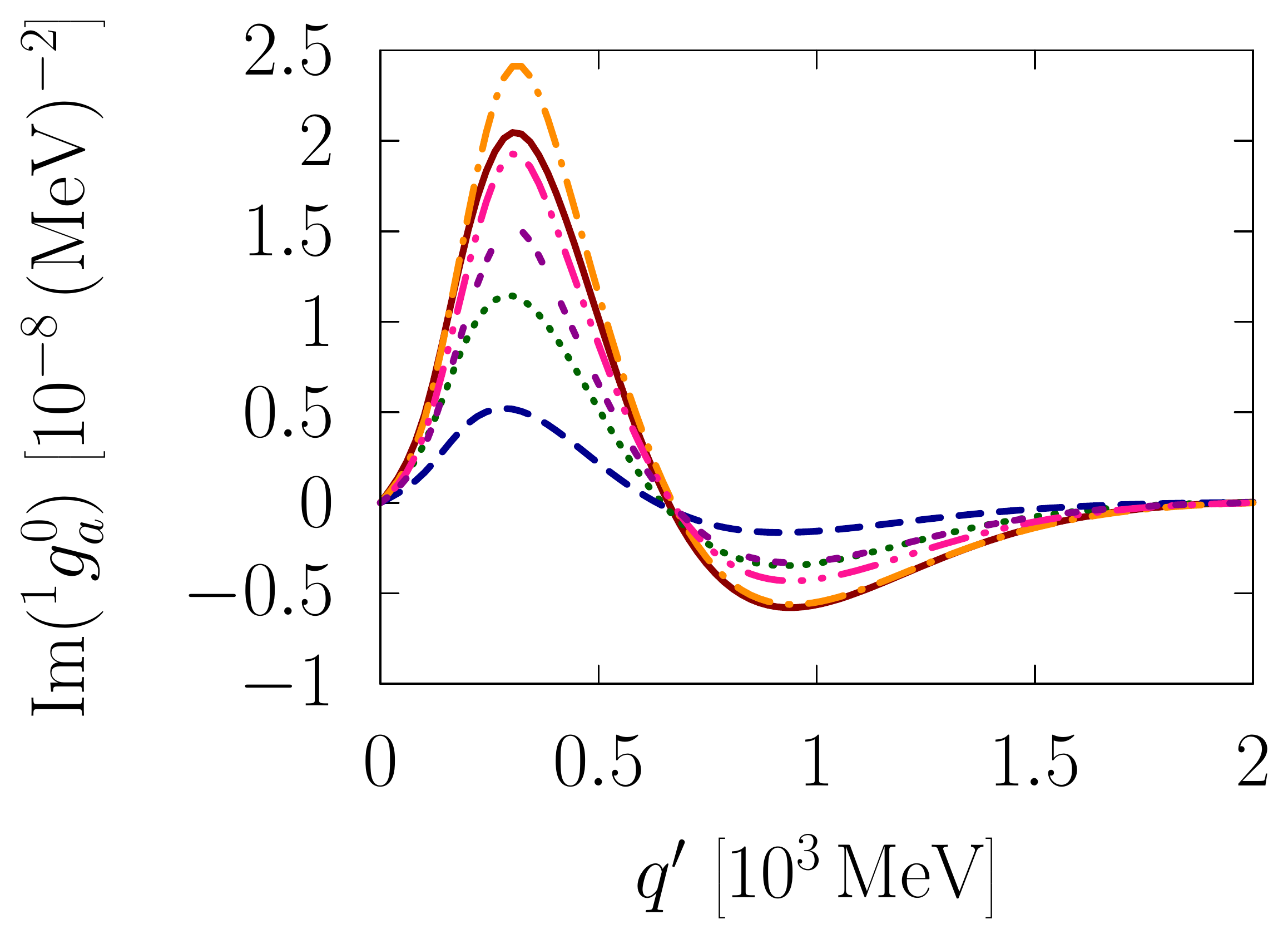}
}
\subfloat{
\includegraphics[width=6cm]{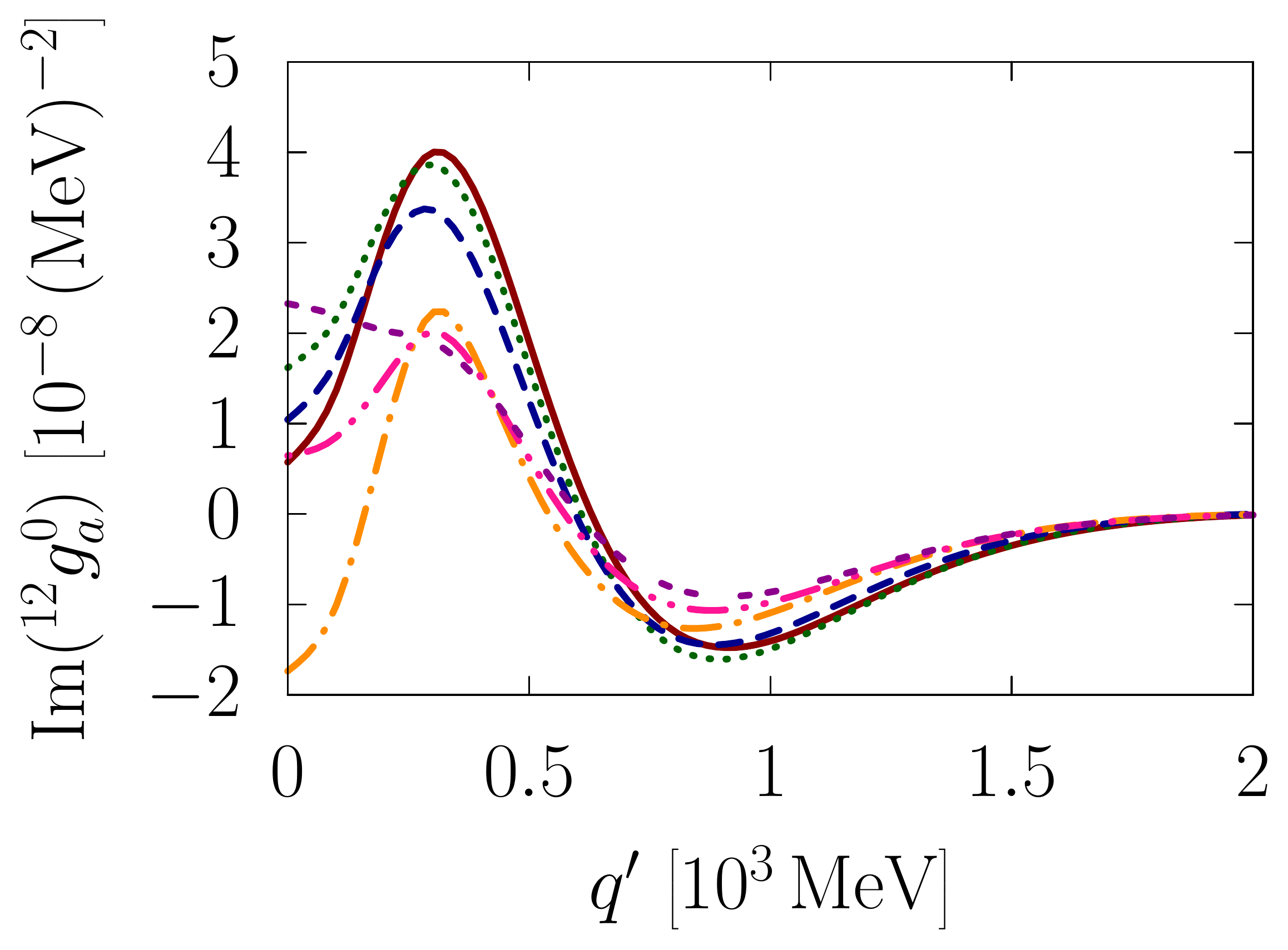}
}
\\
\subfloat{
\includegraphics[width=6cm]{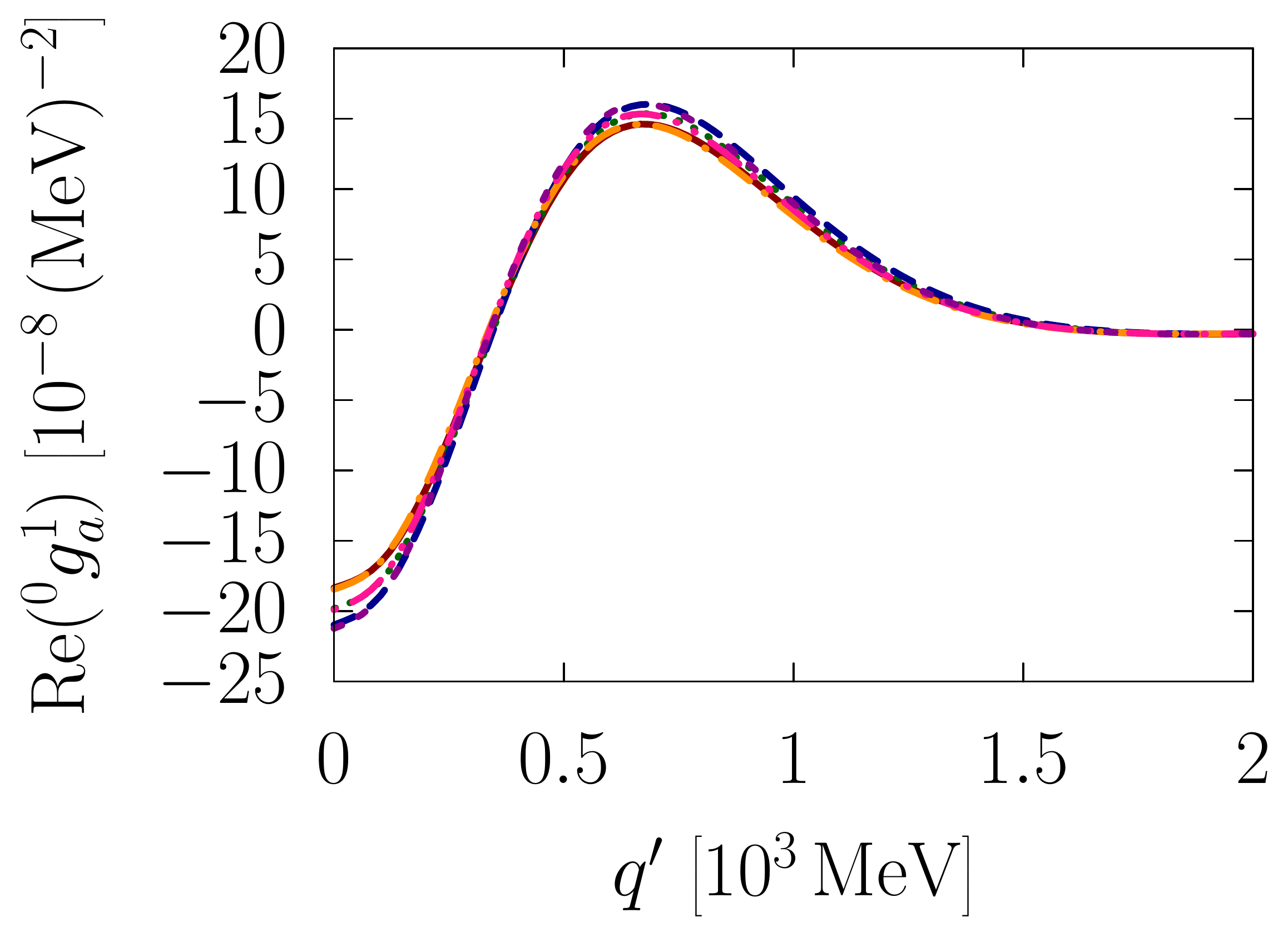}
}
\subfloat{
\includegraphics[width=6cm]{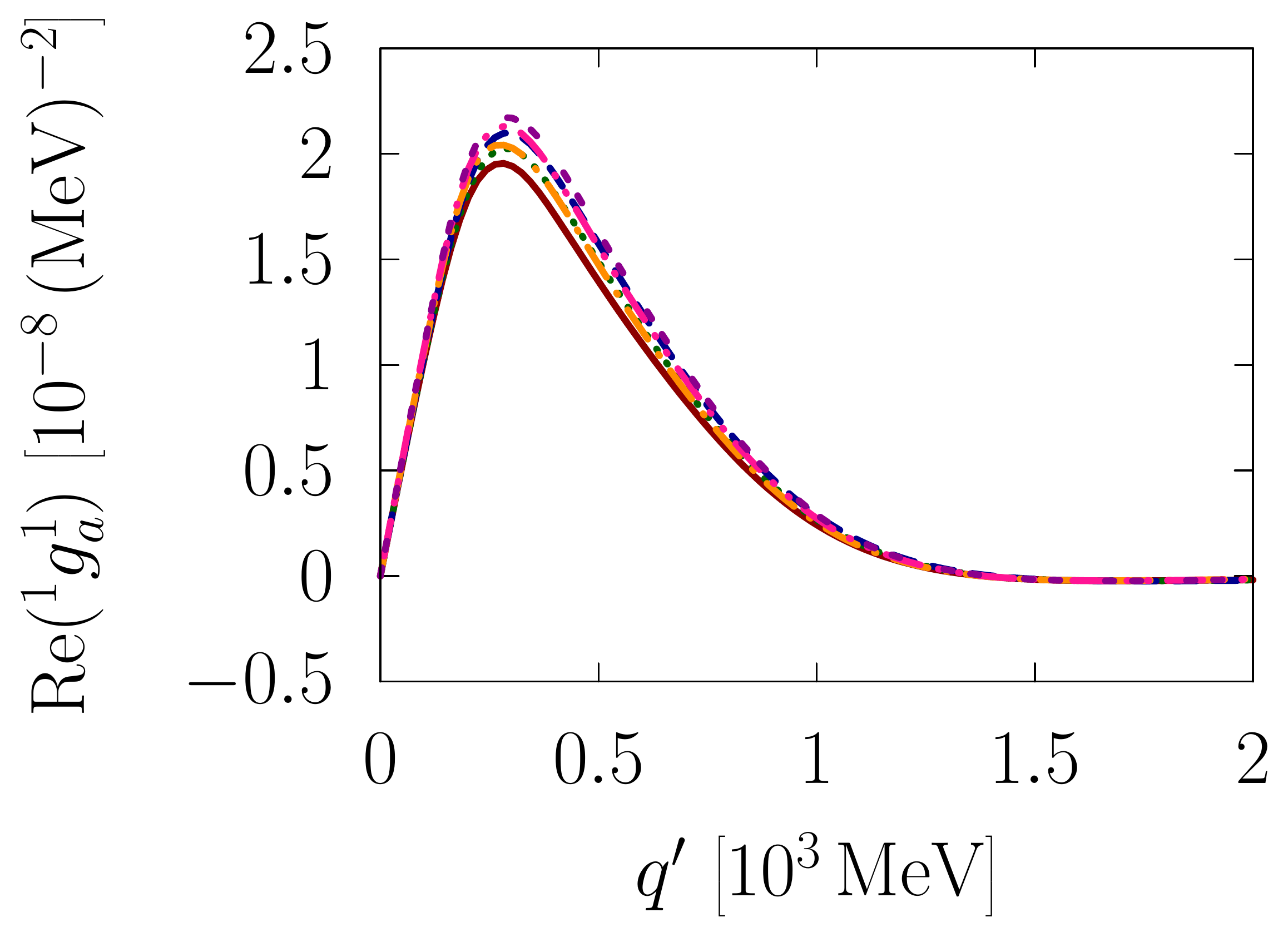}
}
\subfloat{
\includegraphics[width=6cm]{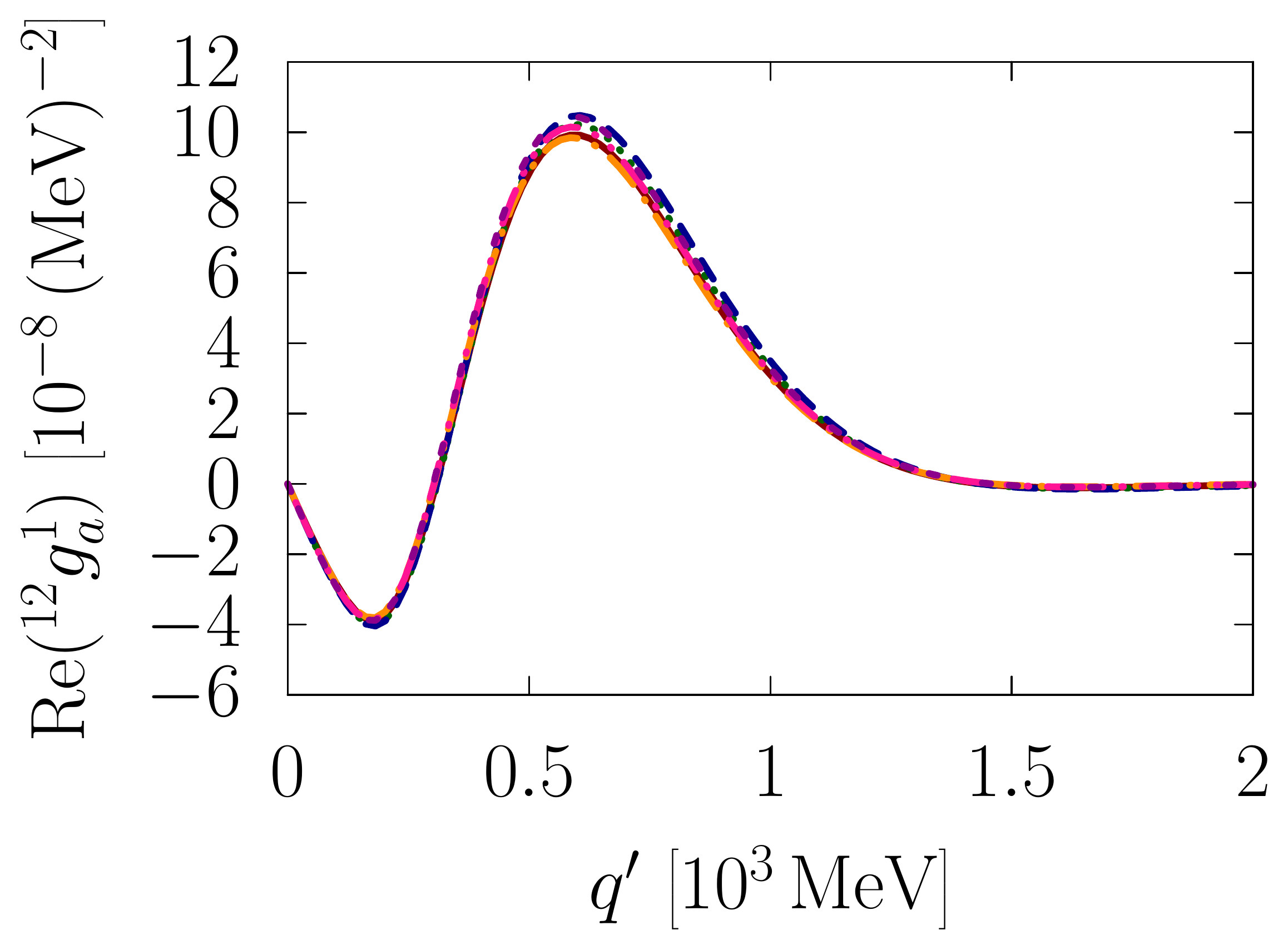}
}
\\
\subfloat{
\includegraphics[width=6cm]{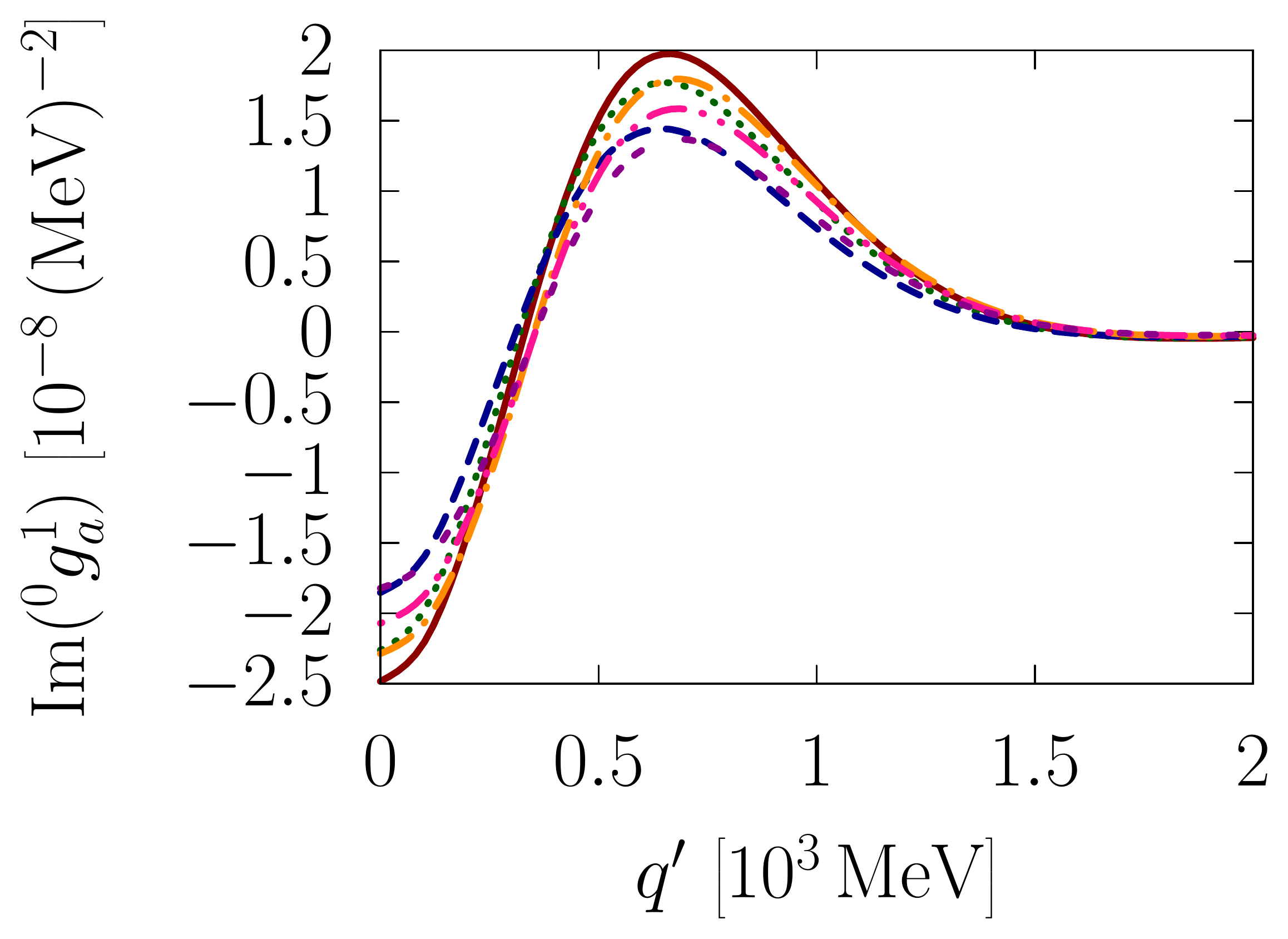}
}
\subfloat{
\includegraphics[width=6cm]{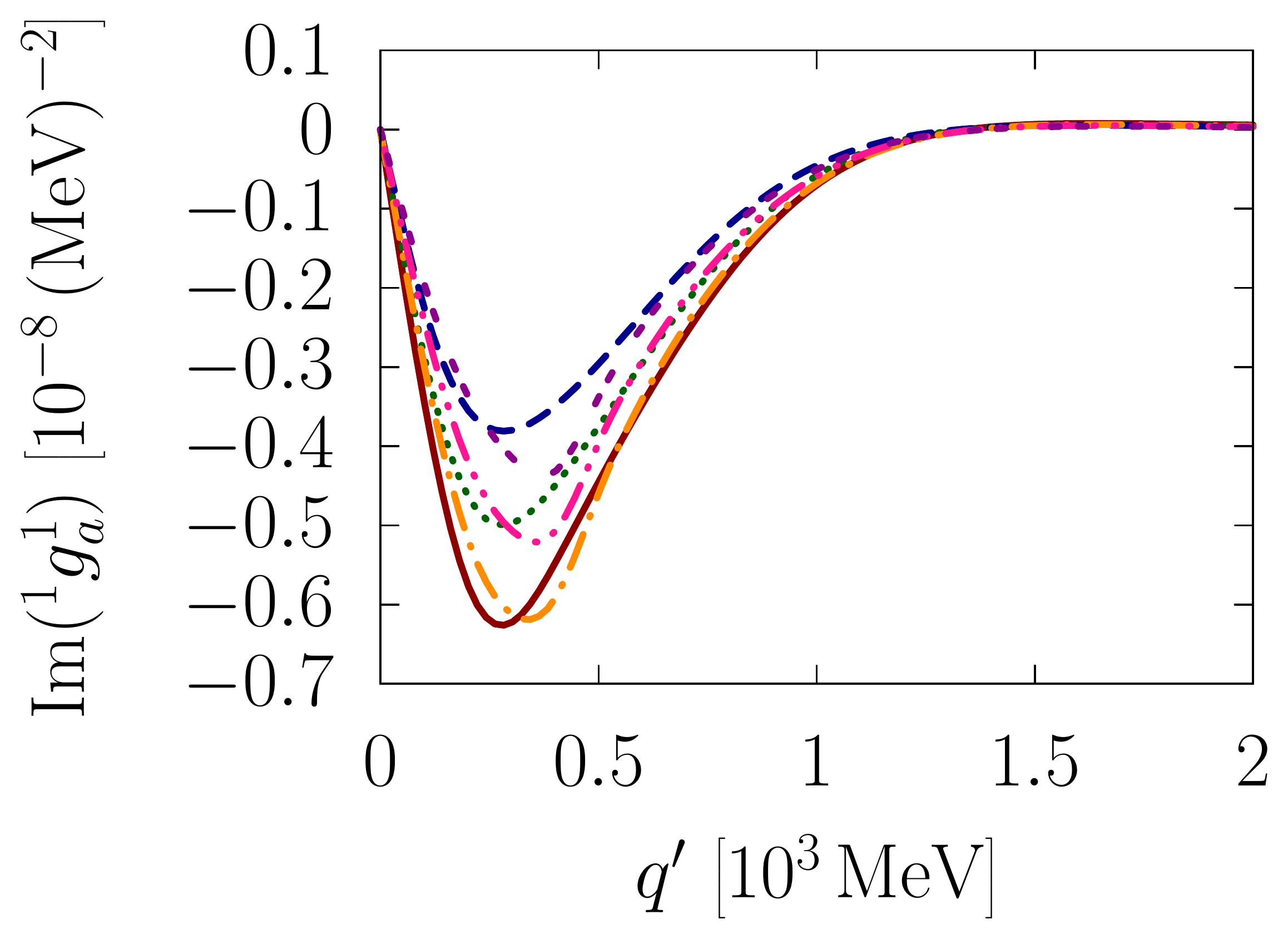}
}
\subfloat{
\includegraphics[width=6cm]{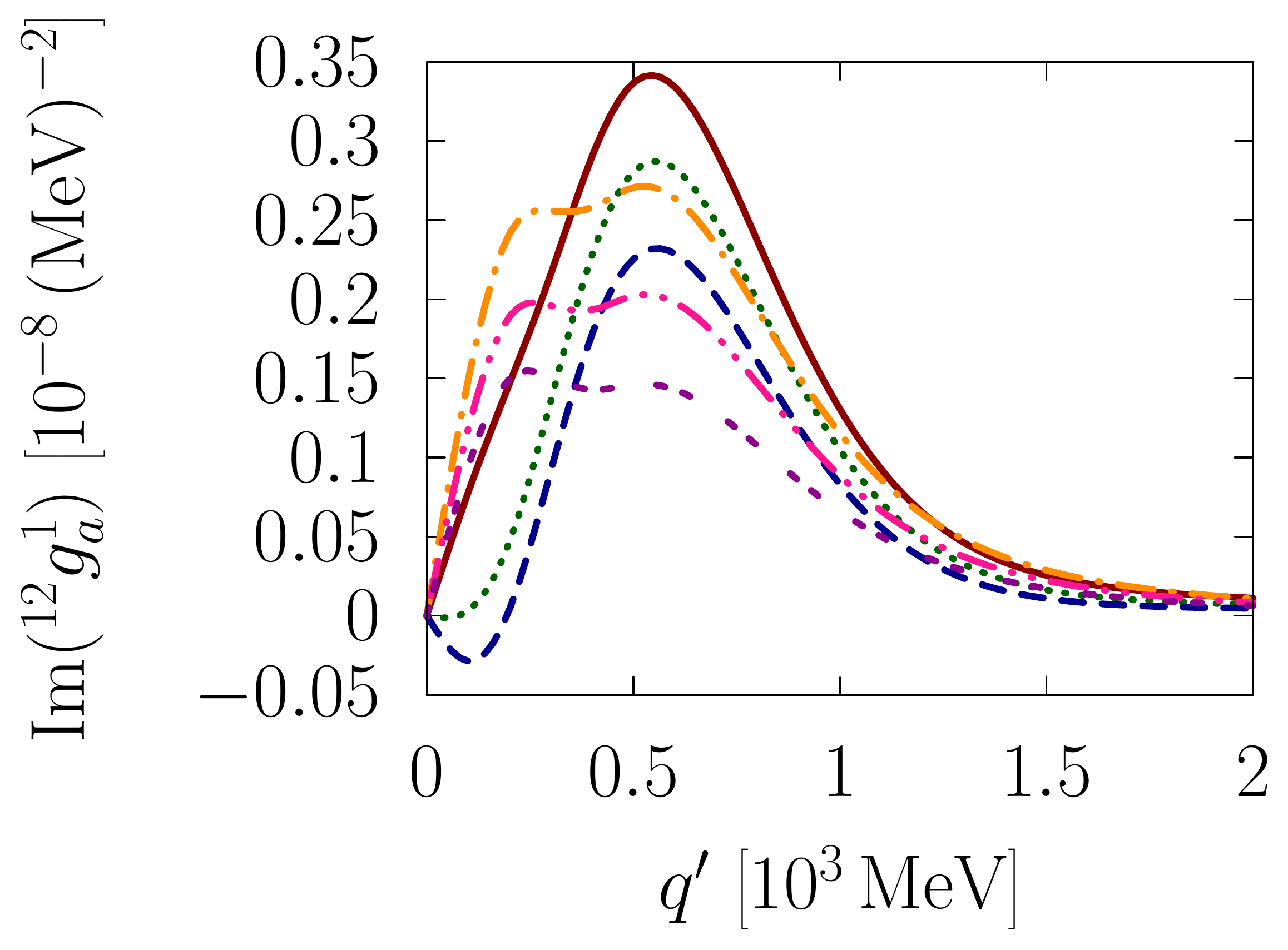}
}
\end{center}
\caption{(Color online) Real and imaginary parts of $\up{n}g^I_a$ (both isospins) for $n=0,1,12$ as a function of $q^\prime$. We set $\theta=\arccos(0.5)$, $q=306.42 \units{MeV}$, and $\theta^\prime =  3$. The solid (red), dotted (green), and dashed (blue) curves are exact Pauli operator calculations preformed at $k_F=1.1$, $1.4$, and $1.6 \units{fm^{-1}}$, respectively. The dashed-dot (orange), dashed-double-dot (pink), and double-dashed (purple) are the corresponding spherical Pauli operator calculations preformed at $k_F=1.1$, $1.4$, and $1.6 \units{fm^{-1}}$, respectively.}
\label{Fig:kf}
\end{figure}

Before closing, some comments are in place concerning the density dependence. In fact, densities lower
than saturation density play an important role in the construction of optical potentials. 
As a demonstration of the density dependence, we take some selected amplitudes and show predictions for Fermi momenta equal to $1.1$, $1.4$, and $1.6 \units{fm^{-1}}$ with both exact $Q$ and angle-averaged $Q$,    
see Fig.~\ref{Fig:kf}. In each frame we plot the real and imaginary parts of amplitudes $\up{0}g^I_a$, $\up{1}g^I_a$, and, $\up{12}g^I_a$ (both isospins) as a function of the off-shell momentum $q^\prime$. We choose specific conditions, namely $q=306.42 \units{MeV}$, $\theta=\arccos(0.5)$, and $\theta^\prime=3$, which are a subset of the most sensitive cases from Figs.\ref{Fig:g_plots50_0}-\ref{Fig:g_plots300_12}. In all frames, the solid (red), dotted (green), and dashed (blue) curves are exact Pauli operator calculations performed at $k_F=1.1$, $1.4$, and $1.6 \units{fm^{-1}}$, respectively. The dashed-dot (orange), dashed-double-dot (pink), and double-dashed (purple) are the corresponding spherical Pauli operator calculations preformed at $k_F=1.1$, $1.4$, and $1.6 \units{fm^{-1}}$, respectively. 

Consistent with our previous findings, at all three densities the real part is less sensitive to model differences than the imaginary part, an observation which applies to all frames in Fig.~\ref{Fig:kf}. Also, the $I=0$ case tends to be more sensitive than the $I=1$ case at all three densities.

By looking, for instance, at the imaginary                            
part of $\up{1}g_a^0$, we see that the differences between the two sets of calculations (solid (red) {\it vs.} 
dashed-dot (orange) for 
$k_F=1.1 \units{fm^{-1}}$, and dashed (blue) {\it vs.} double-dashed (purple) for                              
$k_F=1.6 \units{fm^{-1}}$), are larger at the larger density.                                                  
Furthermore, at the highest density the imaginary parts of 
$\up{12}g_a^0$ and $\up{12}g_a^1$, as calculated with the two methods, show different qualitative trends, whereas,
at the lower densities, all curves tend to display similar trends.                                             

In summary, we identified some remarkable differences between predictions with or without the angle-average
approximation in the Pauli operator, particularly in the imaginary part of the coupled states.
Application of the present $g$-matrix in nuclear systems/reactions which are sensitive to the off-shell nature
of the NN amplitudes have the best potential to reveal sensitivity to the improved description of Pauli blocking.
\begin{table}[H]
\caption{Our calculated and transformed free-space $LSJ$ partial waves (inside square brackets), along with the direct partial wave decomposition solution (outside square brackets). We show on-shell partial waves at $E_{Lab}=50,100 \units{MeV}$ (i.e. $q^\prime=q=153.21,216.67 \units{MeV}$).}
\label{Table:freespaceLSJ_1}
\begin{center}
\begin{tabular}{l \tspace{1.0cm} l \tspace{1.0cm} l}
\hline\hline 
Partial wave & $50 \quad (\units{ 10^{-9} \, MeV^{-2} })$ & $100 \quad (\units{ 10^{-9} \, MeV^{-2} })$ \\
\hline
$\up{1}S_0$ & $-2165.71-i1902.87 [-2165.69-i1902.72]$ & $-1240.24-i637.93 [-1240.20-i637.87]$ \\
$\up{3}P_0$ & $-1002.57-i243.73 [-1002.56-i243.73]$ & $-623.64-i133.38 [-623.63-i133.38]$ \\
\\
$\up{1}P_1$ & $685.36-i110.33 [687.55-i111.06]$ & $607.79-i126.39 [612.35-i128.38]$ \\
$\up{3}P_1$ & $671.08-i105.66 [671.08-i105.66]$ & $701.09-i170.76 [701.09-i170.76]$ \\
$\up{3}S_1$ & $-1885.57-i3277.79 [-1885.67-i3277.58]$ & $-1491.65-i1222.11 [-1491.63-i1222.05]$ \\
$\up{3}S_1$-$\up{3}D_1$ & $-78.35-i105.27 [-78.38-i105.30]$ & $-79.19-i39.96 [-79.21-i39.97]$ \\
$\up{3}D_1$-$\up{3}S_1$ & $-78.35-i105.27 [-78.38-i105.29]$ & $-79.19-i39.96 [-79.21-i39.97]$ \\
$\up{3}D_1$ & $506.59-i63.63 [506.59-i63.63]$ & $642.79-i144.98 [642.79-i144.98]$ \\
\\
$\up{1}D_2$ & $-119.96-i3.30 [-119.96-i3.30]$ & $-175.20-i10.10 [-175.19-i10.10]$ \\
$\up{3}D_2$ & $-730.69-i125.87 [-730.66-i125.86]$ & $-944.44-i327.76 [-944.40-i327.74]$ \\
$\up{3}P_2$ & $-459.85-i53.59 [-459.83-i53.59]$ & $-597.58-i130.22 [-597.55-i130.21]$ \\
$\up{3}P_2$-$\up{3}F_2$ & $139.64+i15.74 [139.64+i15.74]$ & $148.16+i32.51 [148.15+i32.51]$ \\
$\up{3}F_2$-$\up{3}P_2$ & $139.64+i15.74 [139.64+i15.74]$ & $148.16+i32.51 [148.15+i32.51]$ \\
$\up{3}F_2$ & $-26.02-i4.68 [-26.01-i4.68]$ & $-41.08-i8.12 [-41.08-i8.12]$ \\
\\
$\up{1}F_3$ & $92.12-i1.94 [92.10-i1.94]$ & $126.03-i5.22 [126.01-i5.22]$ \\
$\up{3}F_3$ & $56.39-i0.73 [56.38-i0.73]$ & $89.49-i2.63 [89.48-i2.63]$ \\
$\up{3}D_3$ & $-28.23-i4.12 [-28.22-i4.12]$ & $-96.55-i15.89 [-96.53-i15.88]$ \\
$\up{3}D_3$-$\up{3}G_3$ & $-131.03-i0.21 [-131.00-i0.21]$ & $-197.13-i2.74 [-197.10-i2.74]$ \\
$\up{3}G_3$-$\up{3}D_3$ & $-131.03-i0.21 [-131.00-i0.21]$ & $-197.13-i2.74 [-197.10-i2.74]$ \\
$\up{3}G_3$ & $21.36-i4.04 [21.36-i4.04]$ & $54.63-i13.79 [54.62-i13.79]$ \\
\\
$\up{1}G_4$ & $-11.89-i0.03 [-11.89-i0.03]$ & $-22.00-i0.16 [-21.99-i0.16]$ \\
$\up{3}G_4$ & $-58.64-i0.79 [-58.61-i0.79]$ & $-122.59-i4.94 [-122.53-i4.93]$ \\
$\up{3}F_4$ & $-7.71-i0.07 [-7.70-i0.07]$ & $-22.37-i0.46 [-22.36-i0.46]$ \\
$\up{3}F_4$-$\up{3}H_4$ & $15.40+i0.03 [15.40+i0.03]$ & $30.10+i0.28 [30.08+i0.28]$ \\
$\up{3}H_4$-$\up{3}F_4$ & $15.40+i0.03 [15.40+i0.03]$ & $30.10+i0.28 [30.08+i0.28]$ \\
$\up{3}H_4$ & $-2.00-i0.06 [-2.00-i0.06]$ & $-5.81-i0.31 [-5.81-i0.31]$ \\
\\
$\up{1}H_5$ & $13.34-i0.04 [13.32-i0.04]$ & $30.26-i0.30 [30.21-i0.30]$ \\
$\up{3}H_5$ & $6.69-i0.01 [6.68-i0.01]$ & $17.13-i0.10 [17.10-i0.10]$ \\
$\up{3}G_5$ & $4.04-i0.07 [4.04-i0.07]$ & $9.70-i0.58 [9.68-i0.58]$ \\
$\up{3}G_5$-$\up{3}I_5$ & $-16.65+i0.02 [-16.63+i0.02]$ & $-40.76+i0.22 [-40.71+i0.22]$ \\
$\up{3}I_5$-$\up{3}G_5$ & $-16.65+i0.02 [-16.63+i0.02]$ & $-40.76+i0.22 [-40.71+i0.22]$ \\
$\up{3}I_5$ & $1.83-i0.06 [1.83-i0.06]$ & $7.08-i0.56 [7.07-i0.56]$ \\
\hline\hline
\end{tabular}
\end{center}
\end{table}
\begin{table}[H]
\caption{Same as Table~\ref{Table:freespaceLSJ_1} but at $E_{Lab}=200,300 \units{MeV}$ (i.e. $q^\prime=q=306.42,375.29 \units{MeV}$).}
\label{Table:freespaceLSJ_2}
\begin{center}
\begin{tabular}{l \tspace{1.0cm} l \tspace{1.0cm} l}
\hline\hline 
Partial wave & $200 \quad (\units{ 10^{-9} \, MeV^{-2} })$ & $300 \quad (\units{ 10^{-9} \, MeV^{-2} })$ \\
\hline
$\up{1}S_0$ & $-315.87-i48.55 [-315.83-i48.54]$ & $125.14-i9.39 [125.23-i9.40]$ \\
$\up{3}P_0$ & $-77.92-i2.89 [-77.91-i2.89]$ & $227.48-i31.43 [227.49-i31.44]$ \\
\\
$\up{1}P_1$ & $513.00-i133.59 [522.53-i138.97]$ & $452.83-i132.73 [467.74-i142.51]$ \\
$\up{3}P_1$ & $693.46-i260.99 [693.46-i260.99]$ & $668.79-i332.51 [668.81-i332.54]$ \\
$\up{3}S_1$ & $-547.70-i156.54 [-547.67-i156.52]$ & $-14.13-i3.07 [-13.74-i3.08]$ \\
$\up{3}S_1$-$\up{3}D_1$ & $-69.25+i3.77 [-69.25+i3.77]$ & $-65.77+i24.97 [-65.85+i25.02]$ \\
$\up{3}D_1$-$\up{3}S_1$ & $-69.25+i3.77 [-69.25+i3.77]$ & $-65.77+i24.97 [-65.85+i25.02]$ \\
$\up{3}D_1$ & $641.71-i221.33 [641.71-i221.33]$ & $565.21-i223.02 [565.23-i223.05]$ \\
\\
$\up{1}D_2$ & $-224.60-i24.26 [-224.59-i24.26]$ & $-220.56-i29.52 [-220.55-i29.52]$ \\
$\up{3}D_2$ & $-869.31-i459.72 [-869.28-i459.70]$ & $-711.51-i394.54 [-711.47-i394.51]$ \\
$\up{3}P_2$ & $-559.24-i166.64[-559.21-i166.63]$ & $-452.36-i134.09 [-452.21-i134.00]$ \\
$\up{3}P_2$-$\up{3}F_2$ & $95.61+i29.80 [95.60+i29.80]$ & $46.29+i14.26 [46.31+i14.26]$ \\
$\up{3}F_2$-$\up{3}P_2$ & $95.61+i29.80 [95.60+i29.80]$ & $46.29+i14.26 [46.31+i14.26]$ \\
$\up{3}F_2$ & $-42.65-i5.65 [-42.65-i5.65]$ & $-22.70-i1.71 [-22.70-i1.71]$ \\
\\
$\up{1}F_3$ & $134.38-i8.62 [134.35-i8.62]$ & $134.40-i10.84 [134.37-i10.83]$ \\
$\up{3}F_3$ & $111.75-i5.95 [111.73-i5.95]$ & $119.08-i8.50 [119.06-i8.49]$ \\
$\up{3}D_3$ & $-200.29-i42.51 [-200.26-i42.50]$ & $-231.33-i56.65 [-231.27-i56.64]$ \\
$\up{3}D_3$-$\up{3}G_3$ & $-217.68-i10.32 [-217.64-i10.31]$ & $-195.35-i12.68 [-195.31-i12.67]$ \\
$\up{3}G_3$-$\up{3}D_3$ & $-217.68-i10.32 [-217.64-i10.31]$ & $-195.35-i12.68 [-195.32-i12.67]$ \\
$\up{3}G_3$ & $103.94-i28.09 [103.92-i28.08]$ & $128.30-i33.32 [128.28-i33.31]$ \\
\\
$\up{1}G_4$ & $-32.53-i0.50 [-32.51-i0.50]$ & $-39.59-i0.93 [-39.56-i0.93]$ \\
$\up{3}G_4$ & $-192.78-i17.82 [-192.68-i17.81]$ & $-223.02-i30.19 [-222.89-i30.17]$ \\
$\up{3}F_4$ & $-51.83-i2.16 [-51.80-i2.16]$ & $-73.35-i4.49 [-73.31-i4.49]$ \\
$\up{3}F_4$-$\up{3}H_4$ & $43.10+i1.30 [43.08+i1.30]$ & $46.13+i2.43 [46.11+i2.43]$ \\
$\up{3}H_4$-$\up{3}F_4$ & $43.10+i1.30 [43.08+i1.30]$ & $46.13+i2.43 [46.11+i2.43]$ \\
$\up{3}H_4$ & $-11.68-i0.95 [-11.68-i0.95]$ & $-14.76-i1.40 [-14.75-i1.40]$ \\
\\
$\up{1}H_5$ & $45.59-i0.99 [45.53-i0.99]$ & $49.27-i1.45 [49.21-i1.45]$ \\
$\up{3}H_5$ & $30.16-i0.43 [30.12-i0.43]$ & $36.12-i0.78 [36.08-i0.78]$ \\
$\up{3}G_5$ & $11.41-i2.30 [11.39-i2.29]$ & $5.19-i3.85 [5.19-i3.84]$ \\
$\up{3}G_5$-$\up{3}I_5$ & $-68.52+i0.98 [-68.43+i0.97]$ & $-80.09+i1.57 [-79.98+i1.57]$ \\
$\up{3}I_5$-$\up{3}G_5$ & $-68.52+i0.98 [-68.43+i0.97]$ & $-80.09+i1.57 [-79.98+i1.57]$ \\
$\up{3}I_5$ & $18.46-i2.40 [18.44-i2.39]$ & $27.61-i4.29 [27.57-i4.28]$ \\
\hline\hline
\end{tabular}
\end{center}
\end{table}
\section{Summary and Conclusion}
\label{IV} 
We have solved the integral equation for scattering of two nucleons in the medium without the use
of partial wave expansion. As part of our three-dimensional formalism, we
provided explicit formulas for the three-dimensional relativistic OBE amplitudes, which are more general
than those already in literature.

First, we verify the accuracy of our calculation by reproducing closely existing free-space results. We then proceed to apply Pauli blocking effects in the integral equation and compare our predictions with those obtained with the popular spherical
approximation. Although the implementation of the exact Pauli operator is straightforward in the three-dimensional formalism, care must be exercised when extracting the physical states in the medium. 

We observe potentially significant differences, particularly in the imaginary part of specific
combinations of off-shell helicity amplitudes. Coupled states, which are driven by the tensor force, appear to be most impacted by the presence of a
non-spherical Pauli operator. It will be interesting and informative to explore to which extent these differences may impact
physically observable systems, a focal point of our future research.  
\begin{acknowledgments}
Support from the U.S. Department of Energy under Grant No. DE-FG02-03ER41270 is acknowledged.
\end{acknowledgments}
\appendix
\appendixpage
\section{One-boson-exchange potentials in plane-wave helicity formalism}
\label{OBEP} 
The momentum space one-boson-exchange potentials (OBEP) presented in this section are a modification of those found in Machleidt \textit{et al}.~\cite[Appx. E]{Mac87}. The following modifications are preformed:
\begin{enumerate}
\item Full three-dimensional treatment of momenta and rotated helicity wavefunctions.                   
\item The formulas apply to two baryons with different masses.                                                  
\item The Thompson propagator is used in place of the Blankenbecler and Sugar propagator. This allows the 
transition from Eq.~(\ref{Eq:Thompson_full}) to Eq.~(\ref{Eq:Thompson_mod}).    
\end{enumerate}
\subsection{Interaction Lagrangians and Dirac spinors}
Guided by symmetry principles, simplicity, and physical intuition, the most commonly used interaction Lagrangians~\cite{Mac89} for meson-nucleon coupling are the scalar($s$), pseudovector($pv$), and vector($v$)
\begin{equation}
\mathcal{L}_{s} = g_s \bar{\psi} \psi \varphi_{(s)} \; ,
\quad
\mathcal{L}_{pv} = -\frac{f_{ps}}{m_{ps}} \bar{\psi} \gamma^5 \gamma^{\mu} \psi \partial_{\mu} \varphi_{(ps)} \; ,
\quad
\mathcal{L}_{v} = g_v \bar{\psi} \gamma_{\mu} \psi \varphi_{(v)}^{\mu} + \frac{f_v}{4m} \bar{\psi} \sigma_{\mu \nu} \psi \rb{\partial^{\mu} \varphi_{(v)}^{\nu} - \partial^{\nu} \varphi_{(v)}^{\mu}} \; .
\end{equation}
We adhere to the conventions and notations of Machleidt e.g., $\psi$ ($m$) is the nucleon and $\varphi_{k}$ ($m_k$) for $k=s,ps,v$ the meson field (mass). In relativistic nuclear structure calculations the $pv$ Lagrangian is used in place of the $ps$ Lagrangian. This is because the contribution from the nucleon-antinucleon pair diagram becomes very large when using the $ps$ coupling, leading to unrealistically large pion-nucleon scattering lengths, whereas the same contributions are strongly reduced when using the $pv$ Lagrangian~\cite{GEB}. 

In addition to the interaction Lagrangians, we also need Dirac spinors in a helicity basis (similar expressions for primed coordinates)
\begin{alignat}{2}
u(\bvec{q},\lambda_1) & = \sqrt{\frac{W_1}{2m_1}} \begin{pmatrix} \bvec{1} \\ \frac{2 \lambda_1 \abs{\bvec{q}}}{W_1} \end{pmatrix} \ket{\lambda_1} \; , & \quad u(-\bvec{q},\lambda_2) & = \sqrt{\frac{W_2}{2m_2}} \begin{pmatrix} \bvec{1} \\ \frac{2 \lambda_2 \abs{\bvec{q}}}{W_2} \end{pmatrix} \ket{\lambda_2} \; .     
\label{Eq:helicity}
\end{alignat}
The most general spinor rotated into a direction with polar angles $\theta$ and $\phi$ through the usual Euler 
rotations can be written as~\cite{BrJack}
\begin{equation}  
\ket{\lambda} \equiv \ket{\theta,\phi,\lambda} = \mathcal{R}_{\phi,\theta,-\phi} \, \chi_{\lambda} = 
e^{i\phi \lambda} \mathcal{R}_{\phi,\theta,0} \, \chi_{\lambda} \;, 
\label{Eq:helicityrot}
\end{equation}  
with 
$\mathcal{R}_{\phi,\theta,0} = \exp(-\frac{i}{2} \sigma_z \phi) \exp(-\frac{i}{2} \sigma_y \theta)$ operating on the conventional Pauli spinor $\chi_{\lambda}$. Notice that $\chi_{\lambda_1}$ and $\chi_{-\lambda_2}$ must be used for 
particle $1$ and $2$ respectively. This is due to the opposite direction of motion in the center-of-mass frame. 

The spinors are normalized covariantly e.g., $u^\dagger(\bvec{p},\lambda) \gamma^0 u(\bvec{p},\lambda) = \bar{u}(\bvec{p},\lambda) u(\bvec{p},\lambda) = 1$, and $W_{1(2)} \equiv E_{1(2)} + m_{1(2)}$ where $E_{1(2)} = \sqrt{\bvec{q}^2 + m_{1(2)}^2}$.

\subsection{Relativistic momentum space OBEP}
From the interaction Lagrangian's and Dirac spinors we can derive the modified OBEP. By definition the OBEP is
\begin{equation}
\braket{\lambda_1^\prime \lambda_2^\prime|\hat{V}^I(\bvec{q}^\prime,\bvec{q})|\lambda_1 \lambda_2} \equiv \sum_{\alpha = \sigma,\eta,\omega} \braket{\bvec{q}^\prime \lambda_1^\prime \lambda_2^\prime |V_\alpha| \bvec{q} \lambda_1 \lambda_2} + ( \delta_{I1} - 3 \delta_{I0} ) \sum_{\alpha = \pi,\delta,\rho} \braket{\bvec{q}^\prime \lambda_1^\prime \lambda_2^\prime |V_\alpha| \bvec{q} \lambda_1 \lambda_2} \; ,
\end{equation}
with scalar ($\delta$, $\sigma$), pseudoscalar ($\pi$, $\eta$), and vector ($\rho$, $\omega$) particles. In the above formula $\delta_{ij}$ stands for the Kronecker delta function, and it's utilized to assigns the proper isospin coefficient. 

For scalar particles ($\delta$, $\sigma$)
\begin{equation}
\braket{\bvec{q}^\prime \lambda_1^\prime \lambda_2^\prime |V_s| \bvec{q} \lambda_1 \lambda_2} = -g_s^2 C_s \rb{1 - \frac{4 \lambda_1 \lambda_1^\prime \abs{\bvec{q}}\abs{\bvec{q}^\prime}}{W_1^\prime W_1}} \rb{1 - \frac{4 \lambda_2 \lambda_2^\prime \abs{\bvec{q}}\abs{\bvec{q}^\prime}}{W_2^\prime W_2}} \braket{\lambda_1^\prime \lambda_2^\prime|\lambda_1 \lambda_2} \; .
\label{Eq:V_s}
\end{equation}

For pseudoscalar particles ($\pi$, $\eta$)
\begin{align}
\braket{\bvec{q}^\prime \lambda_1^\prime \lambda_2^\prime |V_{pv}| \bvec{q} \lambda_1 \lambda_2} & = \frac{f_{ps}^2}{m_{ps}^2} C_{ps} (4 m_1 m_2) \Bigg[ \rb{\frac{2 \lambda_1^\prime \abs{\bvec{q}^\prime}}{W_1^\prime} - \frac{2 \lambda_1 \abs{\bvec{q}}}{W_1}} \rb{\frac{2 \lambda_2^\prime \abs{\bvec{q}^\prime}}{W_2^\prime} - \frac{2 \lambda_2 \abs{\bvec{q}}}{W_2}} + \frac{(E_1^\prime - E_1)(E_2^\prime - E_2)}{4 m_1 m_2} \nonumber \\
& \times \rb{\frac{2 \lambda_1^\prime \abs{\bvec{q}^\prime}}{W_1^\prime} + \frac{2 \lambda_1 \abs{\bvec{q}}}{W_1}} \rb{\frac{2 \lambda_2^\prime \abs{\bvec{q}^\prime}}{W_2^\prime} + \frac{2 \lambda_2 \abs{\bvec{q}}}{W_2}} + \frac{E_1^\prime - E_1}{2m_1} \rb{\frac{2 \lambda_1^\prime \abs{\bvec{q}^\prime}}{W_1^\prime} + \frac{2 \lambda_1 \abs{\bvec{q}}}{W_1}} \rb{\frac{2 \lambda_2^\prime \abs{\bvec{q}^\prime}}{W_2^\prime} - \frac{2 \lambda_2 \abs{\bvec{q}}}{W_2}} \nonumber \\ 
& + \frac{E_2^\prime - E_2}{2m_2} \rb{\frac{2 \lambda_1^\prime \abs{\bvec{q}^\prime}}{W_1^\prime} - \frac{2 \lambda_1 \abs{\bvec{q}}}{W_1}} \rb{\frac{2 \lambda_2^\prime \abs{\bvec{q}^\prime}}{W_2^\prime} + \frac{2 \lambda_2 \abs{\bvec{q}}}{W_2}} \Bigg] \braket{\lambda_1^\prime \lambda_2^\prime|\lambda_1 \lambda_2} \; .
\end{align}

For vector particles ($\rho$, $\omega$) the potential is the sum of three terms $V_v = V_{vv} + V_{tt} + V_{vt}$
\begin{align}
\braket{\bvec{q}^\prime \lambda_1^\prime \lambda_2^\prime |V_{vv}| \bvec{q} \lambda_1 \lambda_2} & = g_v^2 C_v \Bigg[ \rb{1 + \frac{4 \lambda_1^\prime \lambda_1 \abs{\bvec{q}^\prime} \abs{\bvec{q}}}{W_1^\prime W_1}}  \rb{1 + \frac{4 \lambda_2^\prime \lambda_2 \abs{\bvec{q}^\prime} \abs{\bvec{q}}}{W_2^\prime W_2}} \braket{\lambda_1^\prime \lambda_2^\prime | \lambda_1 \lambda_2} - 4 \rb{\frac{\lambda_1 \abs{\bvec{q}}}{W_1} + \frac{\lambda_1^\prime \abs{\bvec{q}^\prime}}{W_1^\prime}} \rb{\frac{\lambda_2 \abs{\bvec{q}}}{W_2} + \frac{\lambda_2^\prime \abs{\bvec{q}^\prime}}{W_2^\prime}} \nonumber \\ 
& \times \braket{\lambda_1^\prime \lambda_2^\prime| \bvec{\sigma}^{(1)} \cdot \bvec{\sigma}^{(2)}  | \lambda_1 \lambda_2 } \Bigg] \; ,
\end{align}
\begin{align}
\braket{\bvec{q}^\prime \lambda_1^\prime \lambda_2^\prime |V_{vt}| \bvec{q} \lambda_1 \lambda_2} & = 2 g_v f_v C_v \Bigg[\bigg \{ \rb{\frac{W_1^\prime + W_2^\prime + W_1 + W_2}{2m}} \rb{\frac{16 \lambda_1^\prime \lambda_2^\prime \lambda_1 \lambda_2 {\abs{\bvec{q}^\prime}}^2 \abs{\bvec{q}}^2}{W_1^\prime W_2^\prime W_1 W_2}} - \rb{\frac{E_1^\prime + E_2^\prime + E_1 + E_2 - 2(m_1+m_2)}{2m}} \bigg \} \nonumber \\ 
& \times \braket{\lambda_1^\prime \lambda_2^\prime | \lambda_1 \lambda_2} - \bigg \{ \rb{\frac{m_1+m_2}{2m}} \rb{\frac{2 \lambda_1 \abs{\bvec{q}}}{W_1} + \frac{2 \lambda_1^\prime \abs{\bvec{q}^\prime}}{W_1^\prime}} \rb{\frac{2 \lambda_2 \abs{\bvec{q}}}{W_2} + \frac{2 \lambda_2^\prime \abs{\bvec{q}^\prime}}{W_2^\prime}} + \frac{E_1^\prime - E_1}{m} \rb{\frac{\lambda_1^\prime \abs{\bvec{q}^\prime}}{W_1^\prime} - \frac{\lambda_1 \abs{\bvec{q}}}{W_1}} \nonumber \\
& \times \rb{\frac{\lambda_2^\prime \abs{\bvec{q}^\prime}}{W_2^\prime} + \frac{\lambda_2 \abs{\bvec{q}}}{W_2}} + \frac{E_2^\prime - E_2}{m} \rb{\frac{\lambda_1^\prime \abs{\bvec{q}^\prime}}{W_1^\prime} + \frac{\lambda_1 \abs{\bvec{q}}}{W_1}} \rb{\frac{\lambda_2^\prime \abs{\bvec{q}^\prime}}{W_2^\prime} - \frac{\lambda_2 \abs{\bvec{q}}}{W_2}} \bigg \} \braket{\lambda_1^\prime \lambda_2^\prime| \bvec{\sigma}^{(1)} \cdot \bvec{\sigma}^{(2)}  | \lambda_1 \lambda_2 } \Bigg] \; ,
\end{align}
\begin{align}
\braket{\bvec{q}^\prime \lambda_1^\prime \lambda_2^\prime |V_{tt}| \bvec{q} \lambda_1 \lambda_2} & = f_v^2 C_v \Bigg[ \bigg \{ \frac{m_1 m_2}{m^2} \rb{1 + \frac{4 \lambda_1^\prime \lambda_1 \abs{\bvec{q}^\prime} \abs{\bvec{q}}}{W_1^\prime W_1}} \rb{1 + \frac{4 \lambda_2^\prime \lambda_2 \abs{\bvec{q}^\prime} \abs{\bvec{q}}}{W_2^\prime W_2}} - \frac{E_1^\prime + E_2^\prime + E_1 + E_2}{2m^2} \bigg[ m_1 \rb{1+\frac{4 \lambda_1^\prime \lambda_1 \abs{\bvec{q}^\prime} \abs{\bvec{q}}}{W_1^\prime W_1}} \nonumber \\
& \times \rb{1-\frac{4 \lambda_2^\prime \lambda_2 \abs{\bvec{q}^\prime} \abs{\bvec{q}}}{W_2^\prime W_2}} + m_2 \rb{1-\frac{4 \lambda_1^\prime \lambda_1 \abs{\bvec{q}^\prime} \abs{\bvec{q}}}{W_1^\prime W_1}} \rb{1+\frac{4 \lambda_2^\prime \lambda_2 \abs{\bvec{q}^\prime} \abs{\bvec{q}}}{W_2^\prime W_2}} \bigg] + \frac{1}{2m^2} \rb{1-\frac{4 \lambda_1^\prime \lambda_1 \abs{\bvec{q}^\prime} \abs{\bvec{q}}}{W_1^\prime W_1}} \nonumber \\
& \times \rb{1-\frac{4 \lambda_2^\prime \lambda_2 \abs{\bvec{q}^\prime} \abs{\bvec{q}}}{W_2^\prime W_2}} \Big[4 m_1 m_2 +\frac{1}{2} \Big \{ (E_1^\prime + E_1)(E_2^\prime + E_2) - (E_1^\prime - E_1)^2 - (E_2^\prime - E_2)^2 + {\abs{\bvec{q}^\prime}}^2 + \abs{\bvec{q}}^2 \nonumber \\
& + 2 \bvec{q}^\prime \cdot \bvec{q} \Big \} \Big] \bigg \} \braket{\lambda_1^\prime \lambda_2^\prime | \lambda_1 \lambda_2} - \bigg \{ \frac{m_1 m_2}{m^2} \rb{\frac{2 \lambda_1 \abs{\bvec{q}}}{W_1} + \frac{2 \lambda_1^\prime \abs{\bvec{q}^\prime}}{W_1^\prime}} \rb{\frac{2 \lambda_2 \abs{\bvec{q}}}{W_2} + \frac{2 \lambda_2^\prime \abs{\bvec{q}^\prime}}{W_2^\prime}} + \frac{m_1(E_2^\prime - E_2)}{2m^2} \nonumber \\
& \times \rb{\frac{2 \lambda_1 \abs{\bvec{q}}}{W_1} + \frac{2 \lambda_1^\prime \abs{\bvec{q}^\prime}}{W_1^\prime}} \rb{\frac{2 \lambda_2^\prime \abs{\bvec{q}^\prime}}{W_2^\prime} - \frac{2 \lambda_2 \abs{\bvec{q}}}{W_2}} + \frac{m_2(E_1^\prime - E_1)}{2m^2} \rb{\frac{2 \lambda_1^\prime \abs{\bvec{q}^\prime}}{W_1^\prime} - \frac{2 \lambda_1 \abs{\bvec{q}}}{W_1}} \rb{\frac{2 \lambda_2^\prime \abs{\bvec{q}^\prime}}{W_2^\prime} + \frac{2 \lambda_2 \abs{\bvec{q}}}{W_2}} \nonumber \\
& + \frac{(E_1^\prime - E_1)(E_2^\prime - E_2)}{4m^2} \rb{\frac{2 \lambda_1^\prime \abs{\bvec{q}^\prime}}{W_1^\prime} - \frac{2 \lambda_1 \abs{\bvec{q}}}{W_1}} \rb{\frac{2 \lambda_2^\prime \abs{\bvec{q}^\prime}}{W_2^\prime} - \frac{2 \lambda_2 \abs{\bvec{q}}}{W_2}} \bigg \} \braket{\lambda_1^\prime \lambda_2^\prime| \bvec{\sigma}^{(1)} \cdot \bvec{\sigma}^{(2)}  | \lambda_1 \lambda_2 }  \Bigg] \; .
\label{Eq:V_tt}
\end{align}

In the above formulas
\begin{equation}
C_k \equiv \frac{1}{4 (2 \pi)^3} \frac{\sqb{F_k ( \abs{\bvec{q}^\prime - \bvec{q}}^2 )}^2}{\rb{\abs{\bvec{q}^\prime - \bvec{q}}^2 + m_k^2}} \sqrt{\frac{W_1^\prime W_2^\prime W_1 W_2}{E_1^\prime E_2^\prime E_1 E_2}} \; ,
\quad
F_k(\abs{\bvec{q}^\prime - \bvec{q}}^2) = \rb{\frac{\Lambda_k^2 - m_k^2}{\Lambda_k^2 + \abs{\bvec{q}^\prime - \bvec{q}}^2}}^{n_k} \; ,
\quad
\text{for } k=s,ps,v \; .
\label{Eq:formfactor} 
\end{equation}
Numerical values for the parameters $m_k$, $n_k$, $\Lambda_k$ and the coupling constants $f_k$, $g_k$ can be found on p.347 of Ref.~\cite{Mac89} or in Ref.~\cite{Mac87}.
\subsection{Helicity matrix elements} \label{Appx:helicity}
For completeness, we provide expressions for the helicity state matrix elements with general dependence on $\theta, \theta^\prime, \phi, \phi^\prime$. These can be derived using 
Eq.~(\ref{Eq:helicityrot}).                                                                   
\begin{subequations}
\begin{align}
& \braket{++|++} = \frac{1}{2}(1 + \cos \theta^\prime \cos \theta + \sin \theta^\prime \sin \theta \cos(\phi^\prime - \phi)) \; , \hspace{8.3cm} \nonumber \\
& \braket{++|+-} = \frac{1}{2}(\cos \theta^\prime \sin \theta - \sin \theta^\prime ( \cos \theta \cos(\phi^\prime - \phi) + i \sin(\phi^\prime - \phi))) \; , \nonumber \\
& \braket{++|--} = \frac{1}{2}(-1 + \cos \theta^\prime \cos \theta + \sin \theta^\prime \sin \theta \cos(\phi^\prime - \phi)) \; , \nonumber \\
& \braket{+-|++} = \frac{1}{2}(\sin \theta^\prime \cos \theta - \sin \theta ( \cos \theta^\prime \cos(\phi^\prime - \phi) + i \sin(\phi^\prime - \phi))) \; , \nonumber \\
& \braket{+-|+-} = \frac{1}{2}(\sin \theta^\prime \sin \theta + (1 + \cos \theta^\prime \cos \theta ) \cos(\phi^\prime - \phi) + i(\cos \theta^\prime + \cos \theta) \sin(\phi^\prime - \phi)) \; , \nonumber \\
& \braket{+-|-+} = \frac{1}{2}(-\sin \theta^\prime \sin \theta + (1 - \cos \theta^\prime \cos \theta ) \cos(\phi^\prime - \phi) + i(\cos \theta^\prime - \cos \theta) \sin(\phi^\prime - \phi)) \; , \nonumber \\
& \braket{++| \bvec{\sigma}^{(1)} \cdot \bvec{\sigma}^{(2)}  | ++} = \braket{++|++} - 2 \; , \nonumber \\
& \braket{++| \bvec{\sigma}^{(1)} \cdot \bvec{\sigma}^{(2)}  | +-} = \braket{++|+-} \; , \nonumber \\
& \braket{++| \bvec{\sigma}^{(1)} \cdot \bvec{\sigma}^{(2)}  | --} = \braket{++|--} + 2 \; , \nonumber \\
& \braket{+-| \bvec{\sigma}^{(1)} \cdot \bvec{\sigma}^{(2)}  | ++} = \braket{+-|++} \; , \nonumber \\
& \braket{+-| \bvec{\sigma}^{(1)} \cdot \bvec{\sigma}^{(2)}  | +-} = \braket{+-|+-} \; , \nonumber \\
& \braket{+-| \bvec{\sigma}^{(1)} \cdot \bvec{\sigma}^{(2)}  | -+} = \braket{+-|-+} \; ,
\end{align}
\label{Eq:six_helicities}
\end{subequations}
\begin{subequations}
\begin{align}
& \braket{--|--} = \braket{++|++} \; , \hspace{12.8cm} \nonumber \\
& \braket{--|+-} = \braket{++|+-} \; , \nonumber \\
& \braket{++|-+} = \braket{--|-+} = - \Re(\braket{++|+-}) + i \Im(\braket{++|+-}) \; , \nonumber \\
& \braket{+-|--} = \braket{+-|++}, \nonumber \\
& \braket{-+|++} = \braket{-+|--} = - \Re(\braket{+-|++}) + i \Im(\braket{+-|++}) \; , \nonumber \\
& \braket{--|++} = \braket{++|--} \; , \nonumber \\
& \braket{-+|+-} = \Re(\braket{+-|-+}) - i \Im(\braket{+-|-+}) \; , \nonumber \\
& \braket{-+|-+} = \Re(\braket{+-|+-}) - i \Im(\braket{+-|+-}) \; ,
\end{align}
\label{Eq:remanin_helicities}
\end{subequations}
\begin{subequations}
\begin{align}
& \braket{--| \bvec{\sigma}^{(1)} \cdot \bvec{\sigma}^{(2)}  | --} = \braket{++| \bvec{\sigma}^{(1)} \cdot \bvec{\sigma}^{(2)}  | ++} \; , \nonumber \\
& \braket{--| \bvec{\sigma}^{(1)} \cdot \bvec{\sigma}^{(2)}  | +-} = \braket{++| \bvec{\sigma}^{(1)} \cdot \bvec{\sigma}^{(2)}  | +-} \; , \nonumber \\
& \braket{++| \bvec{\sigma}^{(1)} \cdot \bvec{\sigma}^{(2)}  | -+} = \braket{--| \bvec{\sigma}^{(1)} \cdot \bvec{\sigma}^{(2)}  | -+} = -\Re(\braket{++| \bvec{\sigma}^{(1)} \cdot \bvec{\sigma}^{(2)}  | +-}) + i \Im(\braket{++| \bvec{\sigma}^{(1)} \cdot \bvec{\sigma}^{(2)}  | +-}) \; , \nonumber \\
& \braket{+-| \bvec{\sigma}^{(1)} \cdot \bvec{\sigma}^{(2)}  | --} = \braket{+-| \bvec{\sigma}^{(1)} \cdot \bvec{\sigma}^{(2)}  | ++} \; , \nonumber \\
& \braket{-+| \bvec{\sigma}^{(1)} \cdot \bvec{\sigma}^{(2)}  | ++} = \braket{-+| \bvec{\sigma}^{(1)} \cdot \bvec{\sigma}^{(2)}  | --} = -\Re(\braket{+-| \bvec{\sigma}^{(1)} \cdot \bvec{\sigma}^{(2)}  | ++}) + i \Im(\braket{+-| \bvec{\sigma}^{(1)} \cdot \bvec{\sigma}^{(2)}  | ++}) \; , \nonumber \\
& \braket{--| \bvec{\sigma}^{(1)} \cdot \bvec{\sigma}^{(2)}  | ++} = \braket{++| \bvec{\sigma}^{(1)} \cdot \bvec{\sigma}^{(2)}  | --} \; , \nonumber \\
& \braket{-+| \bvec{\sigma}^{(1)} \cdot \bvec{\sigma}^{(2)}  | +-} = \Re(\braket{+-| \bvec{\sigma}^{(1)} \cdot \bvec{\sigma}^{(2)}  | -+}) - i \Im(\braket{+-| \bvec{\sigma}^{(1)} \cdot \bvec{\sigma}^{(2)}  | -+}) \; , \nonumber \\
& \braket{-+| \bvec{\sigma}^{(1)} \cdot \bvec{\sigma}^{(2)}  | -+} = \Re(\braket{+-| \bvec{\sigma}^{(1)} \cdot \bvec{\sigma}^{(2)}  | +-}) - i \Im(\braket{+-| \bvec{\sigma}^{(1)} \cdot \bvec{\sigma}^{(2)}  | +-}) \; .
\end{align}
\label{Eq:matrix_elements}
\end{subequations}
\section{Converting the Bethe-Goldstone integral equations into matrix equations} \label{systemofeq}
In this section we give details on the numerical solution of Eq.~(\ref{Eq:BG_final}). Clearly, the free-space Thompson equation follows along similar lines.                                                    

Six integral equations are obtained from Eq.~(\ref{Eq:BG_final}) using the linear combinations in Eq.~(\ref{Eq:linearcombinations}) along with information from Eq.~(\ref{Eq:symmetries}) 
\begin{subequations}
\begin{align}
\up{0}g^I(\tilde{q}^\prime,\tilde{q}) & = \up{0}v^I(\tilde{q}^\prime,\tilde{q}) + \pi \int_0^{\pi} \rb{ \mathcal{P} \int_0^\infty \frac{ \up{(0,0)}f^I(q^\pp) {q^\pp}^2 }{E^\ast_q - E^\ast_{q^\pp}} \dd q^\pp - i \pi q E^\ast_q \up{(0,0)}f^I(q) } \sin \theta^\pp \dd \theta^\pp \; , \\
\up{1}g^I(\tilde{q}^\prime,\tilde{q}) & = \up{1}v^I(\tilde{q}^\prime,\tilde{q}) + \pi \int_0^{\pi} \rb{ \mathcal{P} \int_0^\infty \frac{ \up{(1,1)}f^I(q^\pp) {q^\pp}^2 }{E^\ast_q - E^\ast_{q^\pp}} \dd q^\pp - i \pi q E^\ast_q \up{(1,1)}f^I(q) } \sin \theta^\pp \dd \theta^\pp \; , \\
\up{12}g^I(\tilde{q}^\prime,\tilde{q}) & = \up{12}v^I(\tilde{q}^\prime,\tilde{q}) + \pi \int_0^{\pi} \rb{ \mathcal{P} \int_0^\infty \frac{ \sqb{ \up{(12,12)}f^I(q^\pp) + \up{(55,66)}f^I(q^\pp)}{q^\pp}^2 }{E^\ast_q - E^\ast_{q^\pp}} \dd q^\pp - i \pi q E^\ast_q \sqb{ \up{(12,12)}f^I(q) + \up{(55,66)}f^I(q) } } \sin \theta^\pp \dd \theta^\pp \; , \\
\up{34}g^I(\tilde{q}^\prime,\tilde{q}) & = \up{34}v^I(\tilde{q}^\prime,\tilde{q}) + \pi \int_0^{\pi} \rb{ \mathcal{P} \int_0^\infty \frac{ \sqb{ \up{(34,34)}f^I(q^\pp) + \up{(66,55)}f^I(q^\pp)}{q^\pp}^2 }{E^\ast_q - E^\ast_{q^\pp}} \dd q^\pp - i \pi q E^\ast_q \sqb{ \up{(34,34)}f^I(q) + \up{(66,55)}f^I(q) } } \sin \theta^\pp \dd \theta^\pp \; , \\
\up{55}g^I(\tilde{q}^\prime,\tilde{q}) & = \up{55}v^I(\tilde{q}^\prime,\tilde{q}) + \pi \int_0^{\pi} \rb{ \mathcal{P} \int_0^\infty \frac{ \sqb{ \up{(12,55)}f^I(q^\pp) + \up{(55,34)}f^I(q^\pp)}{q^\pp}^2 }{E^\ast_q - E^\ast_{q^\pp}} \dd q^\pp - i \pi q E^\ast_q \sqb{ \up{(12,55)}f^I(q) + \up{(55,34)}f^I(q) } } \sin \theta^\pp \dd \theta^\pp \; , \\
\up{66}g^I(\tilde{q}^\prime,\tilde{q}) & = \up{66}v^I(\tilde{q}^\prime,\tilde{q}) + \pi \int_0^{\pi} \rb{ \mathcal{P} \int_0^\infty \frac{ \sqb{ \up{(34,66)}f^I(q^\pp) + \up{(66,12)}f^I(q^\pp)}{q^\pp}^2 }{E^\ast_q - E^\ast_{q^\pp}} \dd q^\pp - i \pi q E^\ast_q \sqb{ \up{(34,66)}f^I(q) + \up{(66,12)}f^I(q) } } \sin \theta^\pp \dd \theta^\pp \; ,
\end{align}
\label{Eq:FIE}
\end{subequations}
where we defined $\up{(n,m)}f^I(q^\pp) \equiv \up{n}v^I(\tilde{q}^\prime,\tilde{q}^\pp) Q(\tilde{q}^\pp,P,k_F) \up{m}g^I(\tilde{q}^\pp,\tilde{q})$ for $n,m=0,1,12,34,55,66$.

The $i \epsilon$ term present in Eq.~(\ref{Eq:BG_final}) was converted into a principle value integral (denoted by $\mathcal{P}$) plus an imaginary term using the Plemelj formula. To handle the principle value integral, we symmetrically distribute Gauss-Legendre (GL) points about the singularity. This is accomplished by breaking the integral into two parts, $ \mathcal{P} \int_0^\infty \dd q^\pp = \int_0^{2q} \dd q^\pp + \int_{2q}^\infty \dd q^\pp$ and creating an $N^\pp$-point GL rule. Namely, $X_i \equiv \begin{pmatrix} \bvec{X}_1 & \bvec{X}_2 \end{pmatrix}$ and $W_i \equiv \begin{pmatrix} \bvec{W}_1 & \bvec{W}_2 \end{pmatrix}$ for $i=1,2,\dots,N^\pp$. The $N^\pp$-point GL rule is built from two separate GL rules. The first being a $N_1$-point GL rule over $(0,2q)$ (with nodes and weights $\bvec{X}_1, \bvec{W}_1$) and the second a $N_2$-point GL rule over $(2q,\infty)$ (with nodes and weights $\bvec{X}_2, \bvec{W}_2$). Choosing $N_1=$ even will ensure that the points are distributed symmetrically about the singularity. With regard to the integration over $(2q,\infty)$, we found that better stability could be achieved by truncating the integration at a sufficiently large value rather than using one of the standard transformations. Finally, for the $(0,\pi)$ integral a standard $N^\pp_\theta$-point GL rule is used with nodes and weight given as $\bvec{x},\bvec{w}$.

Although various methods exist for solving Fredholm integral equations of the second kind, we prefer the Nystrom method. ``Delves and Mohamed~\cite{DM85} investigated methods more complicated than the Nystrom method. For straightforward Fredholm equations of the second kind, they concluded ` . . . the clear winner of this contest has been the Nystrom routine . . . with the $N$-point Gauss-Legendre rule. This routine is extremely simple . . . Such results are enough to make a numerical analyst weep' ''~\cite{nr}. The details of Nystrom method can be found in Ref.~\cite{nr}, but the general idea is to convert the system of integral equations into a system of matrix equations. From there, we solve them using a LAPACK~\cite{lapack} LU-factorization routine. The matrix equations corresponding to Eq.~(\ref{Eq:FIE}) are
\begin{subequations}
\begin{align}
\up{0}\bvec{K}^I(q)\up{0}\bvec{g}^I(q) & = \up{0}\bvec{v}^I(q) \; , \\
\up{1}\bvec{K}^I(q)\up{1}\bvec{g}^I(q) & = \up{1}\bvec{v}^I(q) \; , \\
\begin{pmatrix}
\up{12}\bvec{K}^I(q) & \up{55}\bvec{K}^I(q) - \bvec{1} \\
\up{66}\bvec{K}^I(q) - \bvec{1} & \up{34}\bvec{K}^I(q)
\end{pmatrix}
&
\begin{pmatrix}
\up{12}\bvec{g}^I(q) & \up{55}\bvec{g}^I(q) \\
\up{66}\bvec{g}^I(q) & \up{34}\bvec{g}^I(q)
\end{pmatrix}
=
\begin{pmatrix}
\up{12}\bvec{v}^I(q) & \up{55}\bvec{v}^I(q) \\
\up{66}\bvec{v}^I(q) & \up{34}\bvec{v}^I(q)
\end{pmatrix} \; .
\end{align}
\end{subequations}
Convenient definitions (for $n=0,1,12,34,55,66$) are the two $N^\pp_\theta(1+N^\pp) \times N_\theta$ matrices $\up{n}v_{jk}^I(q) \equiv \up{n}v^I(\tilde{q}_j^\prime,q,\theta_k)$, $\up{n}g_{jk}^I(q) \equiv \up{n}g^I(\tilde{q}_j^\prime,q,
\theta_k)$, and the $N^\pp_\theta(1+N^\pp) \times N^\pp_\theta(1+N^\pp)$ matrix $\up{n}\bvec{K}^I(q) \equiv \bvec{1} - \begin{pmatrix} \up{n}\bvec{\alpha}^I(q) & \up{n}\bvec{\beta}^I(q) \end{pmatrix}$. The $N^\pp_\theta(1+N^\pp) \times N^\pp_\theta$ $\bvec{\alpha}$ matrix and the $N^\pp_\theta(1+N^\pp) \times N^\pp N^\pp_\theta$ $\bvec{\beta}$ matrix are defined as
\begin{subequations}
\begin{align}
\up{n}\alpha_{jk}^I(q) & \equiv -i \pi^2 q E^\ast_q w_k \sin(x_k) Q(q,x_k,P,k_F) \up{n}v^I(\tilde{q}_j^\prime,q,x_k) \; , \\
\up{n}\beta_{j k}^I(q) & \equiv \pi W_m w_\ell  \frac{ X_m^2 \sin(x_\ell) }{ E^\ast_q - E^\ast_{X_m} } Q(X_m,x_\ell,P,k_F) \up{n}v^I(\tilde{q}_j^\prime,X_m,x_\ell) \; , \quad m \equiv \floor{ \frac{k-1}{N^\pp_\theta} + 1 } \; , \quad \ell \equiv k-N^\pp_\theta(m-1) \; . \label{Eq:betamatrix}
\end{align}
\end{subequations}
In the previous equations we utilize the definition of $\tilde{q}^\prime \equiv (q^\prime,\theta^\prime)$ to create a vector of points
\begin{equation}
\tilde{\bvec{q}}^\prime \equiv 
\begin{pmatrix} 
\bvec{y} \\ 
\bvec{z} 
\end{pmatrix} \; ,
\quad y_j \equiv (q,x_j) \; , \quad z_k \equiv (X_m,x_\ell) \; ,
\end{equation}
where $m$ and $\ell$ in terms of $k$ are given in Eq.~(\ref{Eq:betamatrix}). Also keep in mind that we sometimes denote matrices and vectors by their entries e.g., $\bvec{A}$ as $A_{jk}$.

As we stated in section~\ref{thompson_heli}, in our formalism we solve the solution over the $q^\prime \times \theta^\prime \times \theta$ grid. This extra dimension requires us to introduce an additional set of points which are not needed, for instance, in the scheme of Ref.~\cite{FEG}. A natural set of points are the nodes $\bvec{\theta}$ for a $N_\theta$-point GL rule over $(0,\pi)$. Fortunately, this added dimension has little effect on computational time. This is because the matrices are dependent only on $q$. Thus, once LU-factorization is complete the actual solution for multiple right hand sides (i.e. the $\theta$ dimension) is trivial.

A brief mention of our computational parameters is in order. We use a $30$-point GL rule over $(0,2q)$, a $100$-point GL rule over $(2q,\infty)$, a $80$-point GL rule over $(0,\pi)$, and a $20$-point GL rule over $(0, 2 \pi)$ for the $\phi$-integrated NN potential. Unfortunately, the number of points needed for a stable GL rule is high. The reason is the structure of the three-dimensional NN potentials [see in particular Eq.~(\ref{Eq:formfactor})]. Notice that, when the angle-dependent potentials are diagonal in the three-momenta, the form factor becomes ineffective in its role of cutting out high-momentum component. The largest value of $q^\pp$ needed for a stable integration to infinity was $20,000 \units{MeV}$ and occurred for $\up{0}t^0$. The others required a smaller cutoff $\approx 3,000 \units{MeV}$. Furthermore, initial stability tests can be very time consuming. Although the computational time was dramatically reduced using OpenMP~\cite{openmp} to fill in the entries of the matrices, once agreement with existing $t$-matrices is verified, we simply run the $g$-matrix calculation under the same computational conditions.
\bibliography{bibfile}
\end{document}